\documentclass{aa}
\usepackage{graphics} 
\usepackage{epsf} 
\usepackage{epsfig}
\usepackage{psfig}
\usepackage{amssymb}
\usepackage{calc}
\usepackage{rotating}

\begin{document}
\title{Near-IR spectra of ISOGAL sources in the Inner Galactic Bulge
\thanks{Based on observations collected at the European Southern 
Observatory, La Silla Chile}}

\author{M.~Schultheis\inst{1}
\and A.~Lan\c{c}on\inst{2} 
\and A.~Omont\inst{1}
\and F.~Schuller\inst{1}
\and D.~K.~Ojha\inst{3}
}
\authorrunning{M. Schultheis et al.}
\titlerunning{Near-IR spectroscopy of ISOGAL sources}  
\offprints{schulthe@iap.fr} 
\institute{Institut d'Astrophysique de Paris, CNRS, 98bis Bd Arago, 75014 Paris, France     
\and  Observatoire de Strasbourg, UMR 7550, 11 rue de l'Universite, 6700 Strasbourg, France 
\and Tata Institute of Fundamental Research, Homi Bhabha Road, Colaba, Mumbai,
400 005, India
}  
  
\voffset 1.0truecm

\date{Received ....  / Accepted .. .........}

\sloppy

\abstract{
In this work we present  near-IR spectra (HK-band) of a sample of
107 sources with   mid-IR excesses  at 7 and 15\,$\rm \mu$m detected during the ISOGAL survey.  Making use of  the DENIS interstellar extinction map from Schultheis et al. (1999) we derive luminosities and find that the $\rm M_{bol}$  vs.~$\rm ^{12}CO$ and $\rm M_{bol} \, vs.~H_{2}O$ diagrams are  powerful tools for identifying
supergiants, AGB stars, giants and young stellar objects. The majority of our
sample are AGB  stars ($\sim$ 80\%) while we find four good supergiant
candidates, nine young stellar objects and 12 RGB candidates. We have
used the  most recent $\rm K_{0}-[15]$
relation by Jeong et al. (\cite{Jeong2002}) based on recent
theoretical modeling of dust formation of AGB stars  to determine mass-loss rates.
The mass-loss rates of the supergiants are comparable with those in the solar neighbourhood while
 the  long-period Variables  cover a mass-loss range from $\rm -5 <
log\,\dot{\it{M}} < -7$. The red giant candidates
lie at the lower end of the mass-loss rate range between
 $\rm -6.5 < log\,\dot{{\it{M}}} < -9$.
We used the equivalent width of the CO bandhead at 2.3\,$\rm \mu m$, the NaI doublet and the CaI
 triplet to estimate metallicities using the relation by Ram\'{\i}rez et al. (\cite{Ramirez2000}). The metallicity distribution of the ISOGAL objects shows a
mean [Fe/H] $\sim$ -0.25\,dex with a dispersion of $\rm \pm 0.40\,dex$ which is 
in agreement with the values of Ram\'{i}rez et al. (\cite{Ramirez2000}) for Galactic Bulge fields between $\rm b = -4^{o}$ and  $\rm b = -1.3^{o}$. 
A comparison with the solar neighbourhood sample of Lan\c{c}on \& Wood (\cite{LW}) shows  that our sample is $\sim$ 0.5\,dex more metal-rich on average.
}
\maketitle

  
\section{Introduction} \label{introduction}

The inner galactic Bulge sometimes referred to as the central stellar 
cluster (see e.~g. Serabyn \& Morris \cite{Serrabyn96}), or as the nuclear
Bulge (Mezger et al. \cite{Mezger96}), presents quite extreme 
conditions (see also Philipp et al. \cite{Philipp99}, Figer \cite{Figer2003}). Extending only
$\sim$ 200-300 pc in the galactic plane and $\sim$ 30-50\,pc perpendicular
to it, it contains a mass $\rm \sim 4x10^{9}\,M_{\odot}$, with mean stellar
and interstellar densities $\sim$ 500 times larger than in the galactic disk.
The galactic Bulge provides a wide metallicity range with $\rm -1 < [Fe/H] < 1$ which
makes it an ideal place for studying stellar evolution. While in the past, 
several studies claimed that we deal in the Bulge with a supersolar
metal-rich stellar population (see e.~g. Whitford \cite{Whitford78},
Frogel \& Whitford \cite{Frogel87}, Frogel \cite{Frogel88}, Rich
\cite{Rich88}), chemical abundances there have
recently been revised (see e.~g. McWilliam \& Rich \cite{William94}, 
Frogel et al. \cite{Frogel99}) and at present the iron relative
abundance is believed to peak at {\mbox{$\sim$ --0.3\,dex}} (Freeman \&
Bland-Hawthorn \cite{Freeman2002}, van Loon et al. \cite{Loon2003}, Ibata \& Gilmore \cite{Ibata95},
 Minniti et al. \cite{Minniti95}, Houdashelt \cite{Houdashelt96}), with
a wide dispersion. The metallicity distribution is an important
ingredient for the different scenarios of galaxy formation such as
 dissipational collapse or accretion of matter. 
We want to refer to Freeman \& Hawthorn (\cite{Freeman2002}) for the
most recent review about the formation and evolution of our Galaxy.

In most parts of the galactic Bulge, the study of the stellar population is
hampered by its high interstellar absorption (Frogel et al. \cite{Frogel99}, Schultheis et
al. \cite{Schultheis99}); studies in the Infrared 
 are therefore crucial.

	Surveys with the ISO satellite, especially with the ISOCAM instrument (C\'esarsky et al. \cite{Cesarsky96}), whose
 sensitivity is several orders of magnitude greater than IRAS and whose 
angular resolution is ten times better, have led to new possibilities.
 The ISOGAL 7 and 15$\mu$m survey 
(Omont et al. \cite{Omont2002}) in particular has observed $\rm \sim 16\,deg^2$ 
of  the central  obscured regions of the Galaxy. The total number of stars detected ($\sim 10^5$) is
comparable to the number  detected by IRAS in the whole Galaxy. 
The main goals of the ISOGAL survey are to quantify the spatial
 distributions of the various stellar populations and their properties
 in the inner Galaxy, together with the properties of the warm
 interstellar medium. The combination of ISOGAL and
DENIS (or 2MASS) near-infrared data is a powerful means for determining the nature of sources even
in regions of high extinction (A$_{V}$ up to 20--30). The various colour-colour and colour-magnitude
diagrams available with the five ISOGAL-DENIS bands provide  rich information on  extinctions,
distances, intrinsic colours and  absolute magnitudes. 
 Among the M giants, a large proportion of those detected at 15~$\rm \mu$m are AGB-LPVs with mass-loss,
 well traced
by their 15 $\mu$m excesses (Omont et al. \cite{Omont99}, Glass et al. \cite{Glass99}, 
Alard et al. \cite{Alard2001}). Ojha et al. (\cite{Ojha2002}, hereafter referred to as OOS)   studied ISOGAL sources in the outer Bulge ($\rm |b| \gtrsim 1^\circ$) and
discussed their nature as well as their mass-loss rates. 
AGB stars contribute  more than 70\% towards the replenishment of the ISM
in the solar neighbourhood (Sedlmayr 1994). Thus, it is important to study
the mass-loss of these objects in different galactic environments.


However, detailed characterization of the sources faces various difficulties:
  very large interstellar extinction, uncertainty in the mid-IR extinction law, photometry uncertainty in the case of the
weakest ISOGAL sources, etc..
 Spectroscopic follow-up observations have therefore been deemed essential. For example, Schultheis et al.
 (\cite{Schultheis2002})  obtained visible spectra of  nearby sources
with mid-IR excesses and could identify interesting objects such as
  Ae/Be stars, possibly post-AGB stars and stars with red excess. 
Optical spectroscopy 
  in the Inner Galactic Bulge fails as the sources become invisible due to the
 high interstellar extinction ($\rm A_{V} > 20\,mag$). Therefore, only
spectroscopy in the near-IR, where interstellar absorption is about
  ten times smaller (at $\rm \sim 2.2\, \mu m$), can reveal us the nature of the source.

 Up to now the study of stellar
populations in the inner Galaxy have mostly been restricted to low extinction fields, such as Baade's windows
(see e.~g.~Rich \cite{Rich88}, Terndrup \cite{Terndrup88}). 
Recently, however,  Ram\'{\i}rez et al. (\cite{Ramirez2000}, hereafter referred to as RSFD) studied the M giant population in the inner Bulge using
K-band spectra and determined a number of metallicities (see also
Frogel et al. \cite{Frogel99}, Ram\'{\i}rez et
al. \cite{Ramirez2000a}). Wood et al. (\cite{Wood98}), Blommaert
et al. (\cite{Blommaert98}) and Ortiz et
al. (\cite{Ortiz2002}) studied OH/IR stars which are known to be the
most extreme mass-losing stars around the Galactic Center while Glass
et al. (\cite{Glass2001}) performed a monitoring program of
large-amplitude variables in the Galactic Centre. 

In this work, we perform a spectroscopic follow-up study (H and K-bands) of 107 ISOGAL
sources with  IR excess at 7 and 15\,$\rm \mu$m in the Inner Galactic
Bulge to study their nature, mainly to identify young stellar objects
and to establish criteria to distinguish between different classes of objects.
 We  discuss the possibility of separating  the different stellar populations 
such as AGB stars, M giants, supergiants
and young stellar objects in highly obscured regions ($\rm A_{V} \sim 20-30\,mag$), combining near-IR spectroscopy, near and mid-IR
photometry and interstellar extinction data further.  We discuss the mass-loss of AGB stars
as well as of supergiants and red giant stars. Using the spectral features
of CO, NaI and CaI, we estimate the metallicity and make a comparison to
the solar neighbourhood sample of Lan\c{c}on \& Wood (\cite{LW},
hereafter referred to as LW). The derived metallicity distribution
will be compared to  the dissipative collapse model of Molla et al. (\cite{Molla2000}).

\section{Observations and data reduction}

\subsection{Near-IR observations}

\begin{figure*}
\includegraphics[height=18cm,width=8cm,angle=270]{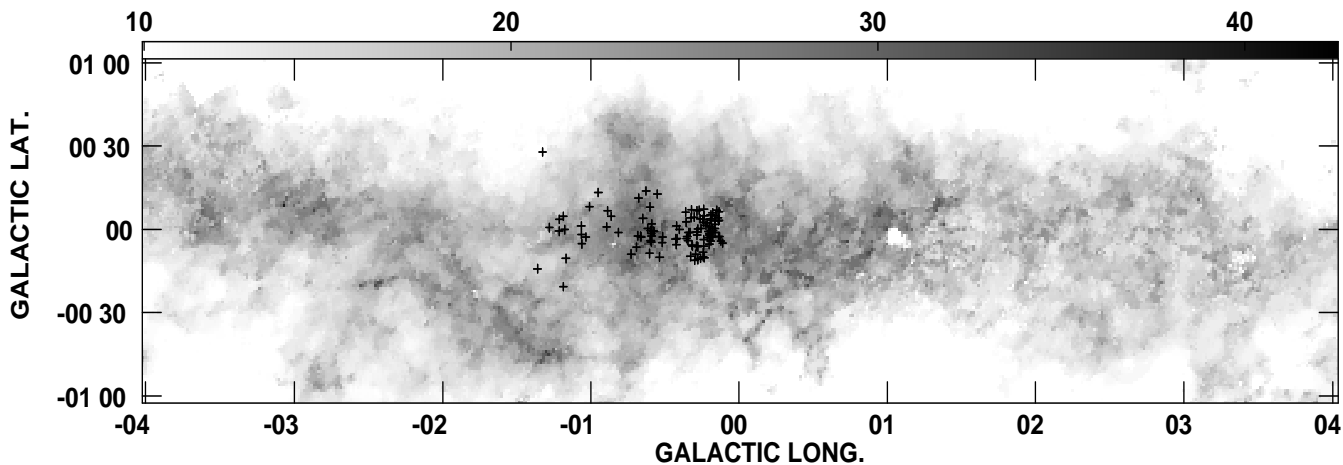}
\caption{Observed near-IR spectra (indicated by plus signs) superimposed on the extinction map
by Schultheis et al. (1999). The greyscale on the top shows the range of $\rm A_{V}$. Galactic
 longitude and latitude are given in units of degrees.
Note the high interstellar extinction with $\rm A_{V} > 20\,mag$. The
comparison fields of RSFD are located at ($\rm l = 0$, $\rm b =
-1.3,-1.8,-2.3,-2.8^{o}$)
and ($\rm b = -1.3^{o} $, $\rm l = 1.0,2.0,3.0,4.0^{o}$)}
\label{extinction}
\end{figure*}

The near-IR spectra were obtained between  16-19 July 2000
 with the NTT telescope at ESO, La Silla, Chile, using the red 
grism of the SOFI spectrograph.
 The spectra were taken under photometric  conditions through a 1\arcsec \,slit providing a 
resolving power of R $\sim$ 1000, and were recorded on a Hawai HgCdTe 1024x1024 array.
Before each spectrum, an  image in the K band was taken in order
to check the identification of the source.

For optimal sky subtraction, ``ABBA'' observing sequences were used.
The star was moved  10\arcsec~along the  slit between the
A and B exposures. The exposure time was 30\,s in each position,  repeated 
2-10 times, depending on the brightness of the object.

 B, A, late F and early G spectrophotometric standard stars
were observed during the night (typically 6-8 stars per night) to 
correct for telluric absorption features. 

\subsection{Data reduction}
The data were reduced using MIDAS, the standard ESO reduction package. After 
removal of cosmic ray events, subtraction of the bias level and the dark, all 
frames were divided by a normalized flat field. The traces of stars at the two
positions along the slit were used to subtract the sky. After extraction and
co-adding of the spectra, a wavelength calibration was performed using the Xe-Ne
lamp which gives an accuracy better  than $\rm 2\,\AA$. The spectra were
rebinned to a linear scale, to obtain a dispersion of $\sim$ 10\,$\rm
\AA$/pixel and a resulting wavelength range from 1.5\,$\rm \mu$m to
2.5\,$\rm \mu$m.

We used B, A, F and G standard stars to correct for  the instrumental and
atmospheric transmissions. For each of the standard stars, a suitable
model spectrum from Kurucz (\cite{Kurucz93}) was selected. Significant stellar absorption features were removed by
interpolation both in the models and the standard star observations ($\rm Br\gamma$ in B, A stars; the strongest
metal lines in F and G stars). The model spectra were then divided
into the standard star spectra to provide the combined instrumental and atmospheric response. The choice of the  standard  used for the correction of a particular
  program star depended on proximity in time and airmass. We did not apply a second order correction for airmass,
 as the residual effects of the telluric bands on the features we were interested in are small compared to intrinsic variations. For a more detailed discussion on spectroscopic data reduction
in general, we refer the reader to  LW. The final spectra are
normalized at 2.28\,$\rm \mu$m and are dereddened (see Sect.~3.1) using the extinction values of Schultheis et al. (\cite{Schultheis99}). They are displayed in Appendix B. Their classification is discussed in Sect.~4.
The spectra are  available electronically at CDS together
 with the corresponding finding charts.

\begin{table}
\caption{Definition of bandpasses for continuum and features}
\begin{tabular}{lcl}
Feature&band passes ($\rm \mu$m)&reference\\
\hline
NaI feature&2.204-2.211&RDF\\
NaI continuum \#1&2.191-2.197&RDF \\
NaI continuum \#2&2.213-2.217&RDF\\
CaI feature&2.258-2.269&RDF\\
CaI continumm \#1&2.245-2.256&RDF\\
CaI continuum \#2&2.270-2.272&RDF\\
$\rm ^{12}CO(2,0)$ bandhead&2.289-2.302&LW\\
$\rm ^{12}CO(2,0)$ continuum\#1 &2.242-2.258&LW\\
$\rm ^{12}CO(2,0)$ continuum\#2 &2.284-2.291&LW\\
$\rm H_{2}O$ continuum &1.629-1.720& \\
$\rm H_{2}O$ absorption wing 1 &1.515-1.525& \\
$\rm H_{2}O$ absorption wing 2 &1.770-1.780& \\
\hline
\end{tabular}
\end{table}

\subsection{Equivalent widths}

Prominent  spectral features in  our data are the NaI, CaI  and CO(2,0) bands which
have  also been discussed by Ram\'{\i}rez et al. (\cite{Ramirez97}, hereafter referred to as RDF), and the
CO(6,3) and  SiI lines (see Origlia et al. \cite{Origlia93} for
details). In addition, the OH radical has many groups of prominent
lines scattered over the whole H window (Origlia et al. \cite{Origlia93}, LW)
 
\begin{figure}
\epsfxsize=8cm
\centerline {\epsfbox[20 220 540 640]{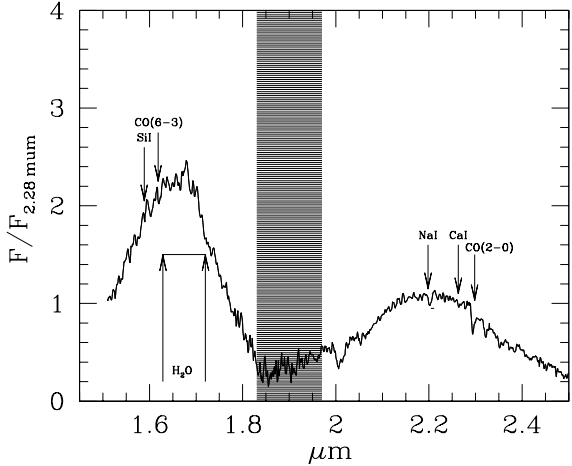}}
\caption{Spectrum of an AGB star superimposed by the most prominent
features of SiI, CO(6-3), NaI, CaI, CO(2-0) and the $\rm H_{2}O$
band. The shaded area indicates the region of very strong telluric absorption}
\label{demonstration}
\end{figure}

In our analysis  we will use the equivalent widths of the $\rm
^{12}CO(2,0)$ bands at 2.3\,$\rm \mu$m (EW(CO)) and the equivalent widths
of the NaI (EW(Na)) and CaI (EW(Ca)) lines. Additionally, the water
absorption (EW($\rm H_{2}O$)) at $\sim$ 1.6\,$\rm \mu$m has been
measured. The adopted index measures the curvature of the
spectrum in the H window due to the wings of the water bands centered
at 1.4 and 1.9\,$\rm \mu$m. The index compares the flux in a central
passband that is only weakly contaminated by $\rm H_{2}O$, to the
fluxes in passbands on either sides. For consistency with the units of
the other feature measurements, its formal expression is that of an
equivalent width (but it takes negative values when $\rm H_{2}O$
absorption is present). The measured features are identified in
Fig.~\ref{demonstration}. The exact bandpasses for all measurements
are provided in Table~1.

\subsection{The ISOGAL catalogue}
The final product of the ISOGAL catalogues (ISOGAL PSC) at present gives  magnitudes, I, J, $\rm K_{S}$, [7], [15], at five wavelengths (0.8, 1.25, 2.15,
7 \& 15\,$\rm \mu$m) with DENIS providing I,J,$\rm K_{S}$ associations when available.
We refer
for a detailed description of data reduction and the cross-identification
 method to Schuller et al. (\cite{Schuller2002}). 
Note that here and elsewhere we
 use $\rm [\lambda]$ to denote the ISOGAL magnitude at wavelength $\rm \lambda \, \mu m$ ([7], [15])

\section{Sample}
Figure~\ref{extinction} shows the location of the ISOGAL sources observed with SOFI
superimposed on the extinction map of Schultheis et al. (\cite{Schultheis99}).
 The sources were selected from the ISOGAL fields located at $\rm l = -0.27^{o}, b = -0.03^{o}$ 
(hereafter  referred to as field A), $\rm l = -0.62^{o}, b =
-0.03^{o}$ (field B), $\rm l = -0.90^{o}, b = -0.03^{o}$ (field C), 
$\rm l = -1.21^{o}, b = -0.03^{o}$ (field D), $\rm l = -0.44^{o}, b = -0.18^{o}$ (field E), 
$\rm l = -1.12^{o}, b=+0.30^{o}$ (field F), $\rm l = -1.12^{o}, b = -0.33^{o}$
 (field G), $\rm l = -5.76^{o}, b = +0.17^{o}$ (field H). Precise details of the fields  can be found
in the Explanatory Supplement of the ISOGAL catalog (Schuller et al. \cite{Schuller2002}).
We selected sources with [7]--[15] $>$ 1.4 and [15] $<$ 8.0 (except
for 4 cases, see Fig.~\ref{ISOGAL}). These colour criteria were adopted to favour the detection
of dusty young objects with mid infrared excesses close to the galactic centre, but they are also
 satisfied by a number of evolved stars, as discussed in Sect.~4. We further restricted the sample to sources
 brighter than $\sim$ 11\,mag in $\rm K_{S}$ in order to avoid
spurious associations (Schuller et al. \cite{Schuller2002}). Such a
value is slightly brighter  than the $\rm K_{S}$ completeness limit of DENIS in
these regions (see also Schultheis \& Glass \cite{Schultheis2001}). From the 1130 ISOGAL targets
obeying the above criteria in these fields, the 107 sources actually observed with
SOFI were selected  approximately at random, with the majority of the
sources (65) in field A.

 After the observations, we cross-identified  our sample again with  the latest version of the ISOGAL
Point Source Catalogue (Version 1) described in Schuller et al. (\cite{Schuller2002}) in order to
 get the final DENIS and ISO photometry. 
4 \& 5; see Schuller et al. \cite{Schuller2002} for details). 5\% of our sample now  show
a more doubtful association between ISOGAL and DENIS (quality flag $<$ 3) and have been dropped.
 Table A1  lists the coordinates and the corresponding ISOGAL and DENIS magnitudes as well as the
measured equivalent widths and the type of each source.

As a comparison sample, we used the low resolution  ($\rm R \sim 1300$) K-band  spectra of M giants of  RSFD 
which  are located at ($\rm l = 0$, $\rm b = -1.3,-1.8,-2.3,-2.8^{o}$) and  at ($\rm b = -1.3^{o} $, $\rm l = 1.0,2.0,3.0,4.0^{o}$) and the solar neighbourhood sample of oxygen-rich stars (M giants, semi-regular Variables and Mira Variables) of LW.

\subsection{Interstellar reddening}
The high interstellar extinction in the Galactic Bulge hampers 
the study of the stellar populations in the inner parts of our Galaxy. As in most
parts of the Galactic Bulge, interstellar absorption is not homogeneous but
occurs in clumps (Glass et al. \cite{Glass87}). Catchpole et al. (\cite{Catchpole90}) mapped the interstellar extinction around the
 Galactic Centre ($\rm \sim 2\,deg^{2}$) using the red giant branch of 47Tuc as a reference while
 Frogel et al. (\cite{Frogel99}) determined the interstellar reddening for a few fields in the 
inner Galactic Bulge using the red giant branch of Baade's window. 
Schultheis et al. (\cite{Schultheis99}) constructed a high
resolution map of the whole inner part using DENIS data together with isochrones
calculated for the RGB and AGB phases. We used their extinction table to
deredden our objects (see Table A1) according to  the interstellar extinction law
of Glass (\cite{Glass99a}) with $\rm A_{V}:A_{J}:A_{K} = 1:0.256:0.089$. For
7 and 15\,$\mu$m photometry we used $\rm A_{7}/A_{V} = 0.020$ and $\rm A_{15}/A_{V} = 0.025$
 (Hennebelle et al. \cite{Hennebelle2001}). However, the extinction
curve particularly at 7 and 15\,$\rm \mu$m is rather uncertain.
One has also  to keep in mind  that such extinction values correspond
to the peak value of $\rm A_{V}$ on the line of sight; but the actual
value for each individual star may differ by several magnitudes. In addition,
 the extinction values in  the very high extincted regions ($\rm A_{V}
> 25)$ are only lower limits. Nevertheless there seems to be on average
a reasonable agreement with the $\rm A_{V}$ values derived by Wood et
al. (\cite{Wood98}) for OH/IR stars as discussed by Ortiz et al. (\cite{Ortiz2002}).

\subsection{Bolometric magnitudes} 
Bolometric magnitudes for our ISOGAL objects were obtained by using
the multi-band bolometric correction for AGB stars (Loup et al. \cite{Loup2002}). It employs
multi-band photometry with passbands as accurate as possible, taking into account a 
mean atmospheric absorption at the site for ground-based observations. The bolometric
corrections are derived numerically  using 69 models  for O-rich stars (see Loup et al. \cite{Loup2002}
 for a detailed description). The models are based on
Groenewegen's radiative transfer code, spanning a large range of dust opacities.
The advantage of multi-band bolometric corrections is that the determination of $\rm M_{bol}$
is more accurate than with traditional one-band bolometric corrections.
We used for the input parameter the  DENIS J and $\rm K_{S}$ counterparts, the LW2 and LW3 magnitudes of our objects and  the $\rm A_{V}$ values  of Schultheis et al. (\cite{Schultheis99}). The main errors
in  the bolometric corrections result from the interstellar extinction  values  giving
an error in $\rm m_{bol}$ of at least $\rm \sim 0.2-0.3\,mag$ and the
intrinsic depth of the ``Bulge'' on the line of sight (Alard et at
al. \cite{Alard2001}, OOS).  Variability of the stars will introduce additional errors in the
determination of $\rm M_{bol}$ because of non-simultaneity of the photometry. 
In Table 2 we list the derived $\rm M_{bol}$
using a distance modulus of $\rm \Delta m = 14.5$ (or a distance of $\sim$ 8\,kpc). 

\subsection{Cross-identification with SIMBAD}

 All the objects have
 been searched for in the  SIMBAD data base using a search radius of 
3\arcsec.  As a result, 
fifteen objects have been identified as OH/IR stars from the sample
of Wood et al. (\cite{Wood98}, see also Ortiz et al. \cite{Ortiz2002}), and twelve as Long Period
variables from Glass et al. (\cite{Glass2001}). These subsamples are
 shown, respectively, in Fig.~B3 and Fig.~B4.
 Four objects (A33, A40, D6, D7) are known radio sources with IRAS
 fluxes characteristic of young stellar objects(see also Sect.~4.3).

\begin{figure}[h!]
\epsfxsize=8cm
\centerline {\epsfbox[50 150 510 710]{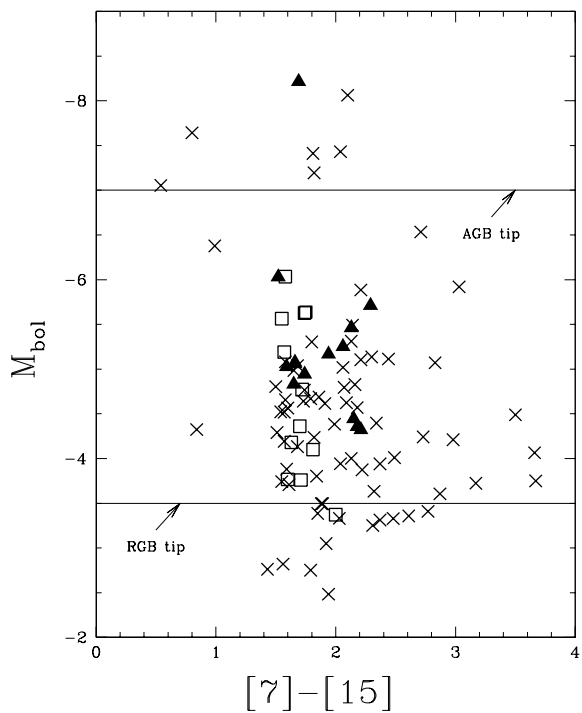}}
\caption{$\rm M_{bol}$ vs. [7]-[15] diagram for the ISOGAL+SOFI sample. 
Known OH/IR stars (Wood et al. \cite{Wood98}) are indicated by filled 
triangles, LPVs (Glass et al. \cite{Glass2001}) by open squares, while
the remaining ISOGAL objects of our sample are shown by crosses. The
two lines indicate the tip of the RGB and AGB.}
\label{Mbol_LW2LW3}
\end{figure}

\section{Classification}

Figure~\ref{Mbol_LW2LW3} shows the bolometric magnitudes vs. the [7]--[15]
colour ([7] and [15] denote the magnitude at 7 and 15\,$\rm \mu$m), which is nearly insensitive to interstellar extinction. 
Indicated
also are the known OH/IR stars and LPVs in our sample (see Sect.~4.2). 
It is obvious from Fig.~\ref{Mbol_LW2LW3} that photometric
information alone is not sufficient to separate the different categories: additional spectroscopic observations are necessary.

\begin{figure*}
\epsfxsize=7.5cm  
\centerline {\epsfbox[50 150 520 710]{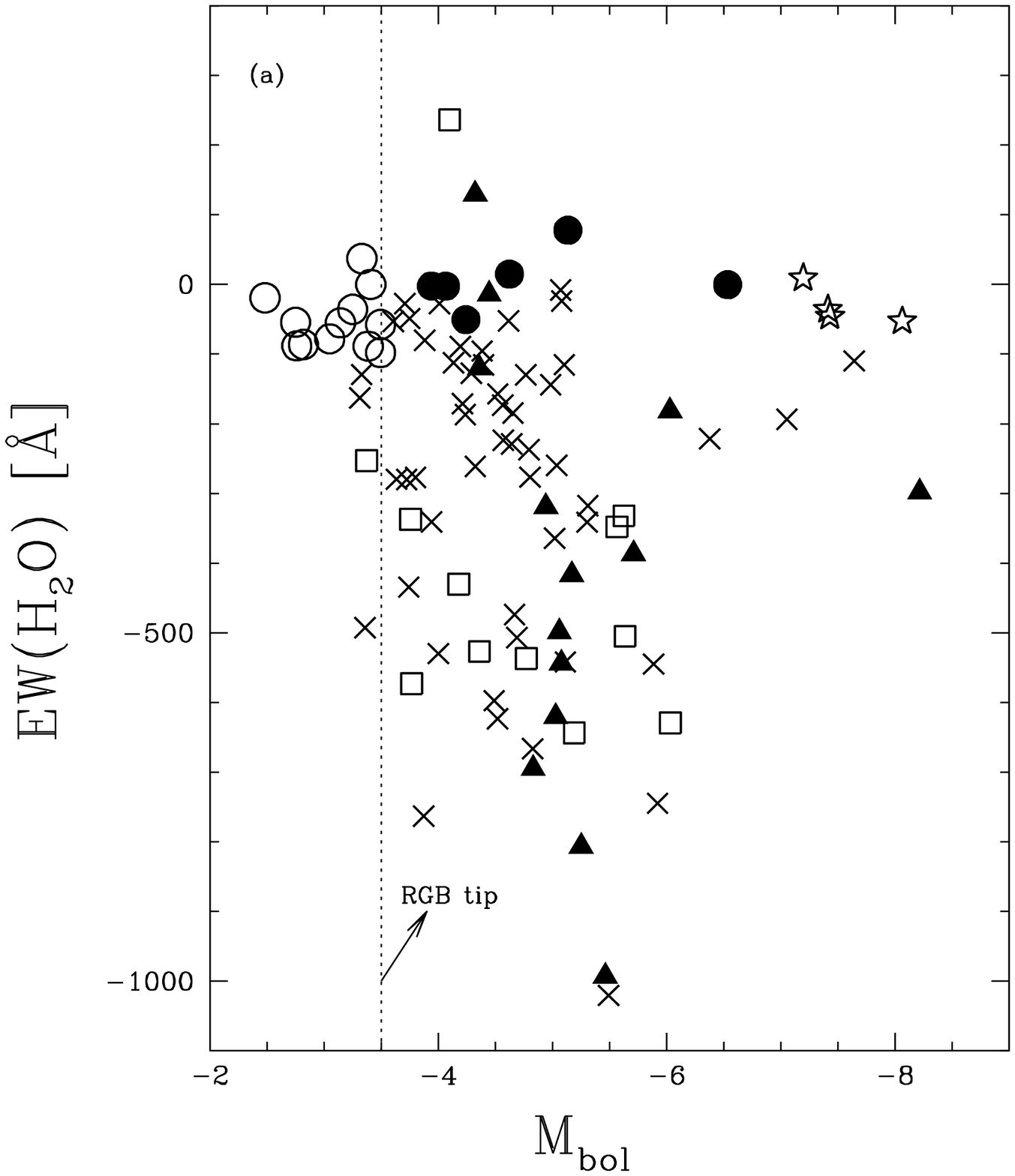}
\epsfxsize=7.5cm  \epsfbox[50 150 520 710]{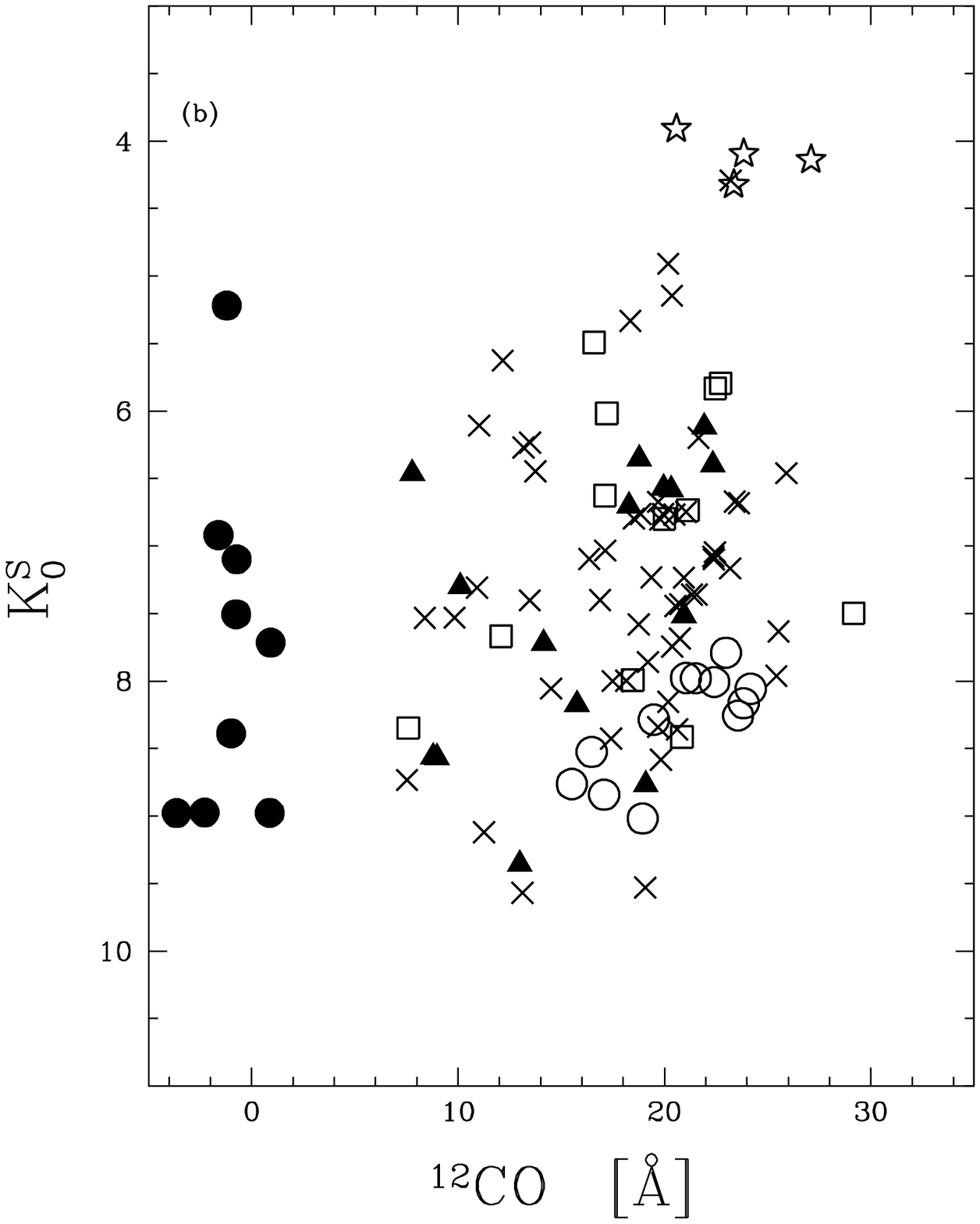}}  
\caption{Equivalent width of $\rm H_{2}O$ vs. $\rm M_{bol}$ (a) and $\rm K_{0}$ vs  equivalent width of CO  (b). YSOs are indicated by filled circles, OH/IR stars (Wood et al. 1998) by filled triangles, LPVs (Glass et al. 2001) by open squares, candidates of red giants by open circles, supergiant candidates by  stars  and 
AGB Variables by crosses. The dotted line indicated the tip of the
RGB. The two AGB stars with $\rm M_{bol} \leq -7$ are probably blue
supergiant candidates (see text). }  
\label{K0CO}  
\end{figure*}

\begin{figure*}  
\epsfxsize=7.5cm  
\centerline {\epsfbox[50 150 520 710]{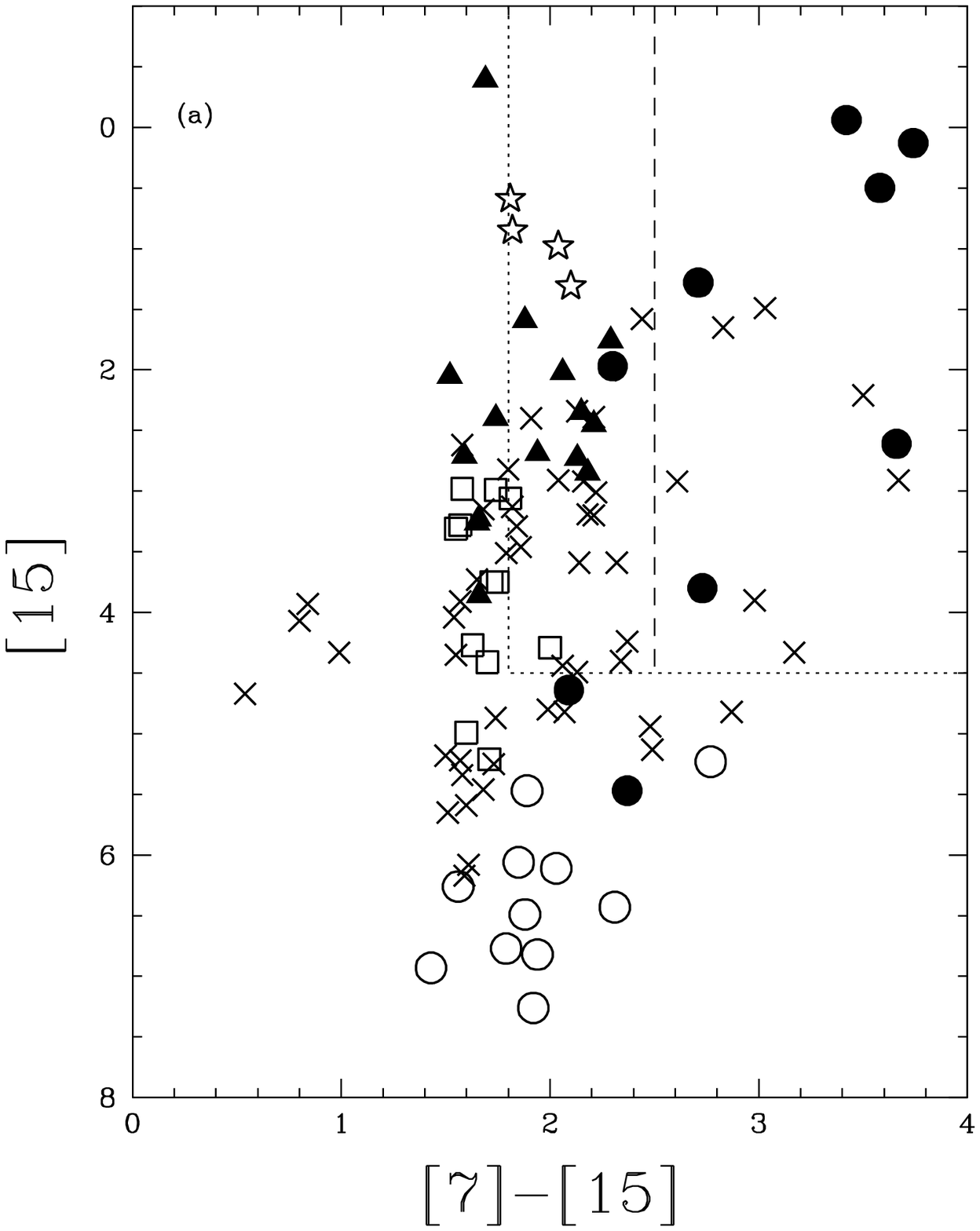} \epsfxsize=7.5cm  \epsfbox[50 150 520 710]{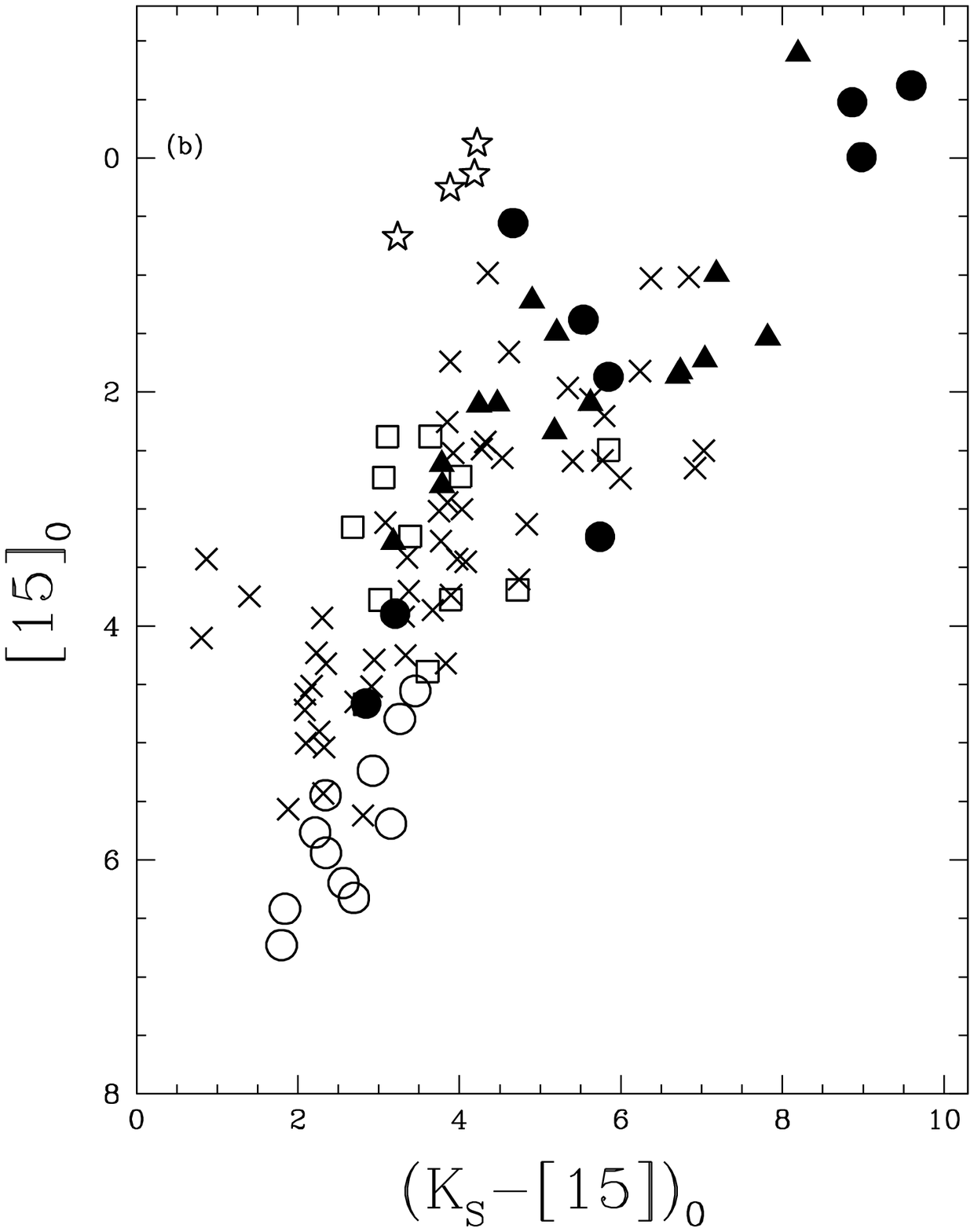}}  
\caption{[15] vs. [7]--[15] diagram (left panel) and [15] vs $\rm
K_{0}-[15]$ diagram (right panel). Symbols are the same as in
Fig.~\ref{K0CO}. The dotted line shows the region  searched by Felli
et al. (\cite{Felli2000}) for YSO candidates (see text). The
long-dashed line indicates a more conservative criterion for
identifying YSOs (Felli et al. \cite{Felli2002} )}  
\label{ISOGAL}  
\end{figure*}

\subsection{Supergiants}
As shown in Fig.~\ref{K0CO}, we find six very luminous objects (A17, B14, B15, B19, E5, E13), besides one OH/IR star,  with 
$\rm M_{bol} \lesssim -7.0$ (or $\rm \gtrsim 50,000\,L_{\odot}$),
assuming  a distance of 8\,kpc to the Galactic Center), which is
approximately the AGB tip luminosity (Vassiliadis \& Wood
\cite{Vassiliadis93}). They show very strong CO bands at 2.3\,$\rm \mu$m with 
equivalent widths $\rm > 20\,\AA$ (see Sect.~2.3 for the band
passes). In addition, four of them (A17,B14,B15,B19)  do not show any
significant $\rm H_{2}O$ absorption ($\rm EW(H_{2}O)< 100\,\AA$) which
is an indication of supergiants (Bessell et al. \cite{Bessell91},
Lan\c{c}on \& Rocca-Volmerange \cite{Lancon92}).  The objects E5 and
E13 show very blue colours in $\rm K_{0}-[15]$ and [7]-[15] and
could be late K or early M-type  supergiant candidates.  Schuller
(\cite{Schuller2003})  found from a systematic search of very
luminous ISOGAL sources ($\rm M_{bol} < -6$) several blue supergiant
candidates with $\rm K_{0}-[15] < 1.0$.

In the [15] vs [7]--[15] diagram (see Fig.~4a), the supergiants
 (stars) are rather luminous at 15\,$\rm \mu$m\  and show  rather blue $\rm K_{0}-[15]$ colours
 compared to the AGB Variables (see Fig.~4b).

\subsection{Long Period variables}
\begin{figure} 
\epsfxsize=8.0cm  
\centerline {\epsfbox[50 150 520 710]{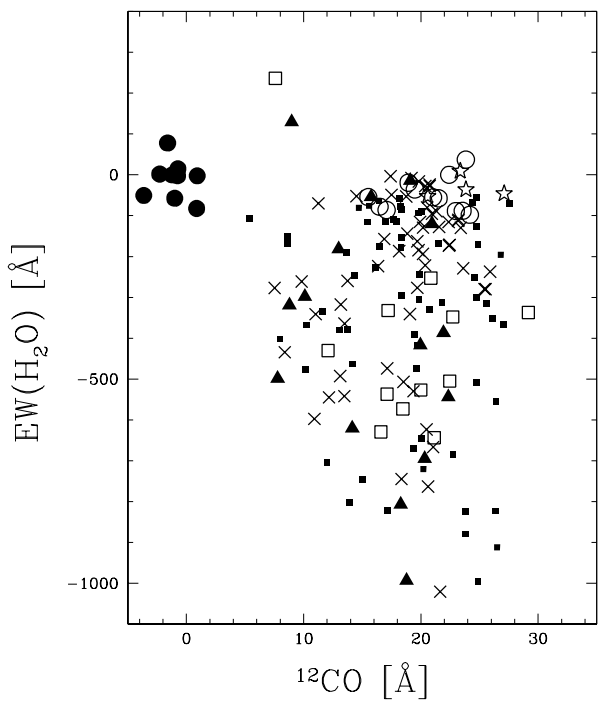}}  
\caption{Equivalent width of $\rm H_{2}O$ vs. CO, using the
sample of oxygen-rich LPVs of LW for comparison.
YSOs are indicated by filled circles,
OH/IR stars by filled triangles, LPVs by open squares, candidates of
giants by open circles, supergiant candidates  by  stars and AGB variables by crosses.
Filled squares indicate the comparison sample of semi-regular variables
and Miras of LW.}  
\label{ariane}  
\end{figure}

The majority of our sources we find to be Long-Period Variables
 (LPVs) of the AGB, mainly based on their
   very deep water absorption features (LW and Fig.~3a). Strong $\rm H_{2}O$ bands
 are associated with large period variability  and  depend on the 
atmospheric structure of the star (see Bessell et al. \cite{Bessell96}).
 Their luminosities are in the range
  $\rm -4 <  M_{bol} < -7$ and their strong CO bands,
 characteristic of stars on the AGB, also imply that they are  likely to be Miras and semi-regular variables (SRVs). Fig.~\ref{ariane} shows a comparison between the water bands
and the CO bands of the ISOGAL sample and a solar neighbourhood sample
of semi-regular variables and Miras (LW). The samples agree surprisingly
well within the broad range of the band strength values of both bands. We identify the ISOGAL sources having strong $\rm H_{2}O$
absorption as AGB variables (semi-regular variables or Miras).
Greene \& Lada (\cite{Greene96}) and G\'omez \& Mardones
 (\cite{Gomez2003}) showed that  YSOs of class III can show 
 rather  strong $\rm H_{2}O$ and CO bands too. However, they
  have rather low luminosities with $\rm M_{bol} >  2$ (see G\'omez \&
 Mardones \cite{Gomez2003}), much too faint to be on the AGB ($\rm M_{bol} < -3.5$)
 Therefore, only additional information about the
 luminosity  give us the real confirmation that our spectra of
 luminous stars with strong water absorption are indeed
 AGB star candidates. (see Fig.~4).

In Fig.~4a one can see, except for supergiants and the other very
bright sources ($\rm M_{bol}  \lesssim -6.0$),  some correlation between the strength
of the water bands and the bolometric magnitude in the sense that
more luminous objects show deeper water absorption. The [15] vs. $\rm K_{0}-[15]$ diagram (Fig.~\ref{ISOGAL}) shows that the LPVs have  a wide $\rm K_{0}-[15]$ colour range ($\rm 0 < K_{0}-[15] < 7$) indicating a large range of  mass-loss rates (see discussion below).

Twelve LPVs  from our sample
and five OH/IR stars (see Table A1) were observed by Glass et al. (\cite{Glass2001}) and we have therefore additional information about their periods and their K amplitudes. 
The period range is rather narrow, starting from $\sim$ $\rm 400^{d}$ up to
$\rm 800^{d}$.  The  sources within the sample follow a period-luminosity relation but we do not find
any relation between the CO or water band strengths and  the amplitude.




\subsection{Young stellar objects}
 A33, A40, D6 and D7 are already known to be young stellar
objects. They  have been detected by IRAS with very red colours and also show radio emission.  They have
more-or-less featureless spectra (see Fig.~B.1) with no CO absorption at 2.3\,$\mu$m
but  in some cases possess hydrogen absorption lines (Brackett series). 
The objects B23, B27, B35, B37, C19 are newly found young stellar objects showing the same
spectral features as the four known ones.

Greene \& Lada (\cite{Greene96}) presented the first systematic spectroscopic survey of
YSOs, comprising a sample from the Ophiucus molecular cloud. They found that the strengths of atomic 
and CO absorption features are closely related to the evolutionary state. The line
strengths decrease from the Class III phase through  Class II  to the self-embedded
 Class I,  where the absorption features are absent. All nine YSOs of our sample
 show more-or-less featureless spectra with no CO absorption lines and  no
$\rm Br\gamma$ emission. Thus, we associate them with young stellar objects of 
Class I.

However, Fig.~\ref{ISOGAL} shows that
YSOs cannot be separated unambiguously from other stellar
populations  by using ISOGAL colours alone. 
Felli et al. (\cite{Felli2000}) give a criterion for
selecting YSO candidates  (see Fig.~4)
in the ISOGAL [15] vs. [7]--[15] diagram and  Felli et al. (\cite{Felli2002})
present a catalog of YSO candidates with $\rm [15] < 4.5$ and $\rm [7]-[15] > 1.8$. 
 The criteria defined by Felli et al. (\cite{Felli2002}) are satisfied by seven of the nine identified YSOs in this region (see Fig.~4a). However, as shown
in Fig.~\ref{ISOGAL}, there are also 32 AGB variables, eight OH/IR stars and four supergiants
 in the same region.  Felli et al. (\cite{Felli2002}) point  out also that $\rm [7]-[15] \geq 2.5$ is 
a more conservative criterion for identifying YSOs. Six of the YSOs meet this criterion but  it is still
satisfied by seven objects which are probably AGB stars.
Thus, an unambiguous separation
between YSOs and evolved stars can only be made by using additional spectroscopic information. 

\subsection{M giants}
In contrast to  the typical LPV,  M giants show nearly no water
vapor bands. Using an additional luminosity criterion, namely $\rm
M_{bol}$ fainter than --3.5 (see Fig.~3), which is approximately the tip of the RGB 
(see Tiede et al. \cite{Tiede96}, Omont et al. \cite{Omont99}), we find 12 objects. Figure~\ref{ISOGAL}  shows that indeed the
M giant candidates populate the lower end in both the [15] vs [7]--[15] and
[15] vs. $\rm K_{0}-[15]$ diagram.
However,  as seen in Fig.~\ref{ariane}, the separation between  M giants and variable AGB stars in the CO vs. $\rm H_{2}O$ plane is ambiguous. Some of the stars classified as giants might
 be AGB stars  that happen to be caught at a phase of pulsation 
where $\rm H_{2}O$ is not very strong (see LW). Additional multi-epoch observations are necessary,
 especially because it has recently been shown that  most  M giants with late spectral types are variable
 (Alcock et al. \cite{Alcock2000}, Glass \& Schultheis \cite{Glass2002}).

\begin{figure}[ht] 
\epsfxsize=8.0cm  
\centerline {\epsfbox[50 150 520 710]{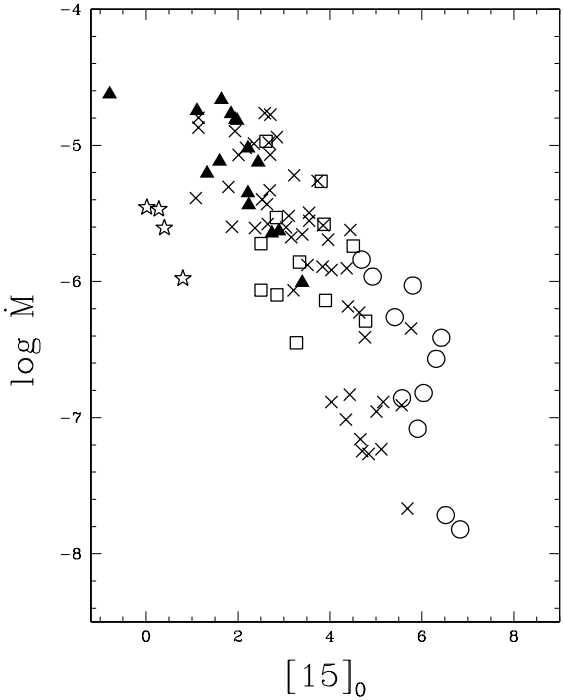}} 
\caption{Mass-loss rate vs.~15\,$\mu$m magnitude vs. for the present sample.
OH/IR stars are indicated by filled triangles, LPVs by open squares, red giant 
 candidates  by open circles and supergiants by stars.} 
\label{massloss}  
\end{figure}

\section{Discussion}

Figure~\ref{ISOGAL} shows that near and mid-IR photometry alone are not sufficient for  distinguishing the different
stellar populations; additional near-IR spectroscopy  is necessary.
Using the $\rm M_{bol}$ vs $\rm H_{2}O$ and the  $\rm ^{12}CO$ vs. $\rm K_{0}$ diagram (see Fig.~\ref{K0CO}) we can identify
  supergiants, AGB Variables, red giants and  young stellar
objects. In addition, mid-IR data at 7 and 15\,$\rm \mu$m
 enables us to estimate mass-loss rates of the stellar population
while the equivalent widths of the CO bandhead, the NaI doublet and the 
CaI triplet give us estimates of the metallicity.

\subsection{Mass loss}

One of the most promising tools for determining mass-loss rates is the 
combination of near-IR and mid-IR colour such as the IRAS $\rm K_{0}$--[12]
or the ISOGAL $\rm K_{0}$--[15] colours (see e.~g. Whitelock et
al. \cite{Whitelock94}, Habing \cite{Habing96}, Le Bertre \&
 Winters \cite{LeBertre98}, Omont et al. \cite{Omont99}, Jeong
et al. \cite{Jeong2002}, OOS etc.).  We will use the most recent
colour-mass loss relation for oxygen-rich AGB stars by Jeong et al. (\cite{Jeong2002}) which is
based on a consistent time dependent treatment of hydrodynamics,
thermodynamics, equilibrium chemistry and dust formation. They give an
explicit relation between $\rm  \dot{M}$ and the $\rm (K-[15])_{0}$
colour.

\begin{equation}
\rm log\,\dot{\it{M}} = -6.83/(K-[15])_{0} - 3.78   \hspace{1.5cm}  [M_{\odot}/yr]
\end{equation}


 One has to be aware that using this  relation between $\rm \dot{M}$ and $\rm K_{0}-[15]$ for our sample includes  uncertainties in the determination
of the mass-loss rate arising from the following causes: (1) Our de-reddend $\rm (K -[15])_{0}$ colours are strongly affected by the uncertainty in the determination of interstellar extinction, in particularly 
in these highly extincted regions (see Schultheis et al. \cite{Schultheis99}). This could
lead easily to errors in $\rm \dot{M}$ of a factor of 2--3; (2)
The relation was derived for oxygen-rich AGB stars in the solar neighbourhood.
 Habing et al. (\cite{Habing94}) argued that metallicity affects the 
dust to gas ratio and the outflow velocity from evolved stars which is
 directly related to the mass-loss and that, therefore, the color-$\rm \dot{M}$ relation might
differ in different galactic environments such as between the Galactic Bulge and the Magellanic
Clouds; (3) 
It is  important to emphasize that the $\rm K_{S}$ magnitudes of DENIS are
single epoch measurements.
 The average  K amplitude  of  our sources associated with known LPVs
(Glass et al. \cite{Glass2001})  is $\sim$ 1.0\,mag, which gives a
factor of $\sim$ 2--5  uncertainty in the determination of $\rm \dot{M}$ for
$\rm -7 < log\,\dot{{\it{M}}} < -5$. 
 We want to emphasize that the colour mass-loss relation by Jeong
et al. (\cite{Jeong2002}) is calculated for variable AGB stars.
 Its adaption to non-variable red giants and
supergiants has to be questioned.
 


From this relation,  the OH/IR stars in our sample cover  mass-loss rates ranging from 
 $\rm -6.0 < log\,\dot{{\it{M}}} < -4.5$ while LPVs range between  $\rm -6.5 <
log\,\dot{{\it{M}}} < -5$. 
 Considering the uncertainties (see above) these
values lie within the ranges  determined by Ortiz et al. (\cite{Ortiz2002}) for OH/IR stars and 
Alard et al. (\cite{Alard2001}) for LPVs. Our red giant candidates
show mass-loss rates in the range between $-8 < log\,\dot{{\it{M}}} < -6$.
 Origlia \& Ferraro (\cite{Origlia2002}) determined mass-loss
rates of red giants in globular clusters using ISOCAM
observations. They find mass-loss rates in the range of $\rm -7 <
log\,\dot{{\it{M}}} < -6$ assuming a gas-to-dust mass ratio of 200 and using
the DUSTY code. Their mass-loss range  is much  narrower than we find.
 However, taking  the color-mass loss relation of Jeong et al. (\cite{Jeong2002})
would result in at least 10 times smaller mass-loss rates for the
sample of Origlia \& Ferraro (\cite{Origlia2002}). It is obvious that
red giants certainly contribute significantly to the integrated
mass-loss (see also OOS). We  want to stress that this is the first
attempt to quantify mass-loss rates of red giants in the central part
of our Galaxy. A systematic study of these objects in the inner Bulge
 is certainly needed to derive accurate number densities and their
contribution to the integrated mass-loss.



Our 4 supergiant candidates show mass-loss rates between $\rm
-6 < log\,\dot{{\it{M}}} < -5.5$ assuming that the relation by Jeong et
al. (\cite{Jeong2002}) is valid for supergiants.
 However, as pointed out by Josselin et al. (\cite{Josselin2000}) the
gas-to-dust ratio for these kind of objects is rather
 uncertain. Therefore a detailed comparison with the solar
neighbourhood sample of Josselin et al. (\cite{Josselin2000}) is at this
 stage rather difficult.
 



\subsection{Spectral determination of luminosity class}
\begin{figure}[h!]
\epsfxsize=7.0cm 
\centerline {\epsfbox[20 20 566 753]{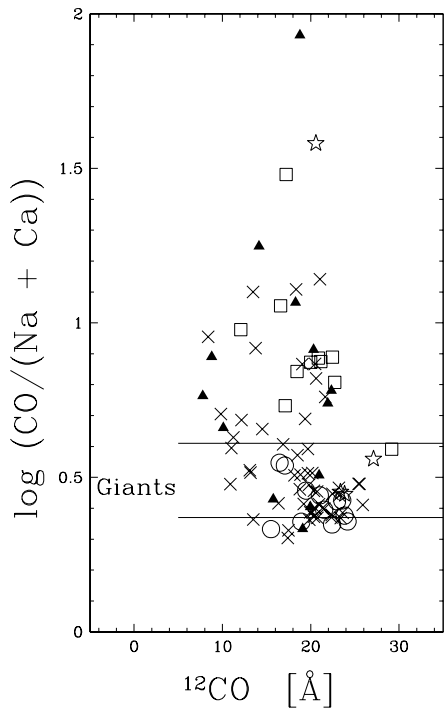}}
\caption{The luminosity indicator of Ram\'{\i}rez et al. (\cite{Ramirez97}) vs. CO band strengths. The two straight
lines show the location of the M giants of Ram\'{\i}rez et al. (\cite{Ramirez97}). The symbols are the same as in Fig.~3  }
\label{luminosity}
\end{figure}

Do spectra in the H and K bands offer  the possibility of determining the
luminosity class?
Ram\'{\i}rez et al. (\cite{Ramirez97}) used the quantity
$\rm log\,[EW(CO)/(EW(Na)+EW(Ca))]$   to distinguish between giants and dwarfs
 over the effective temperature range between 3400 and 4600\,K (see their Fig.~11). They argued that this quantity might be a powerful luminosity diagnostic. 
Figure~\ref{luminosity} shows that the M giants of RSFD (indicated by the two straight lines)  and our sample
agree quite well. For comparison, we calculated also the values for the
LW sample. We find that there is no clear separation between
supergiants and M giants or LPVs. However, as seen in Fig.~\ref{luminosity},
the dispersion in  $\rm log\,[EW(CO)/(EW(Na)+EW(Ca))]$ is much narrower
for M giants than for variable AGB stars. 
Pulsation is responsible for the extended
atmospheres of LPVs (Feuchtinger et al. \cite{Feuchtinger93}) which is seen
in the large dispersion of $\rm log\,[EW(CO)/(EW(Na)+EW(Ca))]$ of the LPVs and OH/IR
stars. We get similar results using the LW
sample.  While the quantity  $\rm log\,[EW(CO)/(EW(Na)+EW(Ca))]$ is a good 
discriminator between dwarfs and giants, it can not be adopted to
separate LPVs from supergiants  and red giants.

\begin{figure} 
\epsfxsize=8.0cm  
\centerline {\epsfbox[0 0 520 763]{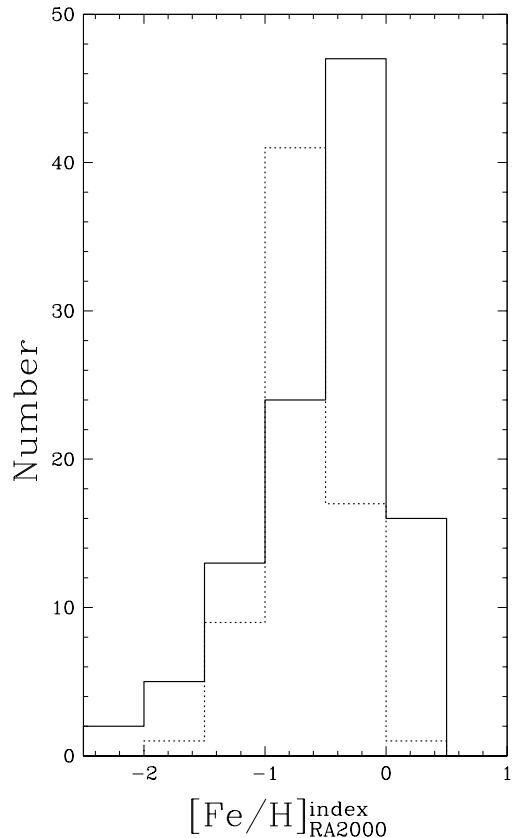}}  
\caption{Distribution of metallicity. The solid lines is our ISOGAL sample located in the
inner Bulge;  the dotted  line is the solar neighbourhood sample of oxygen-rich AGB stars and M giants by LW.}  
\label{metallicity}  
\end{figure}

\subsection{Metallicity}

Recently, RSFD and Frogel et al. (\cite{Frogel2001}) have obtained 
a new metallicity scale for luminous red giants  based on  equivalent
width measurements of the CO bandhead, the NaI doublet and the CaI triplet (see Table 1).
Their calibration is based on giants in globular clusters  for $\rm -1.8 < [Fe/H] < -0.1$ .
We have used the following relation (solution 1, RSFD)

\medskip

 \hspace*{-0.6cm} $\rm [Fe/H]^{index}_{RSFD}$ = -1.782 + 0.352\,EW(Na) - 0.0231\,EW(Na)$^{2}$\\
 \hspace*{1.0cm} -0.0267\,EW(Ca) + 0.0129\,EW(Ca)$^{2}$\\
 \hspace*{1.9cm} + 0.0472\,EW(CO) - 0.00109\,EW(CO)$^{2}$ \hspace*{0.2cm}     (2) \\

where EW(Na), EW(Ca) and EW(CO) are the equivalent widths of NaI, CaI and $\rm ^{12}CO(2,0)$ (see Table~1). As described in RA97, the typical errors
 in the determination of  [Fe/H] are  of the order of $\rm \sim 0.1\,dex$.
RSFD provide a second expression for $\rm [Fe/H]^{index}_{RSFD}$, that reduces the scatter in their data with correction terms
based on $\rm (J-K)_{0}$. As most of our program stars do not have measurable J magnitudes due to heavy
extinction, this second relation could not be used.

In addition, we determined this index from Eq. (2) for the sample of oxygen-rich stars (M giants, semi-regular variables and Miras) in
the solar neighbourhood of LW using the same bandpasses.
We used repeated  observations of LW for semi-regular  and Mira variables, such as BD Hya, R Cha or KV Car.  For the same star the metallicity index of RSFD can
vary by $\sim$ 0.5\,dex showing that it does not actually measure
metallicity in individual spectra of strongly variable stars. 

Fig.~\ref{metallicity} shows the distribution of the RSFD metallicity index  of our ISOGAL objects superimposed on the solar neighbourhood sample of LW.
If we use only the non-varying static stars of LW and our sample of red giant candidates, the mean $\rm [Fe/H]^{index}_{RSFD}$ does not change, although the dispersion gets smaller. This shows that, 
even if the metallicity index of RSFD is not strictly valid
  for single observations of a variable star, it can be used {\em {on average}} over a whole  pulsation cycle. This is important for  stellar population studies, as our results show that
 very  time consuming  multi-epoch observations of LPVs are not
necessary in order to obtain average metallicities. We will only
use average values in the remaining discussion.

The mean $\rm [Fe/H]^{index}_{RSFD}$
is $\rm -0.25\,dex$,  with a dispersion of $\rm \pm 0.40\,dex$, which is in agreement with the values
obtained for the static sample of RSFD. Their mean $\rm
[Fe/H]^{index}_{RSFD}$ is $\rm -0.21\,dex$ with a dispersion of $\rm \pm 0.30\,dex$. This suggests that the ISOGAL sample
and the sample of RSFD have similar metallicities despite their spatial separation.

Superimposed is the solar neighbourhood sample of LW with an average
$\rm [Fe/H]^{index}_{RSFD}$ of $\rm \sim -0.6\,dex$ (see Fig.~9). It is known
  that there is a large spread in the solar neighbourhood metallicity
distribution (see e.~g. Haywood \cite{Haywood2001}) with
$\rm -0.3 <  [Fe/H]^{index}_{RSFD} < 0$;  an offset of $\rm \sim
0.3\,dex$ with respect to the solar neighbourhood sample of LW is apparent. 
The relation of RSFD is based on the calibration of globular
clusters and thus is biased towards metal-poor stars.
 At higher metallicities the CO band reaches a plateau and becomes insensitive to changes
in $\rm [Fe/H]^{index}_{RSFD}$. Since solution 1 has a strong
dependence on EW(CO), it will therefore underestimate $\rm
[Fe/H]^{index}_{RSFD}$ for larger metallicities.
However, Fig.~\ref{metallicity} shows  that our ISOGAL sample as well
as the sample of RSFD is $\rm \sim 0.4\,dex$ metal richer than the solar neighbourhood sample of LW. Haywood (\cite{Haywood2001}) gives a modelled and an observed age-metallicity
relation (see their Fig.~14a). According to that, a [Fe/H] of $\rm -0.6\,dex$ corresponds
to ages of about 8--9\,Gyr  and thus  for AGB stars an initial mass of the order of one solar mass.
 So, according to the age-metallicity relation, the solar neighbourhood sample of LW have
initial masses around 1\,$\rm M_{\odot}$,  in agreement with  mass estimates derived from 
pulsation periods (see Table~8 of LW). 
The difference in the metallicity distribution
in the solar neighbourhood and in the Bulge results probably from both different
mean stellar ages and different mean age-metallicity relations due to different star 
formation histories and chemical evolution histories.
However, the validity of such  age-metallicity relations is still on debate (see e.g. the recent review of Freeman \& Bland-Hawthorn  \cite{Freeman2002})

The accurate determination of mean stellar metallicities is essential
for constraining models of star formation and chemical evolution in
the Bulge.
Our mean [Fe/H] value is consistent with previous chemical abundance studies of the 
galactic Bulge (see e.~g. McWilliam \& Rich \cite{William94}, Minniti et al. \cite{Minniti95},
 Houdashelt \cite{Houdashelt96}, Frogel et al. \cite{Frogel99}, RSFD) where the metallicity peaks 
$\rm \sim -0.3\,dex$. Moreover, Molla et al. (\cite{Molla2000}) model
the evolution of the Galactic Bulge which assumes 
  a dissipative collapse of the gas from a protogalaxy or halo to
form the Bulge and the disk. They predict  a mean stellar Bulge metallicity
[Fe/H] $\rm \sim -0.2\,dex$ with a dispersion of $\rm \pm 0.40\,dex$. 

A chemical abundance gradient in the Bulge is one characteristics
  of a Bulge formed by a dissipative collapse.  The determination of
  the metallicity gradient in the central regions of our Galaxy is
  important  for testing models of Galaxy formation. Up to now, only a
  few  studies of the metallicity gradient exist in the
inner Bulge  ($\rm |b| < 3^{0}$). Frogel et al. (\cite{Frogel99})
  found a metallicity gradient based on near-IR photometric data while
 RSFD could not find any evidence. 
The  dissipative collapse model of Molla et al. (\cite{Molla2000})
predict a steep metallicity gradient ($\rm -0.8\,dex\,kpc^{-1}$) in the inner Bulge ($\rm R < 500\,pc$) which is not supported by the
data of RSFD. Our sample of near-IR spectra presented here, which
agrees with   the metallicity distribution of  RSFD, seems to confirm the absence
of a metallicity gradient in the Bulge.

We want to stress that our data set is not sufficient to do
a large statistical analysis of chemical abundances in the galactic Bulge. Therefore a more detailed
 discussion of the implications of the results is beyond the scope of this paper; a large observational
program is necessary.

It is obvious from the discussion above that  only realistic models of M giants including 
the effects of metallicity can enable us to make  a quantitative 
determination of their  metallicity. First tests of synthetic spectra based on hydrostatic
MARCS model atmospheres (in collaboration with B.~Aringer) and including a complete
atomic line list already show  some promising results (Aringer et al. \cite{Aringer2002}). However, for AGB stars (the majority of our stars) hydrodynamical models would be more appropriate (see e.~g. H\"ofner
 \cite{Hoefner99}, Aringer et al. \cite{Aringer2000}).

\section{Conclusions}

We have studied a sample of 107 near-IR spectra of ISOGAL sources with excesses at 7 and 15\,$\mu$m in the innermost parts of the galactic Bulge where the interstellar extinction
is high and clumpy. We have shown that using the molecular bands of CO  and $\rm H_{2}O$, together with the bolometric
magnitudes and the interstellar extinction values, one can reasonably well separate AGB stars, red giants, supergiants and young stellar objects. We have found  four supergiant candidates,
twelve red giant candidates and nine YSOs, while the rest are probably
variable AGB stars. We have used
the most recent   $\rm K_{0} -[15]$ vs. $\rm \dot{M}$ relation by Jeong
et al. (\cite{Jeong2002}) which is based on a self-consistent time
dependent model of dust formation in AGB stars to determine mass-loss rates. We emphasize
that the color-$\rm \dot{M}$ relation has been determined from a model
of AGB/LPVs and gives  only  an {\em{indication}} of the mass-loss rate.
 From our sample, OH/IR stars cover  mass-loss rates ranging  from $\rm -6 <
log\,\dot{{\it{M}}} < -4.5$ while LPVs range between  $\rm -6.5 < log\,\dot{{\it{M}}}
< -5$. Our red giant candidates show mass-loss rates in the range
between $ -8 < log\,\dot{{\it{M}}} < -6$. However, this selection is biased
in favor of large $\rm \dot{{\it{M}}}$. A comparison with mass-loss rates
of red giants in globular clusters (Origlia \& Ferrero
\cite{Origlia2002}) shows that our mass-loss range is much
broader towards larger values of $\rm \dot{{\it{M}}}$. However, the uncertainty in the determination of $\rm
\dot{{\it{M}}}$ is rather large.

While the quantity \mbox{$\rm log\,[EW(CO)/(EW(Na)+EW(Ca))]$} of RDF is  a good discriminator between giants and dwarfs, it cannot be used to separate supergiants and LPVs from red giants. However, the dispersion 
is much narrower for  M giants than for variable AGB stars.

We have used the metallicity index $\rm [Fe/H]^{index}_{RSFD}$  of
RSFD, determined from the line strengths of CO, NaI and CaI,  to estimate the metallicity of the stellar 
population. Our mean $\rm [Fe/H]^{index}_{RSFD}$ is $\rm -0.25\,dex$ with a dispersion of $\rm \pm 0.40\,dex$ which
 is in agreement with the values obtained by RSFD and supports the argument of RSFD that there is no metallicity
gradient in the Bulge. Our mean $\rm [Fe/H]^{index}_{RSFD}$ is consistent with previous abundance studies of the  galactic Bulge  and with the multiphase evolution 
models of Molla et al. (\cite{Molla2000}) which assume a dissipative collapse of a protogalaxy to form  the Bulge and the galactic disk. Our results confirm that  even if the metallicity index 
of RSFD is only valid  for  M giants, it can be used to estimate an
average metallicity valid over a whole pulsation cycle.  This is an
important result for future stellar population studies using
multi-fiber spectrographs (such as NIRMOS, GIRAFFE,etc.). However,
there is  a pressing need for a grid of realistic models of M giants
with different [Fe/H] to facilitate a quantitative determination of
metallicities, as well as further modeling of LPV/AGB stars.

\noindent

\medskip

{\bf{Acknowledgements:}}

\medskip

We want to thank I.~S.~Glass, J.~Blommaert, J.~ van Loon for their fruitful comments and discussions.
MS is supported by the Fonds zur F\"orderung der wissenschaftlichen
Forschung (FWF), Austria, under the project number J1971-PHY.

This research is supported by the project 1910-1 of Indo-French Center for the
Promotion of Advanced Research (CEFIPRA).

The DENIS project is supported, in France by the Institut National des
Sciences de l'Univers, the Education Ministry and the Centre National de la
Recherche Scientifique, in Germany by the State of Baden-W\"urtemberg, in
Spain by the DGICYT, in Italy by the Consiglio Nazionale delle Ricerche, in
Austria by the Fonds zur F\"orderung der wissenschaftlichen Forschung und
Bundesministerium f\"ur Wissenschaft und Forschung.

This research has made use of the Simbad database, operated at CDS, Strasbourg, France.

{}

\appendix  

\section{Tables}

\begin{table*}[H]
\begin{turn}{90}
\begin{tabular}{ccccccccccccccc}
\multicolumn{15}{c}{Table A1: Coordinates (J2000), magnitudes (DENIS,ISOGAL),$\rm A_{V}$ values,
equivalent widths (in \AA) of $\rm ^{12}CO(2,0)$,NaI,CaI,$\rm
H_{2}O$,$\rm M_{bol}$,$\rm log\,\dot{{\it{M}}}$}\\
\hline
name& right ascension & declination &J&$\rm K_{S}$&[7]&[15]&$\rm
A_{V}$&$\rm 12^{CO}(2,0)$&NaI&CaI&$\rm H_{2}O$&$\rm M_{bol}$&$\rm log\,\dot{{\it{M}}}$& type\\
\hline
\hline
A3 &  17  44  23.8 & -29  8  55.3&14.57 & 10.03 &  8.74 &  6.43 & 19.60 & 19.48 &  4.93& 1.87& -35.78 & -3.25 & -6.69 & RGB \\
A4 &  17  44  30.4 & -29  7  15.5&15.14 &  9.23 &  4.77 &  3.93 & 19.10 &  9.83 &  2.44&-0.50& -261.11 & -4.32 & -5.46 & AGB \\
A5 &  17  44  31.7 & -29  6  20.6&13.80 &  8.50 &  5.08 &  2.92 & 19.70 & 21.04 &  2.26&-0.74& -666.39 & -4.83 & -5.36 & AGB \\
A6 &  17  44  31.8 & -29  17  10.5& -- &  9.58 &  5.58 &  4.04 & 24.50 & 16.88 &  3.94& 0.23& -157.16 & -4.52 & -5.50 & AGB \\
A7 &  17  44  34.4 & -29  10  38.5&13.41 &  8.04 &  4.05 &  1.76 & 21.60 & 21.93 &  4.10&-0.11& -386.72 & -5.71 & -5.17 & OH/IR \\
A8 &  17  44  35.7 & -29  13   5.4&15.20 &  8.68 &  4.83 &  3.15 & 25.10 & 13.75 &  2.17&-0.51& -259.69 & -5.04 & -5.52 & AGB \\
A9 &  17  44  36.4 & -29  9  25.2& -- &  9.67 &  4.20 &  2.62 & 22.30 & 20.74 &  4.88& 2.41& -23.84 & -5.08 & -4.99 & AGB \\
A10 &  17  44  39.7 & -29  16  45.9& -- & 10.99 &  4.50 &  2.35 & 25.00 & 19.09 &  5.25& 3.61& -13.92 & -4.45 & -4.75 & OH/IR \\
A12 &  17  44  44.5 & -29  5  38.3& -- & 11.18 &  3.57 &  2.05 & 20.50 & 12.99 &  1.29&-2.11& -181.29 & -6.03 & -4.65 & OH/IR \\
A13 &  17  44  44.6 & -29  7  33.6&15.84 &  9.26 &  5.90 &  4.35 & 19.40 &  8.38 &  2.30&-1.37& -434.36 & -3.74 & -5.64 & AGB \\
A14 &  17  44  48.0 & -29  6  49.8&15.24 &  9.34 &  5.03 &  2.85 & 20.50 & 20.94 &  4.57& 1.94& -120.23 & -4.36 & -5.10 & OH/IR \\
A16 &  17  44  48.6 & -29  0  13.2&15.62 &  9.89 &  8.16 & 99.99 & 21.50 & 21.04 &  6.26& 1.35& -54.97 & -3.14 & -- & RGB \\
A17 &  17  44  49.1 & -29  19  54.2&11.08 &  6.17 &  3.41 &  1.31 & 25.40 & 20.58 &  0.78&-0.24& -52.70 & -8.06 & -5.89 & supergiant \\
A18 &  17  44  49.5 & -29  3  15.9&15.08 &  9.42 &  6.92 &  5.21 & 21.60 & 29.17 &  6.16& 1.31& -336.94 & -3.76 & -6.20 & LPV \\
A19 &  17  44  50.3 & -29  19  21.5&14.58 &  9.01 &  6.92 &  5.34 & 24.80 & 19.80 &  5.21& 1.41& -184.69 & -4.65 & -7.06 & AGB \\
A20 &  17  44  52.7 & -29  14  11.1& -- &  9.61 &  7.62 &  5.13 & 24.50 & 20.76 &  5.47& 3.40& -27.37 & -4.01 & -6.13 & AGB \\
A21 &  17  44  52.9 & -29  7   6.7& -- &  9.30 &  5.38 &  3.73 & 28.50 & 18.84 &  4.21& 1.63& -144.06 & -4.98 & -5.60 & AGB \\
A22 &  17  44  53.1 & -28  59  46.5& -- & 10.78 &  8.76 &  6.82 & 19.80 & 18.94 &  5.42& 2.91& -19.26 & -2.48 & -6.32 & RGB \\
A23 &  17  44  54.2 & -29  13  44.9&14.84 &  8.58 &  4.89 &  3.23 & 24.50 & 22.34 &  3.70& --& -543.70 & -5.08 & -5.59 & OH/IR \\
A24 &  17  44  54.7 & -29  3  59.0& -- & 10.70 &  5.13 &  3.29 & 22.10 &  7.52 &  0.19&-0.95& -277.07 & -3.80 & -4.92 & AGB \\
A25 &  17  44  55.5 & -29  1  43.4&14.34 &  8.53 &  5.48 &  3.91 & 19.80 & 20.48 &  3.54&-0.76& -623.25 & -4.52 & -5.82 & AGB \\
A26 &  17  44  56.8 & -29  10  20.3&15.77 &  9.05 &  6.11 &  4.41 & 25.30 & 19.99 &  3.76&-1.08& -526.70 & -4.36 & -6.04 & LPV \\
A27 &  17  44  57.0 & -29  5  57.3&15.25 & 10.11 &  4.48 &  1.65 & 25.30 & 19.20 &  4.29& 3.11& -8.01 & -5.07 & -4.78 & AGB \\
A28 &  17  44  57.1 & -29  15  24.7& -- &  8.84 &  5.30 &  3.51 & 20.30 & 17.13 &  1.49&-1.44& -473.84 & -4.67 & -5.47 & AGB \\
A29 &  17  44  57.8 & -29  20  42.5& -- & 10.65 &  4.66 &  2.45 & 23.40 &  8.98 & -0.05&-1.75& 129.52 & -4.32 & -4.80 & OH/IR \\
A30 &  17  44  57.9 & -29  4   6.5&14.64 &  9.15 &  7.14 &  5.46 & 22.30 & 23.18 &  5.60& 2.32& -112.56 & -4.13 & -6.80 & AGB \\
A31 &  17  44  58.9 & -29  9  10.8& -- &  8.19 &  4.73 &  2.99 & 24.40 & 17.21 &  1.34&-0.77& -332.12 & -5.63 & -5.66 & LPV \\
A33 &  17  44  59.5 & -29  16   4.6& -- & 10.78 &  4.08 &  0.50 & 20.30 & -2.28 &  0.42&-1.24&  1.78 & 2504.72 & -- & YSO \\
A34 &  17  44  59.6 & -29  11  15.4&13.43 &  7.95 &  5.50 &  3.75 & 23.80 & 22.46 &  3.96&-1.06& -505.14 & -5.64 & -6.33 & LPV \\
A35 &  17  45   1.0 & -29  1  14.9& -- &  9.43 &  5.90 &  4.27 & 19.80 & 12.08 &  1.27& --& -430.36 & -4.18 & -5.53 & LPV \\
A36 &  17  45   1.7 & -29  2  50.0&13.83 &  8.23 &  4.91 &  3.26 & 18.50 & 20.32 &  2.64&-0.16& -694.69 & -4.83 & -5.58 & OH/IR \\
A38 &  17  45   4.8 & -29  1  24.9&13.94 &  8.52 &  5.52 &  3.86 & 23.10 &  7.78 &  0.65& 0.69& -498.84 & -5.06 & -5.93 & OH/IR \\
A39 &  17  45   4.8 & -29  5  48.5&13.69 &  7.85 &  4.86 &  3.31 & 23.10 & 22.72 &  3.34& 0.20& -347.86 & -5.56 & -6.01 & LPV \\
A40 &  17  45   4.9 & -29  11  46.8&15.24 &  9.01 &  4.27 &  1.97 & 23.50 & -1.61 &  0.58&-0.18& 77.81 & -5.14 & -- & YSO \\
A41 &  17  45   5.3 & -29  1  35.8&13.89 &  8.46 &  5.47 &  3.75 & 20.60 & 17.11 &  2.32& 0.85& -536.87 & -4.77 & -5.79 & LPV \\
A43 &  17  45   7.0 & -29  3  34.2&13.94 &  8.56 &  4.86 &  2.73 & 24.80 & 18.78 & -0.14& 0.36& -993.36 & -5.46 & -5.39 & OH/IR \\
A45 &  17  45   9.8 & -29  5  17.8& -- & 10.87 &  7.82 &  6.26 & 22.80 & 17.07 &  3.82& 1.12& -86.09 & -2.82 & -5.95 & RGB \\
A47 &  17  45  10.8 & -29  7  12.2&15.59 &  9.61 &  7.69 &  4.82 & 22.80 & 18.75 &  4.46& 2.02& -52.85 & -3.61 & -5.83 & AGB \\
A49 &  17  45  12.2 & -29  4  25.2& -- &  8.72 &  4.85 &  3.28 & 22.30 & 21.14 &  3.20&-0.38& -643.49 & -5.19 & -5.48 & LPV \\
A50 &  17  45  13.5 & -29  5  26.7&13.17 &  7.62 &  4.56 &  2.98 & 23.90 & 16.59 &  1.52&-0.06& -629.52 & -6.03 & -5.98 & LPV \\
A51 &  17  45  14.0 & -29  15  27.4&15.93 & 10.31 &  3.47 &  1.59 & 24.00 & 15.75 &  3.32& 2.54& -54.14 & 2504.72 & -4.73 & OH/IR \\
A52 &  17  45  14.3 & -29  7  20.8& -- & 10.60 &  4.14 &  2.40 & 22.90 &  8.78 &  1.01& 0.12& -318.86 & -4.94 & -4.79 & OH/IR \\
\end{tabular}
\end{turn}
\end{table*}

\begin{table*}[hh!]
\begin{turn}{90}
\begin{tabular}{ccccccccccccccc}
\multicolumn{15}{c}{Table A1 (cont.)}\\
\hline
A54 &  17  45  16.7 & -29  7  34.3& -- & 10.80 &  8.56 &  6.77 & 22.90 & 15.51 &  2.97& 4.24& -54.35 & -2.75 & -6.44 & RGB \\
A55 &  17  45  17.5 & -29  10  18.9&14.18 &  9.19 &  7.19 &  5.59 & 23.50 & 22.39 &  6.42& 2.94& -173.11 & -4.56 & -7.04 & AGB \\
A56 &  17  45  18.0 & -29  8  42.9& -- &  9.39 &  6.79 &  5.22 & 22.80 & 21.33 &  5.60& 2.88& -89.14 & -4.19 & -6.30 & AGB \\
A57 &  17  45  18.8 & -29  5   5.4& -- & 10.05 &  7.69 &  6.08 & 25.90 & 20.37 &  4.82& 3.82& -27.65 & -3.71 & -6.73 & AGB \\
A58 &  17  45  19.4 & -29  14   5.9& -- &  8.57 &  4.08 &  2.02 & 21.00 & 18.28 &  2.32&-0.75& -806.41 & -5.25 & -5.09 & OH/IR \\
A59 &  17  45  20.5 & -29  7  19.5& -- & 10.54 &  6.29 &  4.29 & 23.90 & 20.85 &  3.16&-0.45& -252.72 & -3.37 & -5.23 & LPV \\
A60 &  17  45  20.9 & -29  13  42.4& -- &  8.80 &  6.68 &  5.18 & 23.90 & 19.69 &  5.00& 1.04& -276.51 & -4.80 & -7.05 & AGB \\
A61 &  17  45  21.5 & -29  6  36.5& -- & 10.13 &  6.59 &  4.99 & 24.00 & 18.46 &  2.97&-0.32& -573.02 & -3.77 & -5.68 & LPV \\
A62 &  17  45  21.9 & -29  13  44.3&13.75 &  8.62 &  6.61 &  4.87 & 21.90 & 23.40 &  5.88& 4.14& -129.72 & -4.77 & -6.69 & AGB \\
A64 &  17  45  26.4 & -29  8   4.2& -- &  8.67 &  4.63 &  2.69 & 23.60 & 19.96 &  5.88& 1.98& -416.59 & -5.17 & -5.31 & OH/IR \\
A66 &  17  45  27.5 & -29  4  39.9&16.14 &  9.97 &  7.91 &  6.06 & 24.50 & 22.98 &  5.62& 3.06& -88.49 & -3.39 & -6.70 & RGB \\
A67 &  17  45  28.4 & -29  4  22.9&15.60 &  9.27 &  6.88 &  3.90 & 25.00 & 22.45 &  5.22& 2.80& -171.28 & -4.21 & -5.59 & AGB \\
A68 &  17  45  28.5 & -29  18   5.0& -- & 10.30 &  8.35 & 99.99 & 19.30 & 19.83 &  4.46& 4.13& -16.33 & 2504.72 & -3.70 & AGB \\
A69 &  17  45  29.5 & -29  9  16.6& -- & 10.35 &  4.87 &  3.06 & 22.50 &  7.59 &  0.23&-0.91& 236.30 & -4.10 & -4.95 & LPV \\
A70 &  17  45  30.5 & -29  15  50.9&14.76 &  9.54 &  7.16 &  5.65 & 24.50 & 21.55 &  6.38& 2.22& -127.51 & -4.29 & -6.72 & AGB \\
A71 &  17  45  34.4 & -29  12  54.2& -- &  9.90 &  4.30 &  2.71 & 24.50 & 14.15 &  1.43&-0.63& -620.11 & -5.03 & -4.99 & OH/IR \\
A72 &  17  45  36.0 & -29  4  38.1& -- & 10.36 &  7.42 &  4.94 & 24.80 & 20.18 &  5.78& 2.50& -129.24 & -3.33 & -5.56 & AGB \\
A74 &  17  45  38.2 & -29  17  20.1&13.69 &  9.05 &  6.79 &  4.80 & 20.40 & 20.94 &  5.41& 2.68& -95.58 & -4.38 & -6.10 & AGB \\
A75 &  17  45  40.0 & -29  16   2.8&14.22 &  8.11 &  4.62 &  2.82 & 22.50 & 11.02 &  2.48& 0.32& -341.28 & -5.30 & -5.55 & AGB \\
A76 &  17  45  40.7 & -29  4  27.7&15.27 &  9.59 &  7.76 &  6.17 & 24.10 & 20.53 &  5.19& 2.98& -80.12 & -3.88 & -7.42 & AGB \\
A77 &  17  45  40.7 & -29  14  54.7& -- & 10.12 &  4.31 &  2.40 & 23.20 & 14.51 &  2.26& 0.94& -52.42 & -4.62 & -4.88 & AGB \\
A78 &  17  45  41.4 & -29  13  43.6&15.99 & 10.08 &  8.36 &  6.93 & 20.50 & 23.56 &  5.46& 3.39& -88.19 & -2.76 & -7.50 & RGB \\
A79 &  17  45  43.4 & -29  13  29.1&15.16 & 10.42 &  9.18 &  7.26 & 21.30 & 16.47 &  2.48& 2.19& -78.17 & -3.05 & -7.58 & RGB \\
B7 &  17  43  48.6 & -29  27  18.8& -- & 11.54 &  8.16 & 99.99 & 27.20 & 11.25 &  0.95& 1.69& -70.05 & 2504.72 & -3.70 & AGB \\
B13 &  17  44   4.7 & -29  25  18.7& -- & 10.51 &  6.58 &  2.91 & 28.20 & 17.49 &  4.07& 4.14& -48.80 & -3.75 & -4.96 & AGB \\
B14 &  17  44   7.5 & -29  27  37.6&13.94 &  6.64 &  2.40 &  0.59 & 28.60 & 23.84 &  4.72& 3.82& -36.04 & -7.41 & -5.40 & supergiant \\
B15 &  17  44   8.1 & -29  32  22.5&13.99 &  6.86 &  2.67 &  0.85 & 28.50 & 23.36 &  4.88& 3.35&  9.05 & -7.19 & -5.41 & supergiant \\
B17 &  17  44  11.5 & -29  31  31.1&15.16 &  9.30 &  5.41 &  3.20 & 28.50 & 19.93 &  5.49& 2.68& -115.19 & -5.10 & -5.38 & AGB \\
B18 &  17  44  12.7 & -29  31  24.6&15.91 &  9.30 &  6.98 &  5.25 & 29.40 & 23.61 &  4.07& 5.65& -229.09 & -4.64 & -6.93 & AGB \\
B19 &  17  44  12.8 & -29  26  55.5&13.22 &  6.72 &  3.02 &  0.98 & 29.00 & 27.10 &  3.79& 3.68& -45.72 & -7.43 & -5.54 & supergiant \\
B20 &  17  44  16.5 & -29  26   7.7& -- & 10.56 &  8.37 &  6.49 & 29.00 & 21.52 &  3.85& 4.81& -57.55 & -3.50 & -6.86 & RGB \\
B22 &  17  44  17.6 & -29  25  58.9& -- & 10.99 & 99.99 &  6.34 & 28.80 & 17.41 &  3.79& 4.86& -3.47 & 2504.72 & -6.21 & AGB \\
B23 &  17  44  17.8 & -29  27  13.0&14.07 &  7.80 &  3.99 &  1.28 & 29.00 & -1.21 & -1.06&-0.15& -0.34 & -6.53 & -- & YSO \\
B27 &  17  44  21.7 & -29  27  36.2& -- & 10.35 &  6.27 &  2.61 & 29.60 &  0.92 &  1.26& 0.52& -2.38 & -4.06 & -- & YSO \\
B28 &  17  44  22.8 & -29  34  55.0& -- & 10.46 &  7.36 &  5.47 & 27.00 & 24.17 &  5.40& 5.22& -97.87 & -3.49 & -5.87 & RGB \\
B31 &  17  44  27.4 & -29  38   5.8&15.80 &  8.70 &  4.47 &  2.34 & 27.30 & 13.17 &  1.78& 2.24& -317.56 & -5.31 & -5.26 & AGB \\
B32 &  17  44  27.5 & -29  28  52.8& -- & 11.27 &  8.14 &  6.11 & 34.90 & 23.83 &  5.62& 4.40& 37.23 & -3.33 & -6.11 & RGB \\
B35 &  17  44  29.9 & -29  28  57.3& -- & 10.38 &  7.84 &  5.47 & 32.30 & -0.75 &  0.09& 0.76& -2.22 & -3.94 & -- & YSO \\
B37 &  17  44  33.8 & -29  23  57.8&15.80 &  9.75 &  6.73 &  4.64 & 29.80 & -0.73 &  0.69& 1.10& 14.83 & -4.62 & -- & YSO \\
B40 &  17  44  44.5 & -29  31  36.8& -- & 10.41 &  8.00 &  5.23 & 27.00 & 22.41 &  5.60& 4.47&  0.04 & -3.41 & -5.76 & RGB \\
C3 &  17  42  58.8 & -29  49  59.9&15.57 &  9.60 &  5.91 &  3.59 & 18.40 & 25.42 &  5.21& 3.22& -279.57 & -3.63 & -5.19 & AGB \\
C6 &  17  43   3.4 & -29  38   1.5& -- &  9.36 &  4.02 &  1.58 & 22.00 & 13.47 &  0.84& 0.23& -541.95 & -5.11 & -4.85 & AGB \\
C10 &  17  43  11.2 & -29  51  29.8&13.31 &  7.14 &  4.52 &  1.49 & 20.30 & 18.34 &  0.49& 0.94& -744.63 & -5.92 & -5.35 & AGB \\
C14 &  17  43  14.7 & -29  37  49.9&15.32 &  9.26 &  6.62 &  4.49 & 22.80 & 19.37 &  2.59& 1.37& -529.80 & -4.00 & -5.84 & AGB \\
\hline
\end{tabular}
\end{turn}
\end{table*}

\begin{table*}[hh!]
\begin{turn}{90}

\begin{tabular}{ccccccccccccccc}
\multicolumn{15}{c}{Table A1 (cont.):}\\ 
\hline
C19 &  17  43  19.3 & -29  50  28.9&13.92 & 10.98 &  6.53 &  3.80 & 22.50 & -3.64 & -0.58& 0.12& -50.59 & -4.24 & -- & YSO \\
C23 &  17  43  24.6 & -29  53  12.4&14.33 &  8.56 &  6.89 &  4.82 & 23.60 & 25.90 &  5.16& 4.89& -237.00 & -4.79 & -6.84 & AGB \\
C25 &  17  43  25.0 & -29  41  23.0& -- &  9.32 &  5.37 &  3.19 & 25.00 & 16.35 &  3.10& 3.17& -223.55 & -4.57 & -5.29 & AGB \\
C45 &  17  43  44.2 & -29  38  26.1&15.61 &  9.75 &  7.50 &  4.33 & 23.80 & 25.52 &  5.19& 3.28& -279.99 & -3.72 & -5.53 & AGB \\
D5 &  17  42  27.3 & -29  54  17.8&13.28 &  8.04 &  6.50 &  4.44 & 20.30 & 13.49 &  3.05& 2.79& -364.41 & -5.02 & -6.75 & AGB \\
D6 &  17  42  28.0 & -29  56  14.6&13.90 & 10.55 &  3.87 &  0.13 & 24.30 & -1.00 &  0.03& 1.44& -57.42 & 2504.72 & -- & YSO \\
D7 &  17  42  29.9 & -30  1  15.9& -- & 10.97 &  3.36 & -0.06 & 22.40 &  0.87 &  1.20& 1.76& -82.31 & 2504.72 & -- & YSO \\
D11 &  17  42  44.4 & -29  58  36.1&14.03 &  7.94 &  4.60 &  2.39 & 26.00 & 12.16 &  1.85& 0.66& -544.74 & -5.89 & -5.54 & AGB \\
D13 &  17  42  47.7 & -29  56  25.5& -- &  9.56 &  6.74 &  4.40 & 27.90 & 22.36 &  5.19& 4.34& -115.25 & -4.40 & -5.80 & AGB \\
D22 &  17  43  29.8 & -30  1  27.3&12.17 &  7.89 &  5.73 &  3.59 & 19.00 & 21.64 &  1.69& 2.07& -1020.49 & -5.49 & -5.99 & AGB \\
E4 &  17  43  13.0 & -29  21   1.4& -- &  7.22 &  5.32 &  4.33 & 23.30 & 20.36 &  4.41& 2.70& -221.63 & -6.38 & -8.66 & AGB \\
E5 &  17  43  15.5 & -29  24  58.7&11.51 &  6.58 &  4.87 &  4.07 & 25.70 & 23.21 &  5.36& 3.04& -109.75 & -7.64 & -11.67 & AGB \\
E13 &  17  43  28.2 & -29  17  41.7&11.33 &  6.93 &  5.21 &  4.67 & 22.70 & 20.18 &  4.05& 2.13& -193.37 & -7.05 & -12.24 & AGB \\
E20 &  17  43  39.2 & -29  22  44.5& -- & 10.61 &  6.61 &  4.24 & 25.50 & 19.69 &  3.39& 1.65& -162.79 & -3.31 & -5.22 & AGB \\
E51 &  17  44  35.0 & -29  4  35.5& -- &  9.09 &  1.30 & -0.39 & 20.10 & 10.11 &  1.60& 0.61& -297.85 & -8.22 & -4.61 & OH/IR \\
F13 &  17  42  28.0 & -29  37  46.4&15.21 &  9.87 &  5.23 &  3.01 & 17.00 & 20.61 &  2.00& 1.12& -763.02 & -3.87 & -4.96 & AGB \\
F14 &  17  42  39.5 & -29  43  27.3&15.36 &  9.91 &  4.95 &  3.13 & 21.50 & 18.15 &  2.57& 3.08& -186.75 & -4.24 & -5.04 & AGB \\
F23 &  17  40  36.3 & -29  49  13.7&11.51 &  8.18 &  5.71 &  2.21 &  9.80 & 10.91 &  0.91& 2.72& -597.37 & -4.49 & -5.06 & AGB \\
G35 &  17  43  16.4 & -30  13  10.7&14.14 & 10.98 &  4.95 &  2.91 & 16.30 & 19.06 &  1.77& 0.82& -340.69 & -3.94 & -4.75 & AGB \\
G46 &  17  44   7.5 & -30  7  41.3&14.09 &  8.63 &  5.32 &  3.46 & 20.60 & 18.51 &  3.51& 1.45& -506.89 & -4.69 & -5.55 & AGB \\
H01 &  17  30  53.2 & -33  40  21.7&15.72 & 10.53 &  5.53 &  2.92 & 10.80 & 13.11 &  2.50& 1.43& -492.72 & -3.36 & -4.77 & AGB \\
\hline
\end{tabular}
\end{turn}
\end{table*}



\section{Near-IR spectra}
The spectra were dereddened using the extinction curve by Mathis et al. (\cite{Mathis90}) and the DENIS extinction map
(Schultheis et al. \cite{Schultheis99}), see Section 3.1. The flux densities
per unit wavelength are normalized at 2.28\,$\mu$m. The spectra are electronically available at CDS. Strong telluric
bands are seen between 1.8 and 1.9\,$\rm \mu$m. 

\vfill \eject

\begin{figure*}[H!]
\caption{Young stellar objects}
{\epsfig{file=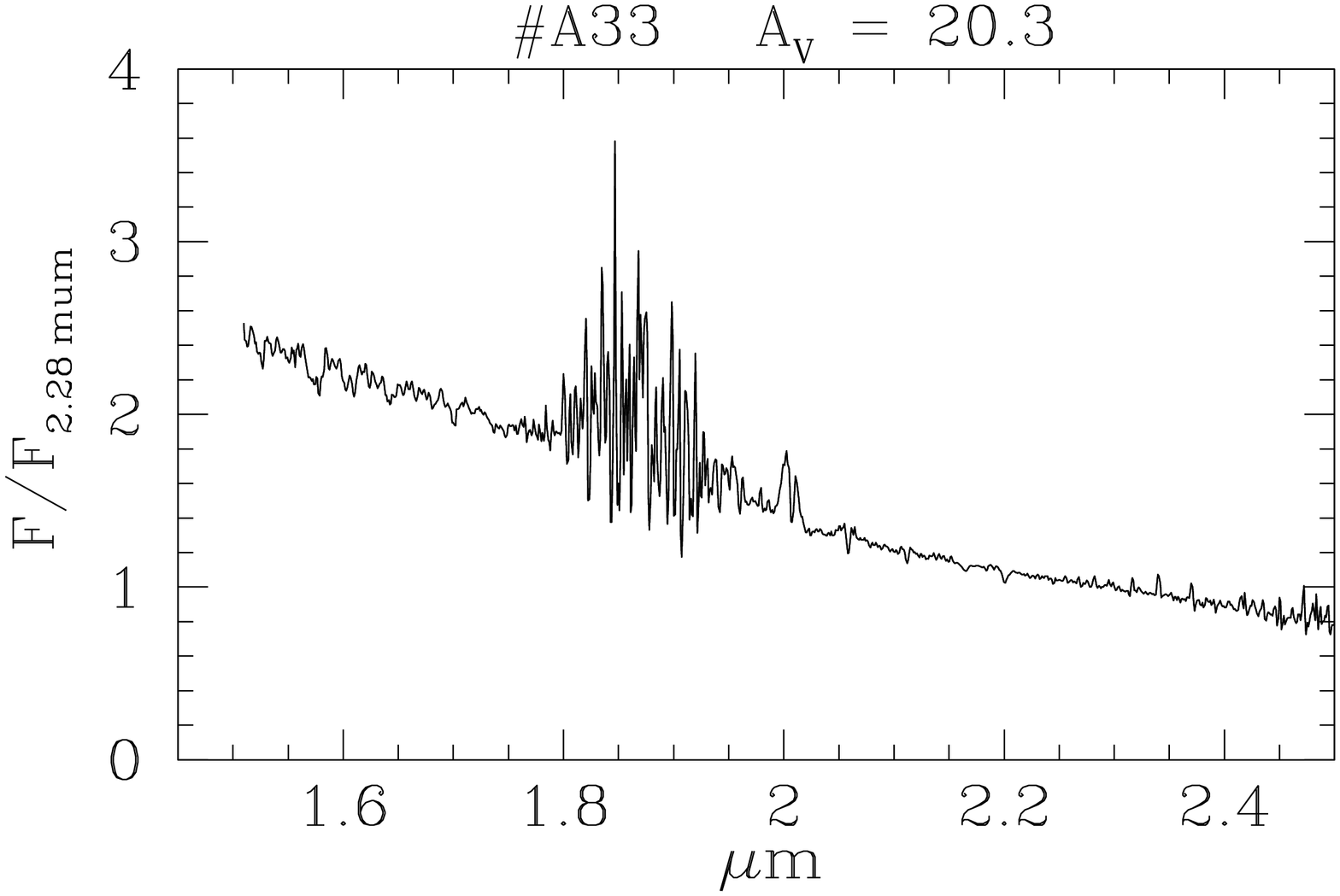,width=4.3cm,angle=270} \epsfig{file=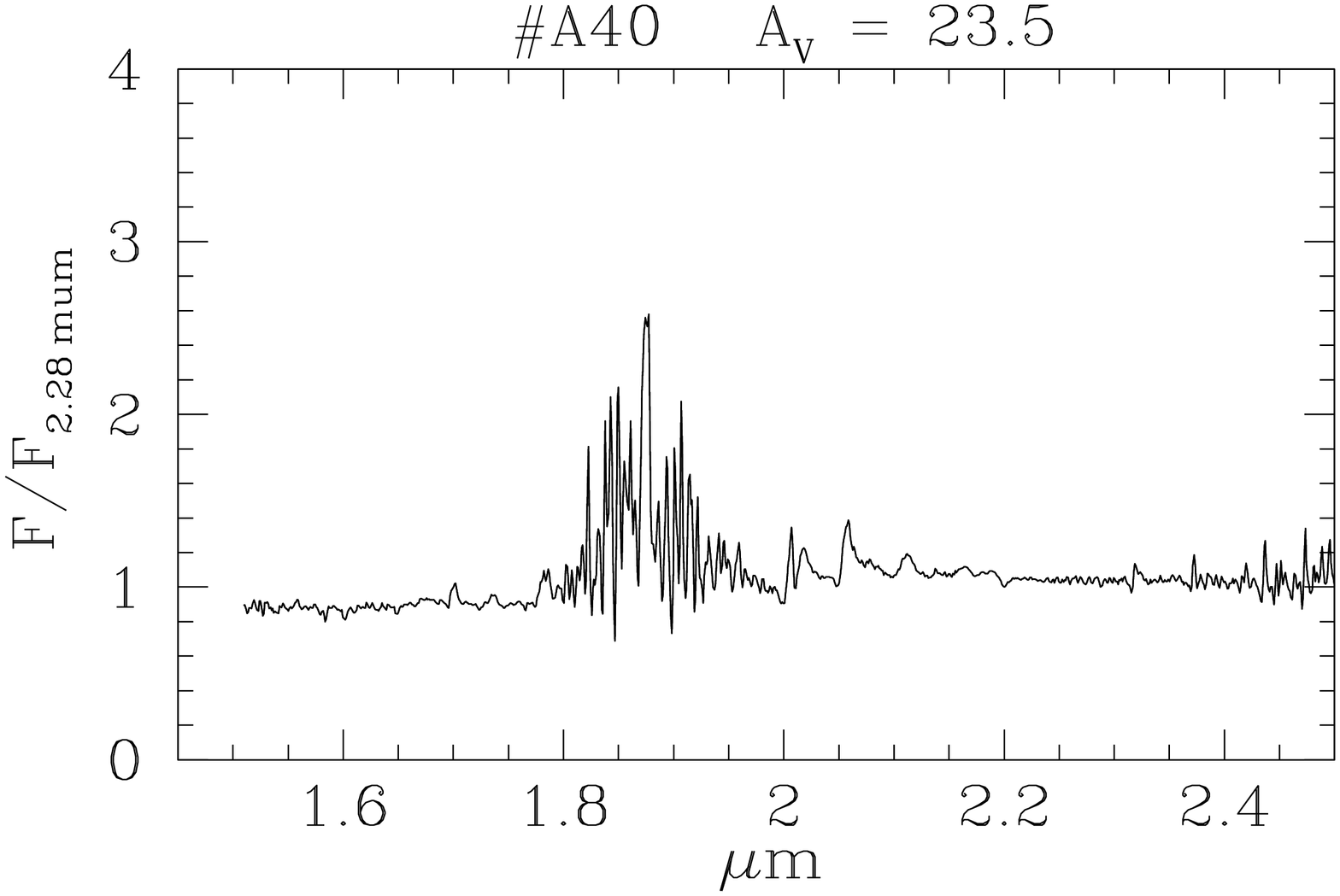,width=4.3cm,angle=270}   \epsfig{file=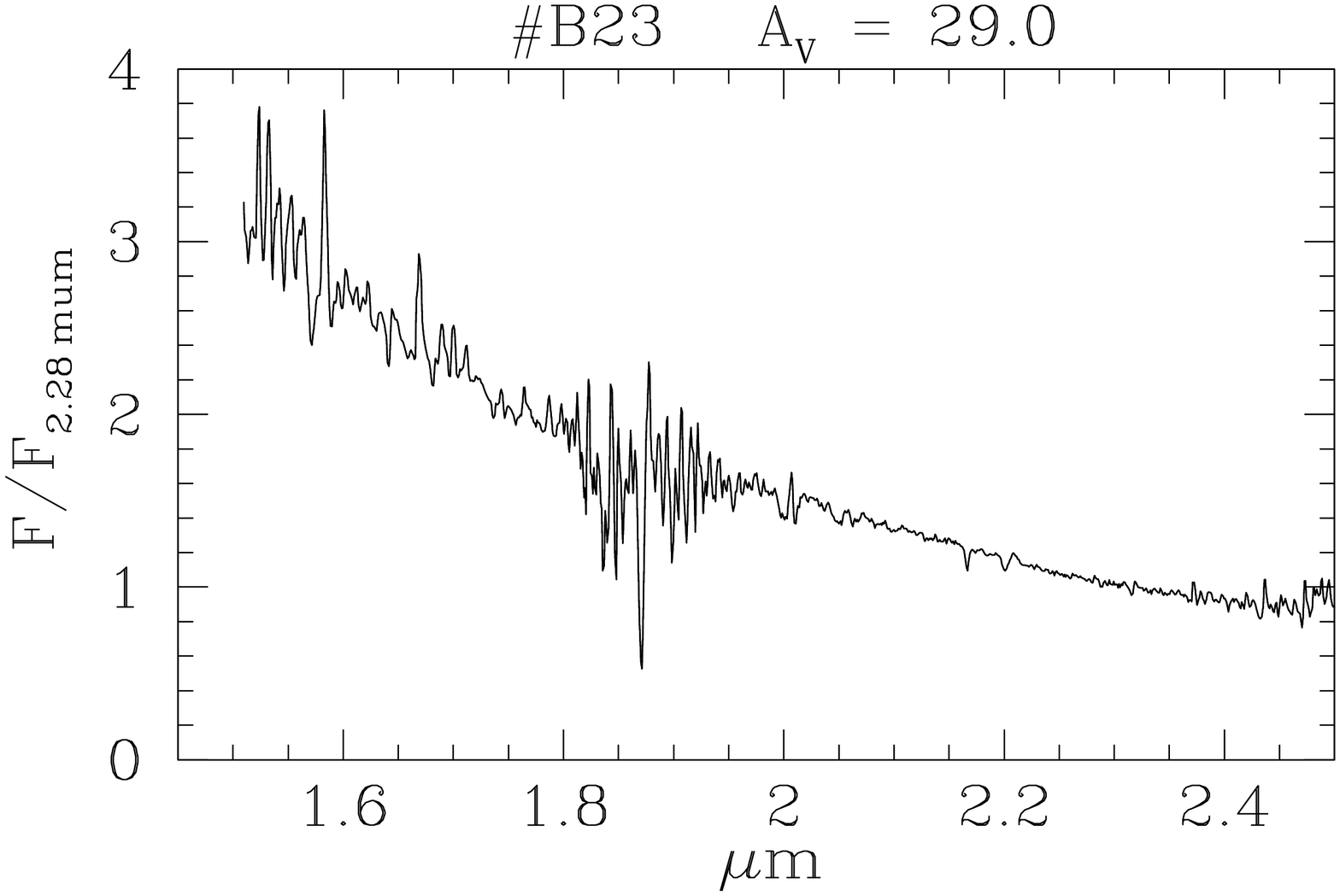,width=4.3cm,angle=270}

\epsfig{file=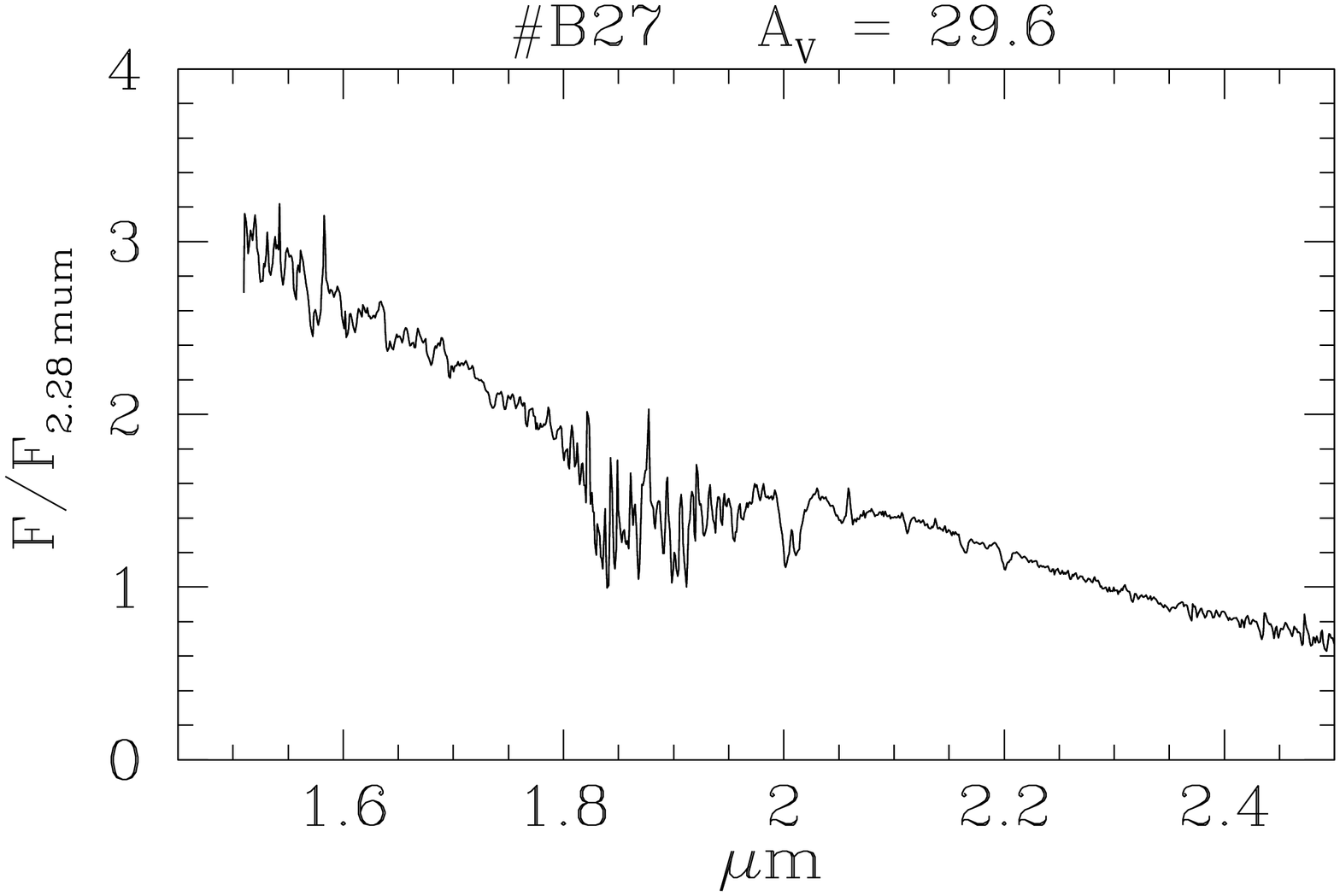,width=4.3cm,angle=270} \epsfig{file=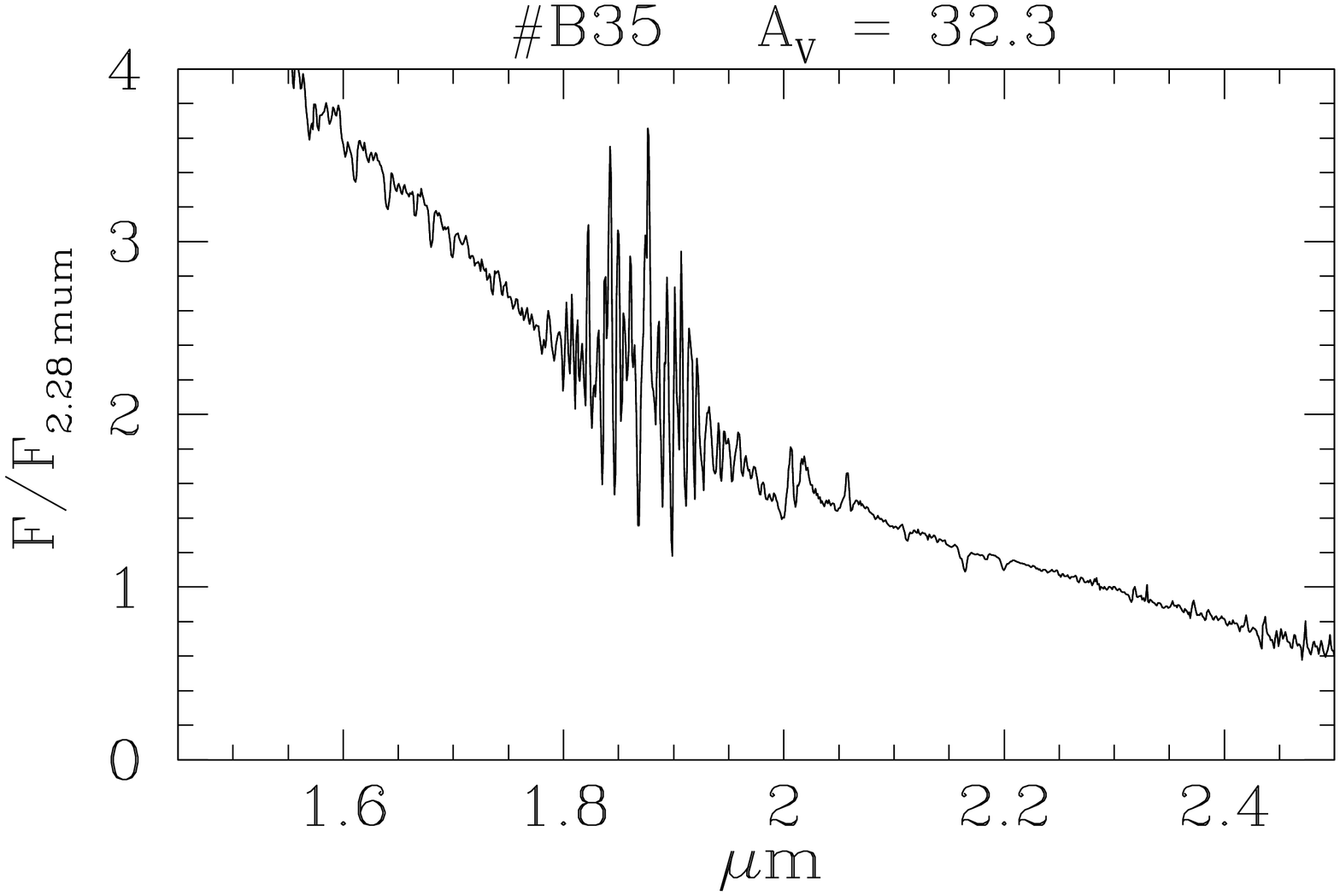,width=4.3cm,angle=270}   \epsfig{file=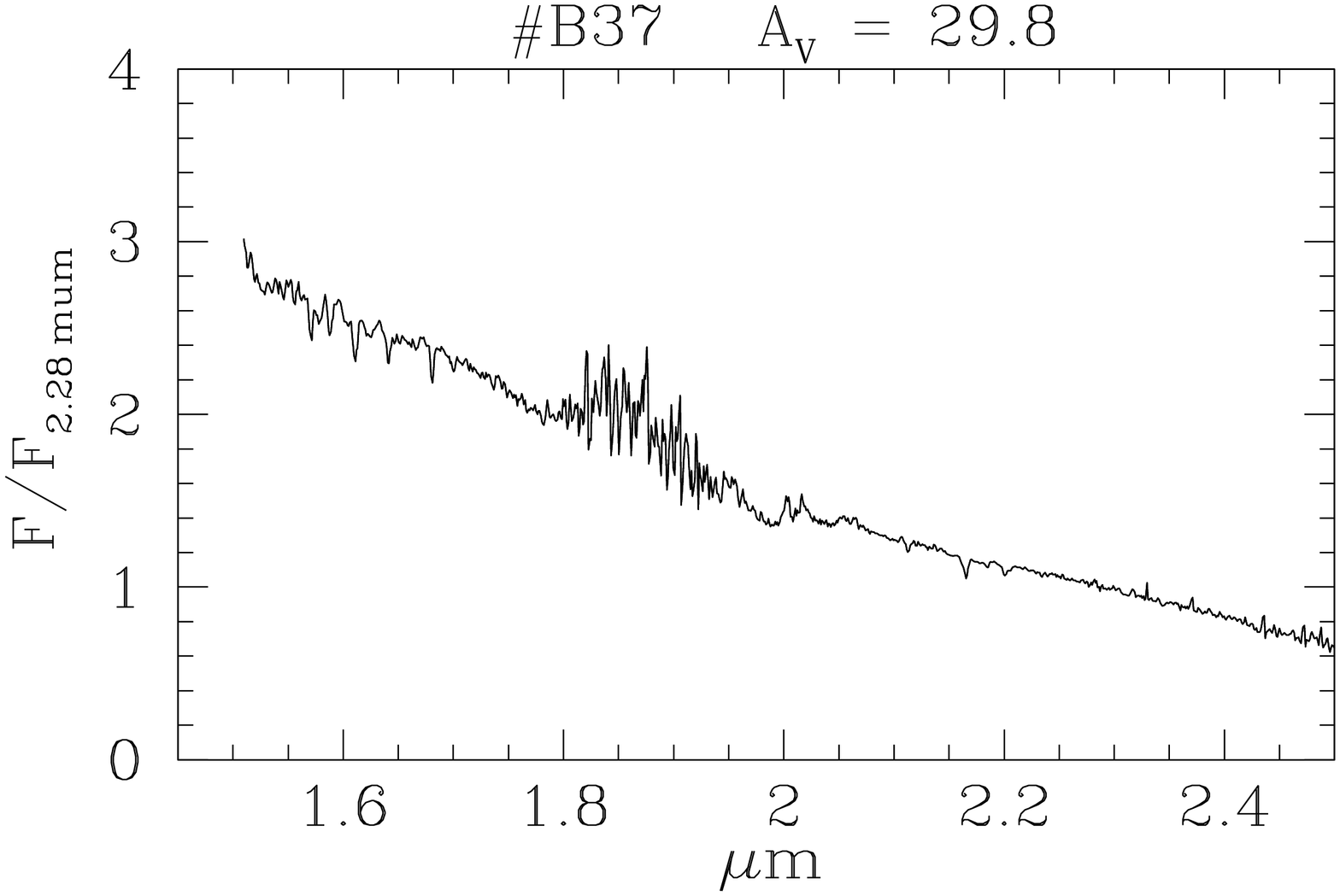,width=4.3cm,angle=270} 

\epsfig{file=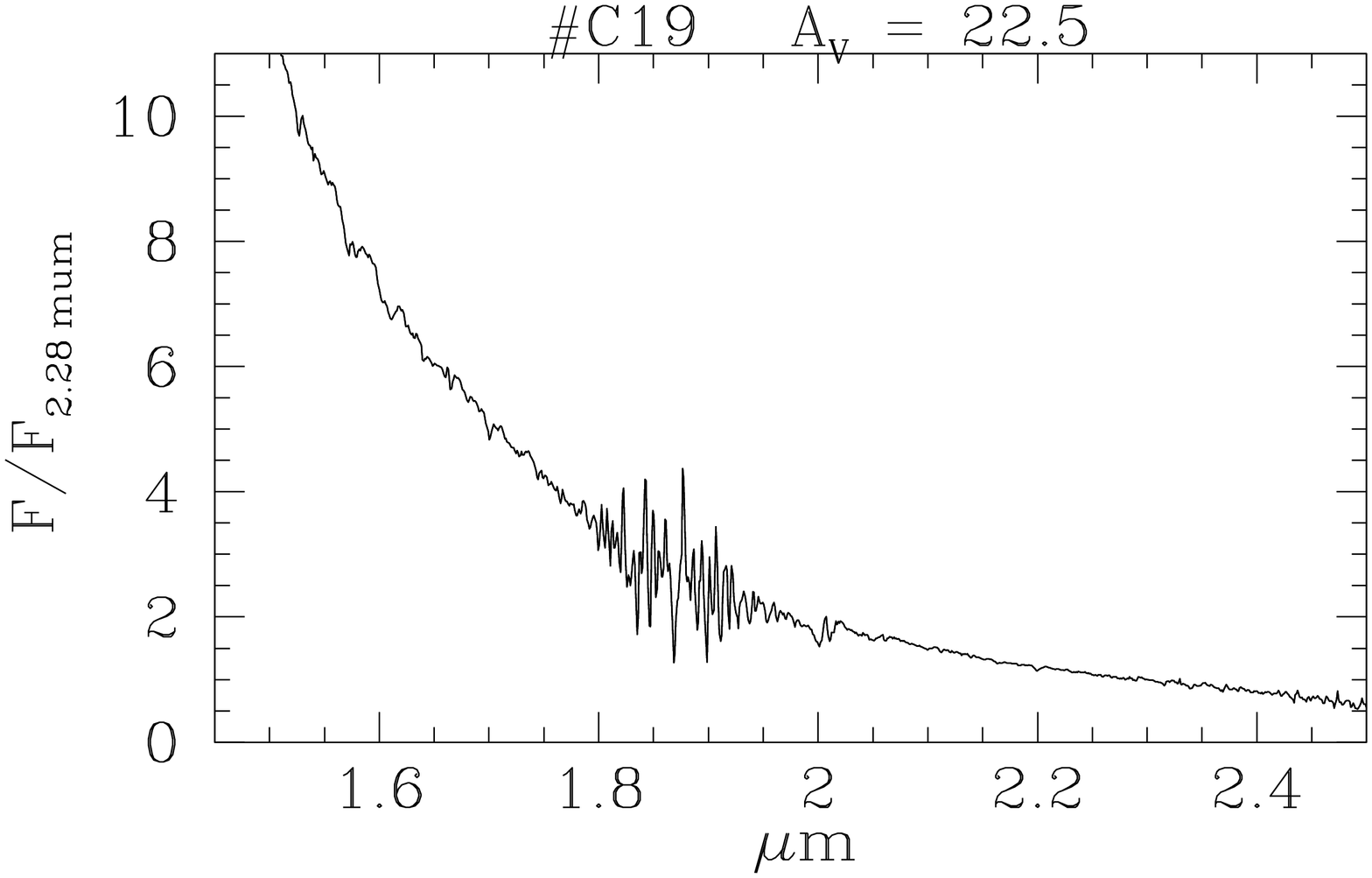,width=4.3cm,angle=270} \epsfig{file=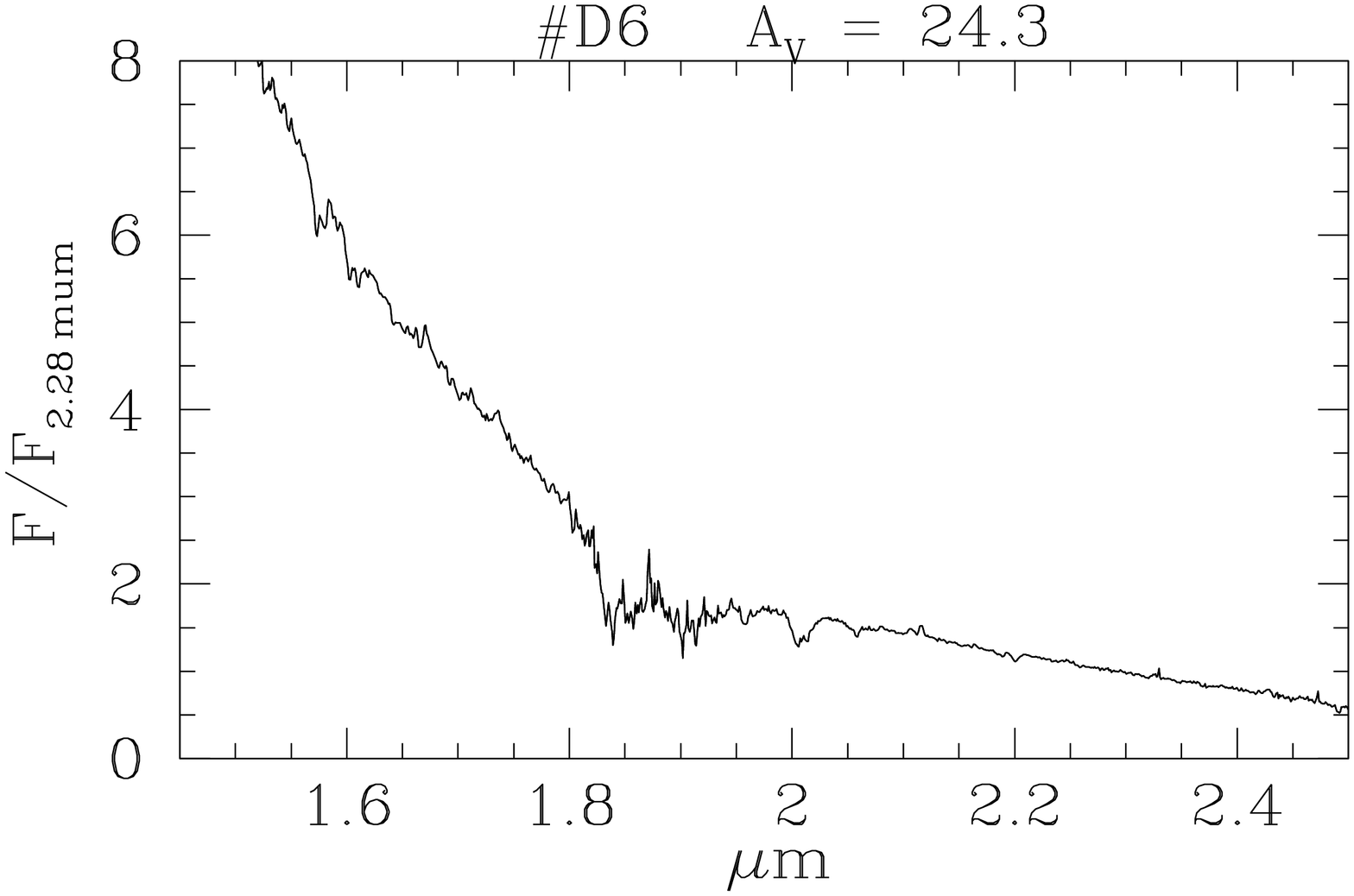,width=4.3cm,angle=270}   \epsfig{file=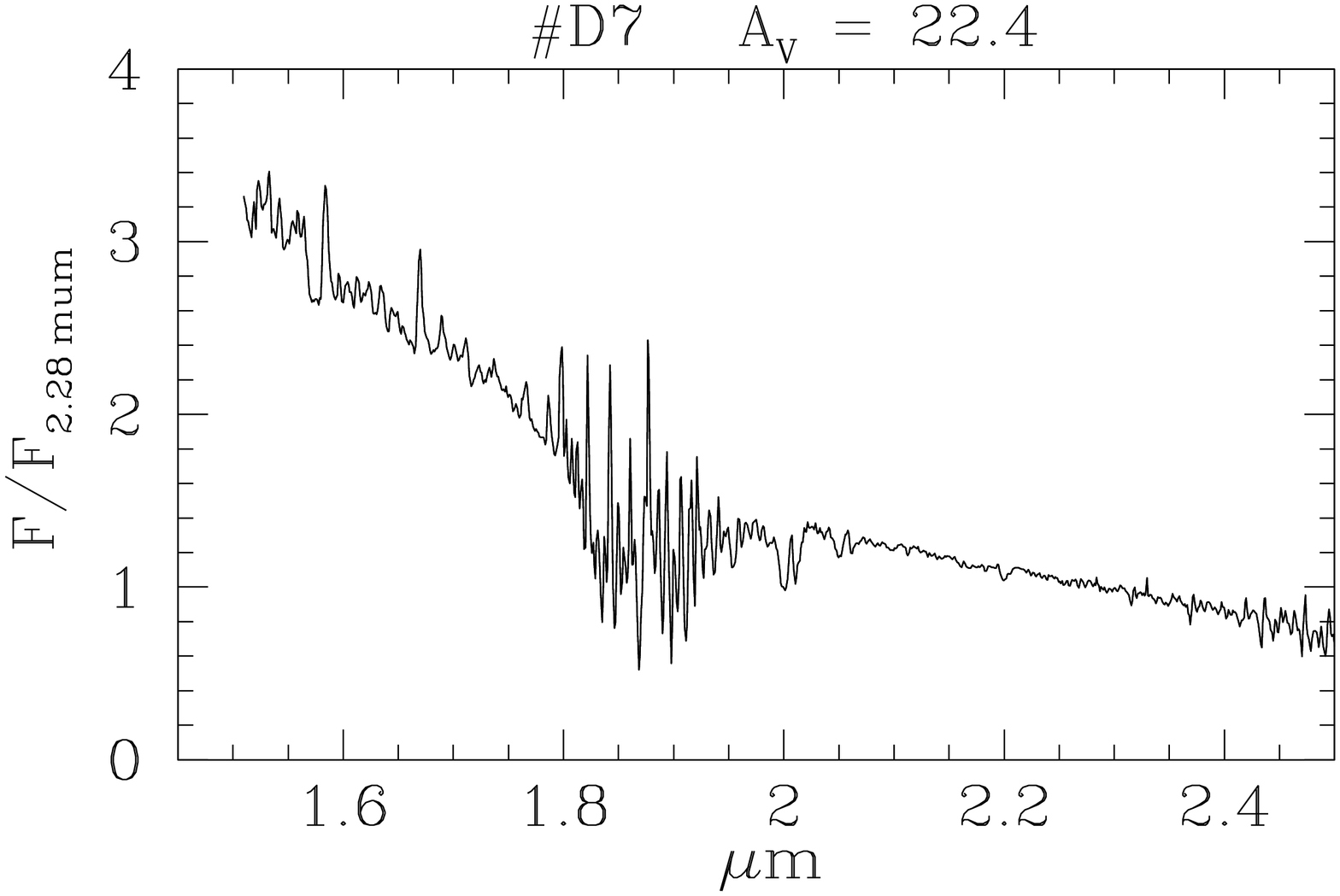,width=4.3cm,angle=270}}

\end{figure*}

\begin{figure*}[H!]
\caption{Supergiants candidates}
 {\epsfig{file=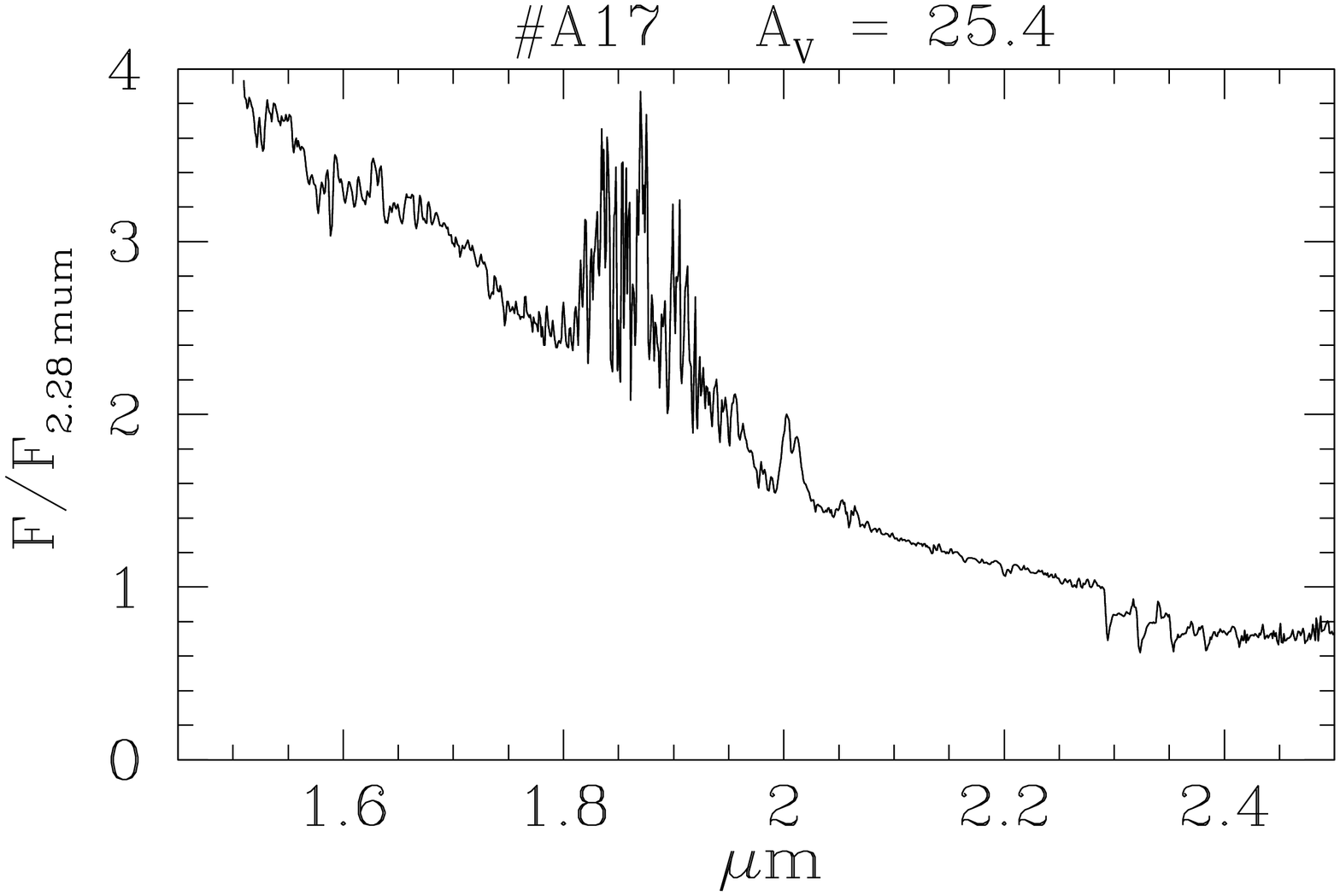,width=4.3cm,angle=270} \epsfig{file=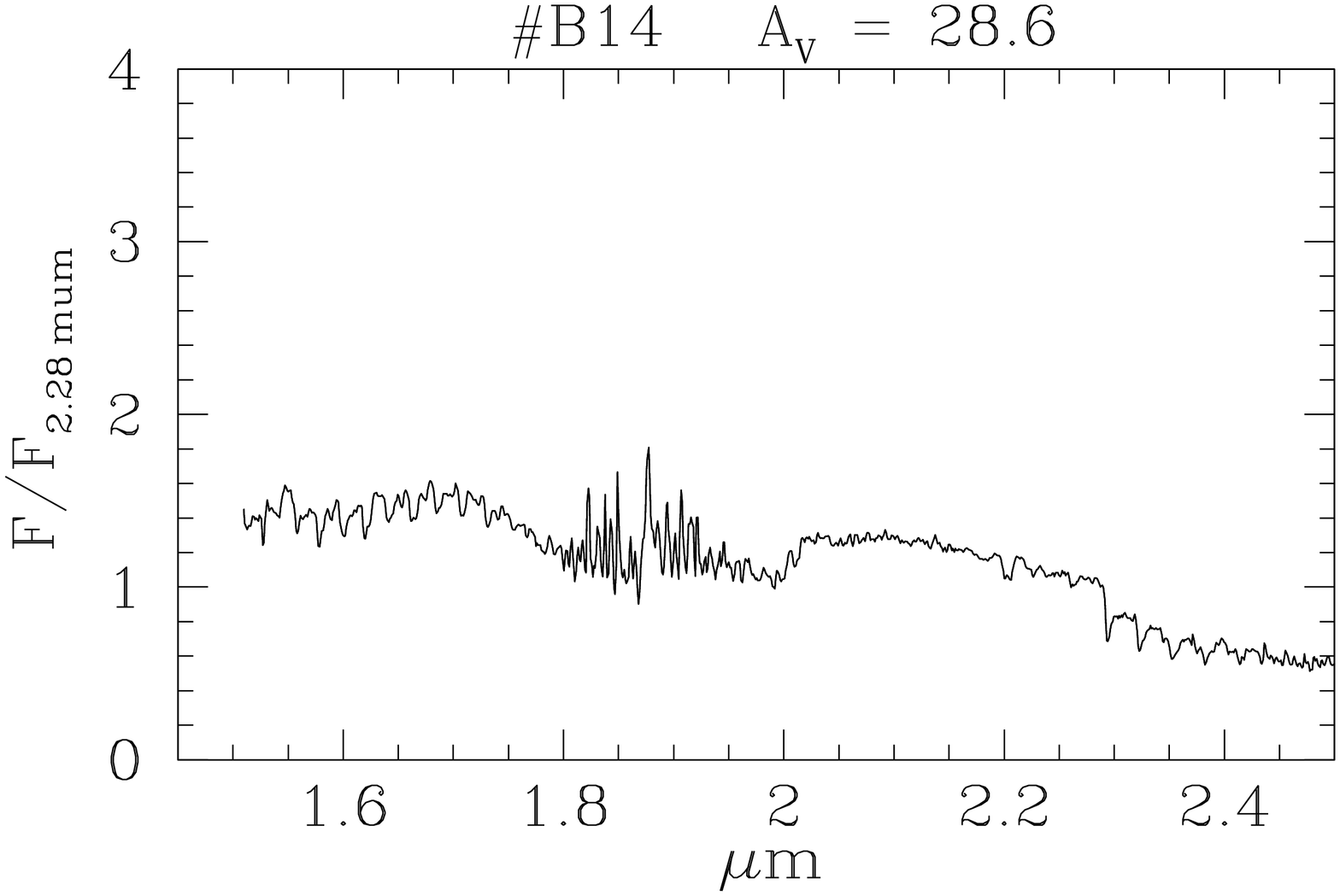,width=4.3cm,angle=270}   \epsfig{file=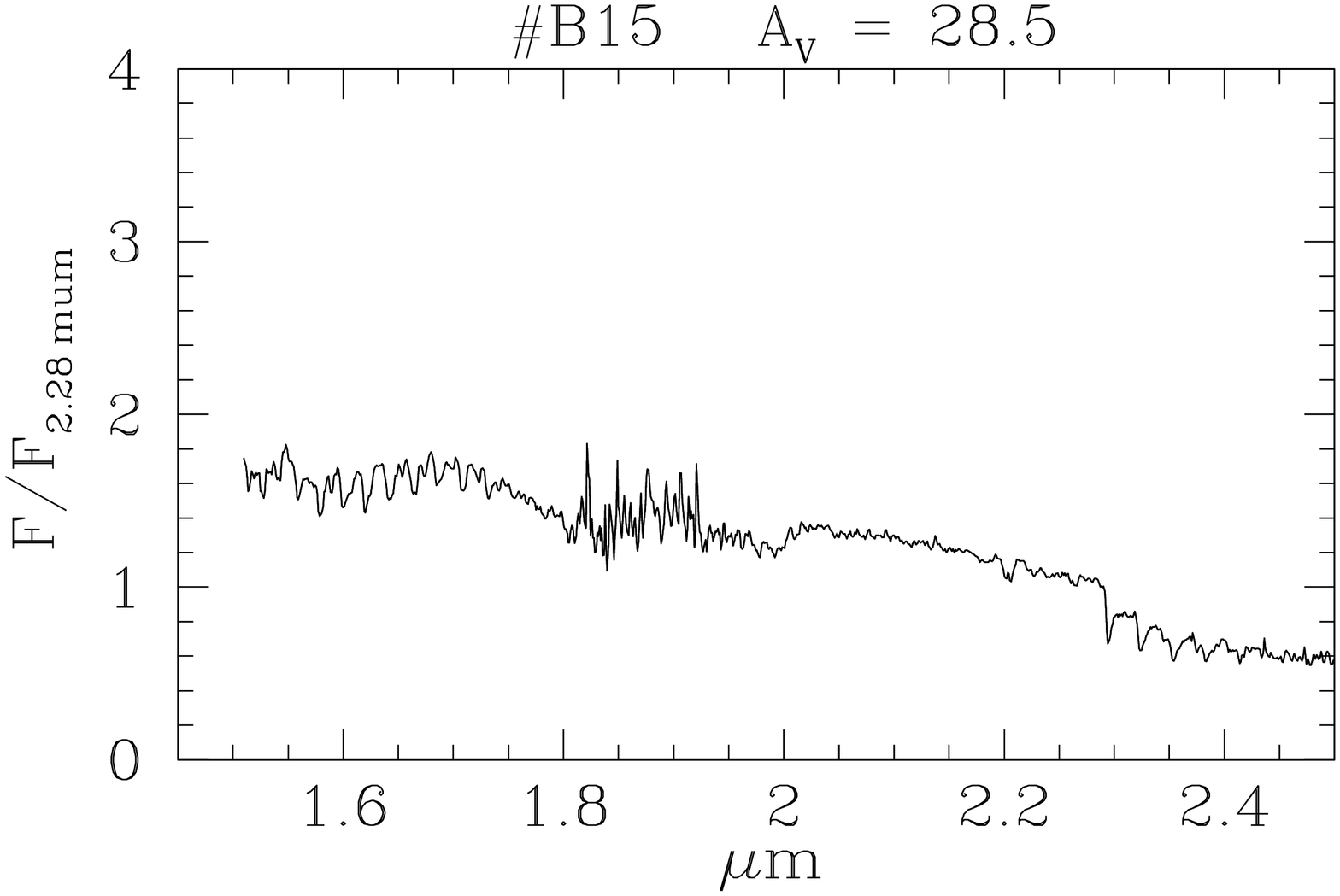,width=4.3cm,angle=270}  \epsfig{file=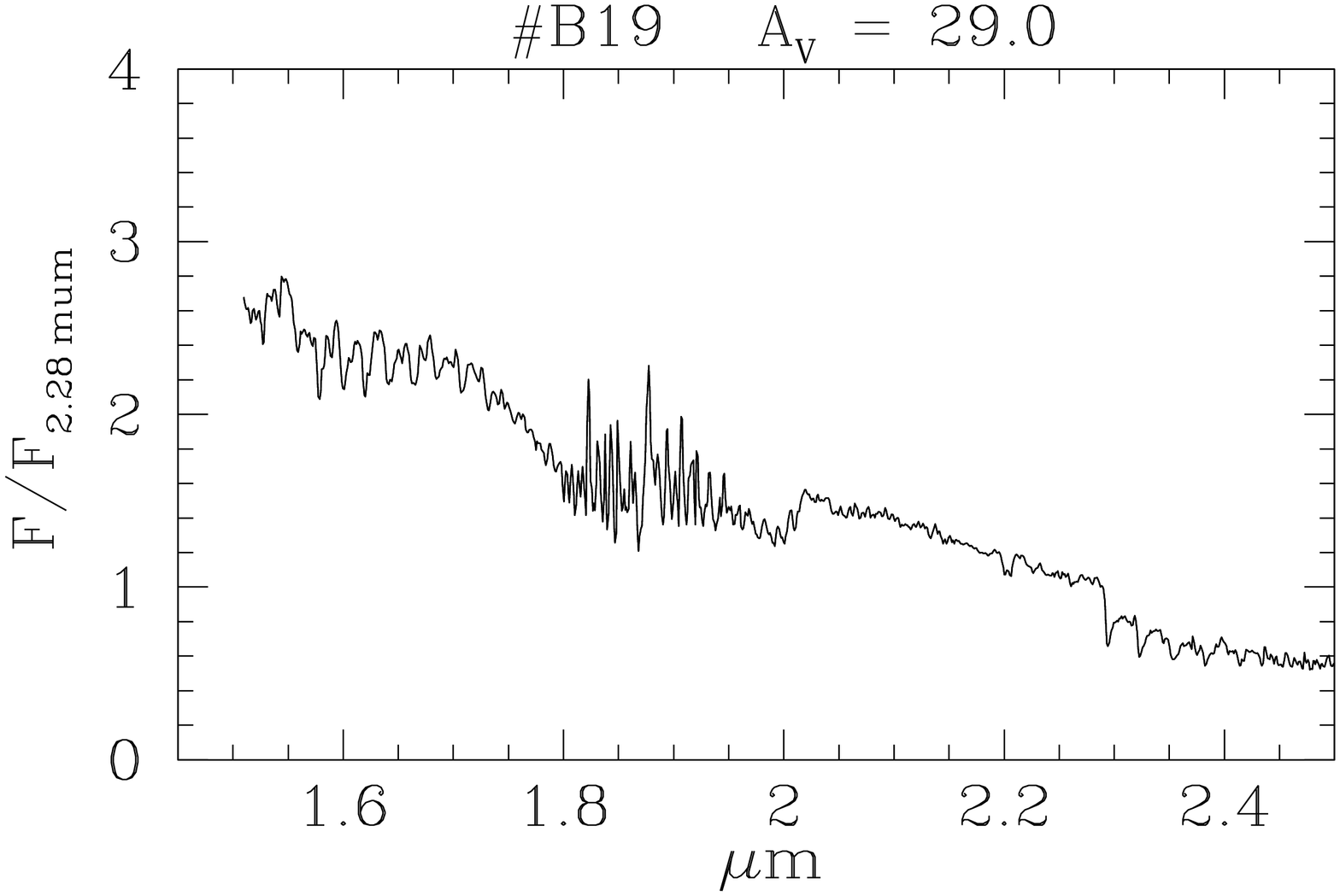,width=4.3cm,angle=270}}

\end{figure*}

\begin{figure*}[H!]
\caption{OH/IR stars (Wood et al. \cite{Wood98}) }
{\epsfig{file=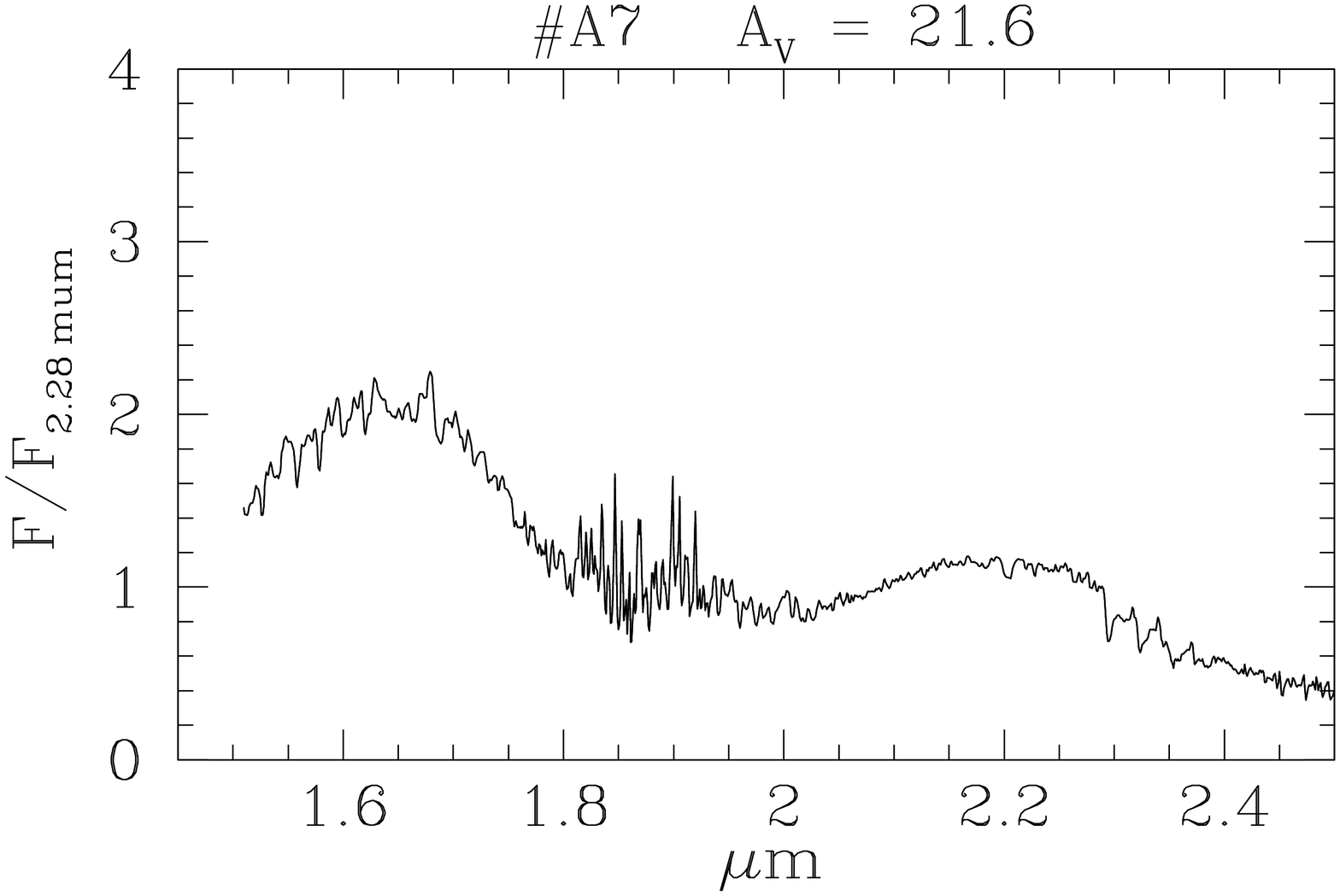,width=4.3cm,angle=270} \epsfig{file=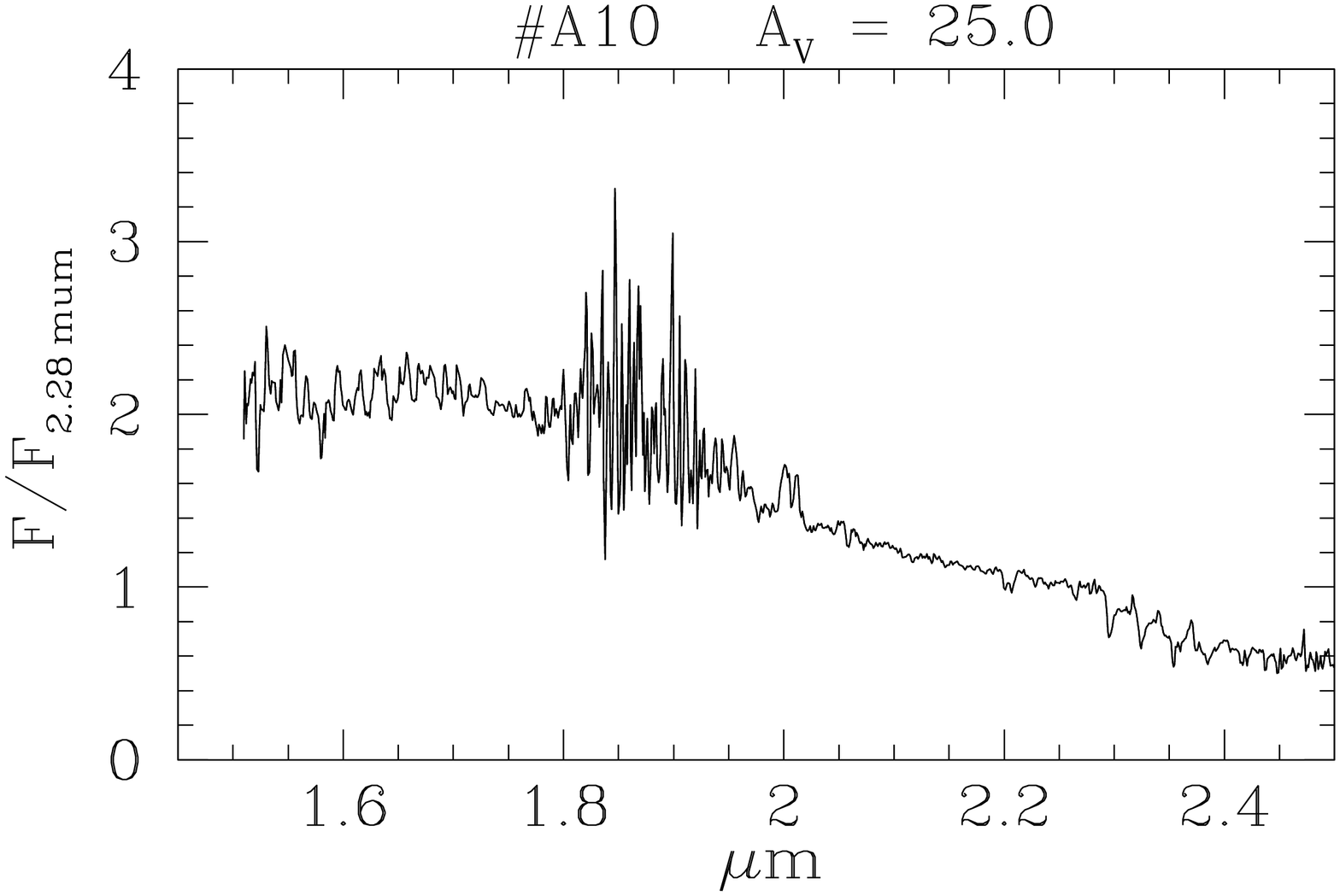,width=4.3cm,angle=270}   \epsfig{file=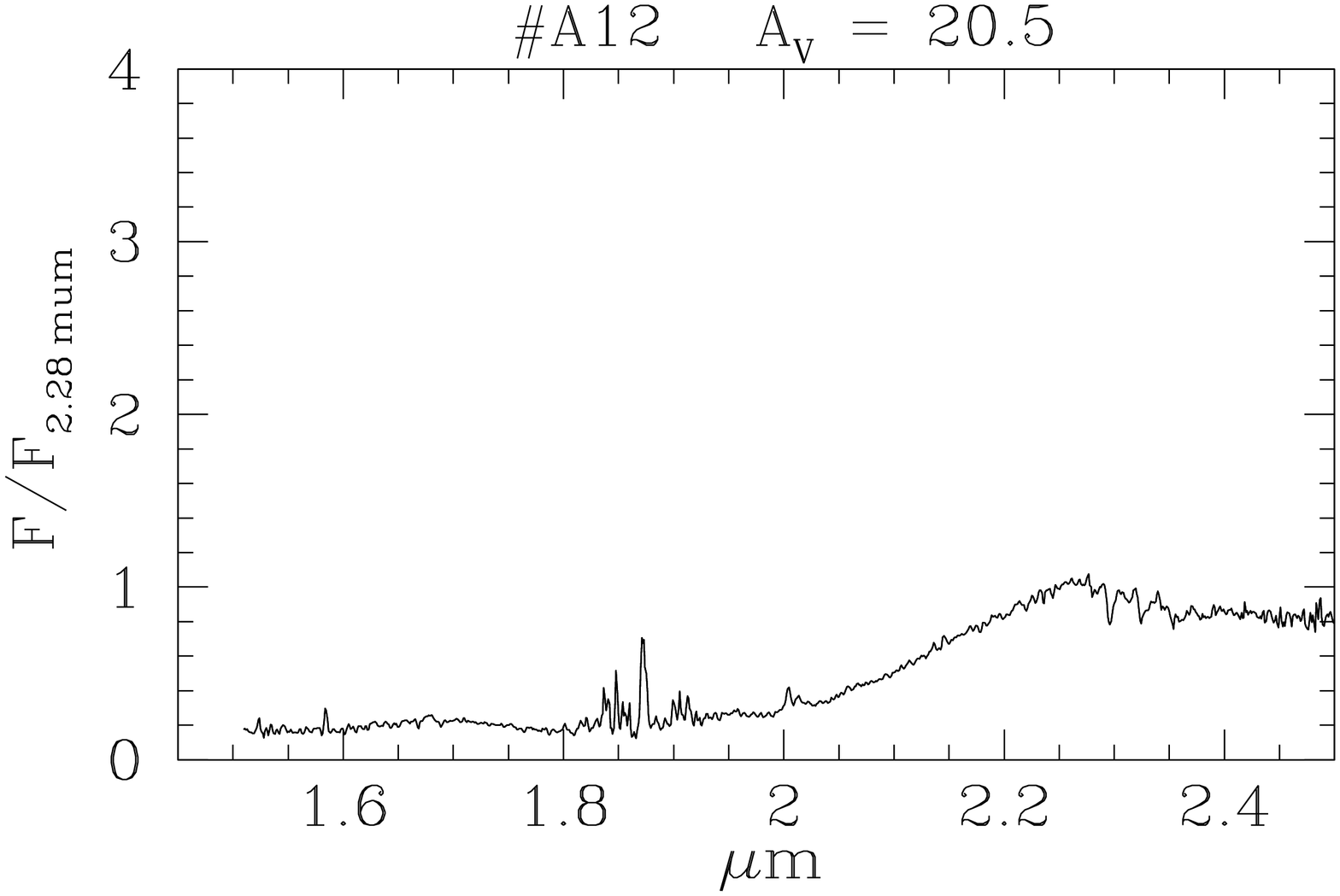,width=4.3cm,angle=270}

\epsfig{file=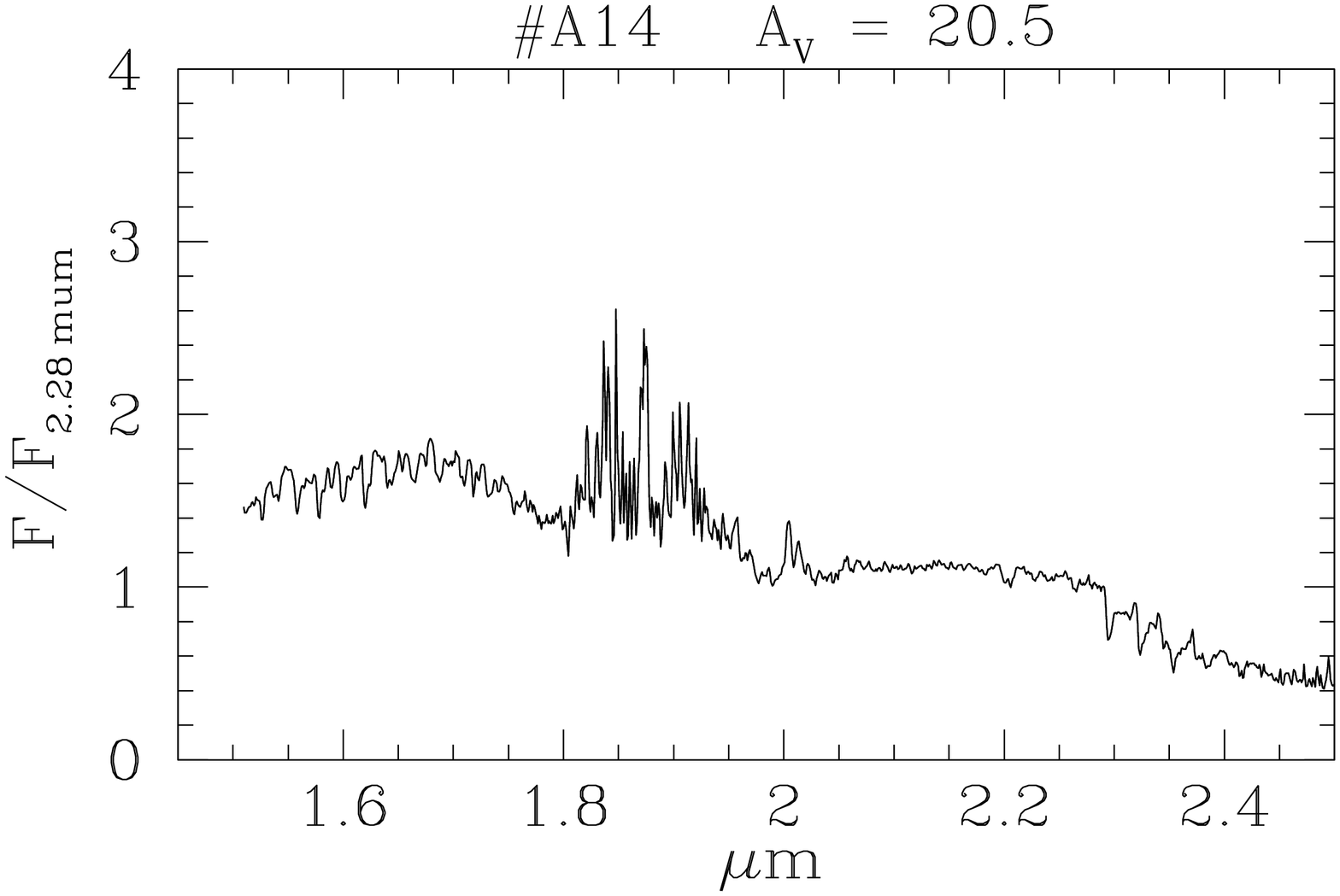,width=4.3cm,angle=270} \epsfig{file=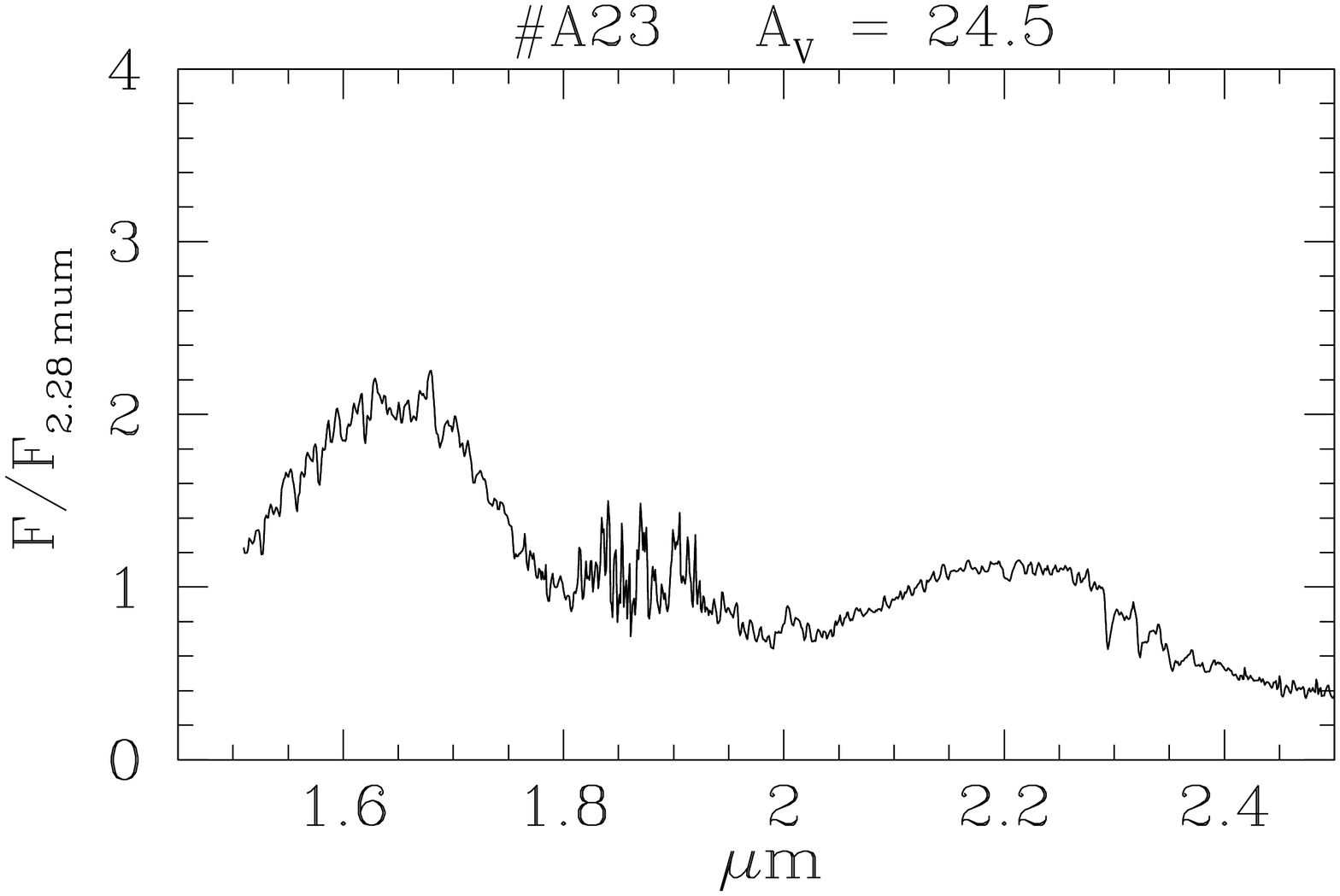,width=4.3cm,angle=270}   \epsfig{file=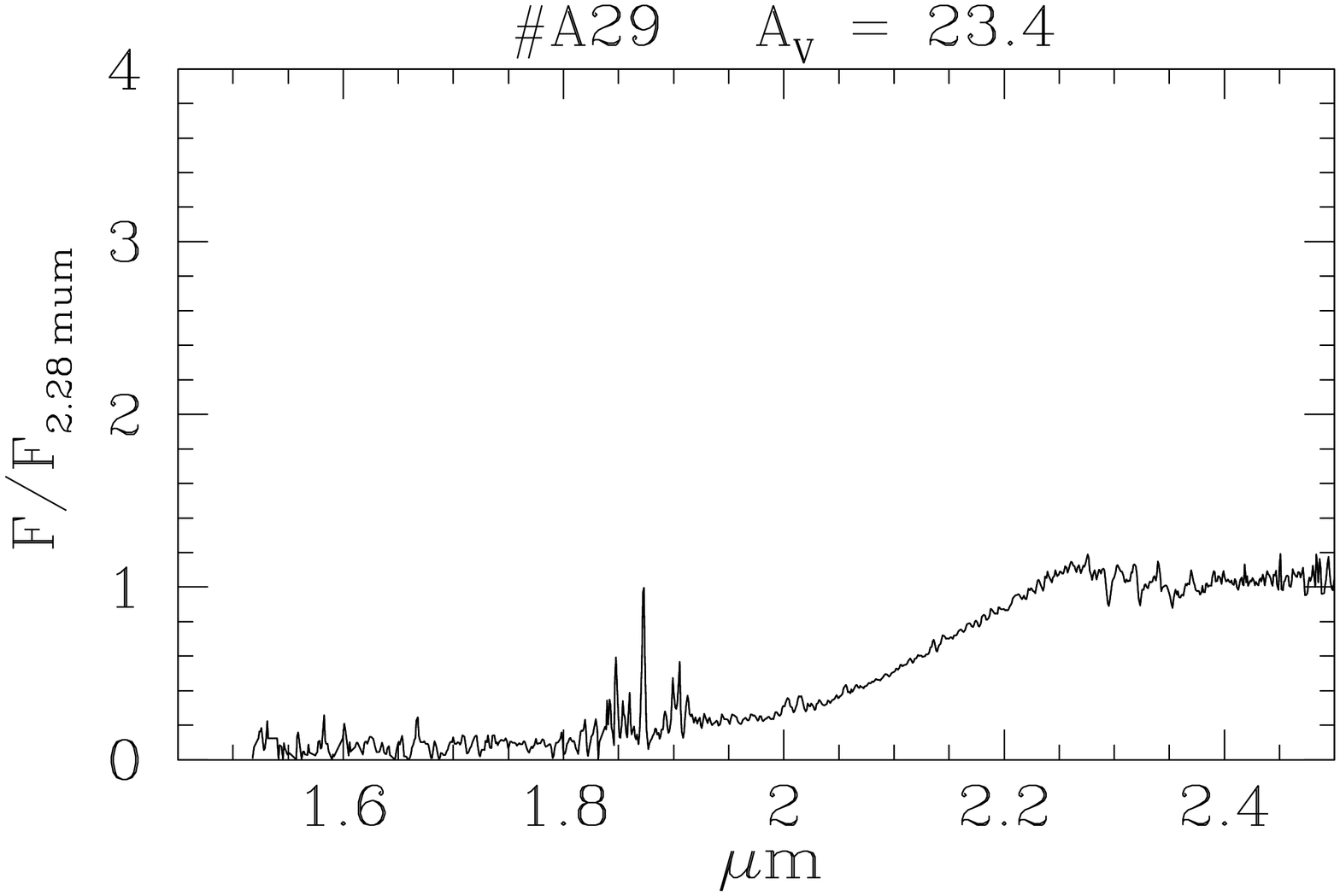,width=4.3cm,angle=270}

\epsfig{file=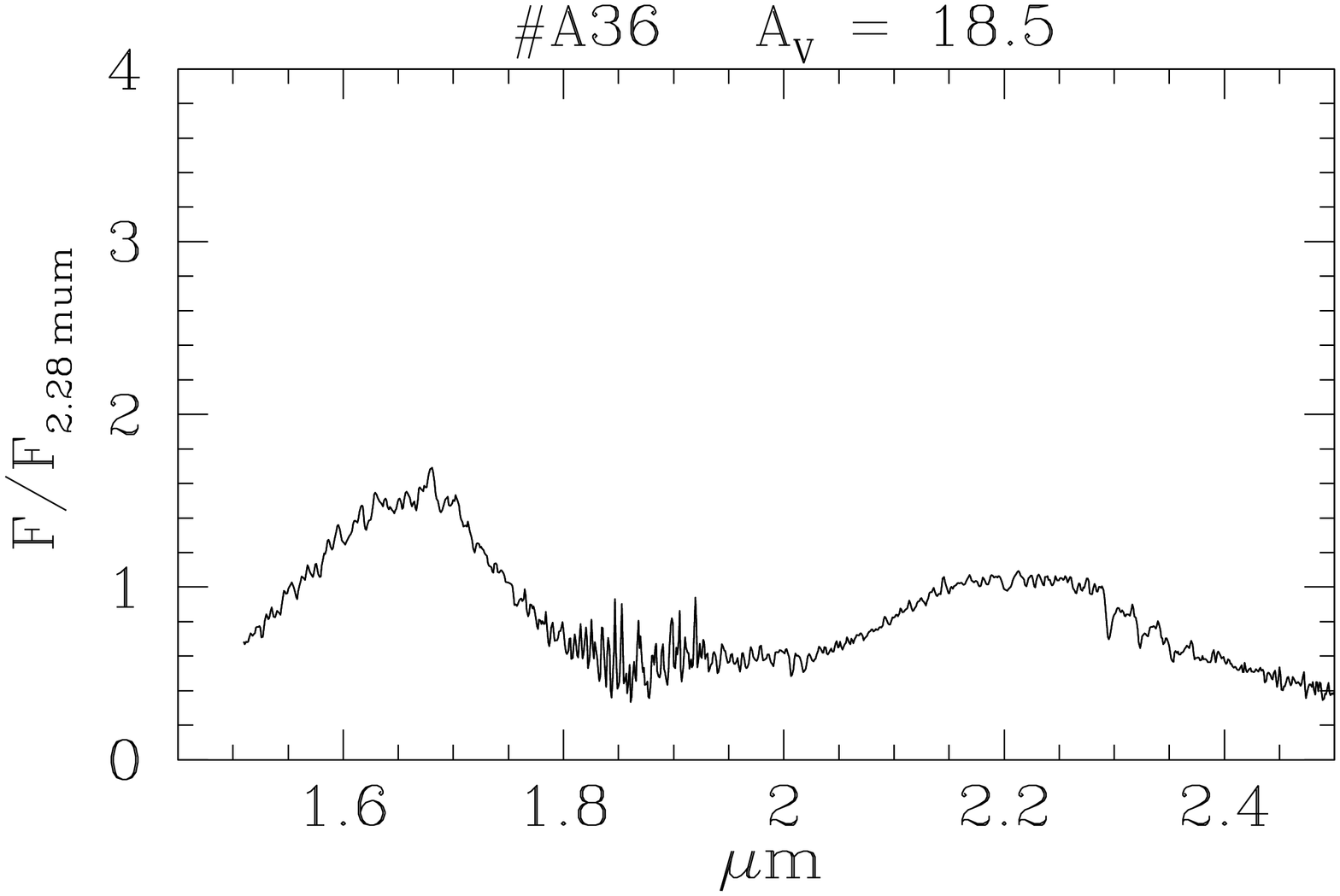,width=4.3cm,angle=270} \epsfig{file=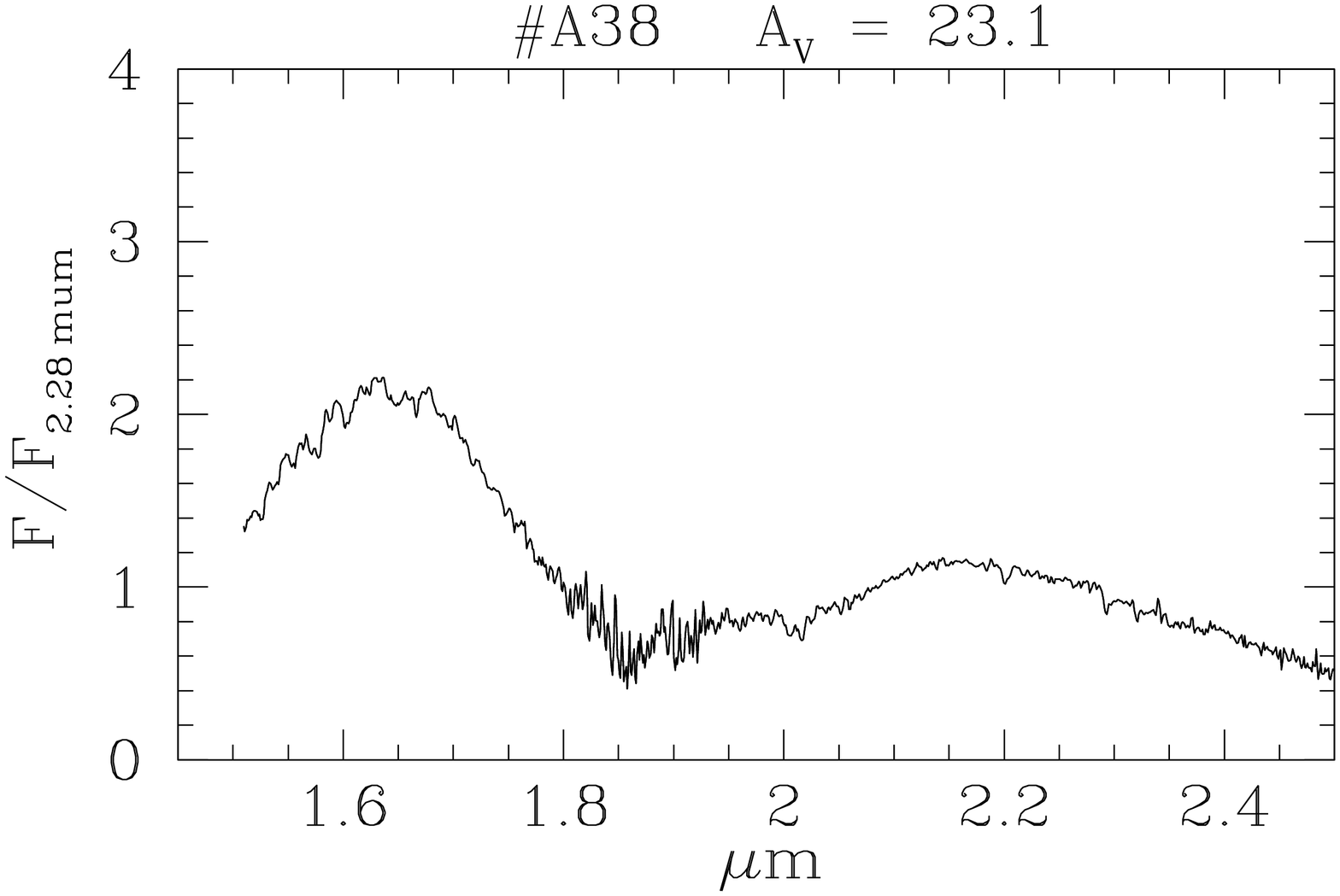,width=4.3cm,angle=270}   \epsfig{file=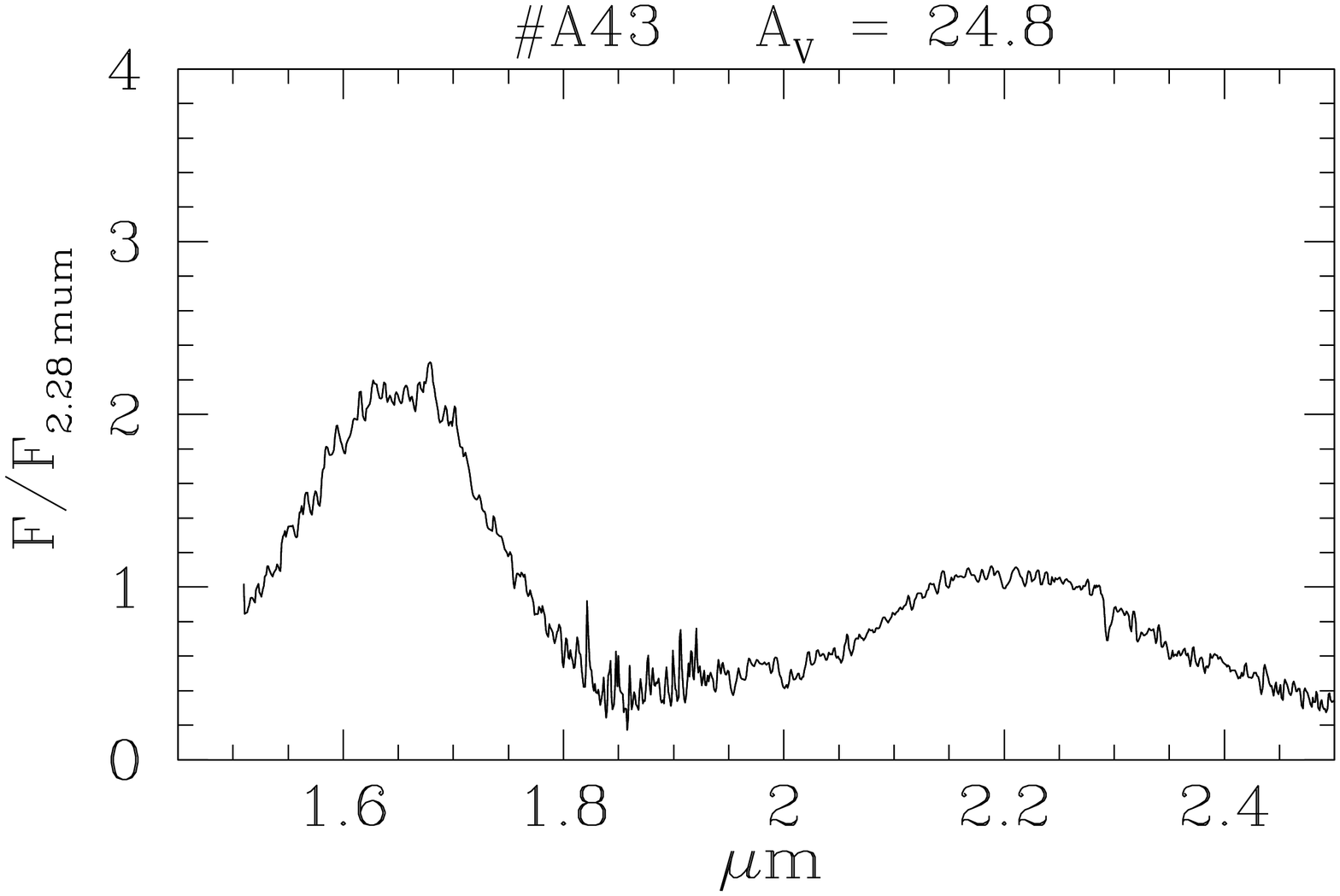,width=4.3cm,angle=270}

\epsfig{file=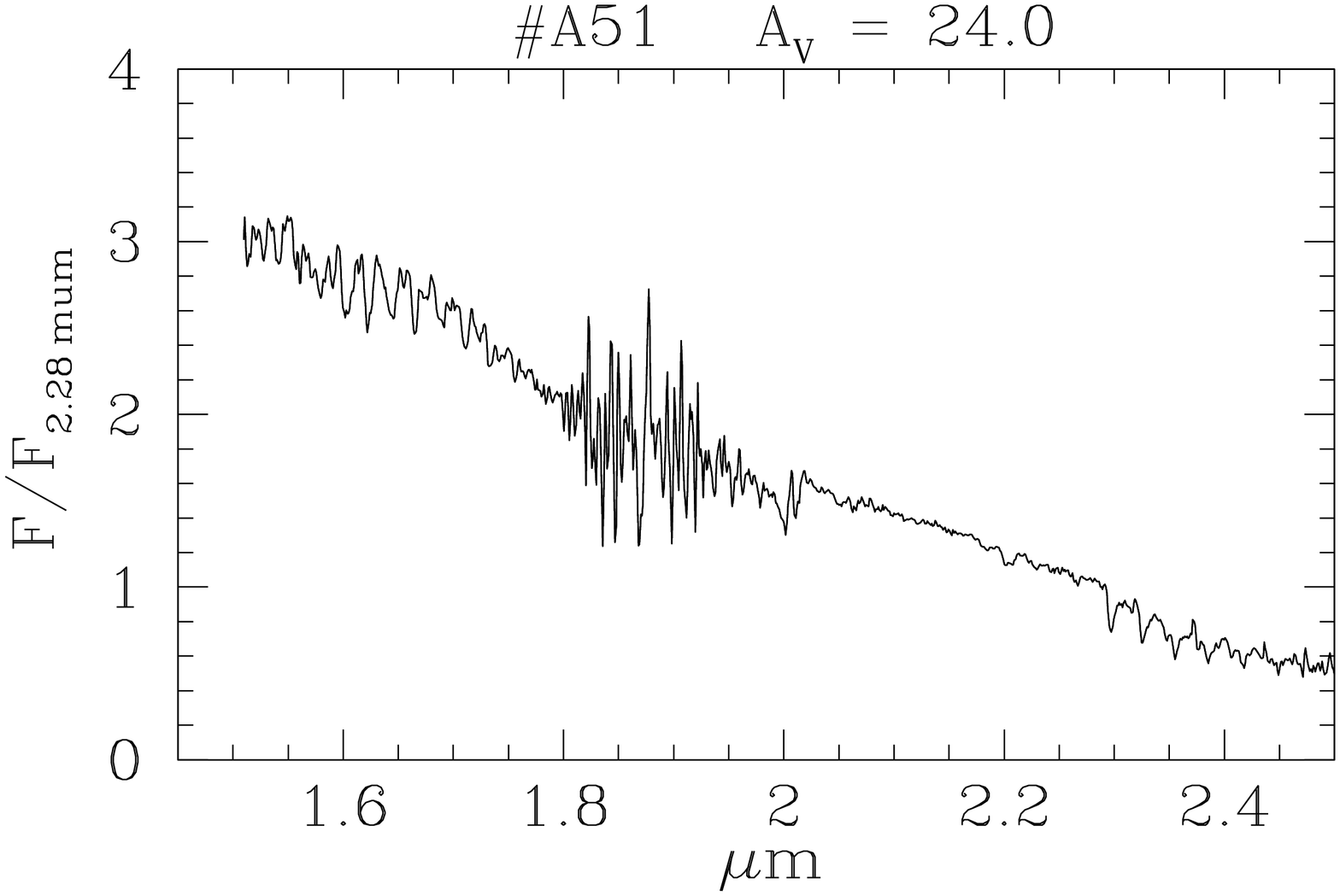,width=4.3cm,angle=270} \epsfig{file=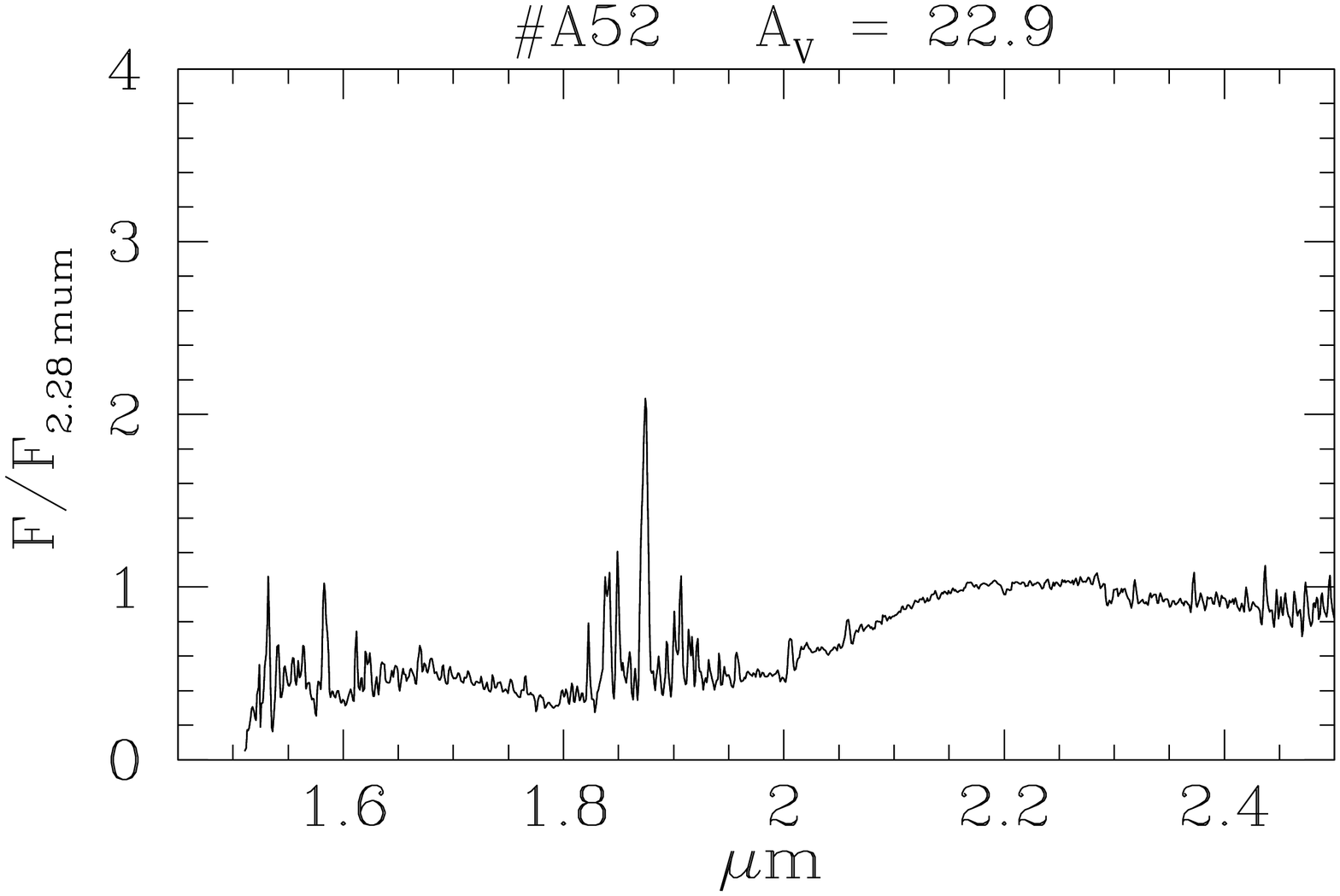,width=4.3cm,angle=270}   \epsfig{file=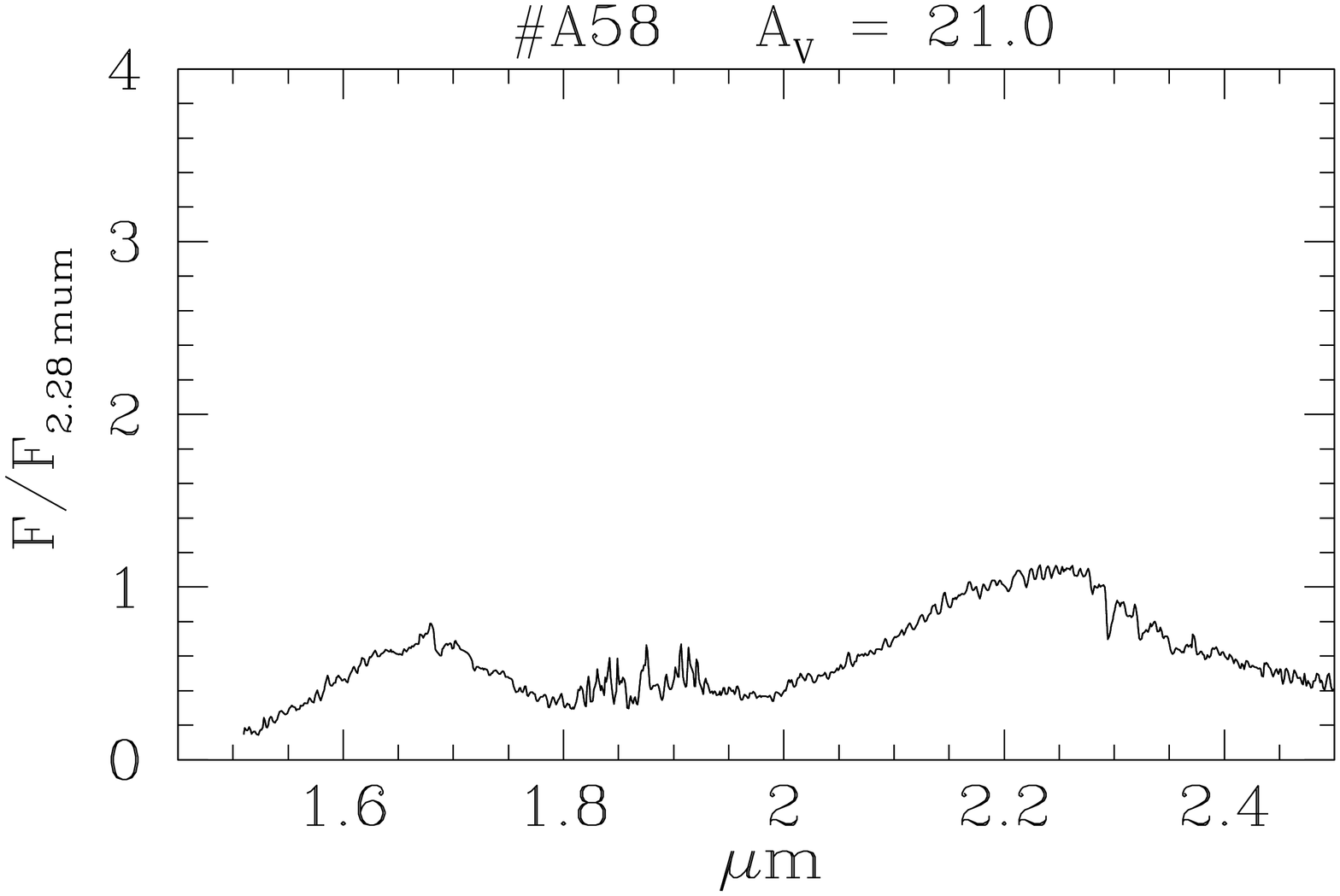,width=4.3cm,angle=270}

\epsfig{file=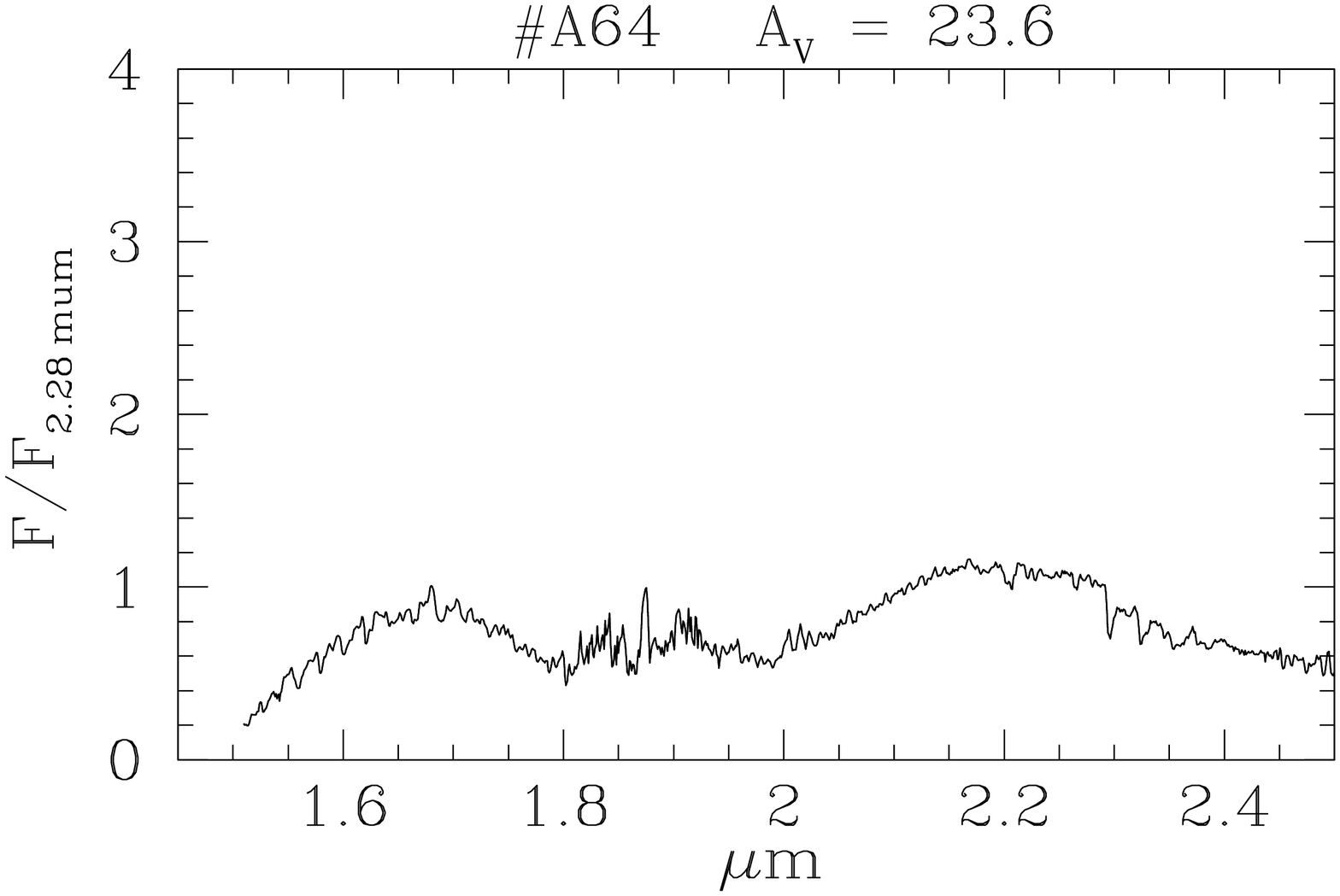,width=4.3cm,angle=270} \epsfig{file=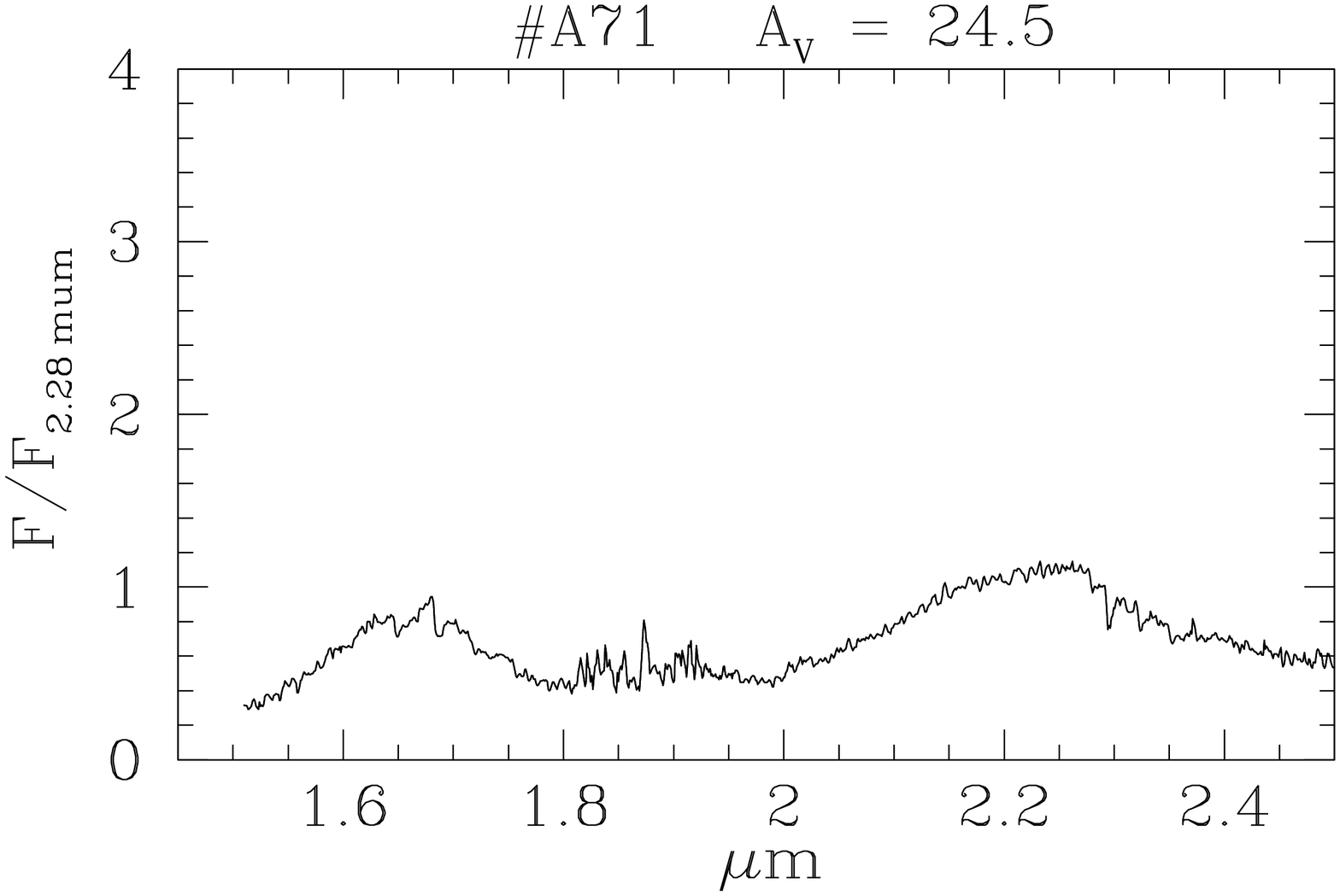,width=4.3cm,angle=270}   \epsfig{file=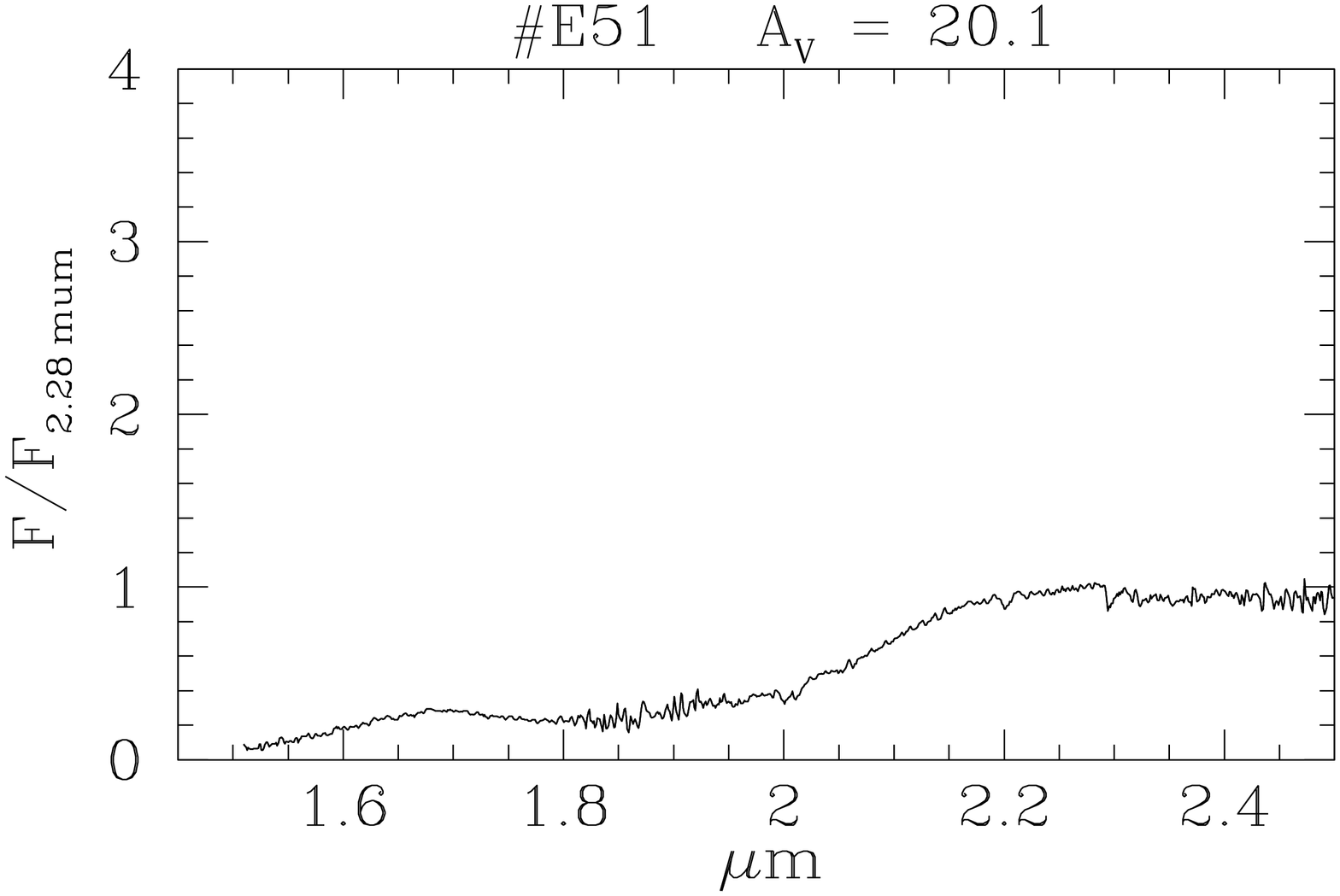,width=4.3cm,angle=270}}

\end{figure*}

\begin{figure*}[H!]
\caption{Long period Variables (Glass et al. \cite{Glass2001}) }
{\epsfig{file=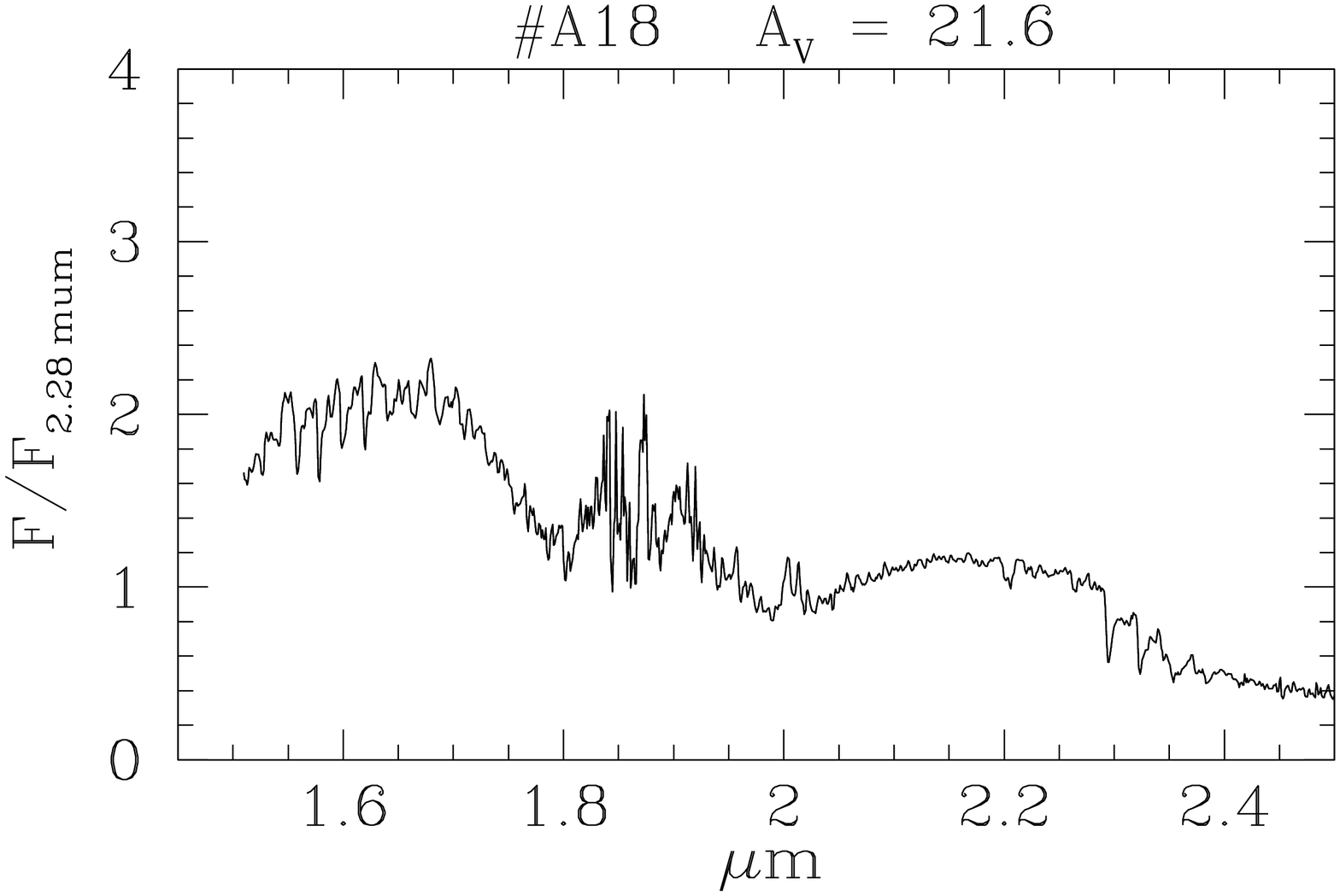,width=4.3cm,angle=270} \epsfig{file=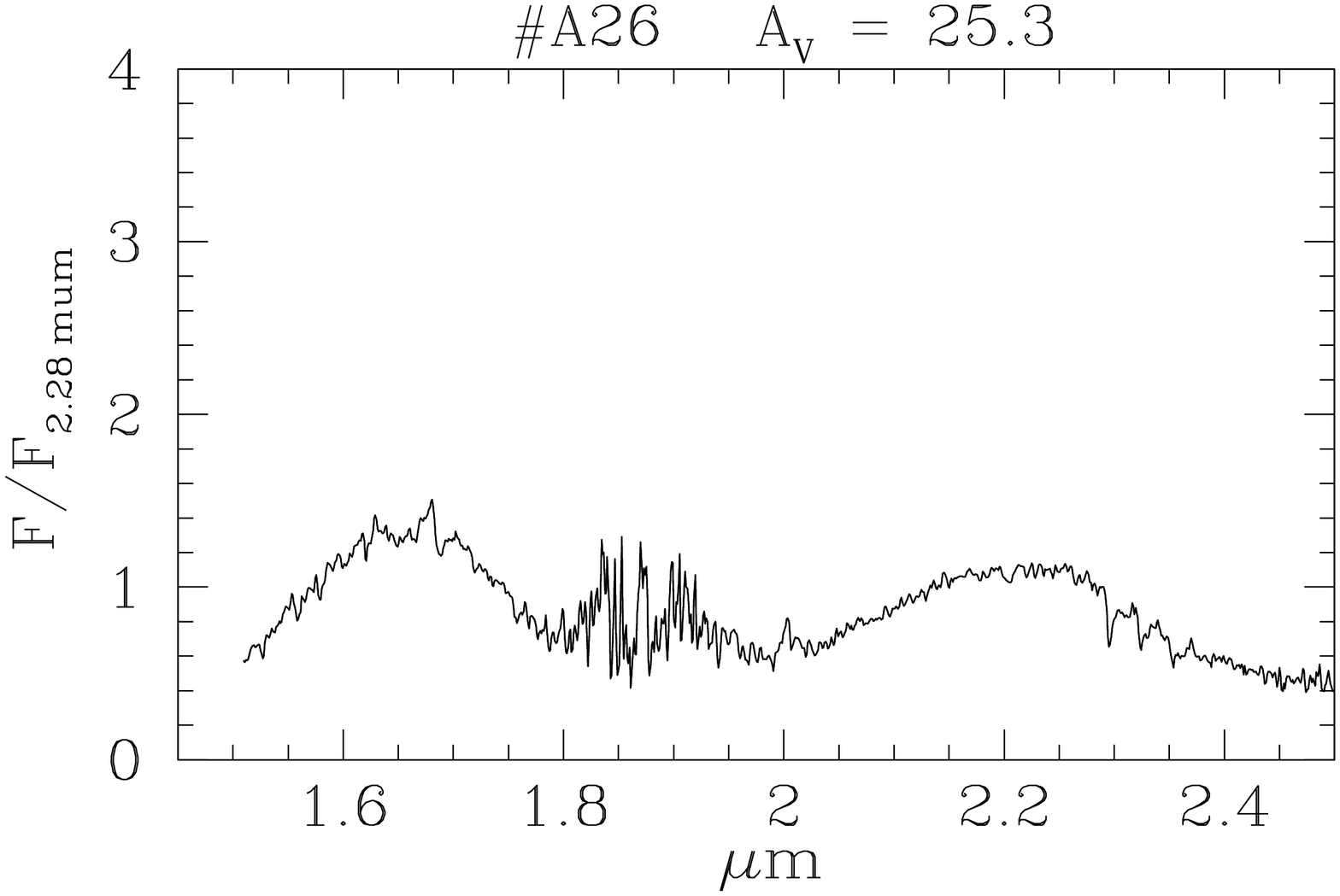,width=4.3cm,angle=270}   \epsfig{file=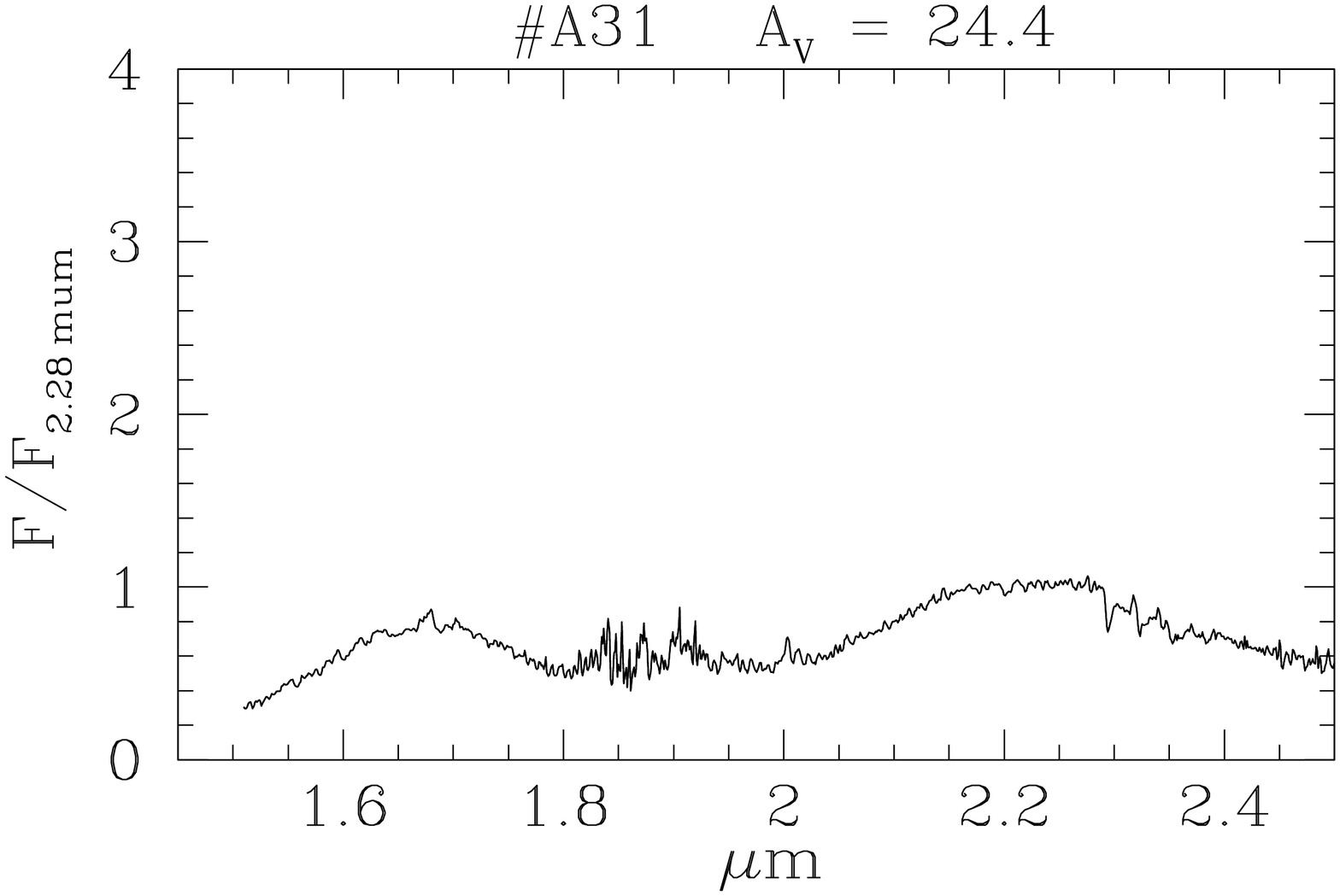,width=4.3cm,angle=270}

\epsfig{file=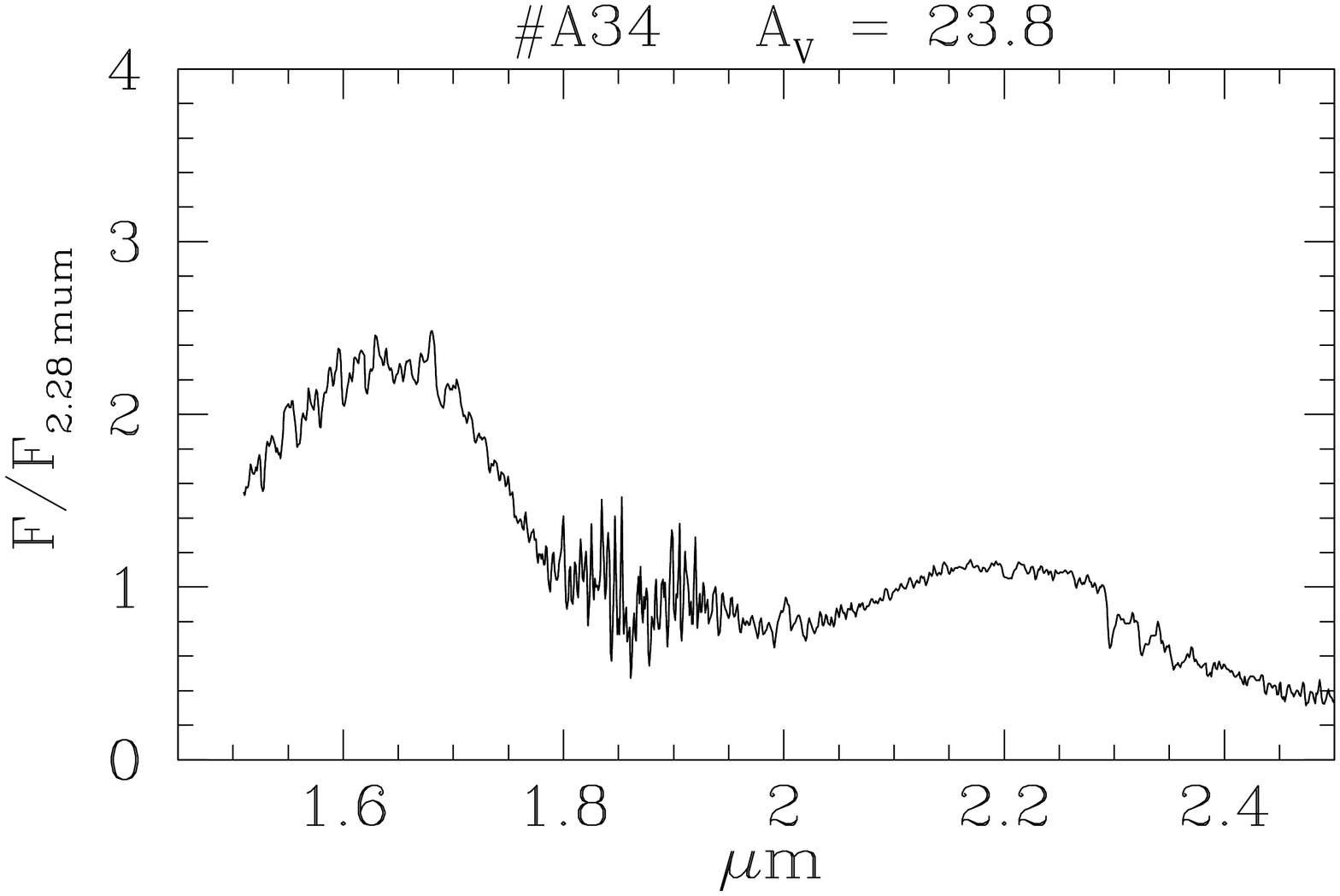,width=4.3cm,angle=270} \epsfig{file=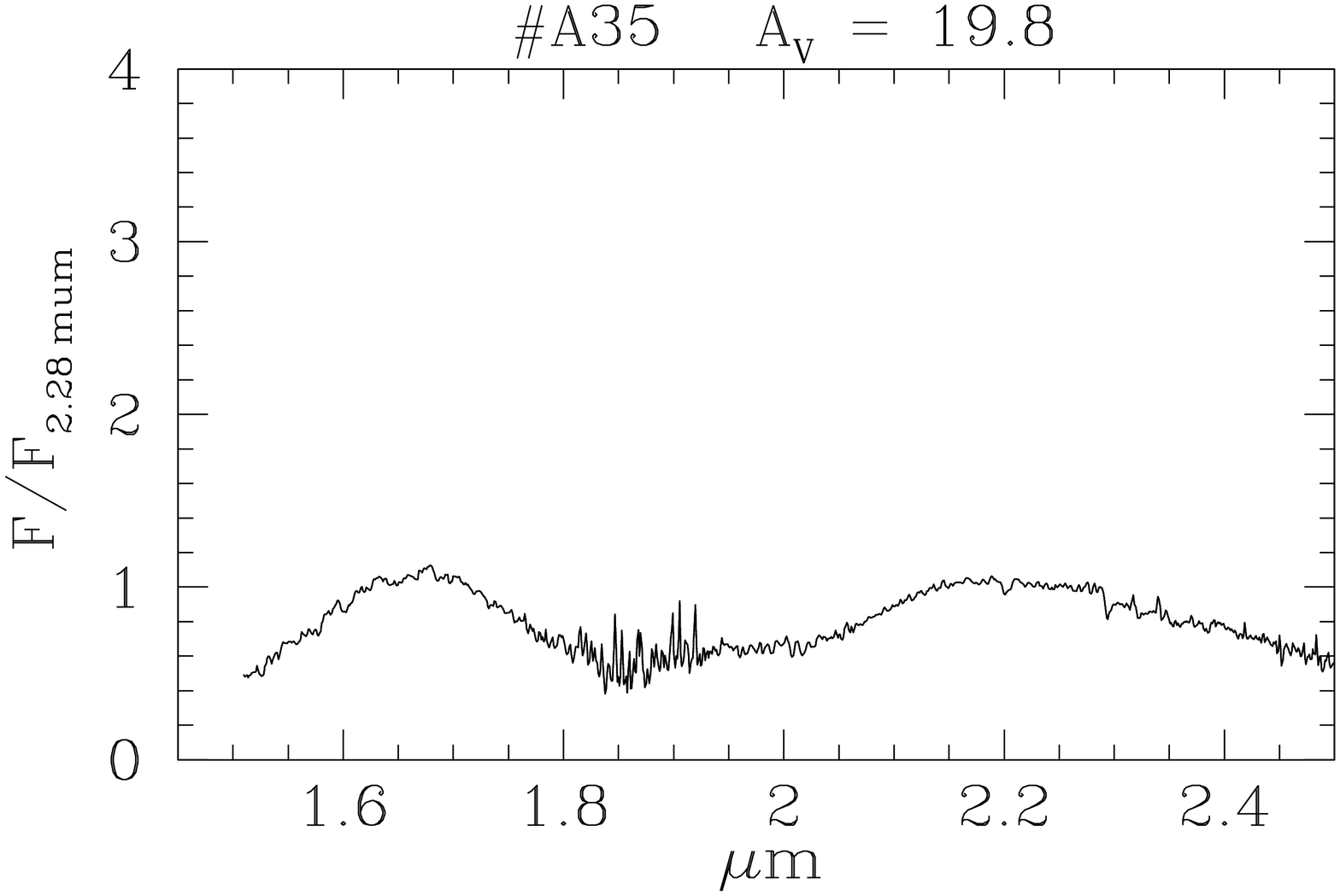,width=4.3cm,angle=270}   \epsfig{file=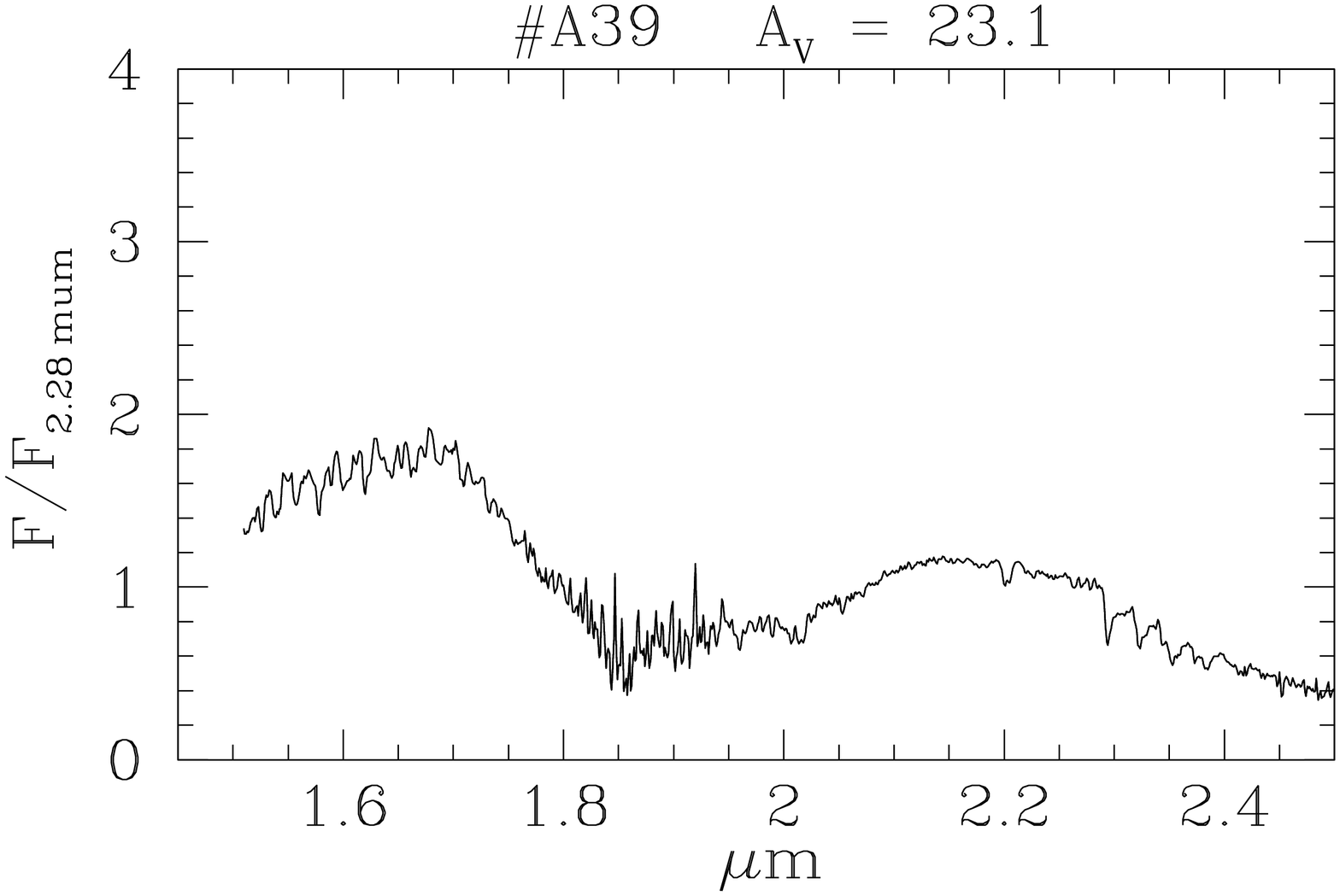,width=4.3cm,angle=270}

\epsfig{file=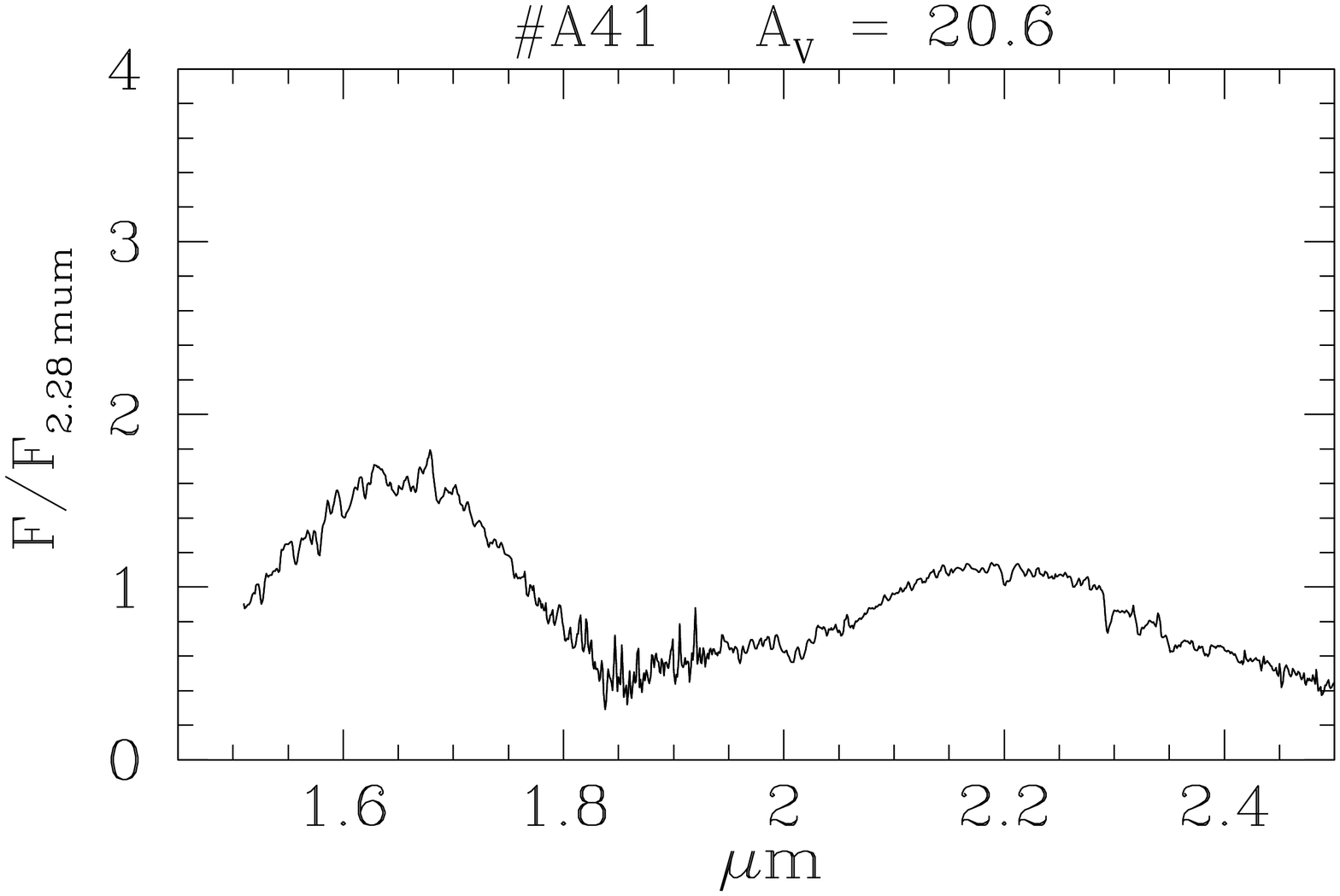,width=4.3cm,angle=270} \epsfig{file=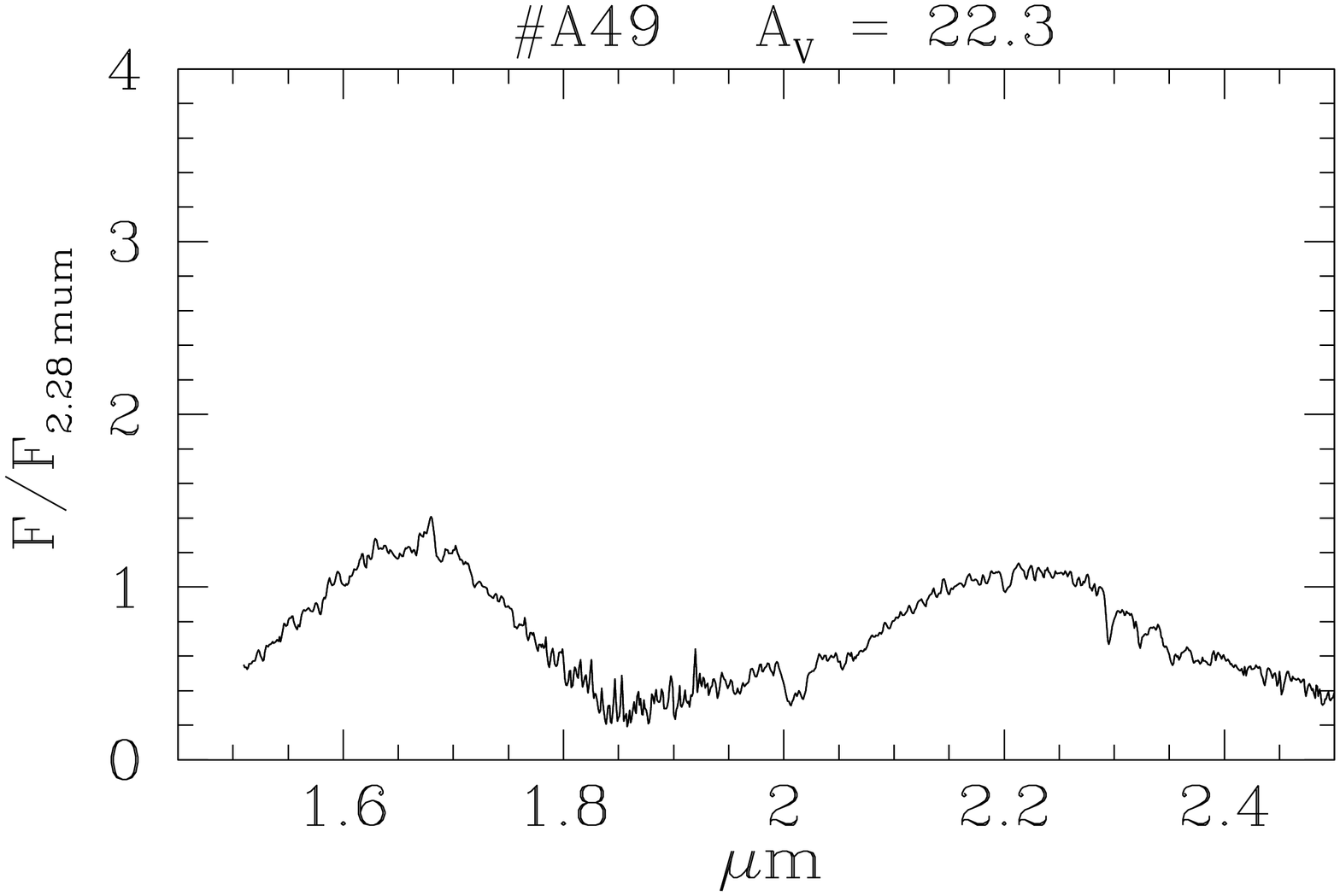,width=4.3cm,angle=270}   \epsfig{file=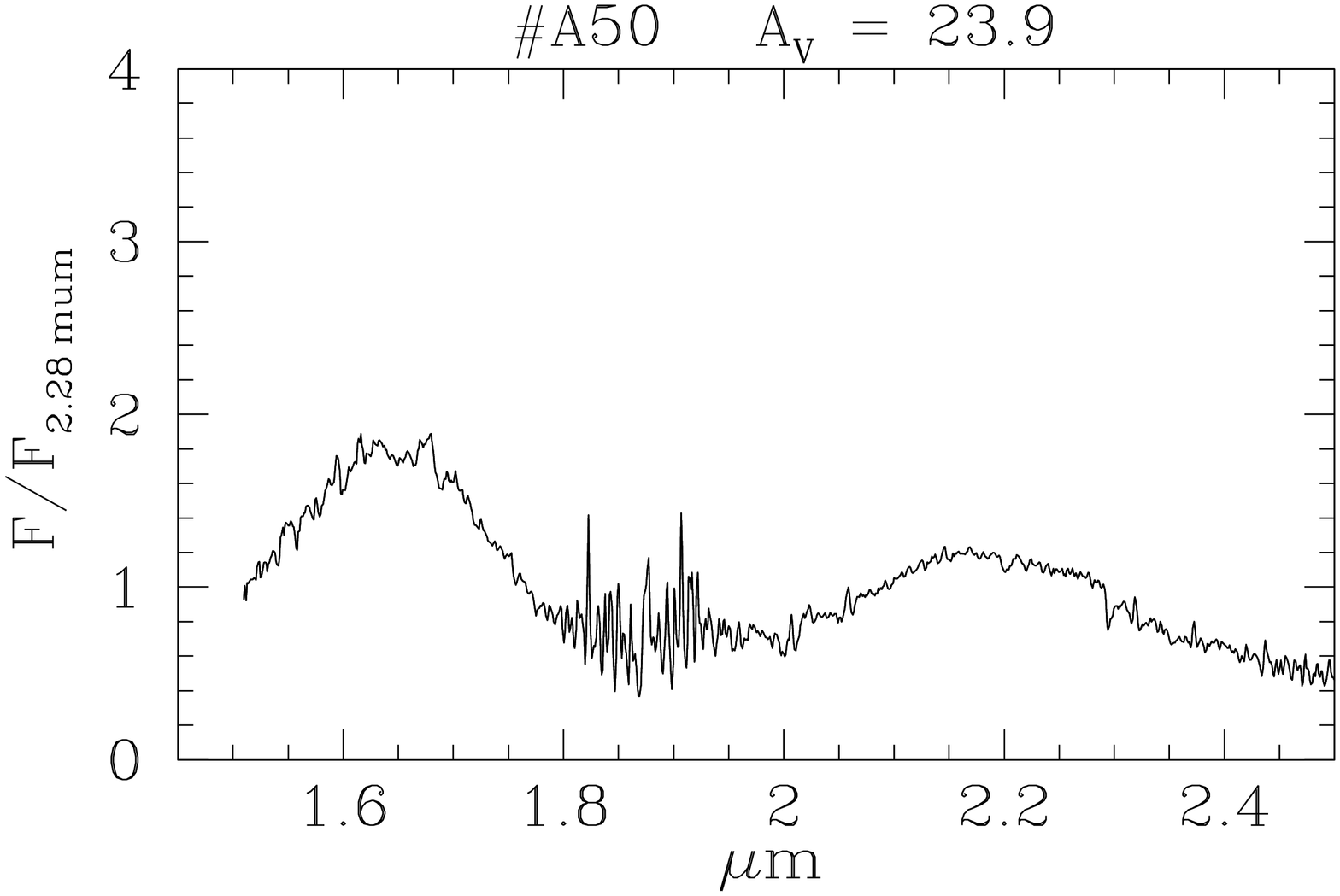,width=4.3cm,angle=270}

\epsfig{file=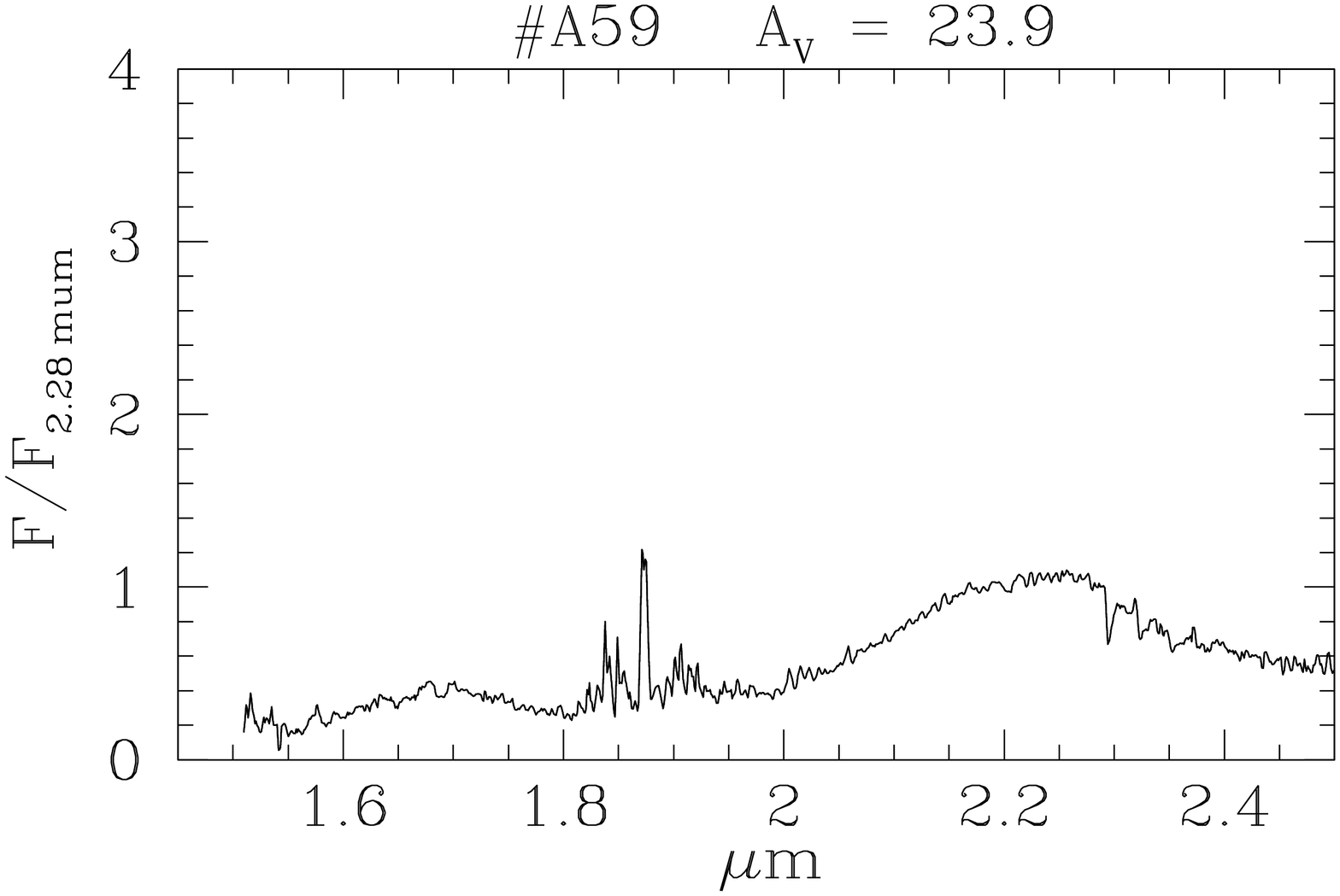,width=4.3cm,angle=270} \epsfig{file=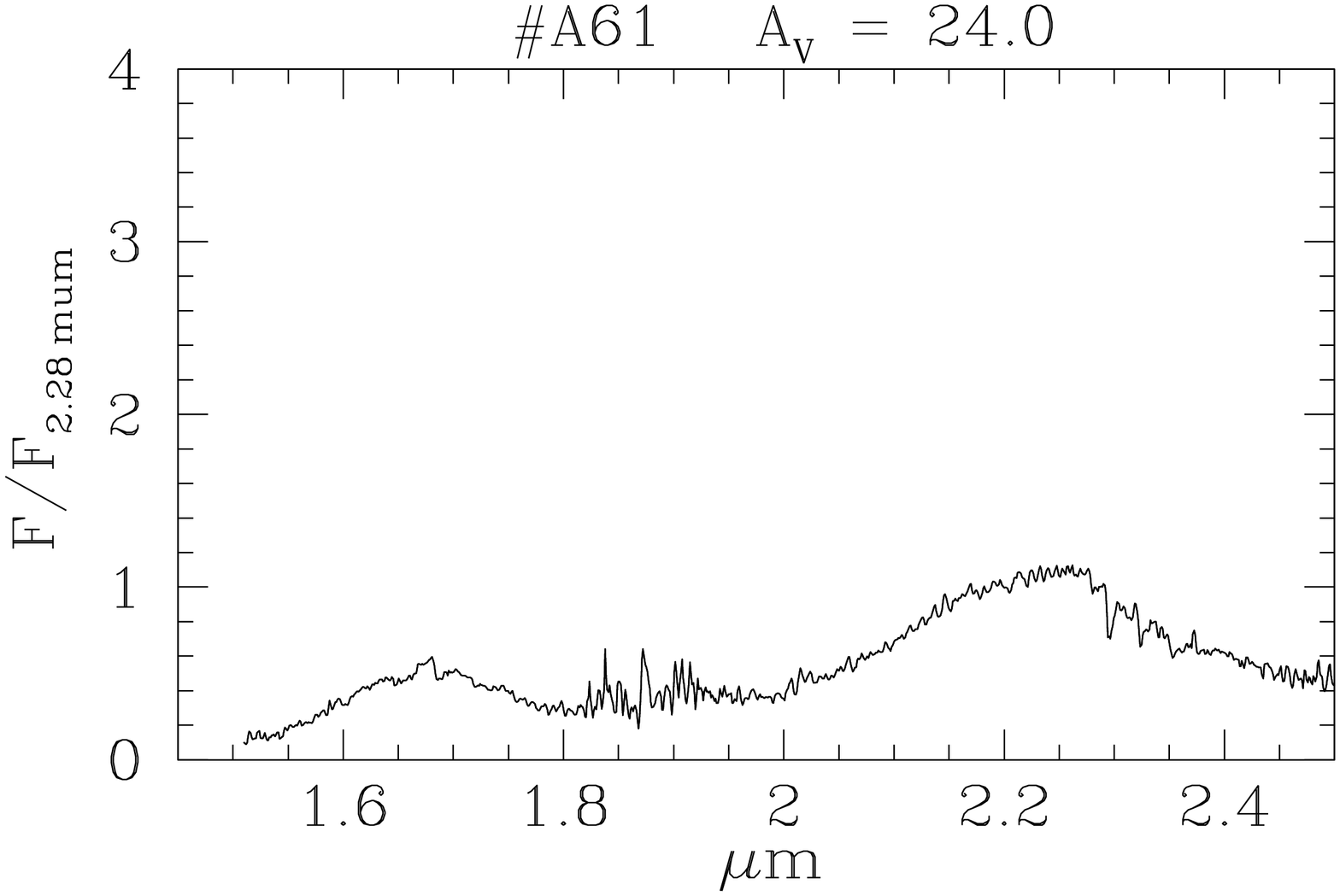,width=4.3cm,angle=270}   \epsfig{file=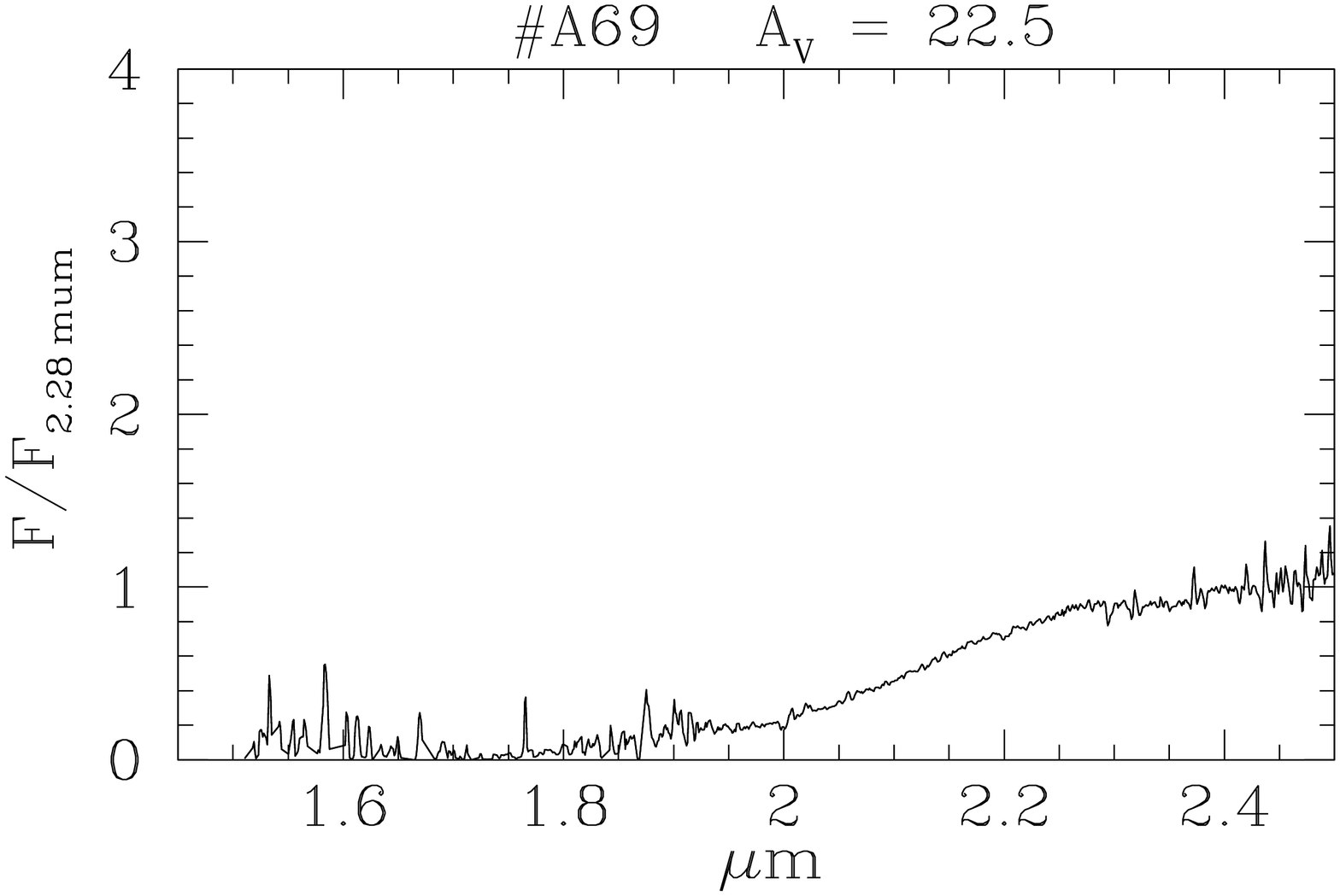,width=4.3cm,angle=270}}

\end{figure*}

\begin{figure*}[H!]
\caption{AGB star candidates}
{\epsfig{file=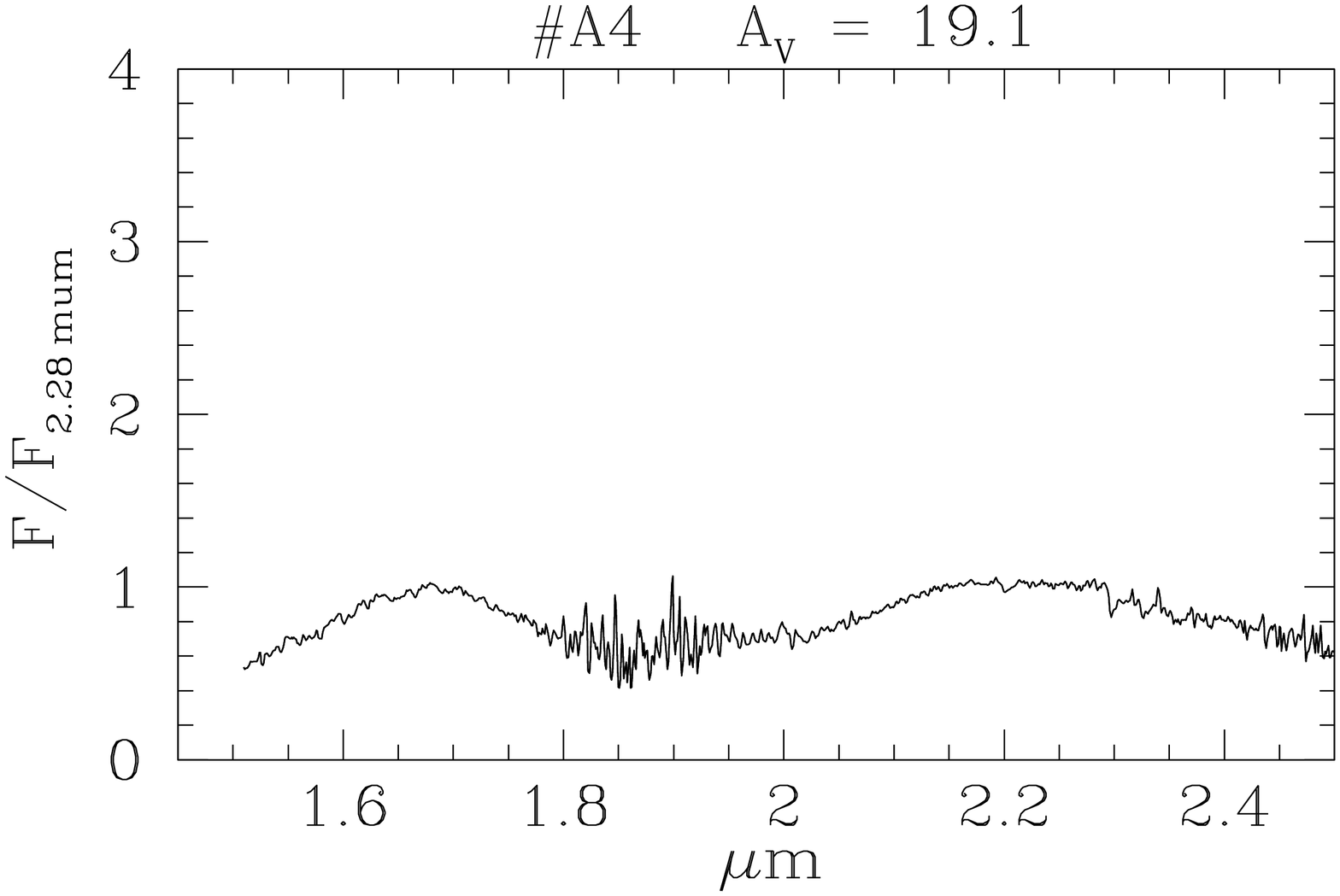,width=4.3cm,angle=270} \epsfig{file=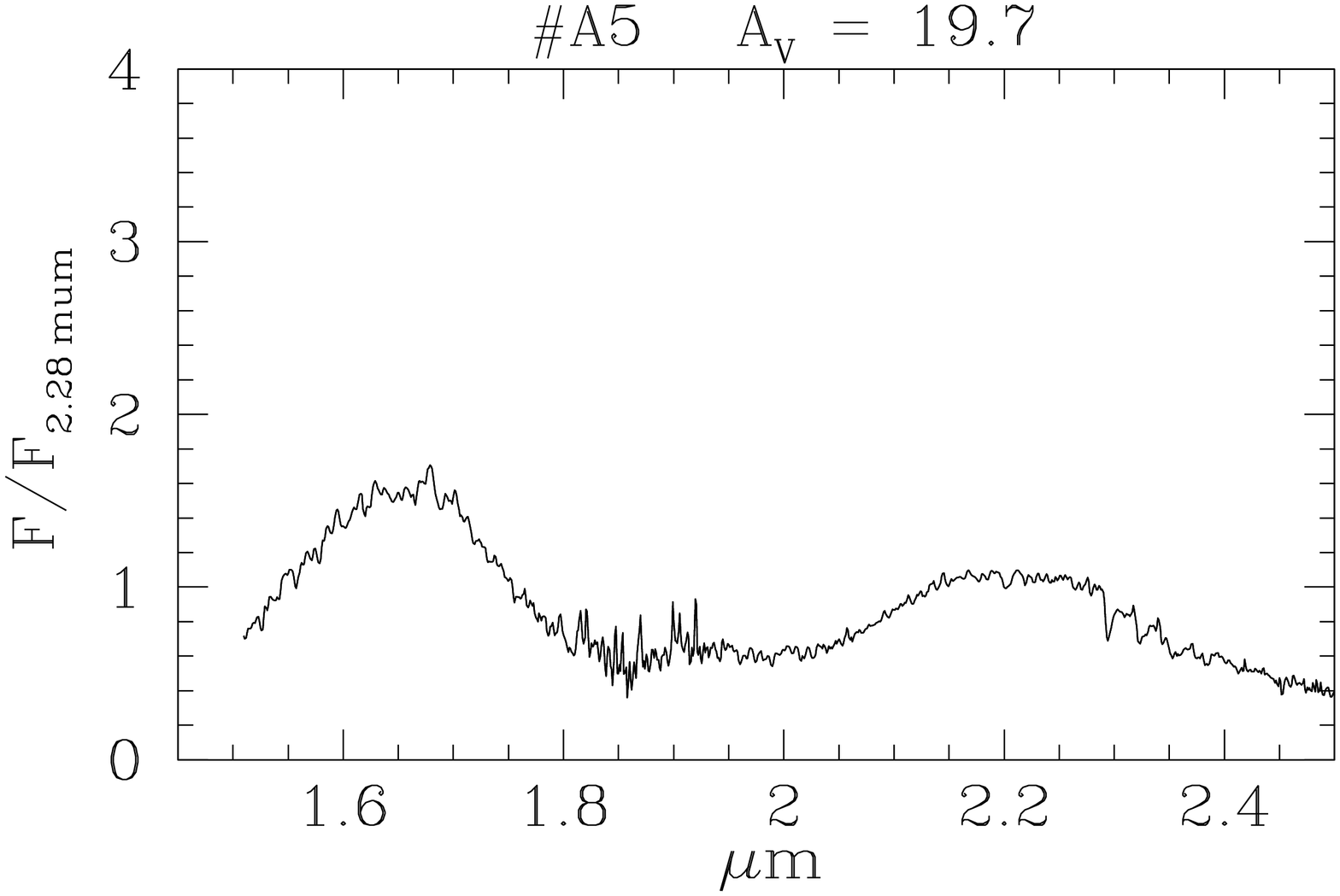,width=4.3cm,angle=270}   \epsfig{file=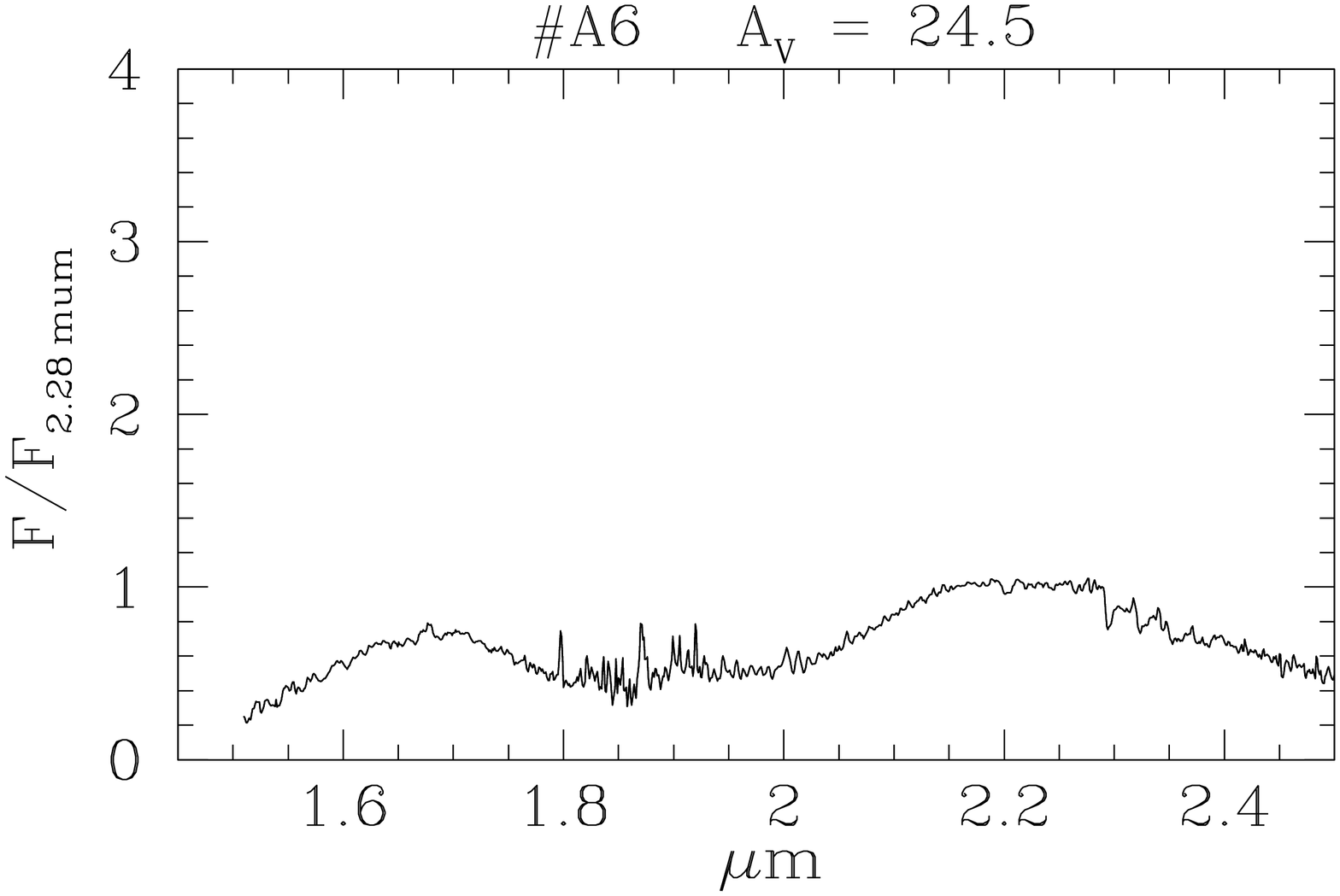,width=4.3cm,angle=270}

\epsfig{file=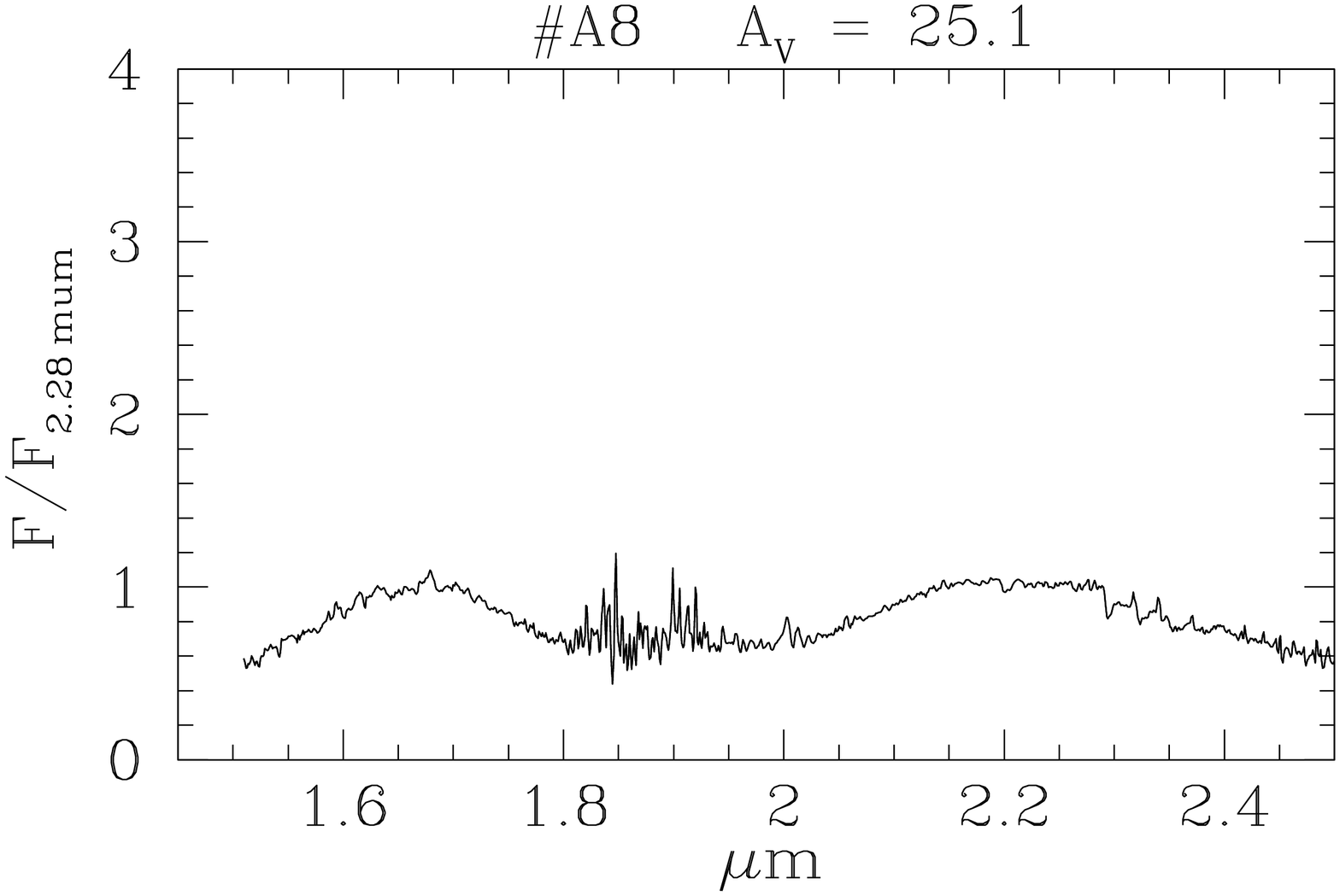,width=4.3cm,angle=270} \epsfig{file=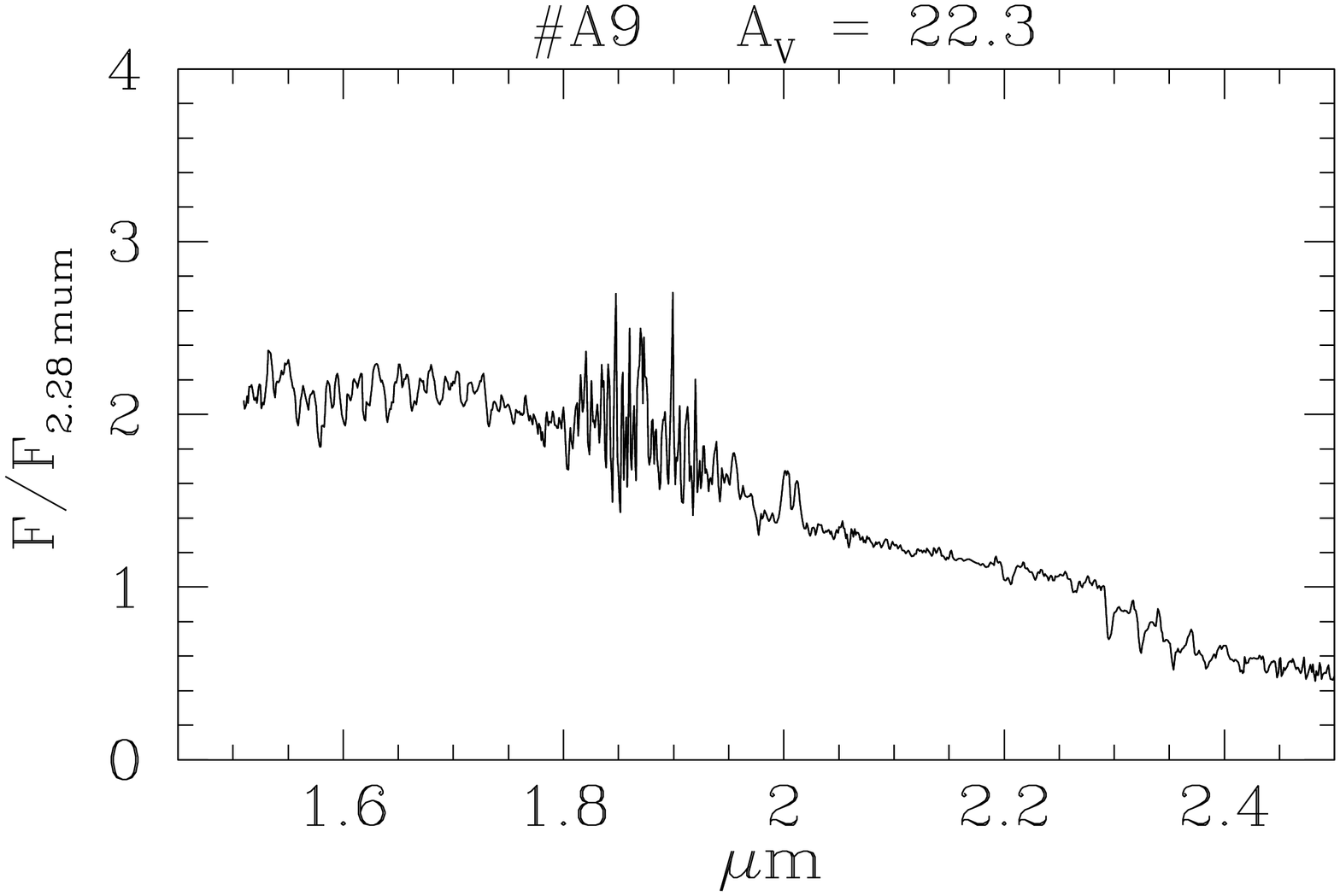,width=4.3cm,angle=270}   \epsfig{file=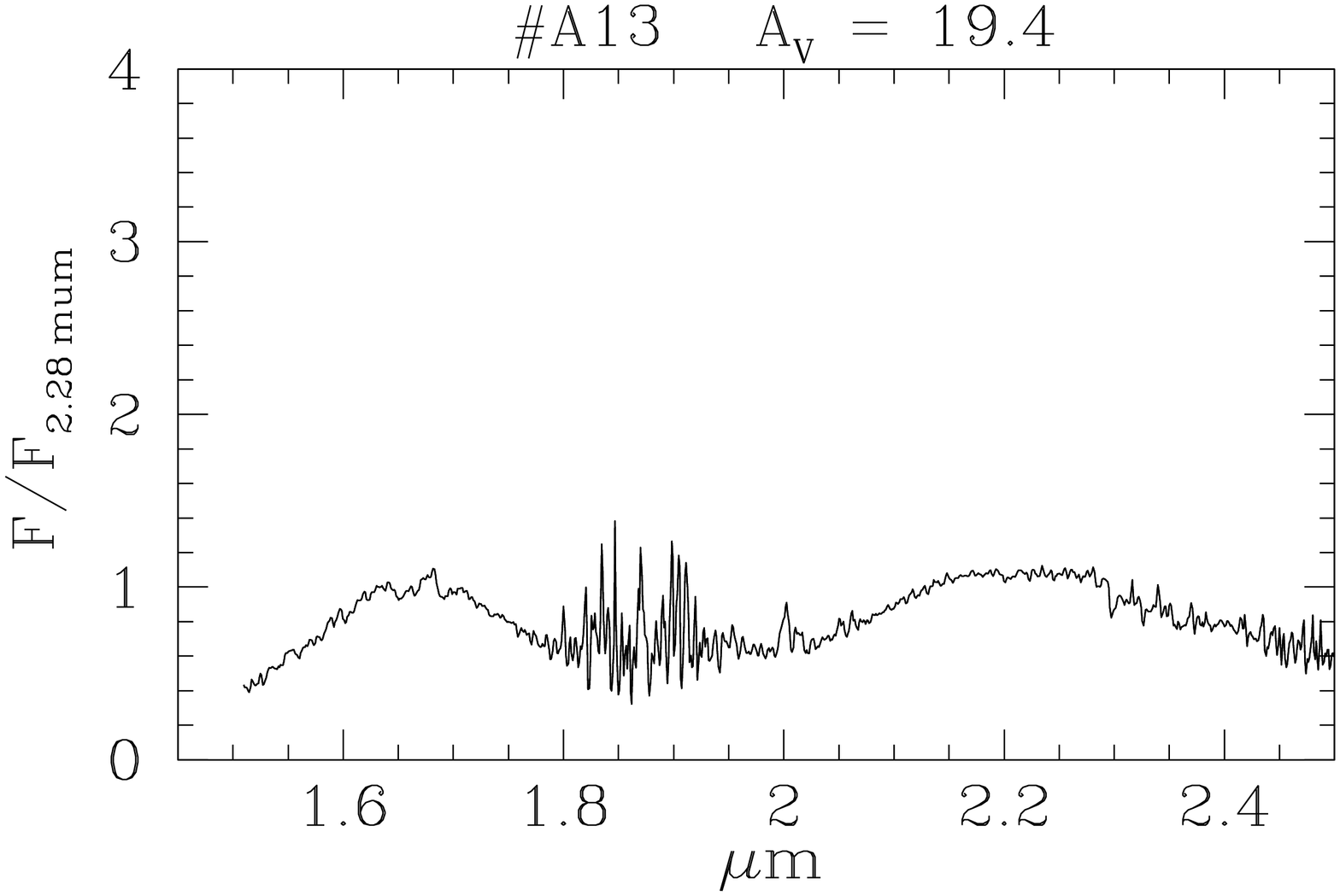,width=4.3cm,angle=270}

\epsfig{file=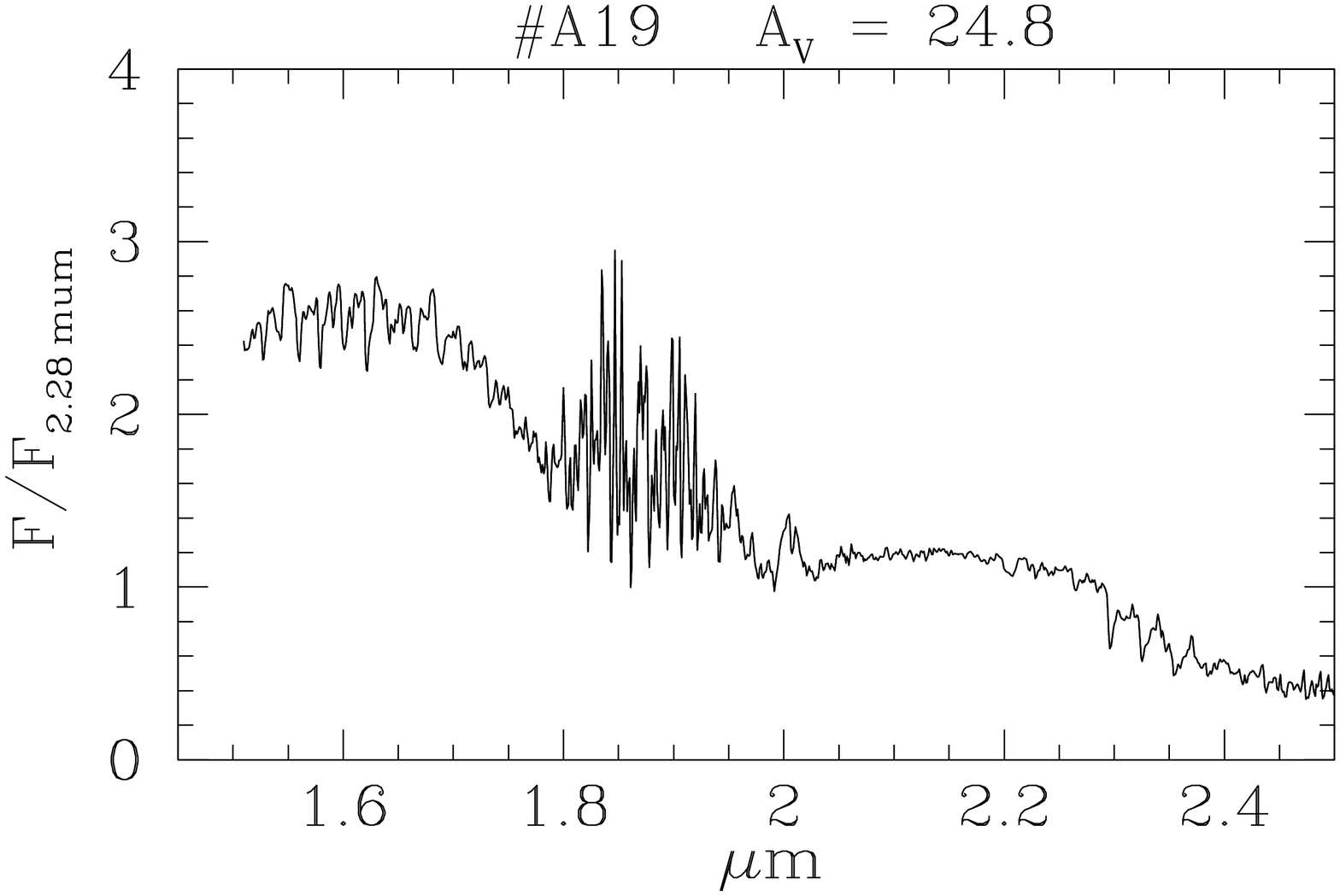,width=4.3cm,angle=270} \epsfig{file=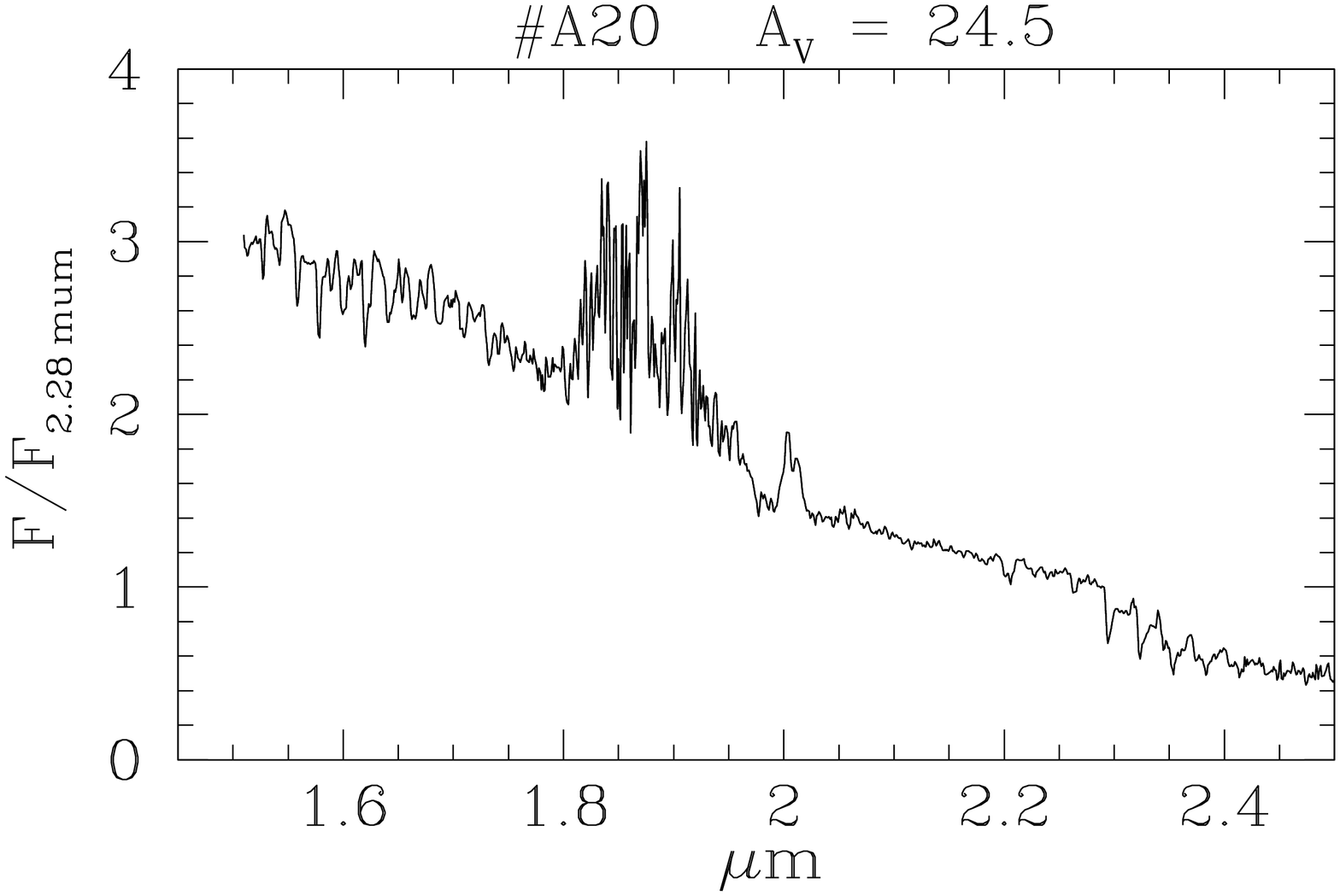,width=4.3cm,angle=270}   \epsfig{file=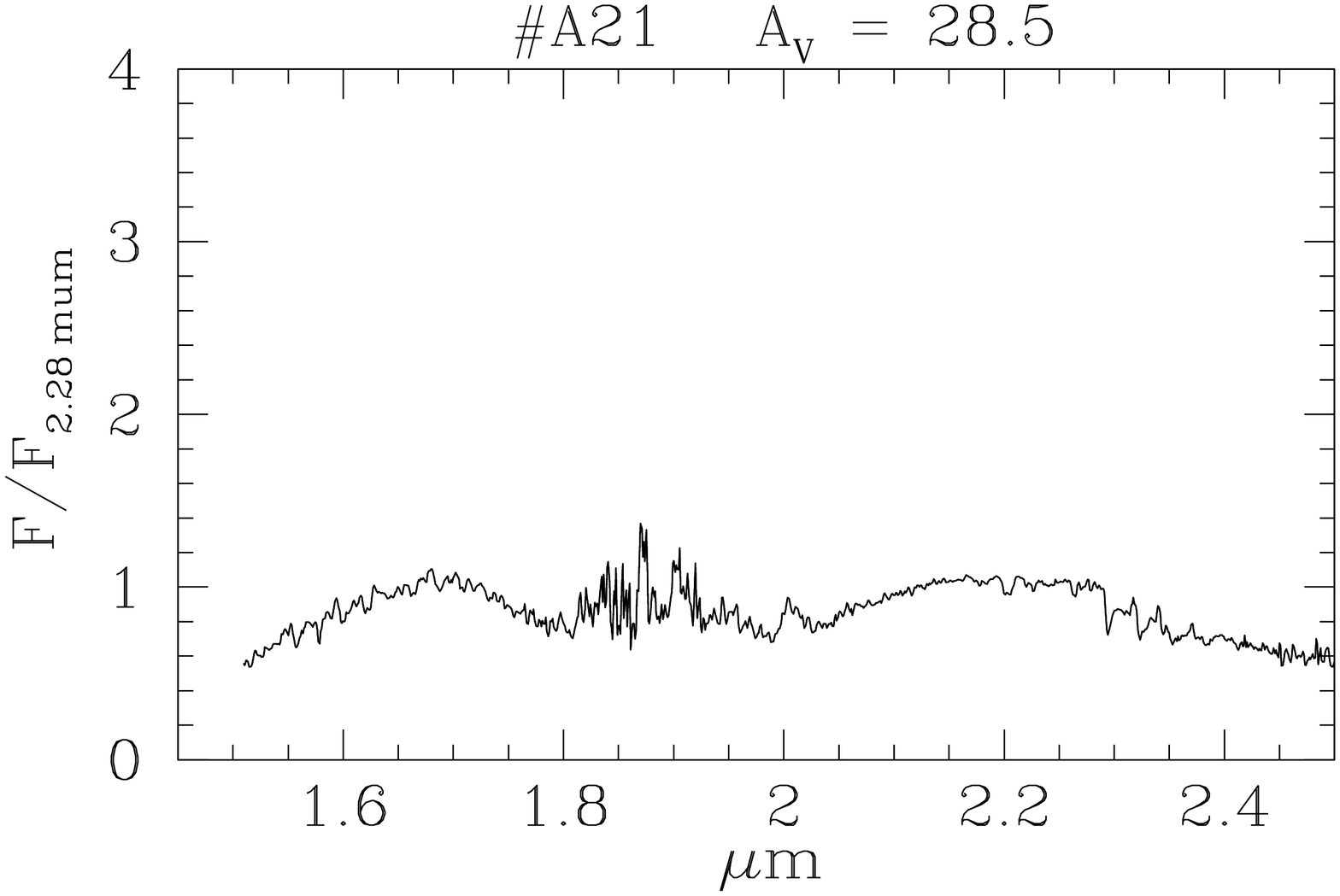,width=4.3cm,angle=270}

\epsfig{file=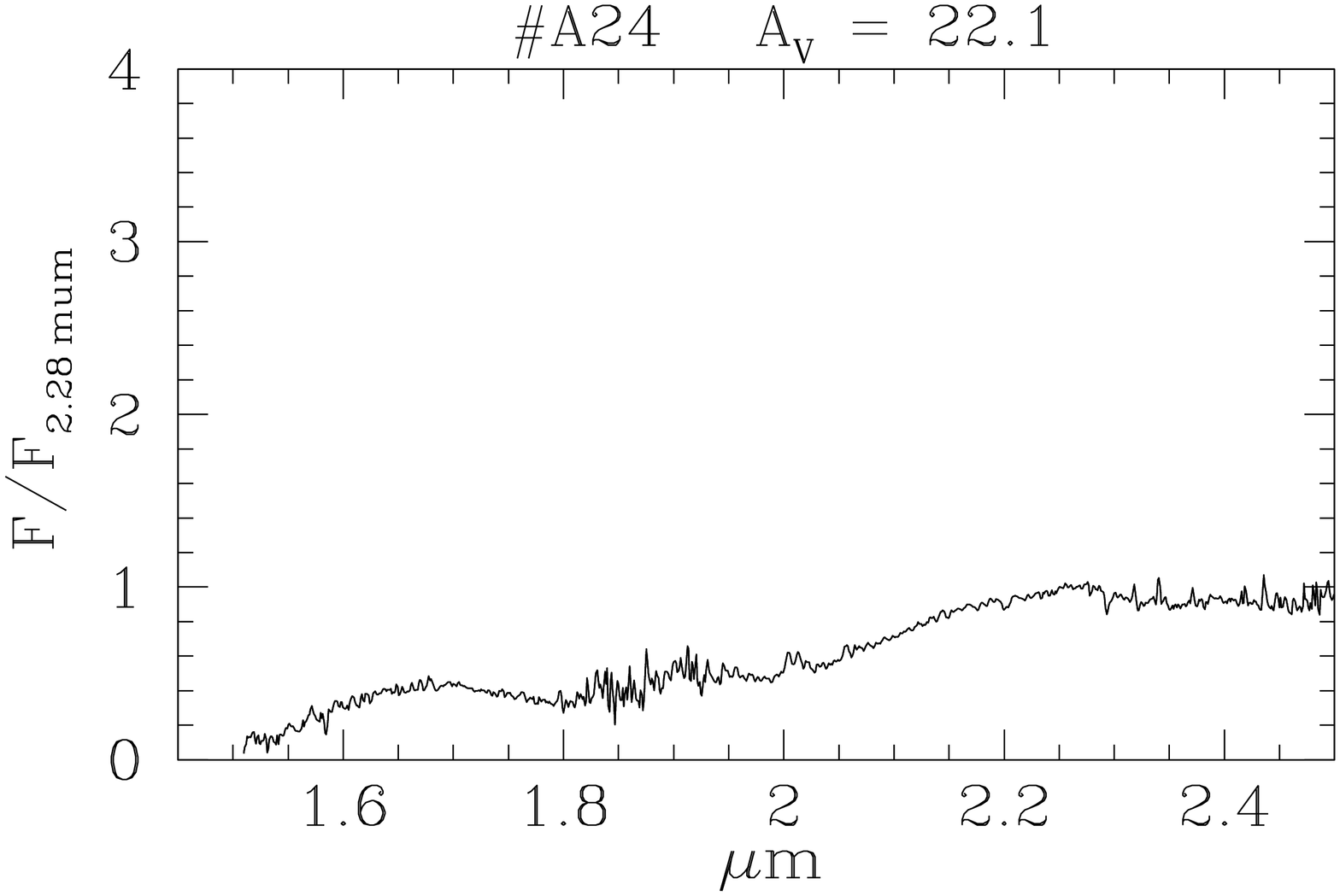,width=4.3cm,angle=270} \epsfig{file=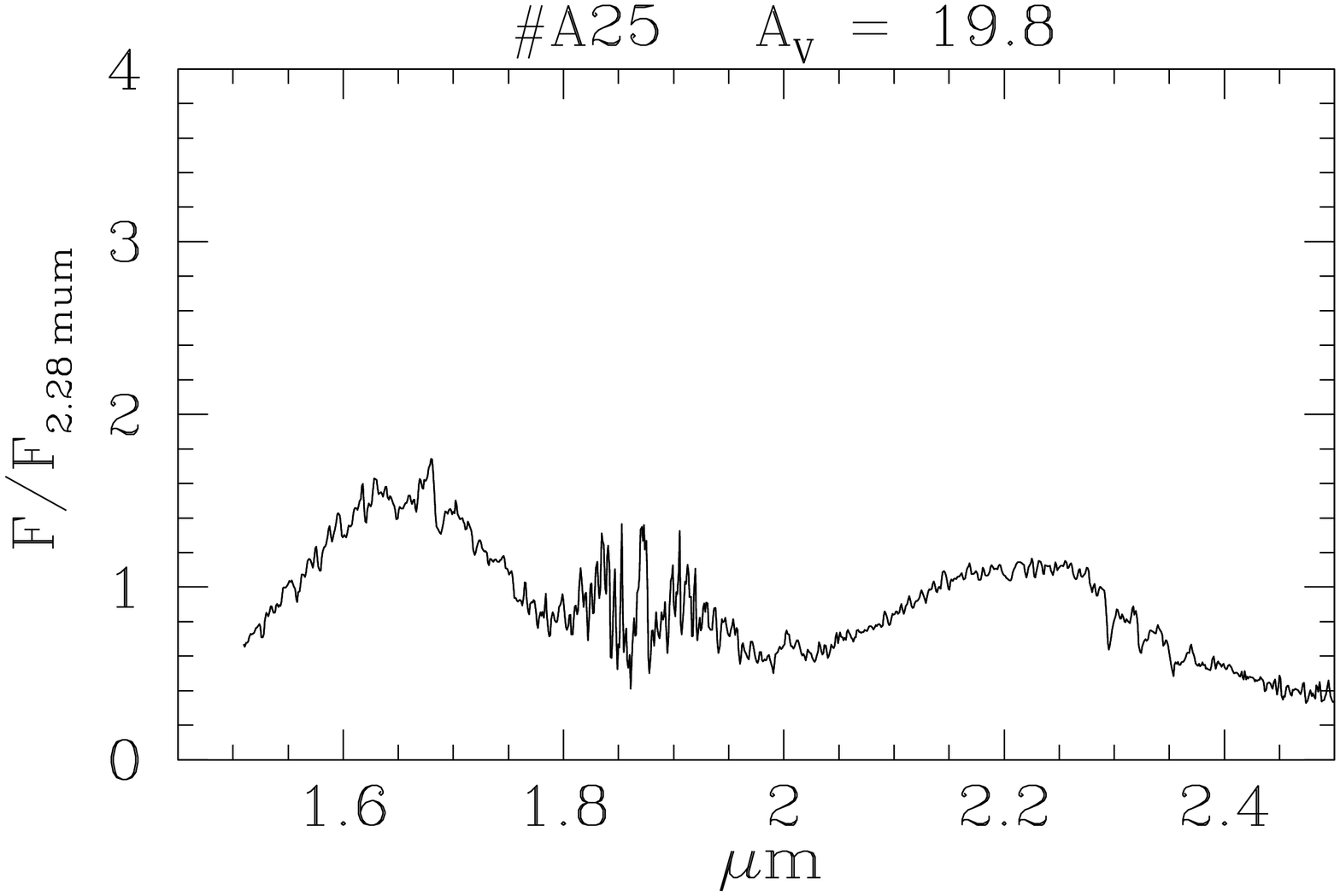,width=4.3cm,angle=270}   \epsfig{file=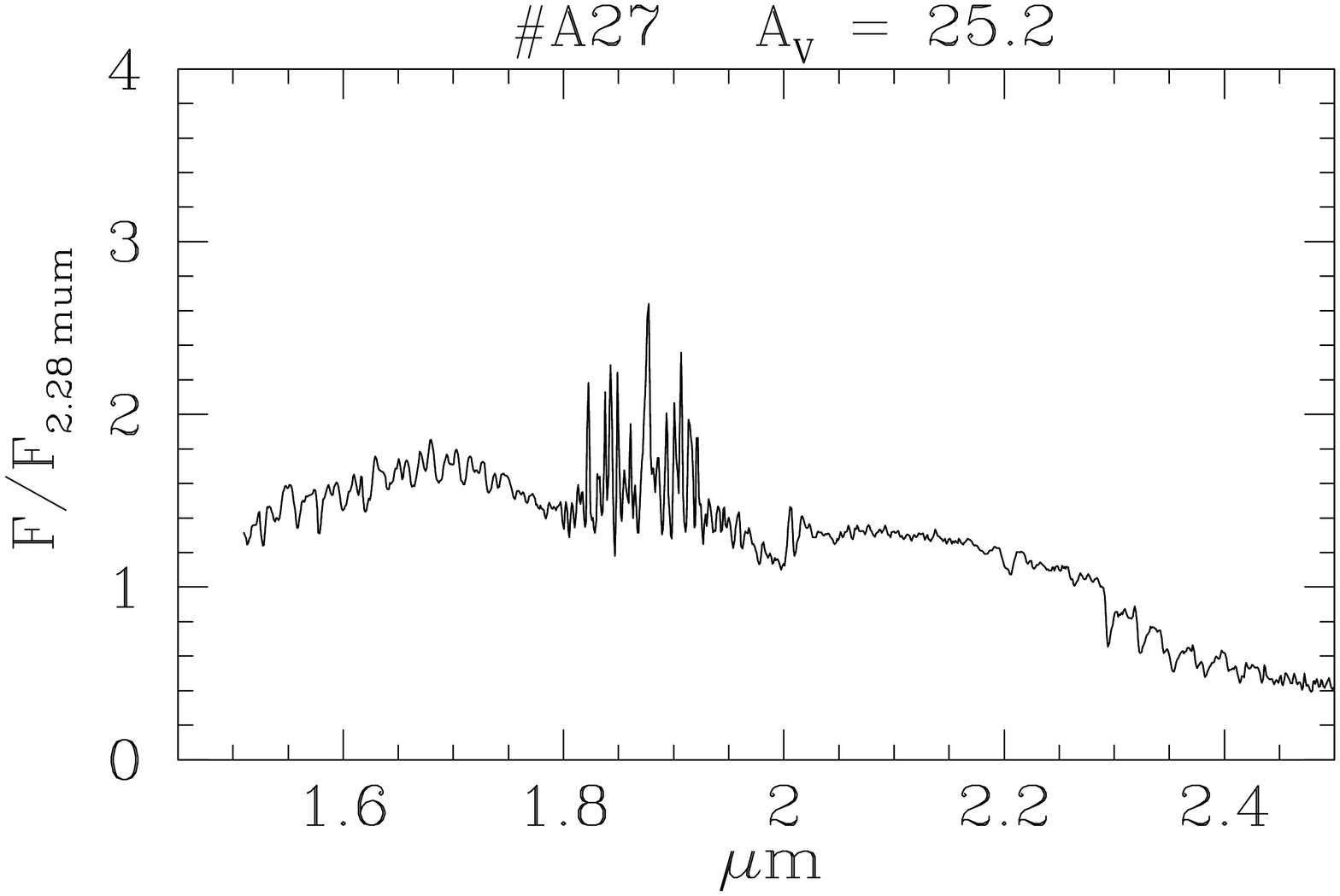,width=4.3cm,angle=270}

\epsfig{file=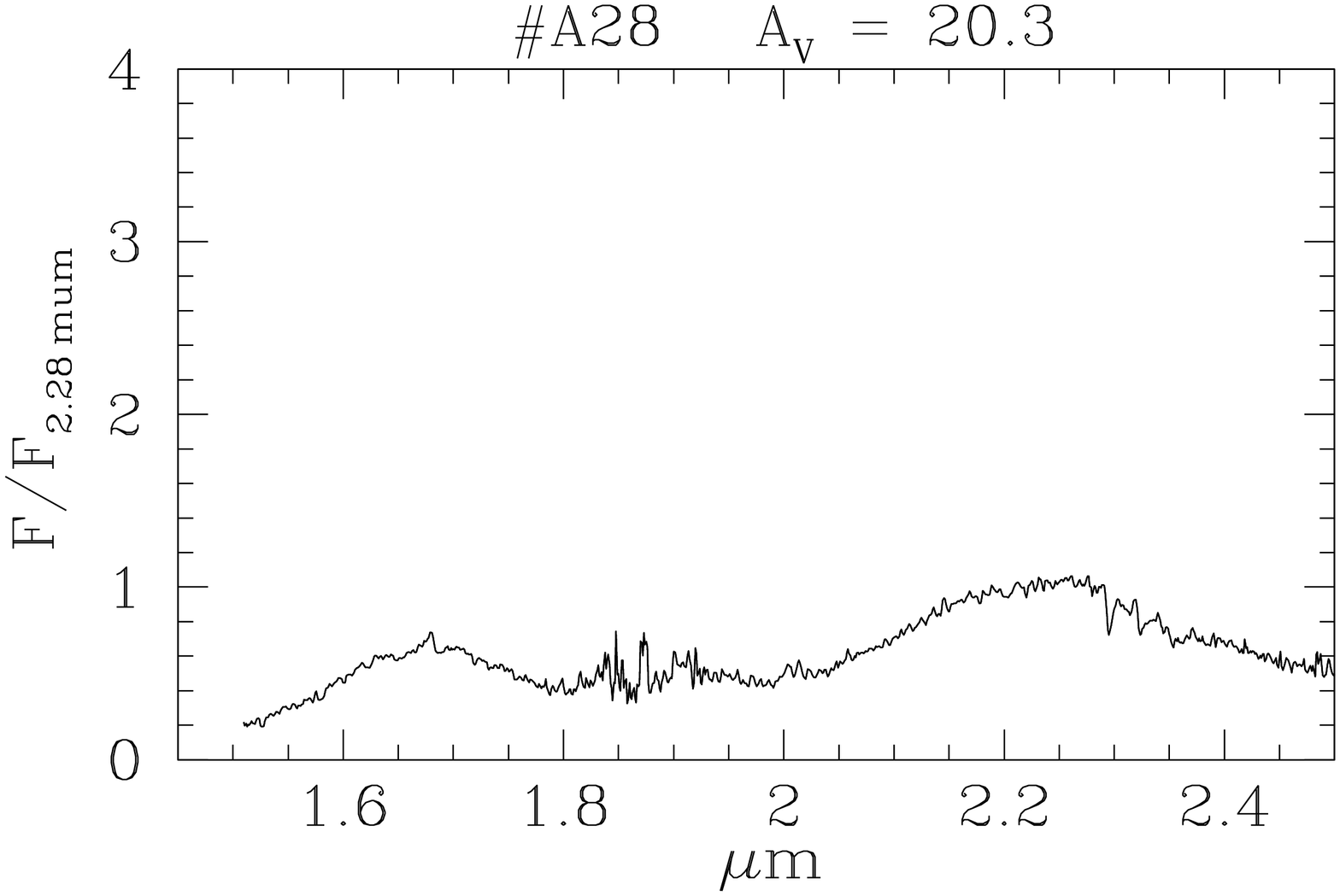,width=4.3cm,angle=270} \epsfig{file=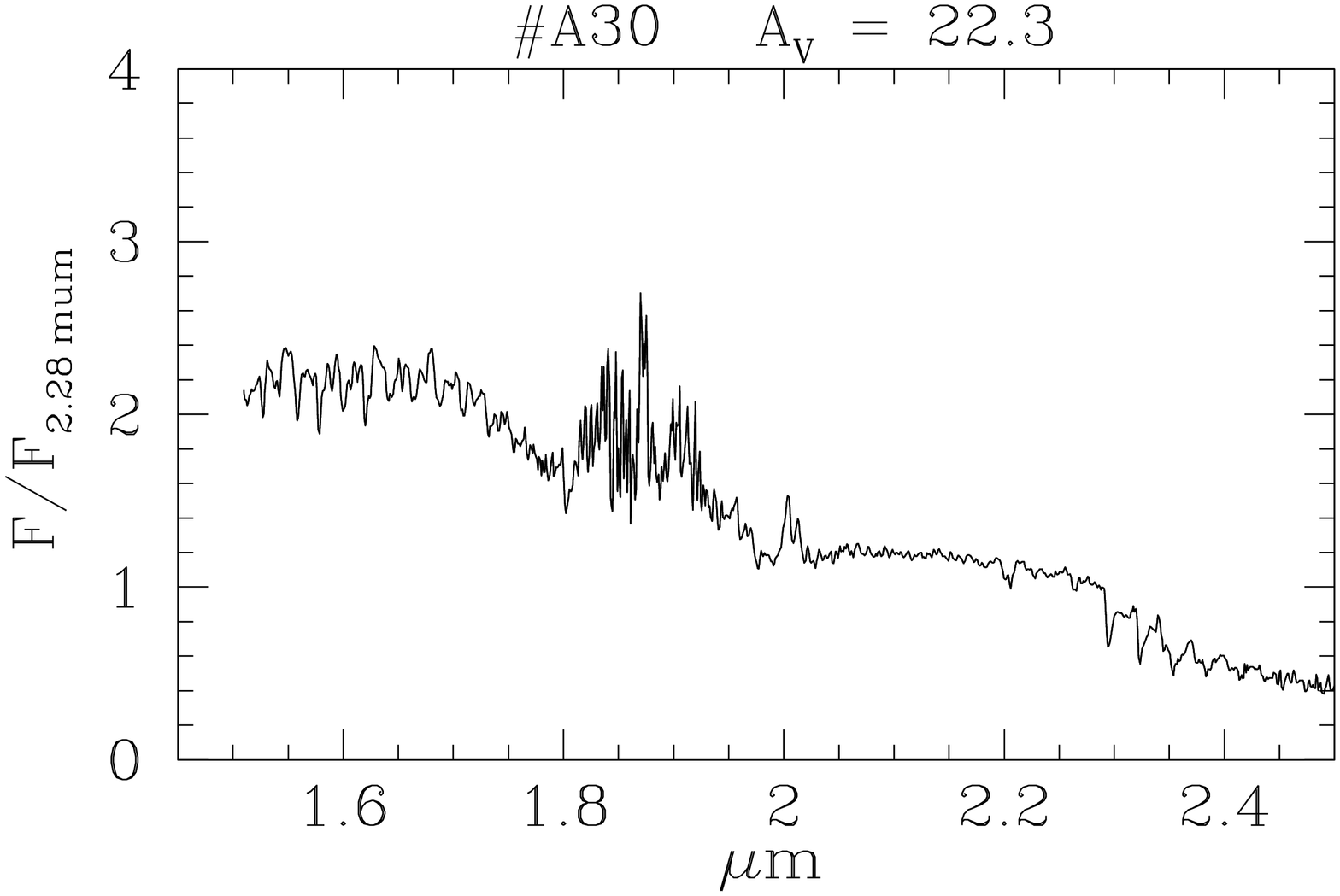,width=4.3cm,angle=270}   \epsfig{file=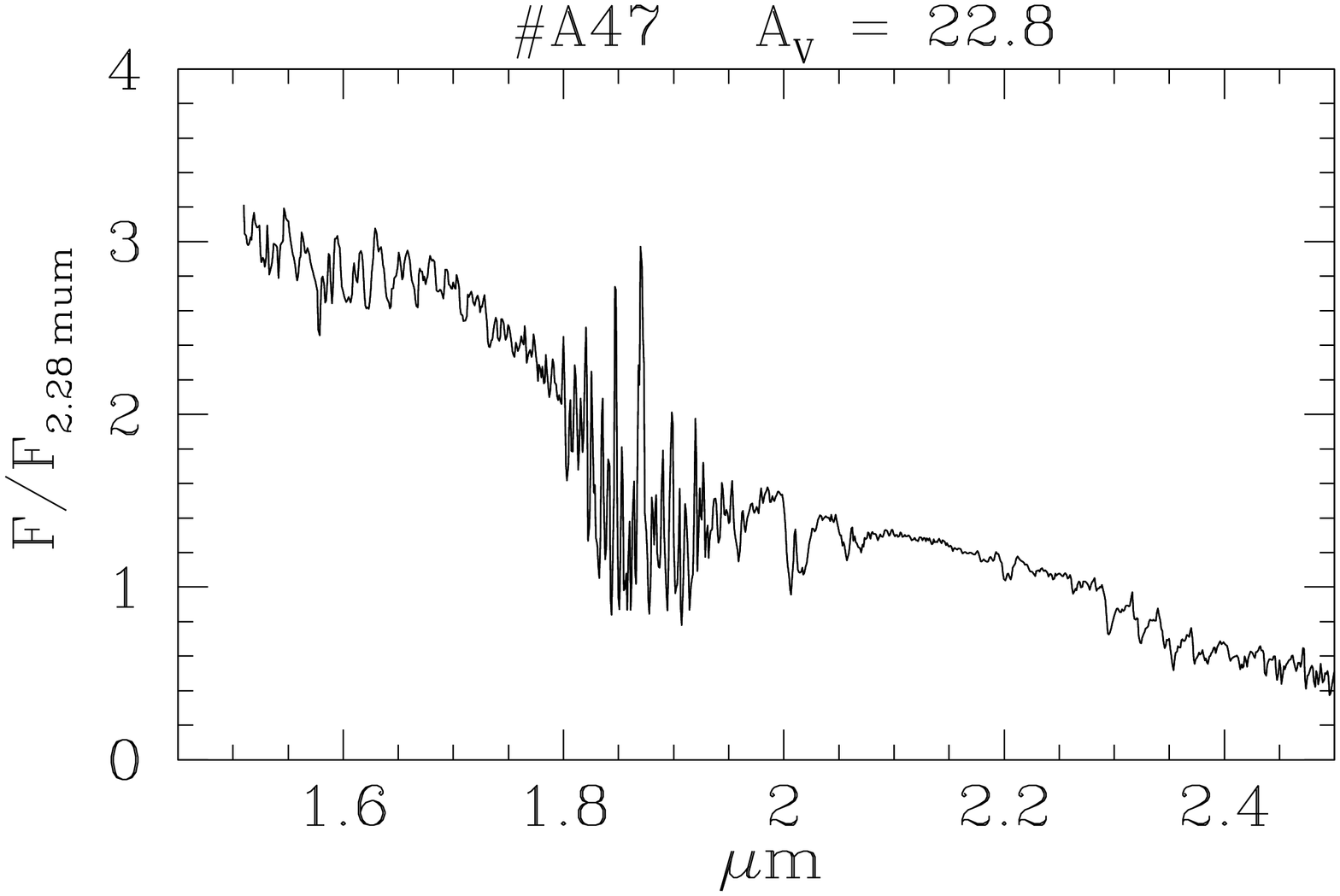,width=4.3cm,angle=270}}

\end{figure*}

\begin{figure*}[H!]
{\bf{Fig.~B.5. (continued)}}

{\epsfig{file=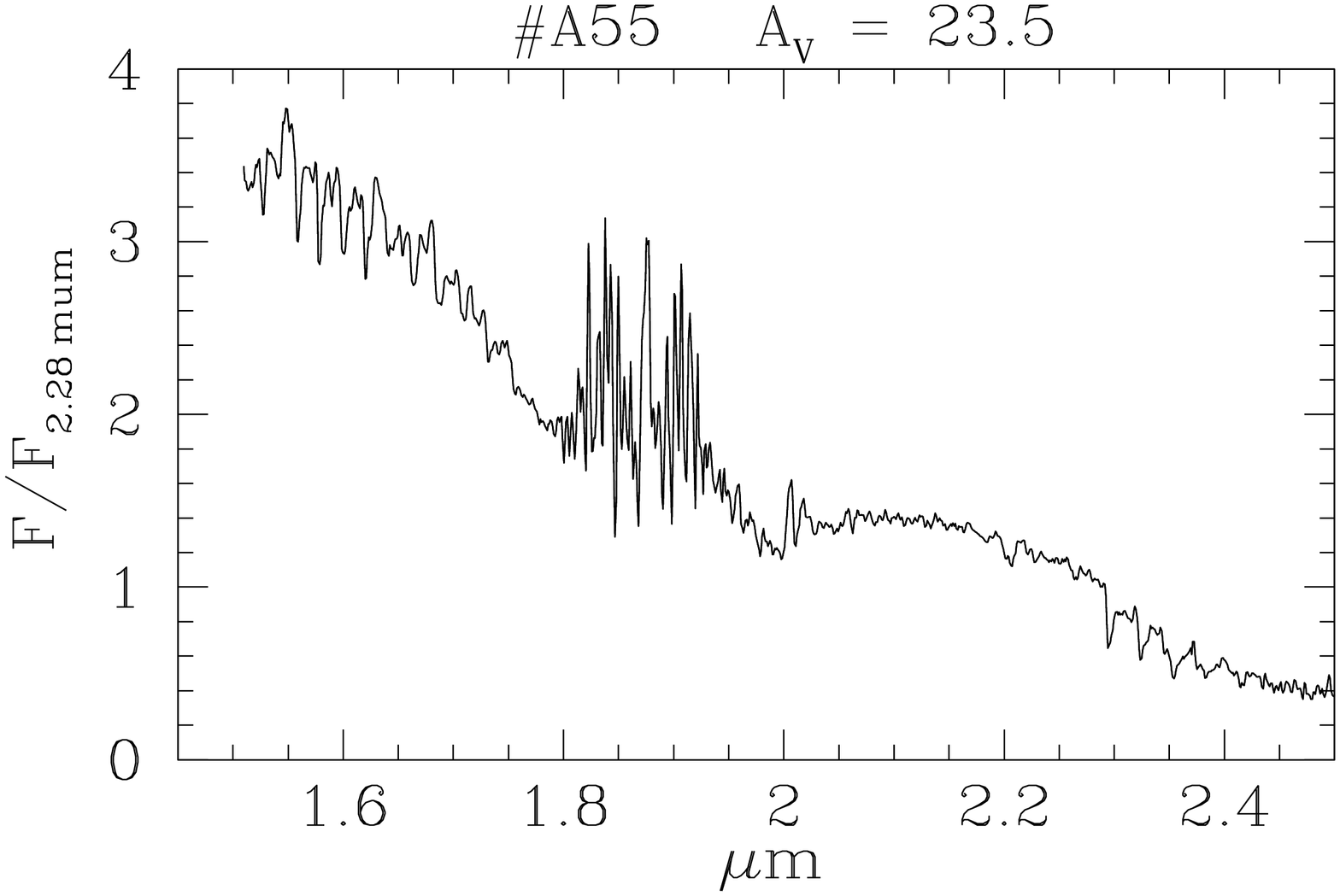,width=4.3cm,angle=270} \epsfig{file=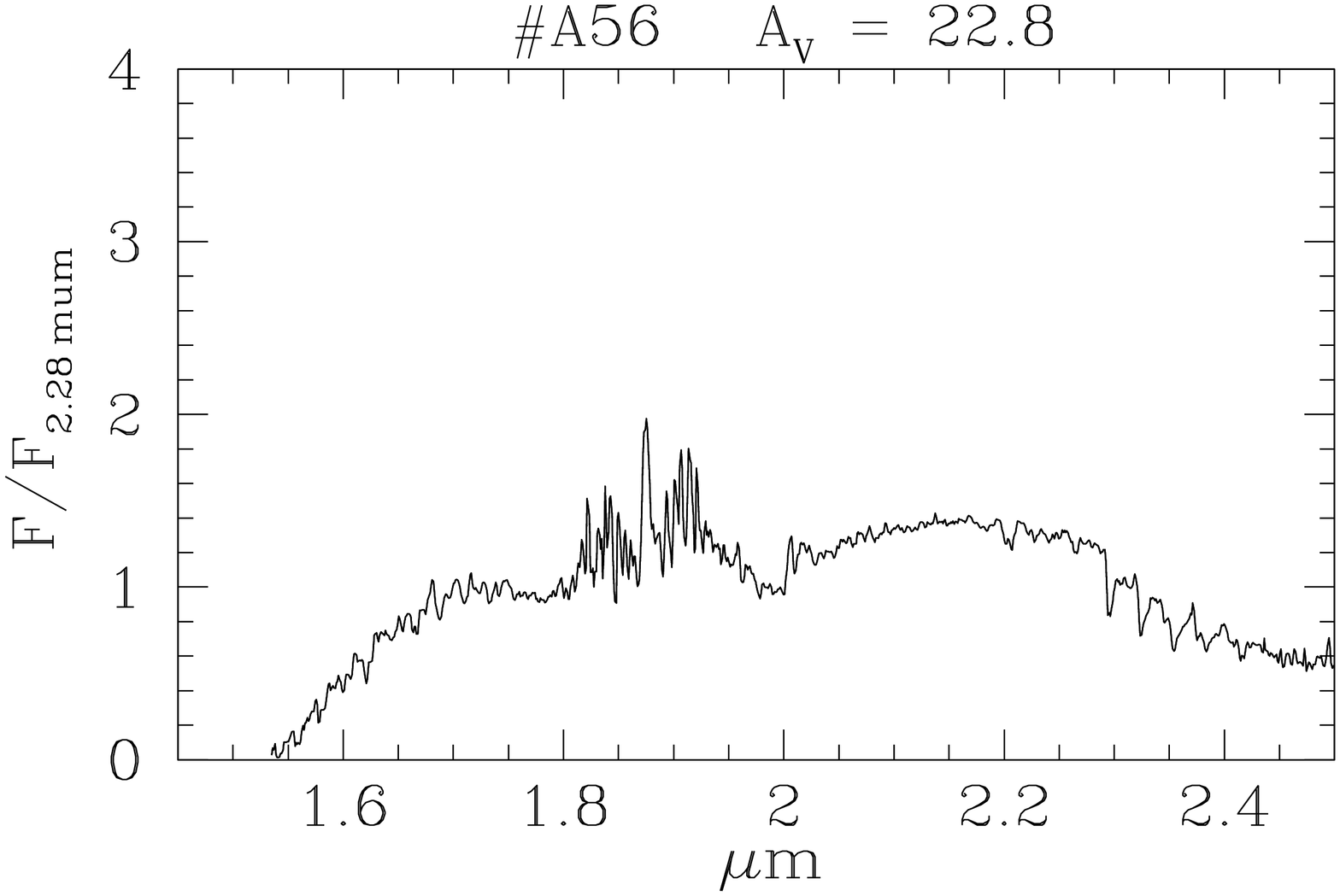,width=4.3cm,angle=270}   \epsfig{file=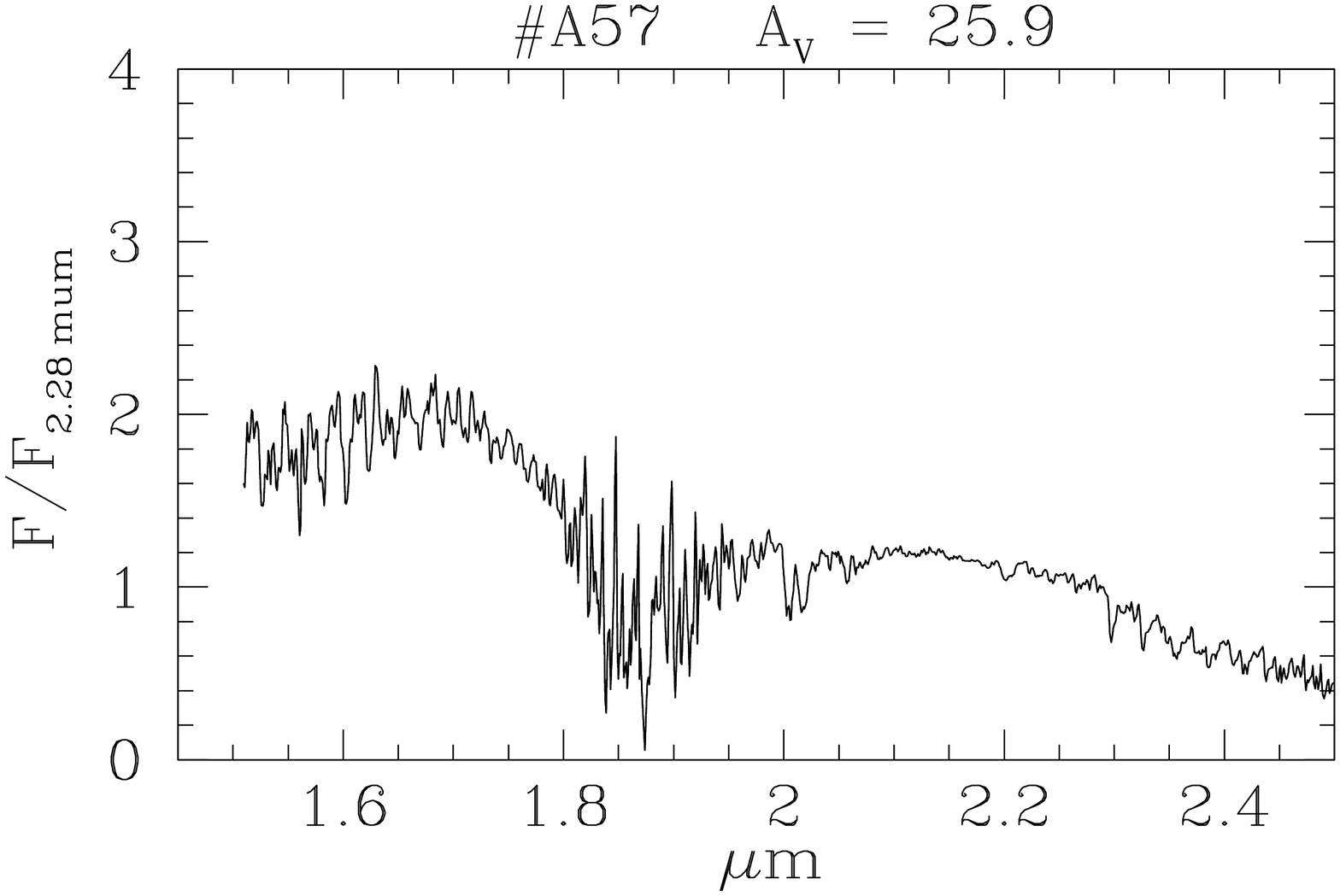,width=4.3cm,angle=270}

\epsfig{file=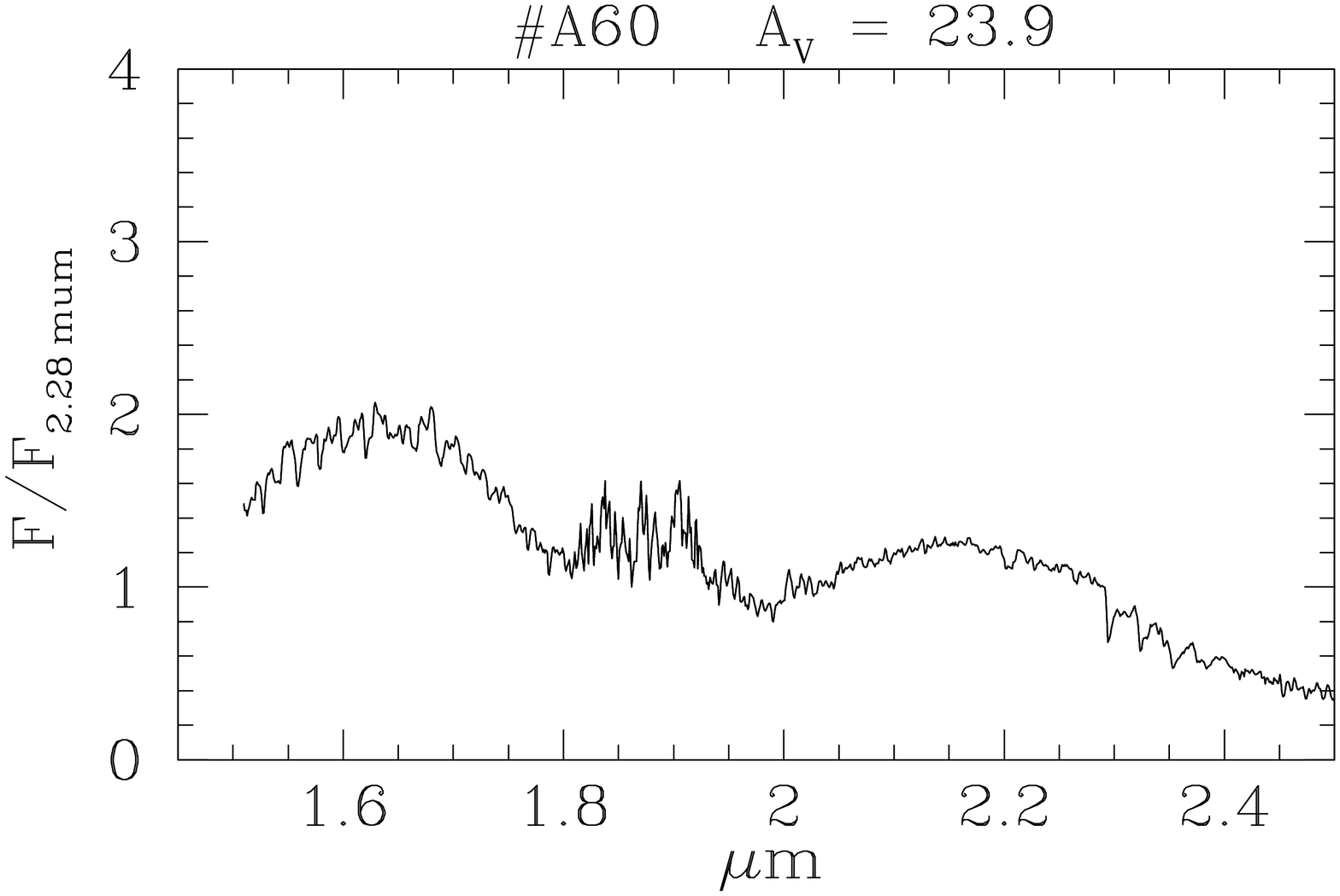,width=4.3cm,angle=270} \epsfig{file=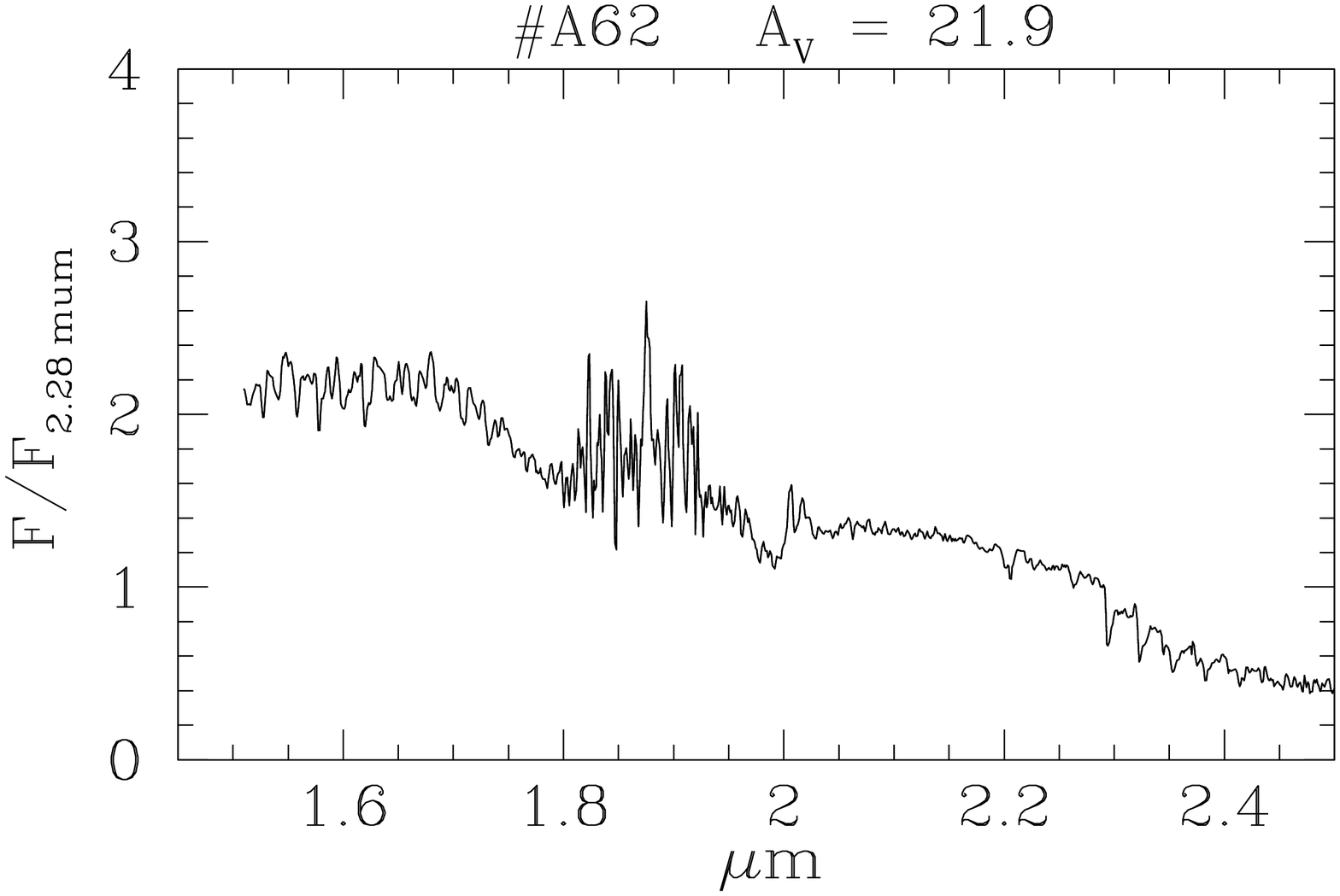,width=4.3cm,angle=270}   \epsfig{file=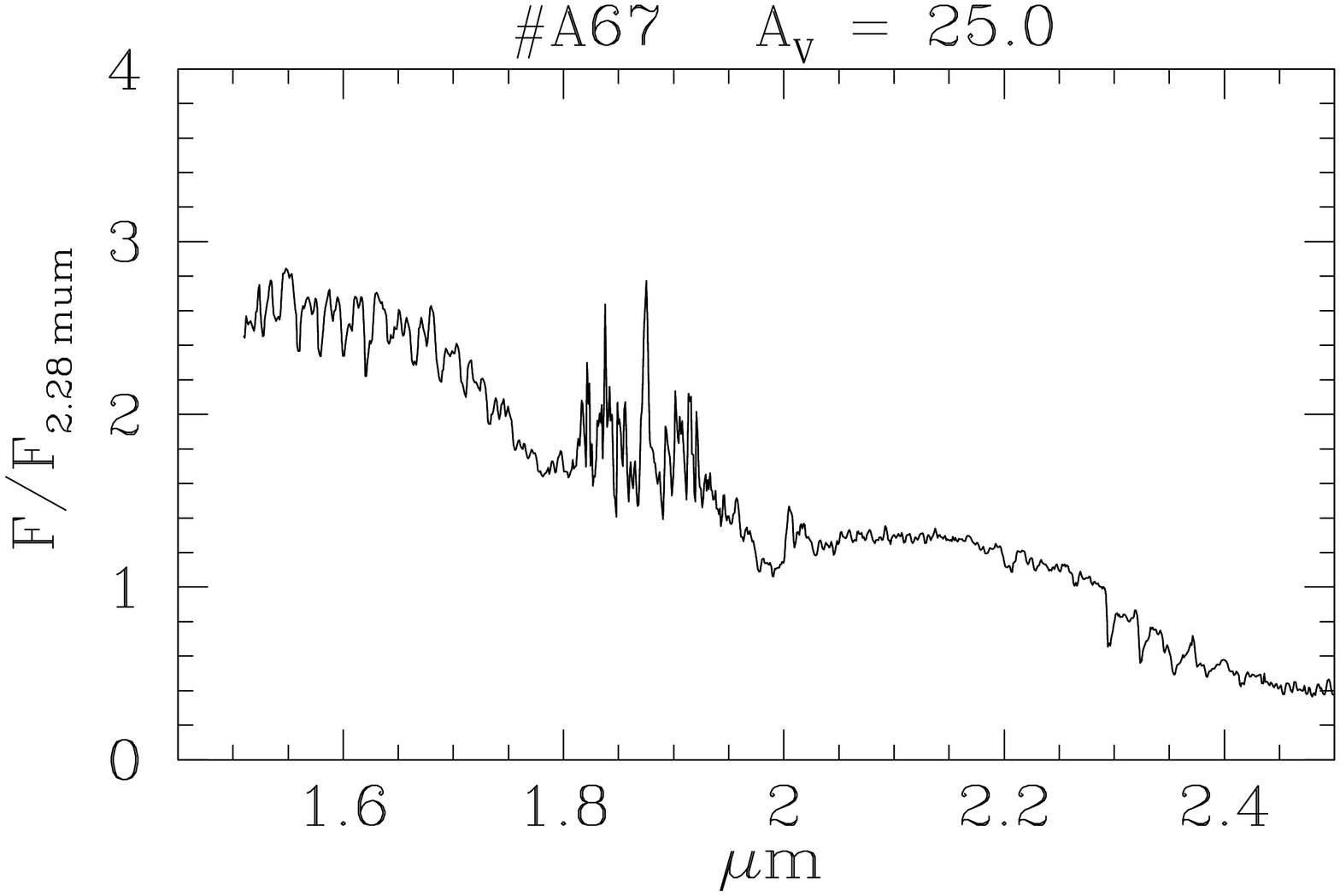,width=4.3cm,angle=270}

\epsfig{file=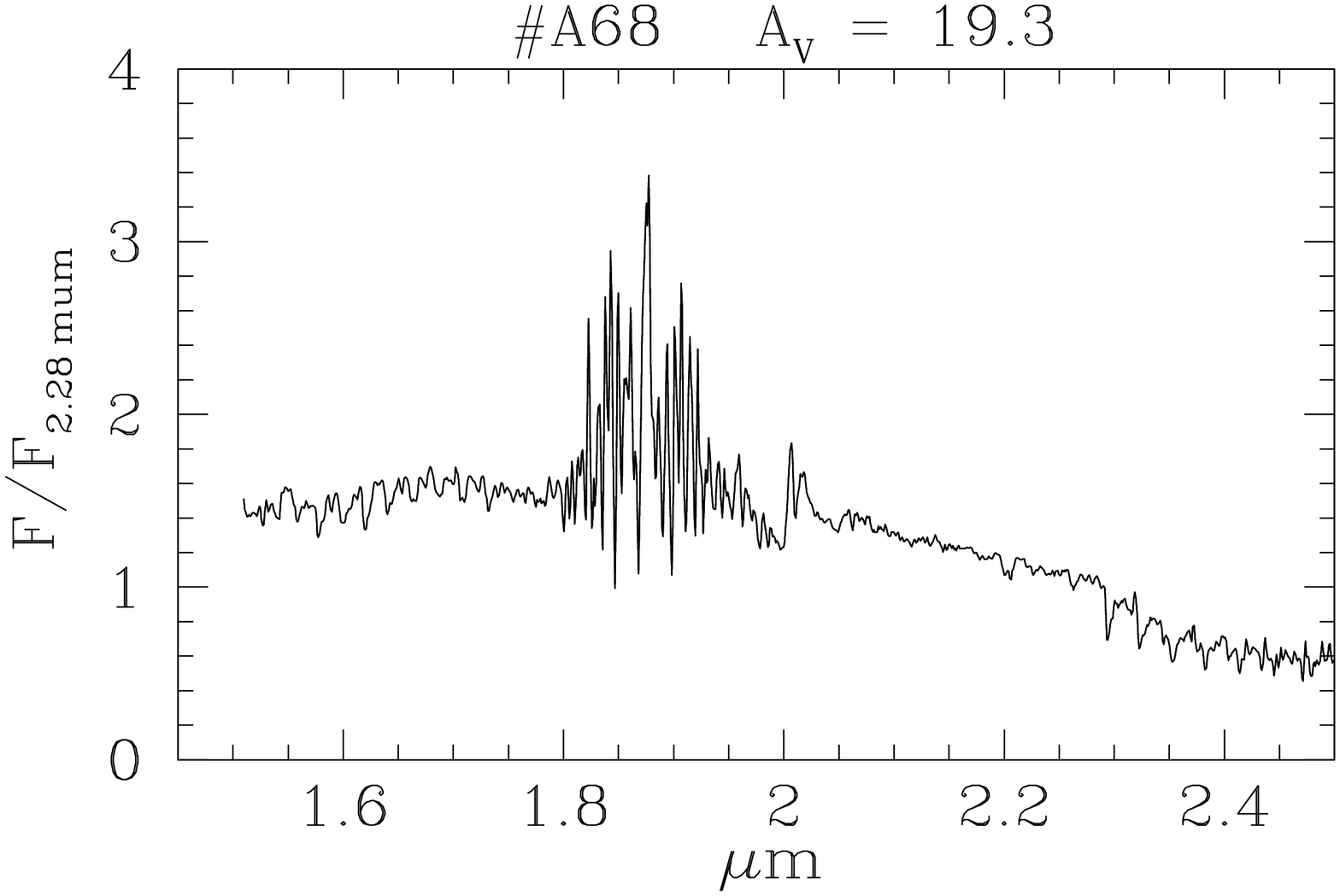,width=4.3cm,angle=270} \epsfig{file=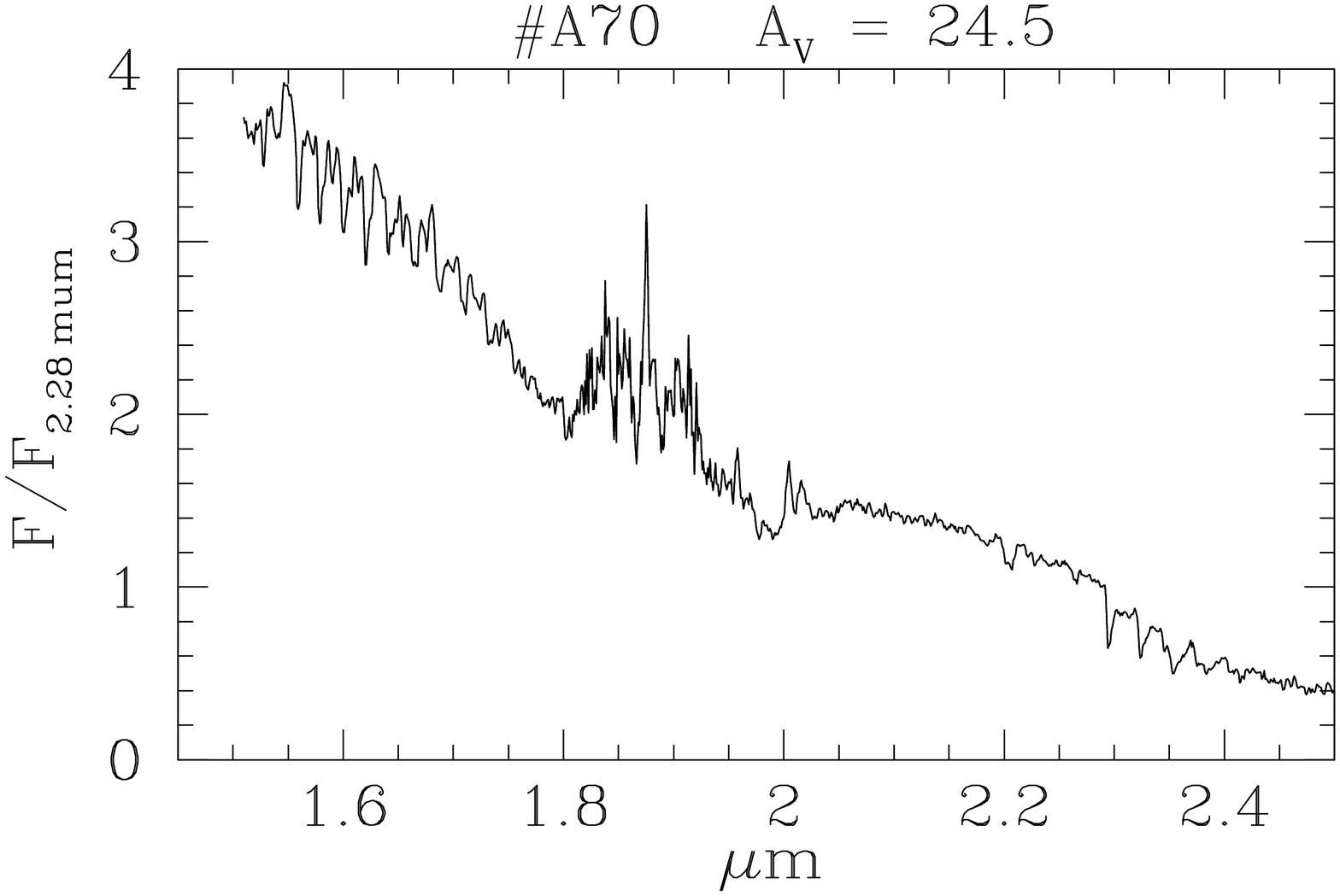,width=4.3cm,angle=270}   \epsfig{file=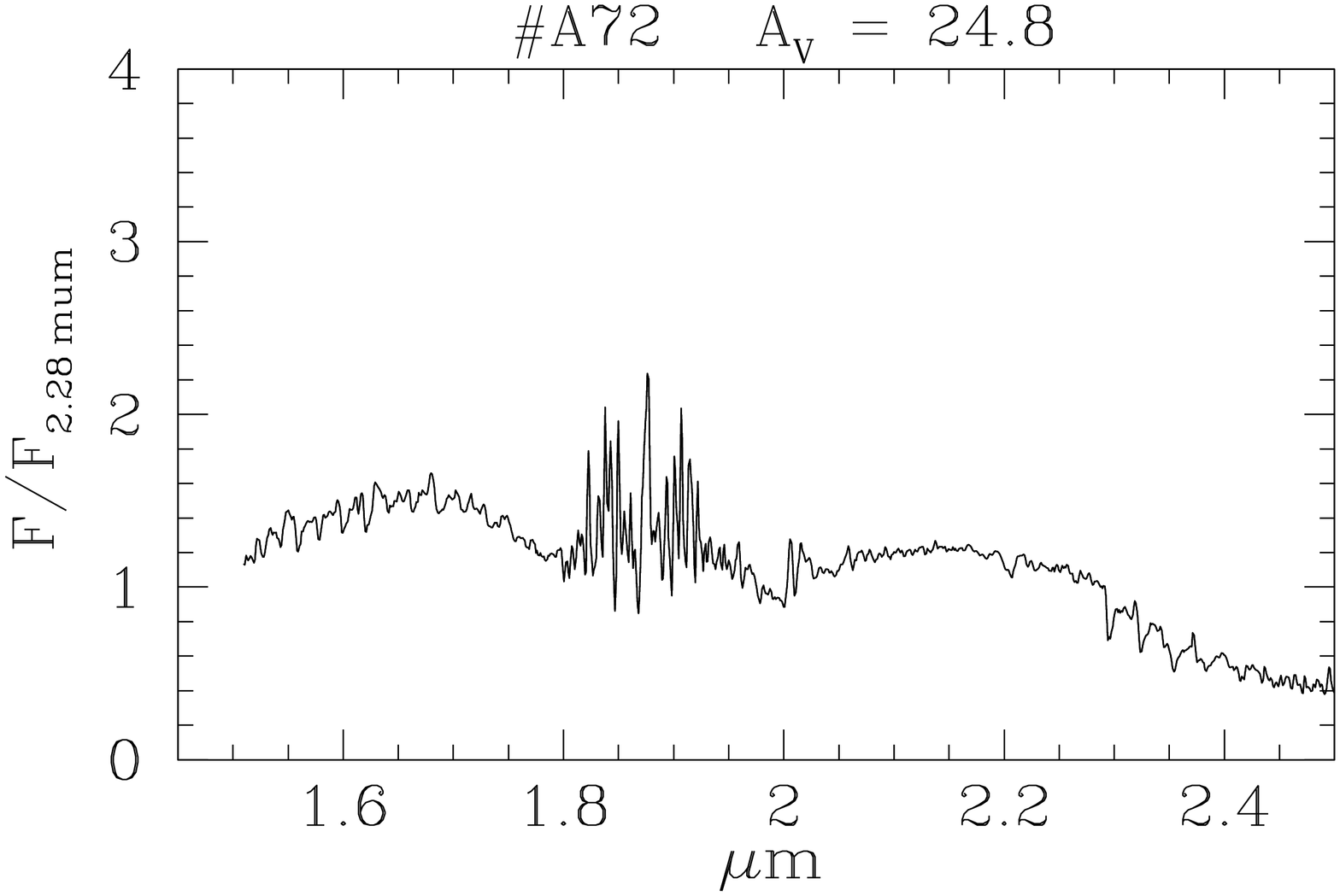,width=4.3cm,angle=270}

\epsfig{file=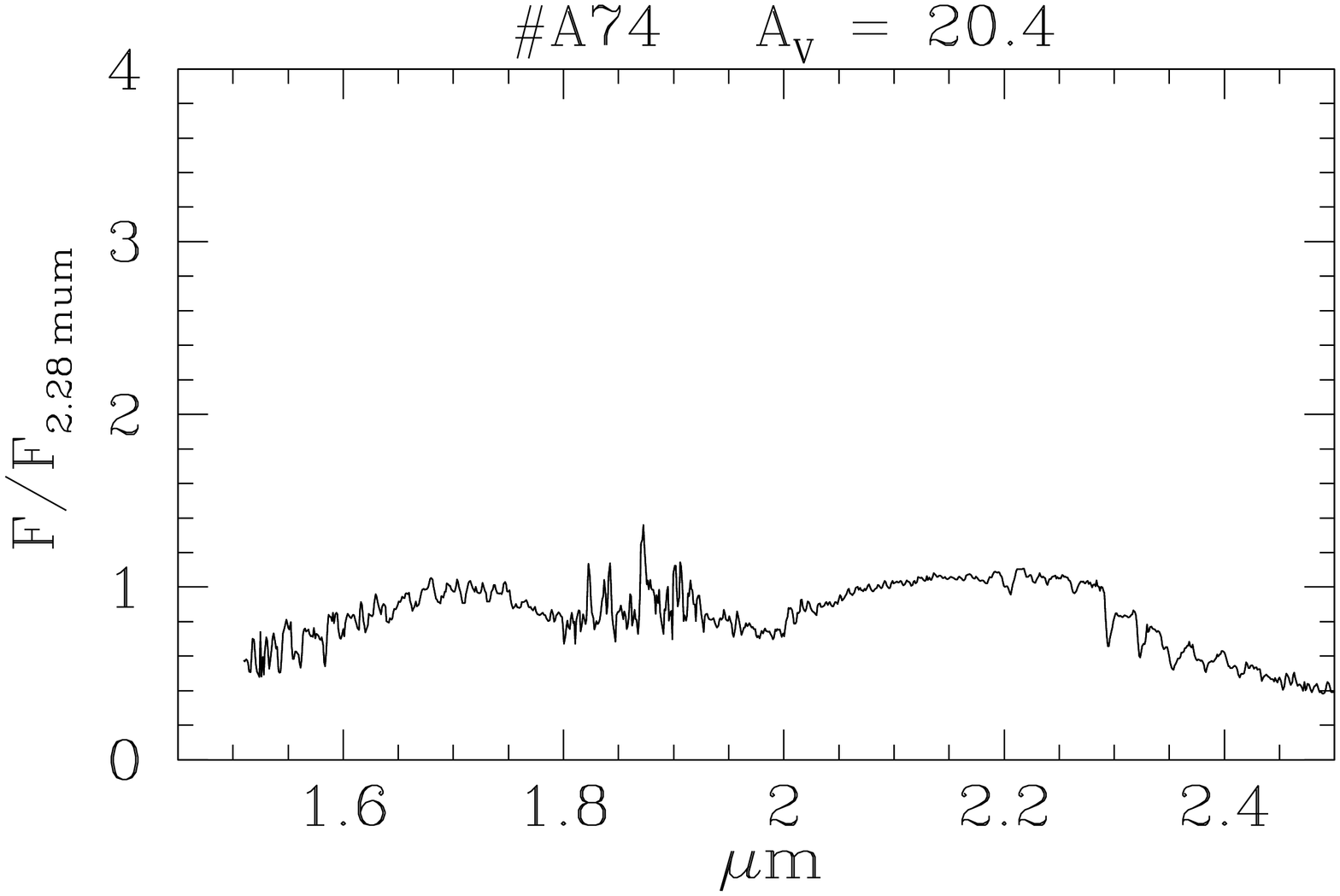,width=4.3cm,angle=270} \epsfig{file=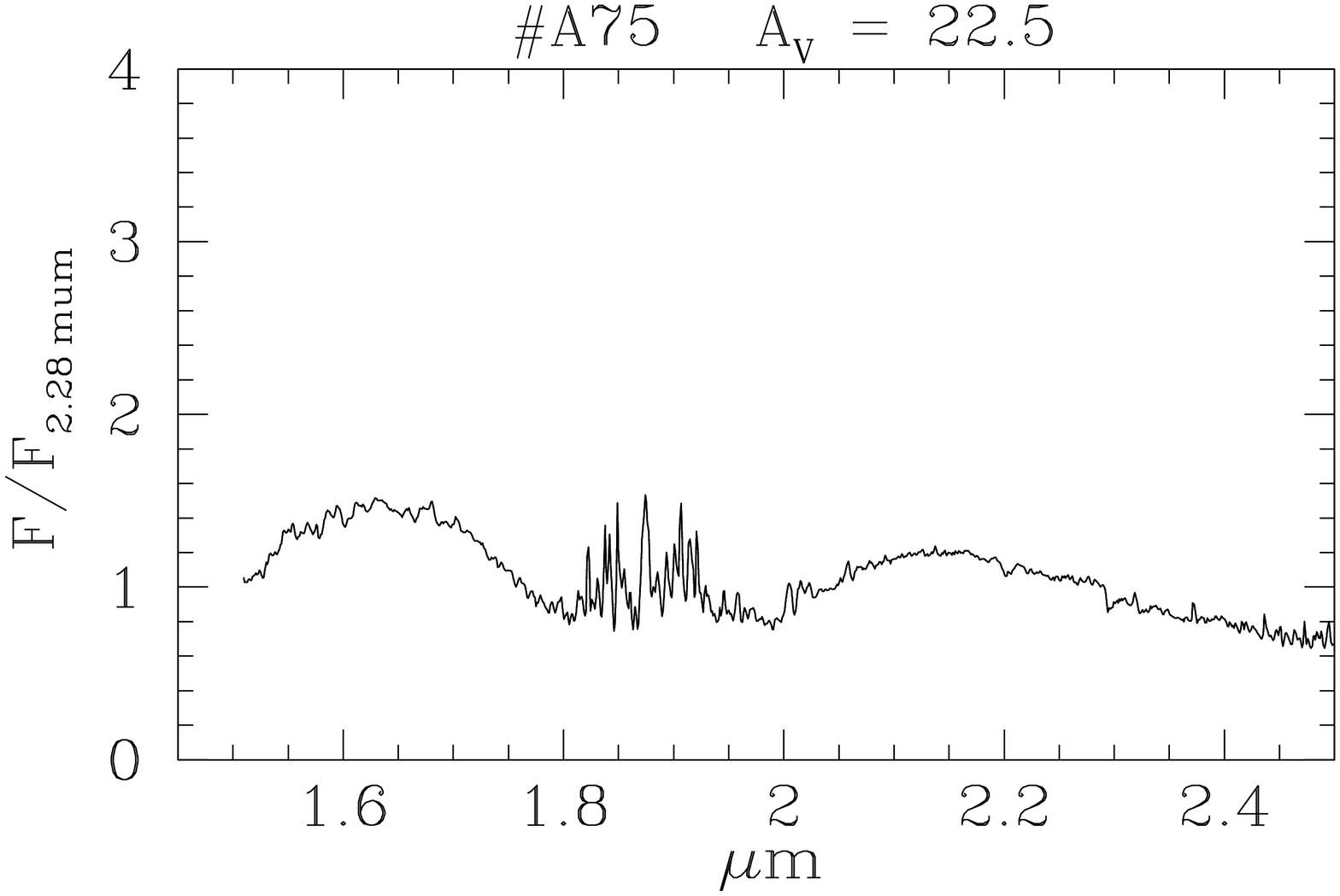,width=4.3cm,angle=270}   \epsfig{file=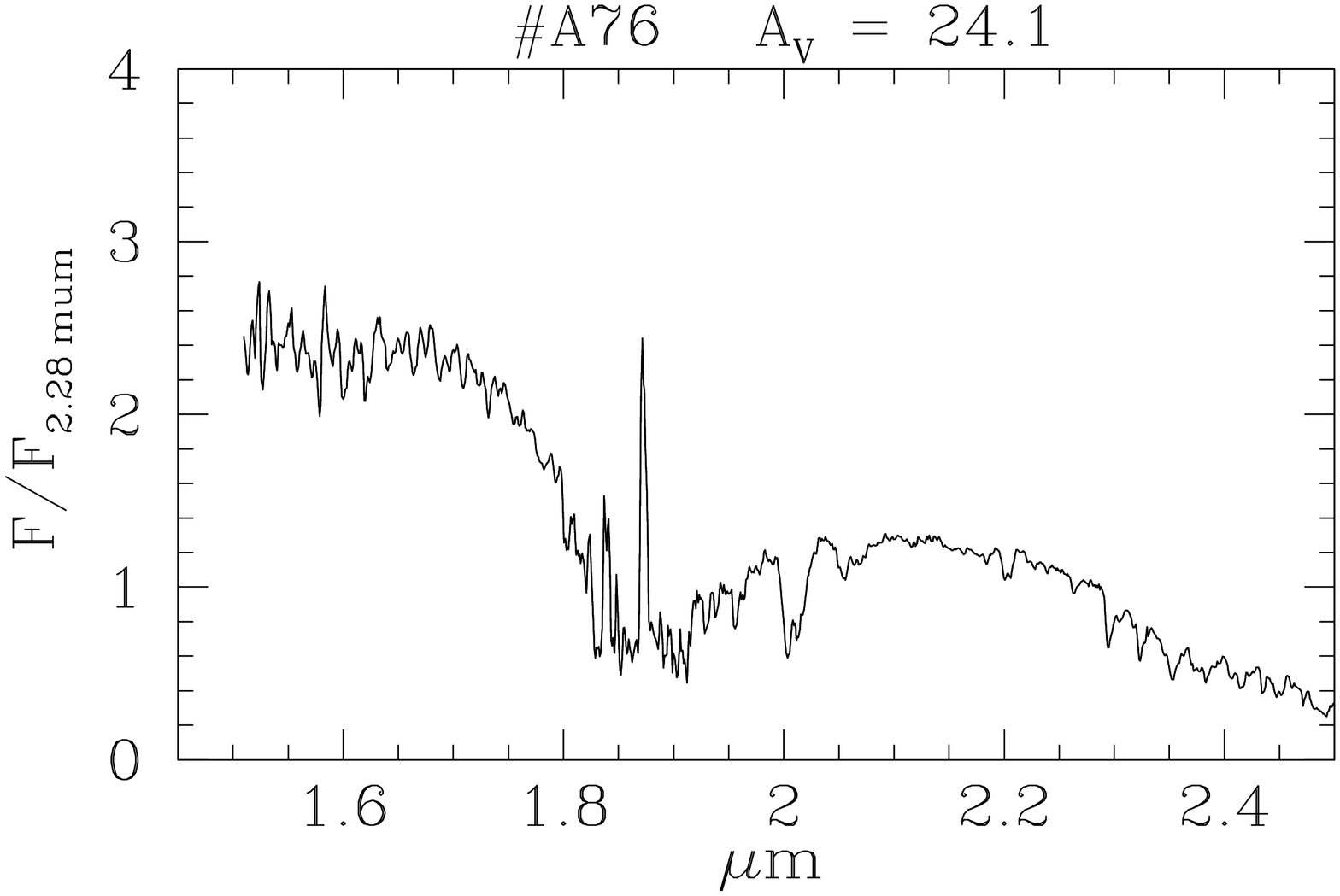,width=4.3cm,angle=270}

\epsfig{file=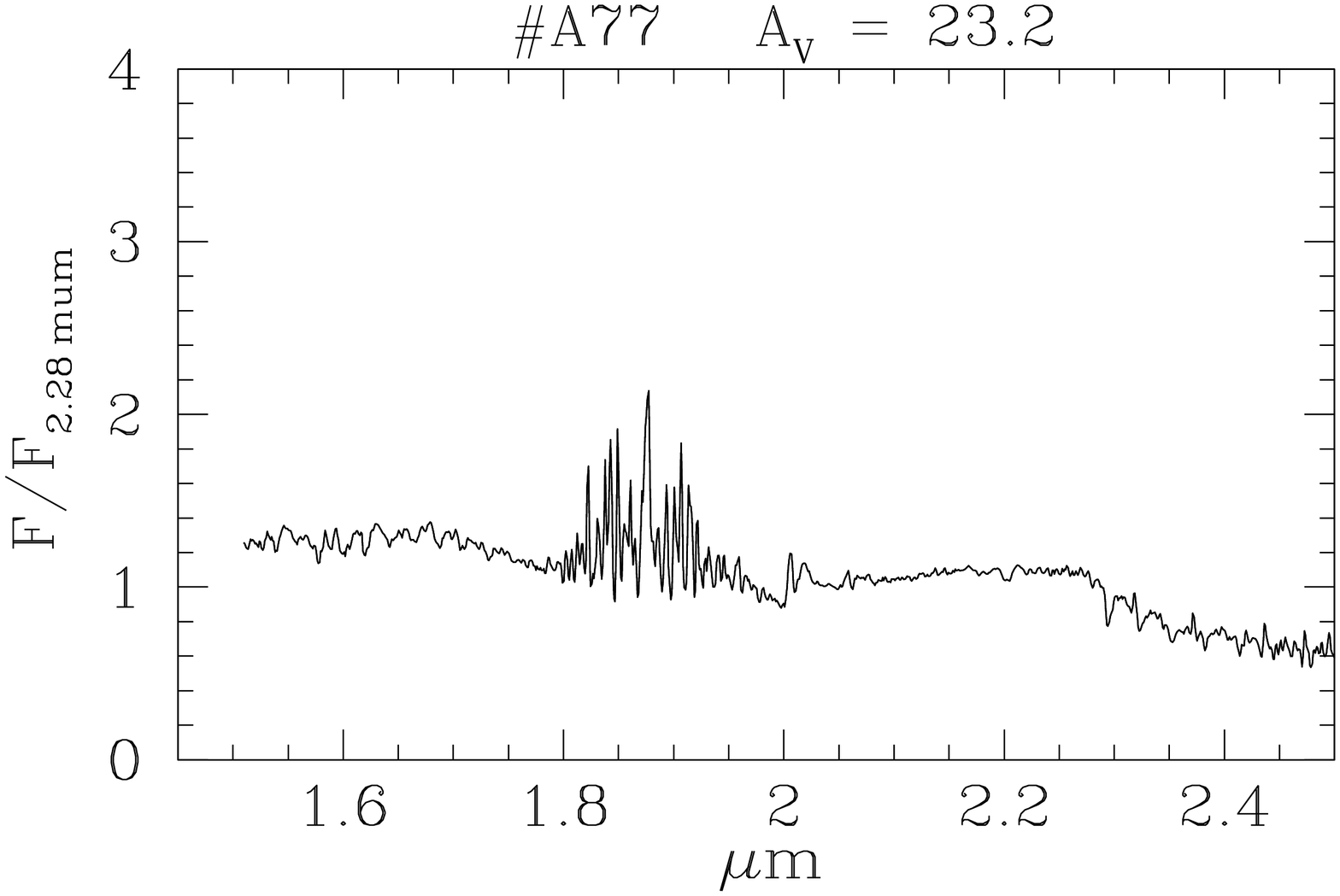,width=4.3cm,angle=270} \epsfig{file=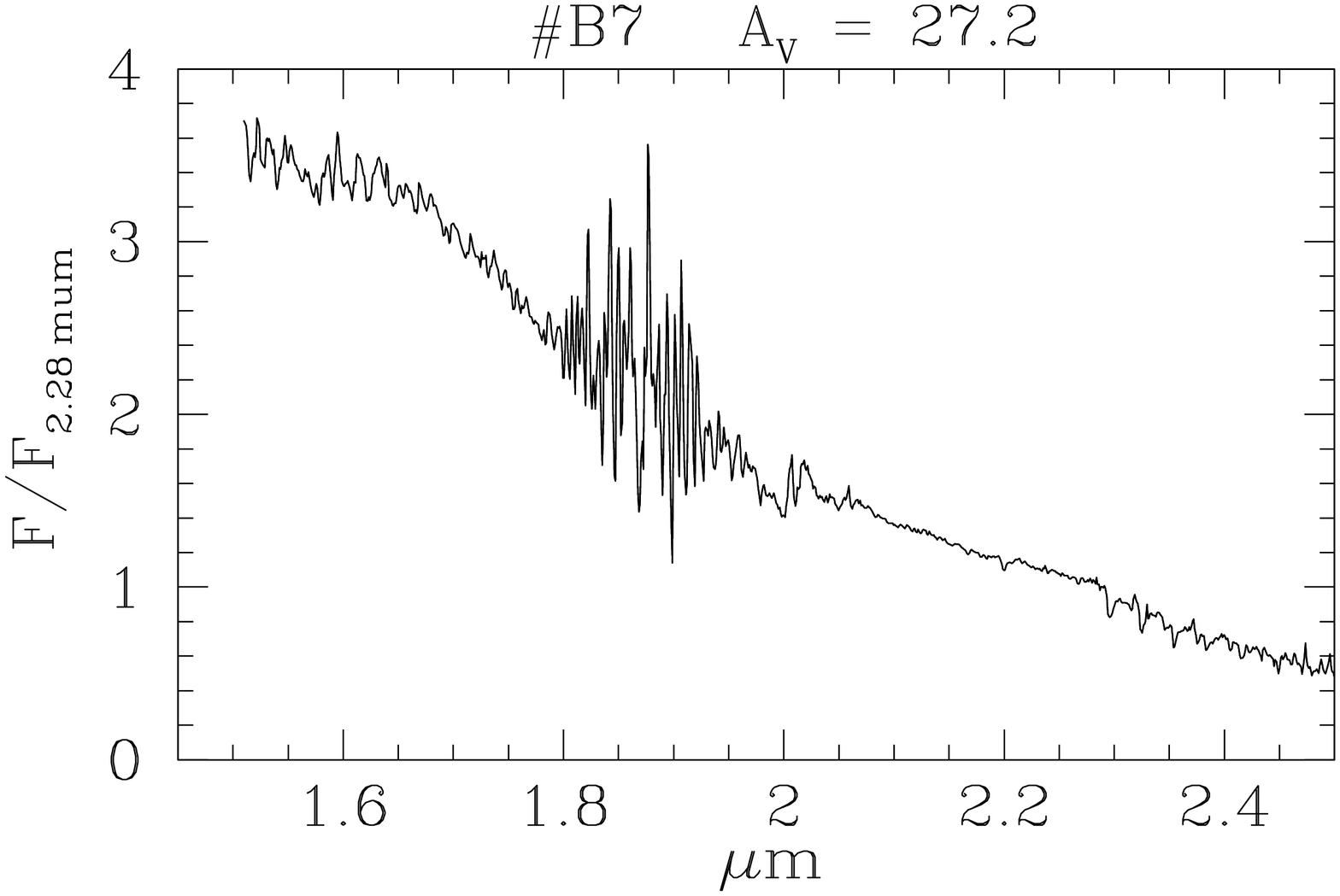,width=4.3cm,angle=270}   \epsfig{file=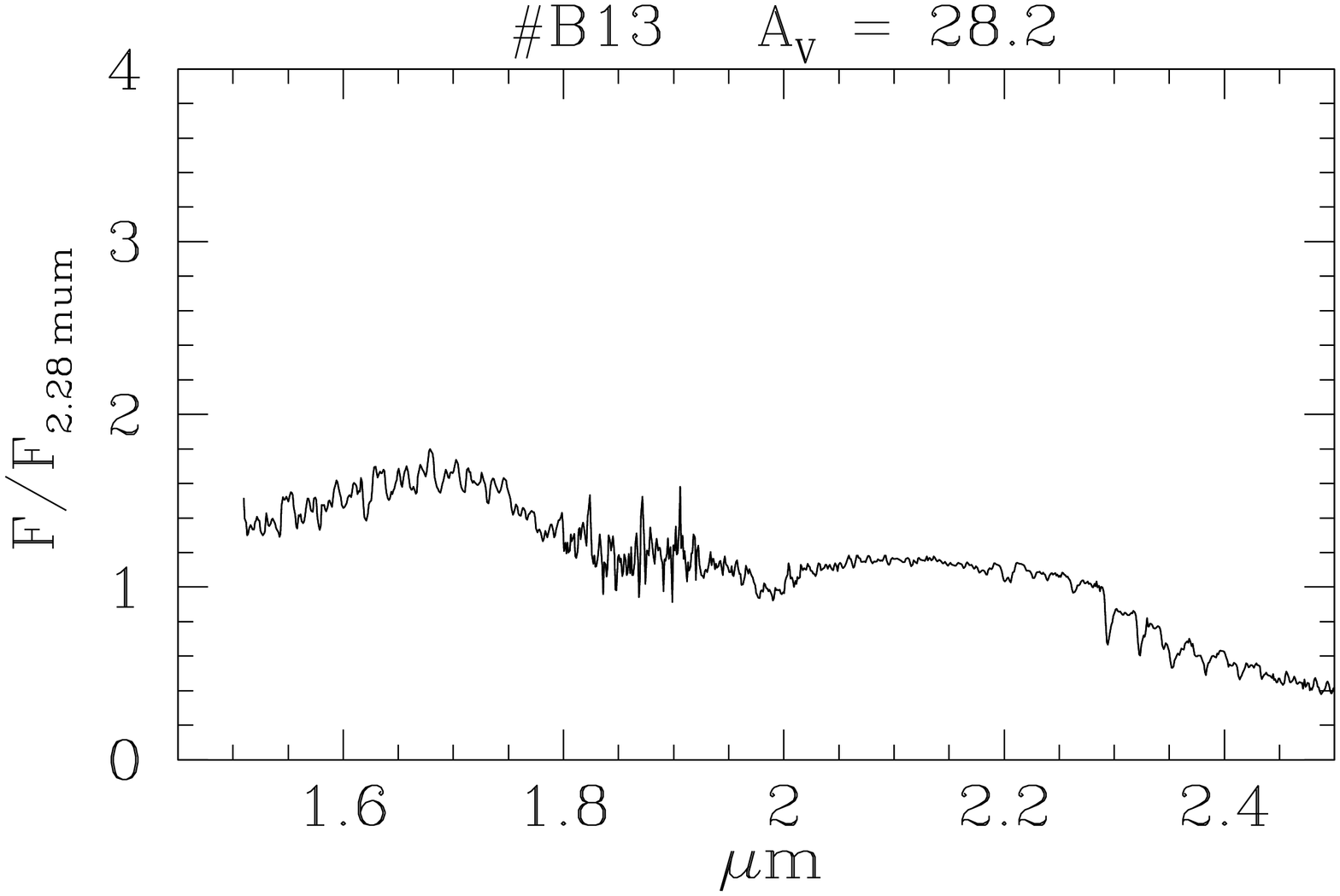,width=4.3cm,angle=270}}

\end{figure*}

\begin{figure*}[H!]
{\bf{Fig.~B.5. (continued)}}

{\epsfig{file=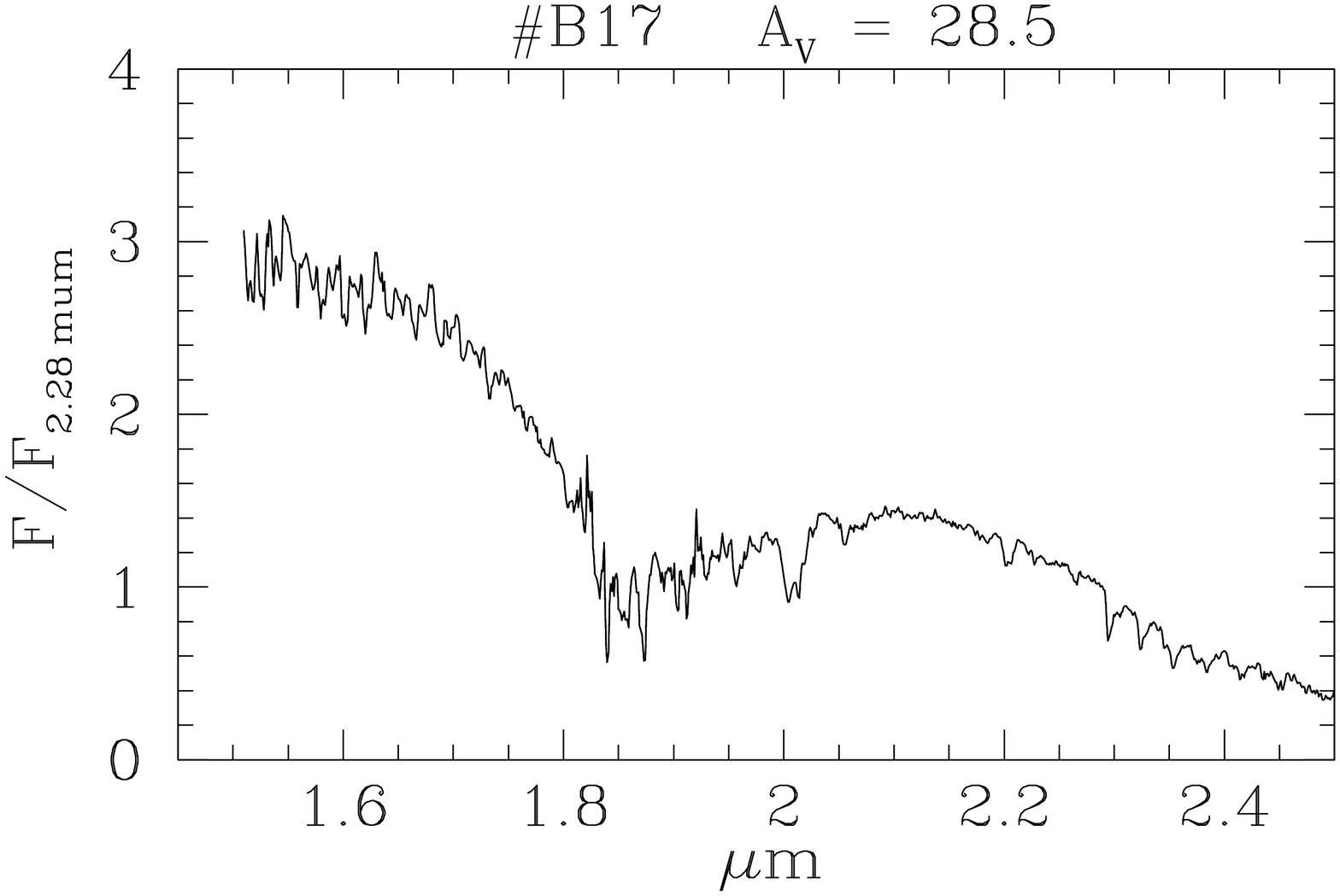,width=4.3cm,angle=270} \epsfig{file=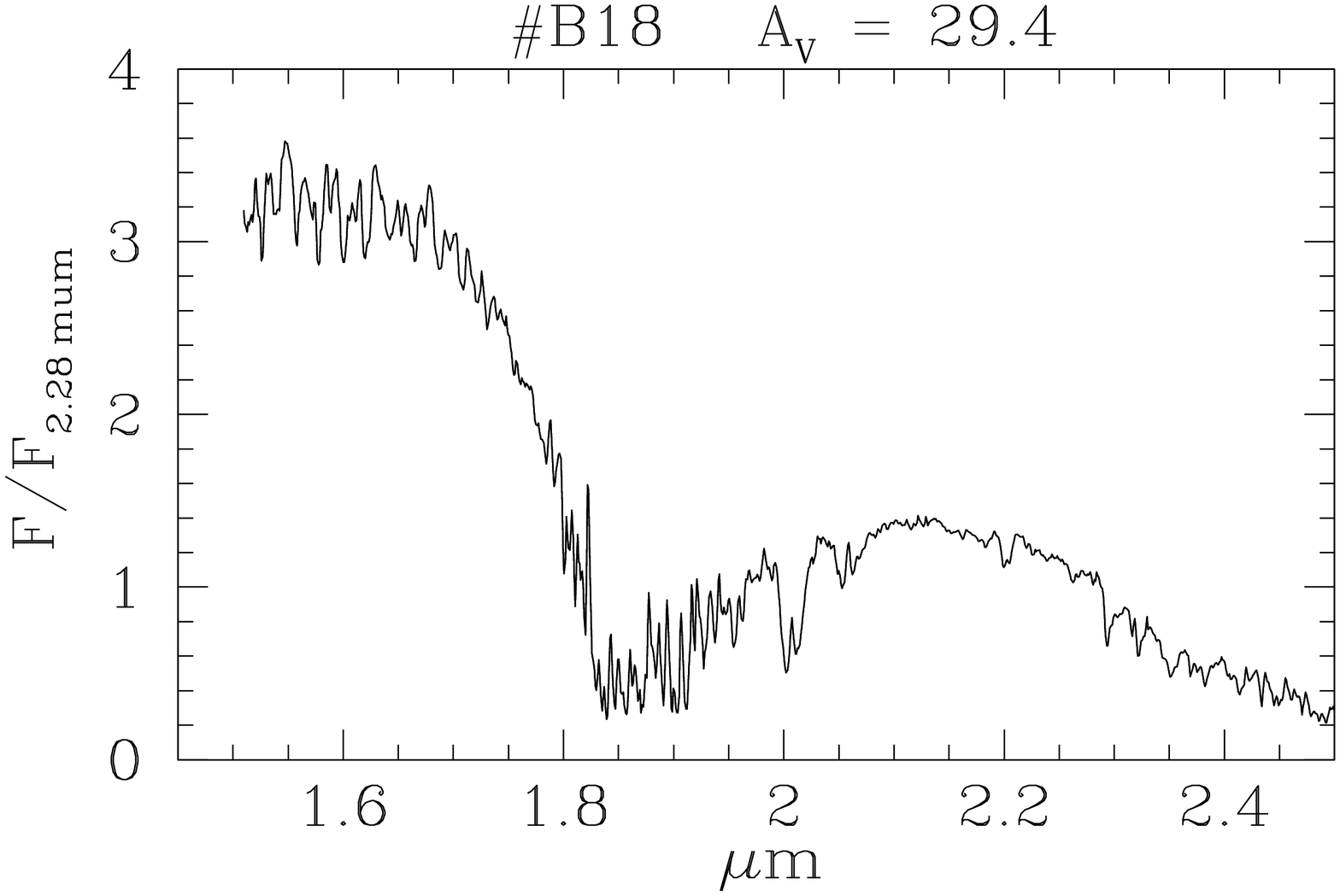,width=4.3cm,angle=270}   \epsfig{file=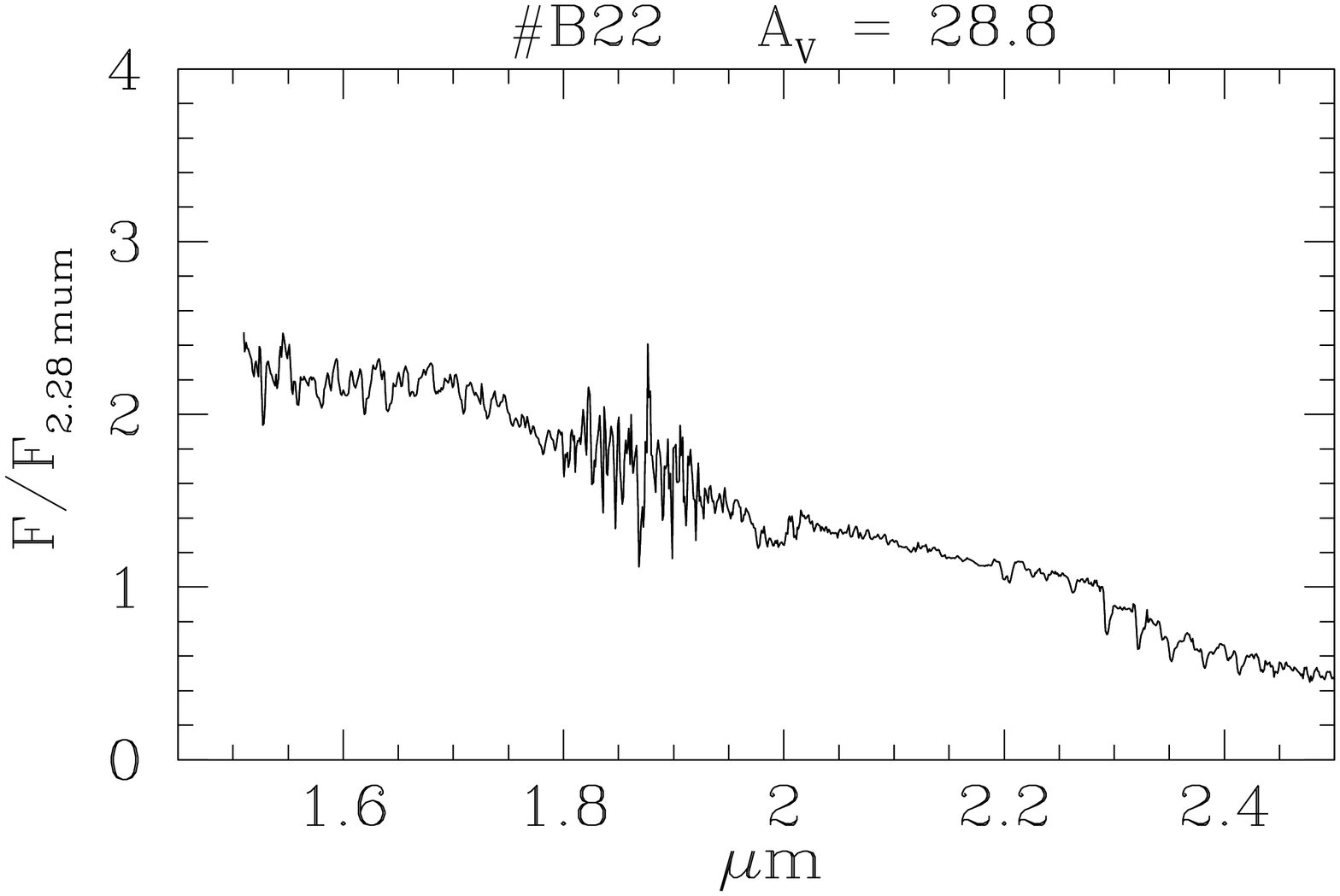,width=4.3cm,angle=270}

\epsfig{file=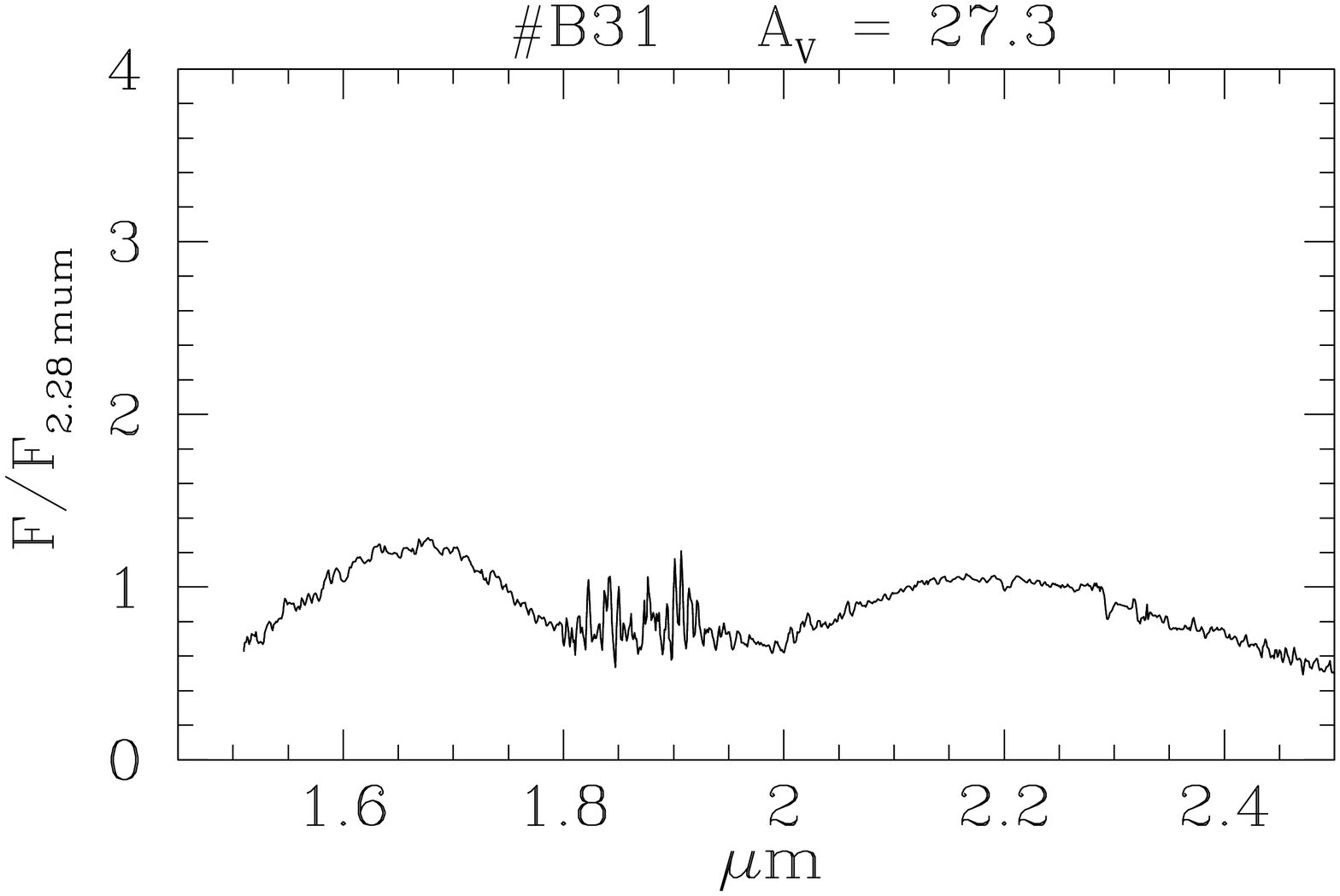,width=4.3cm,angle=270} \epsfig{file=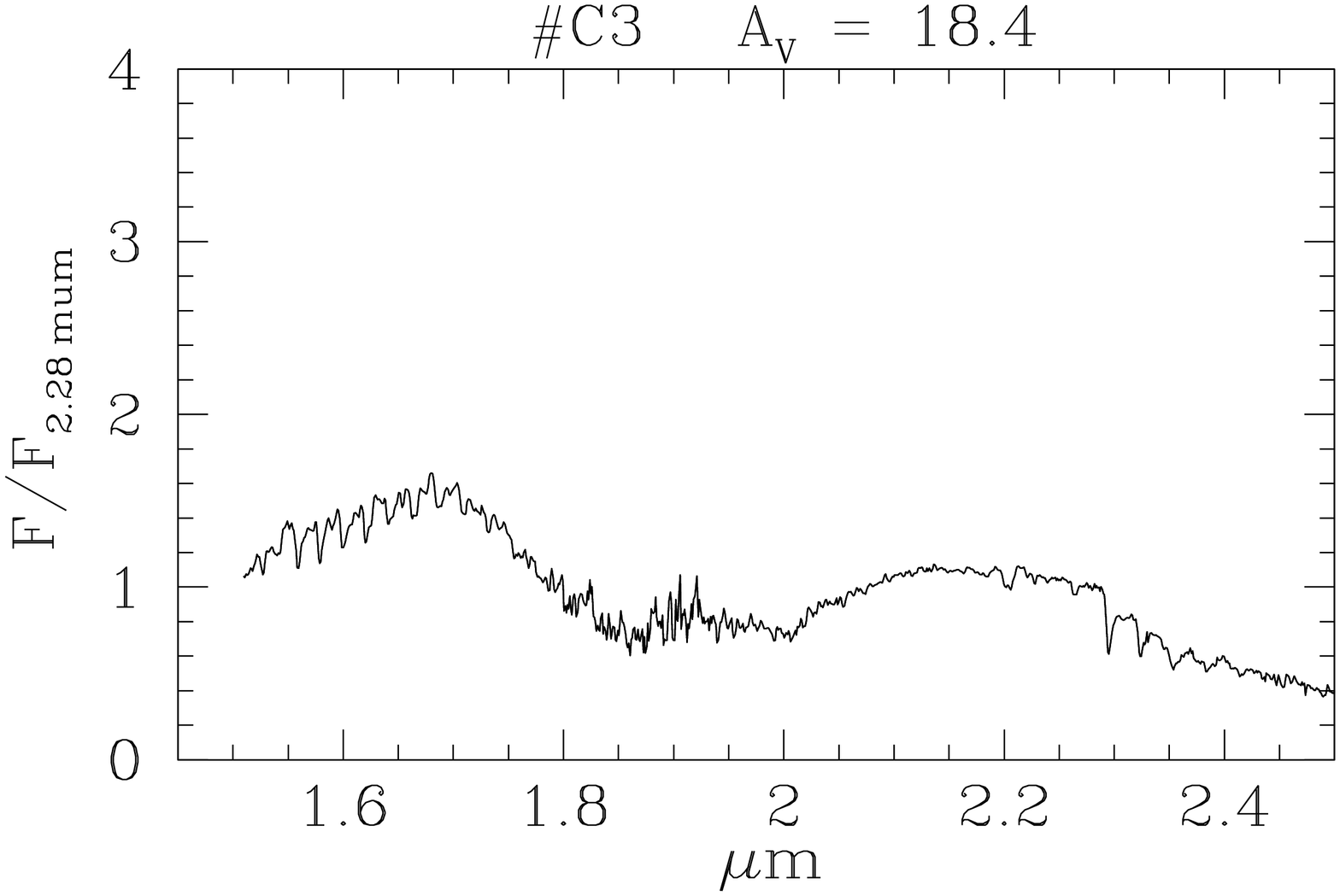,width=4.3cm,angle=270}   \epsfig{file=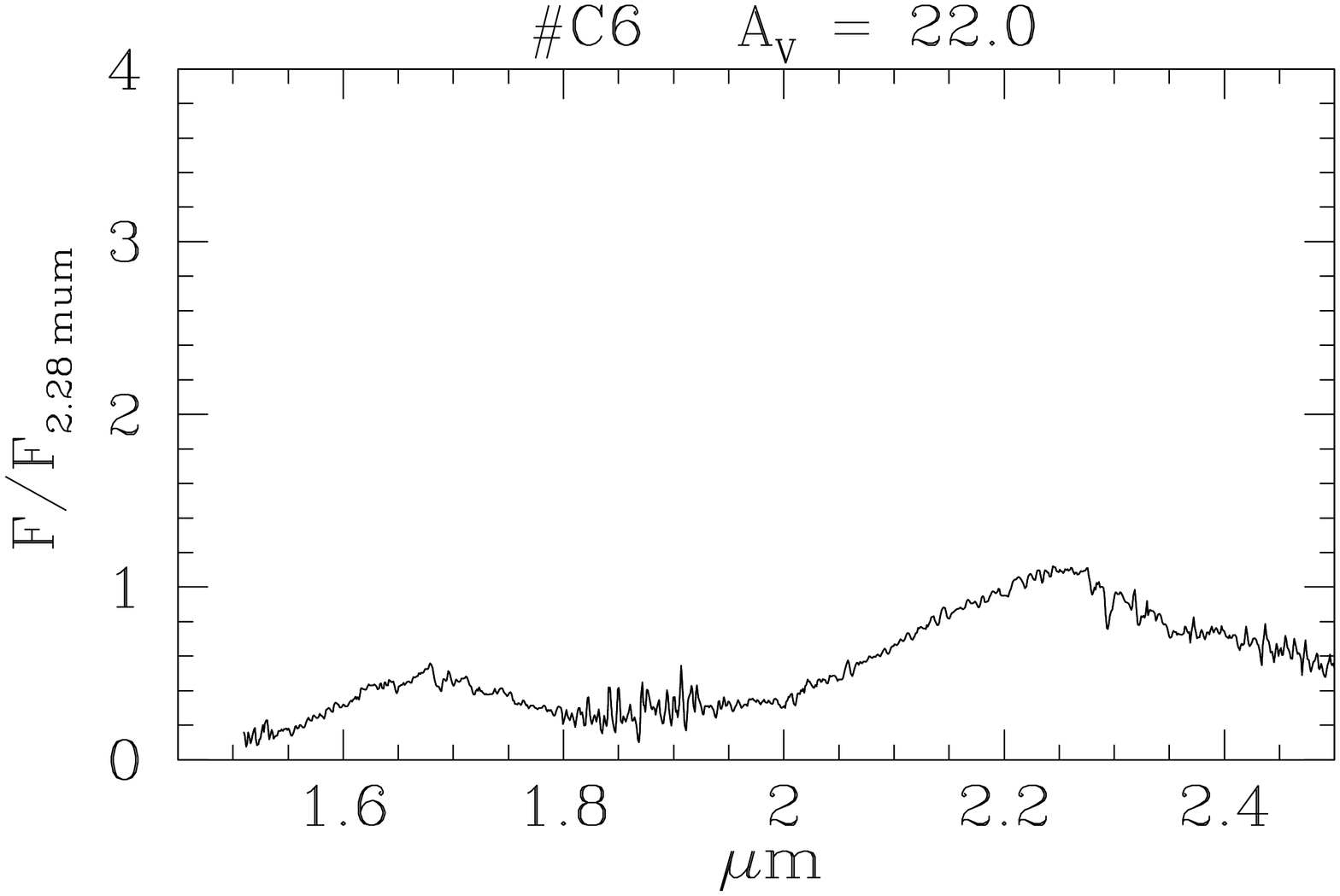,width=4.3cm,angle=270}

\epsfig{file=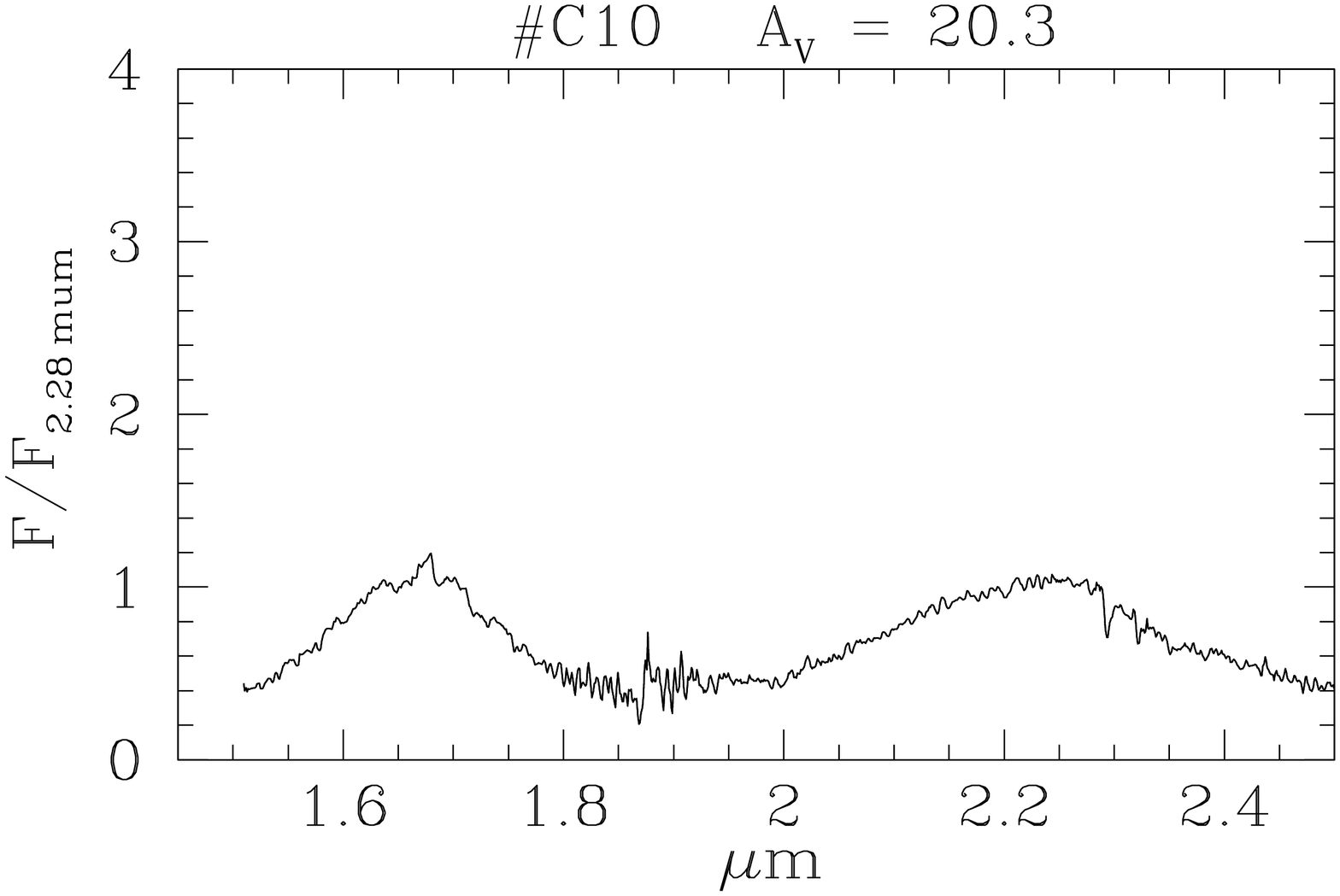,width=4.3cm,angle=270} \epsfig{file=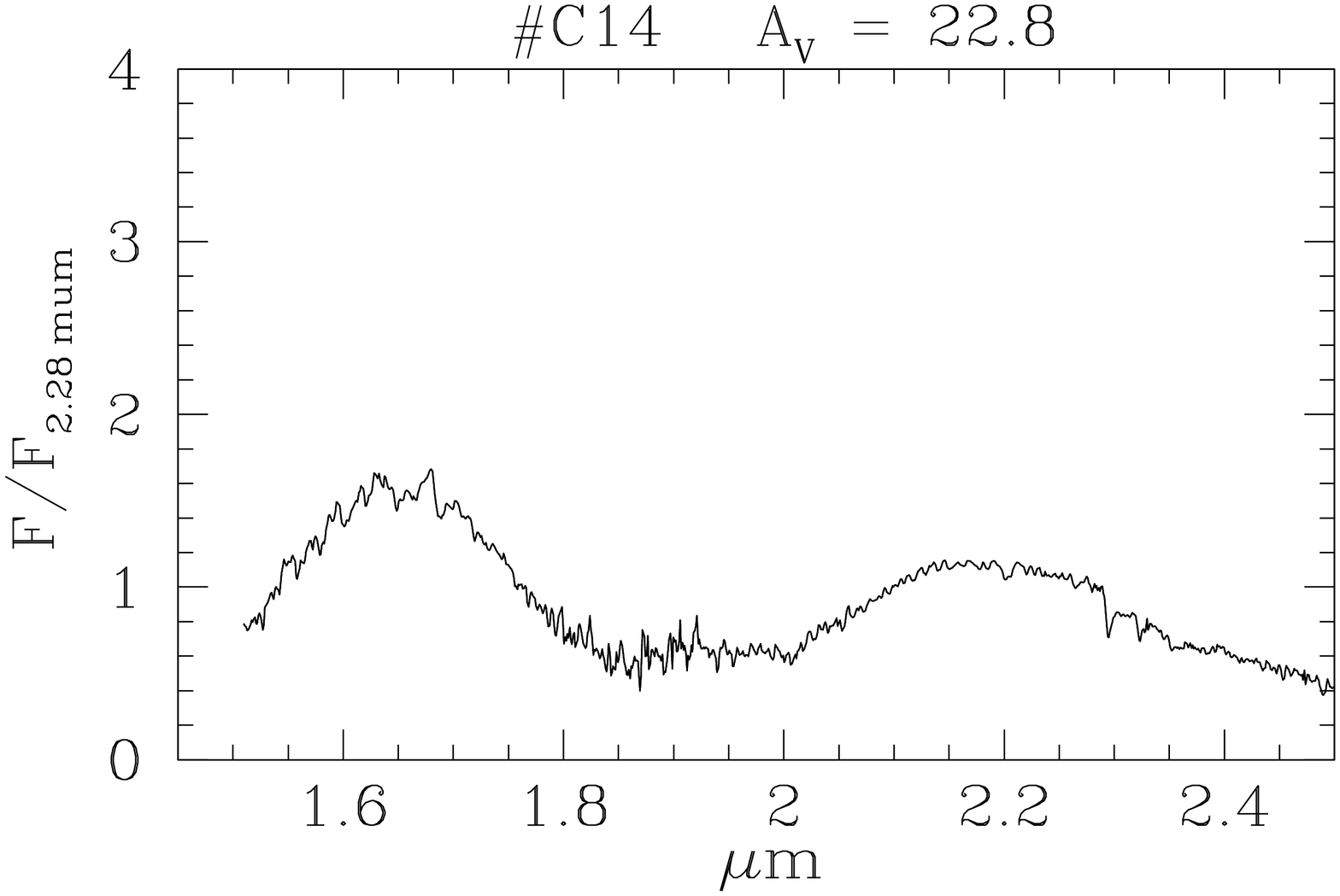,width=4.3cm,angle=270}   \epsfig{file=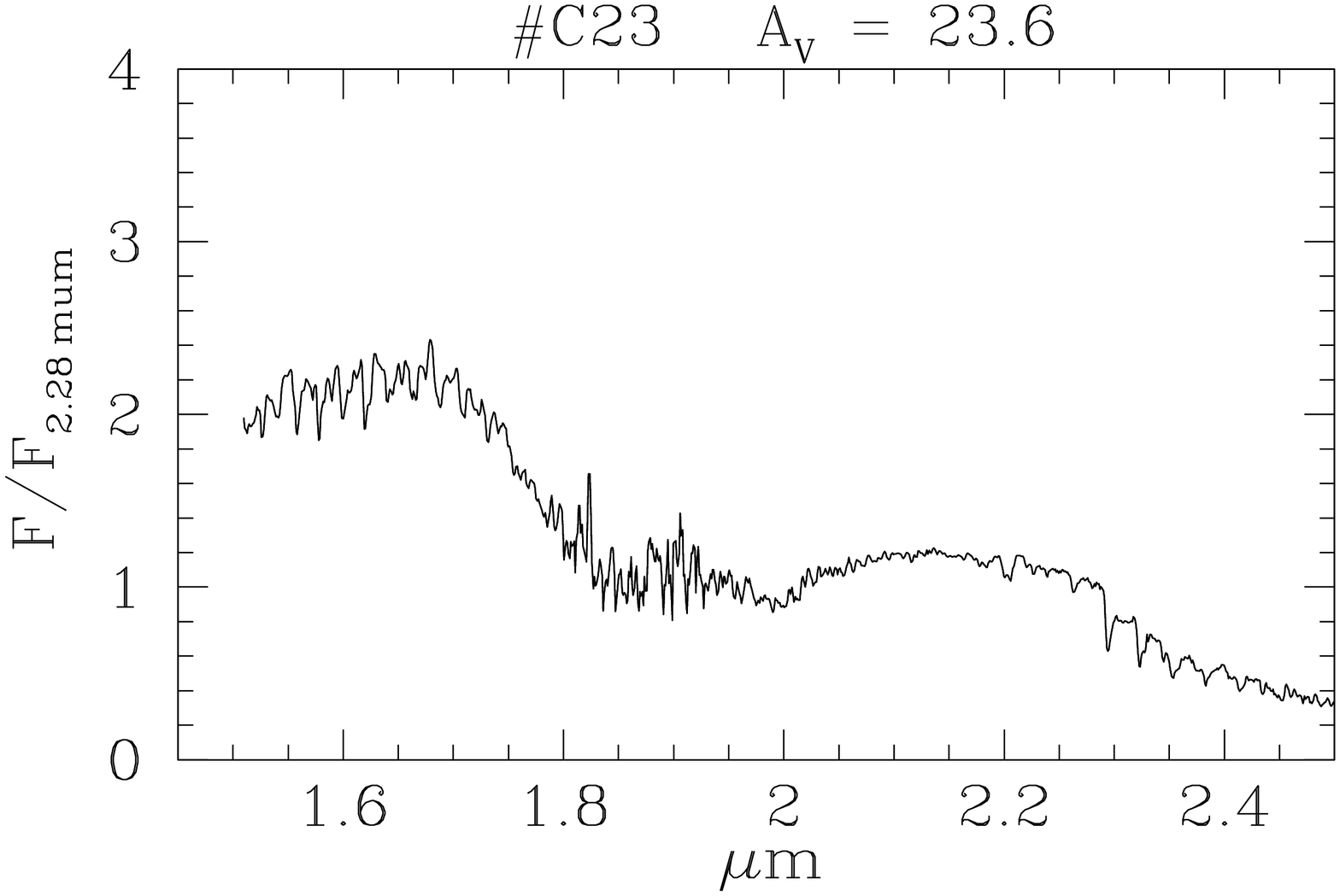,width=4.3cm,angle=270}

\epsfig{file=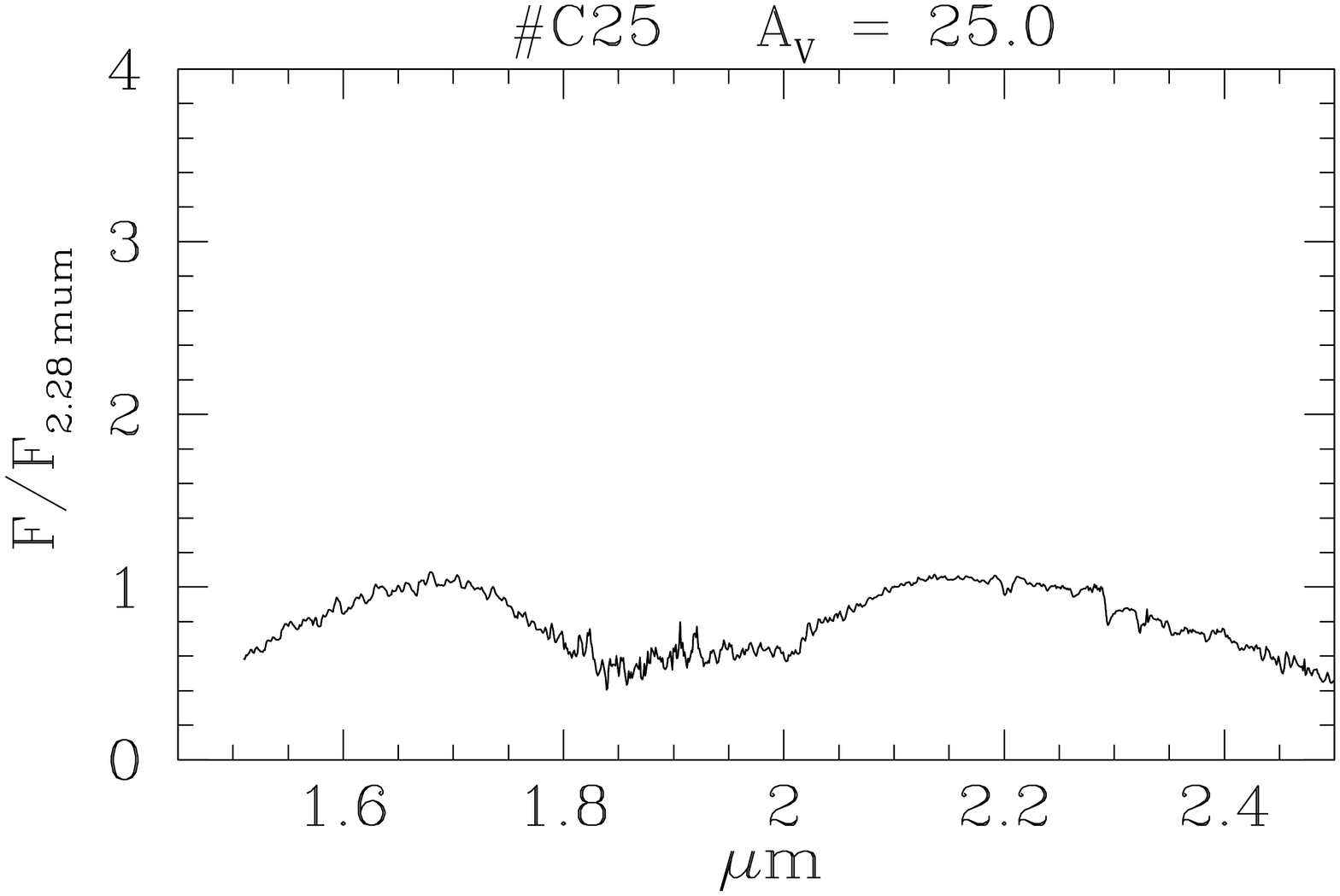,width=4.3cm,angle=270} \epsfig{file=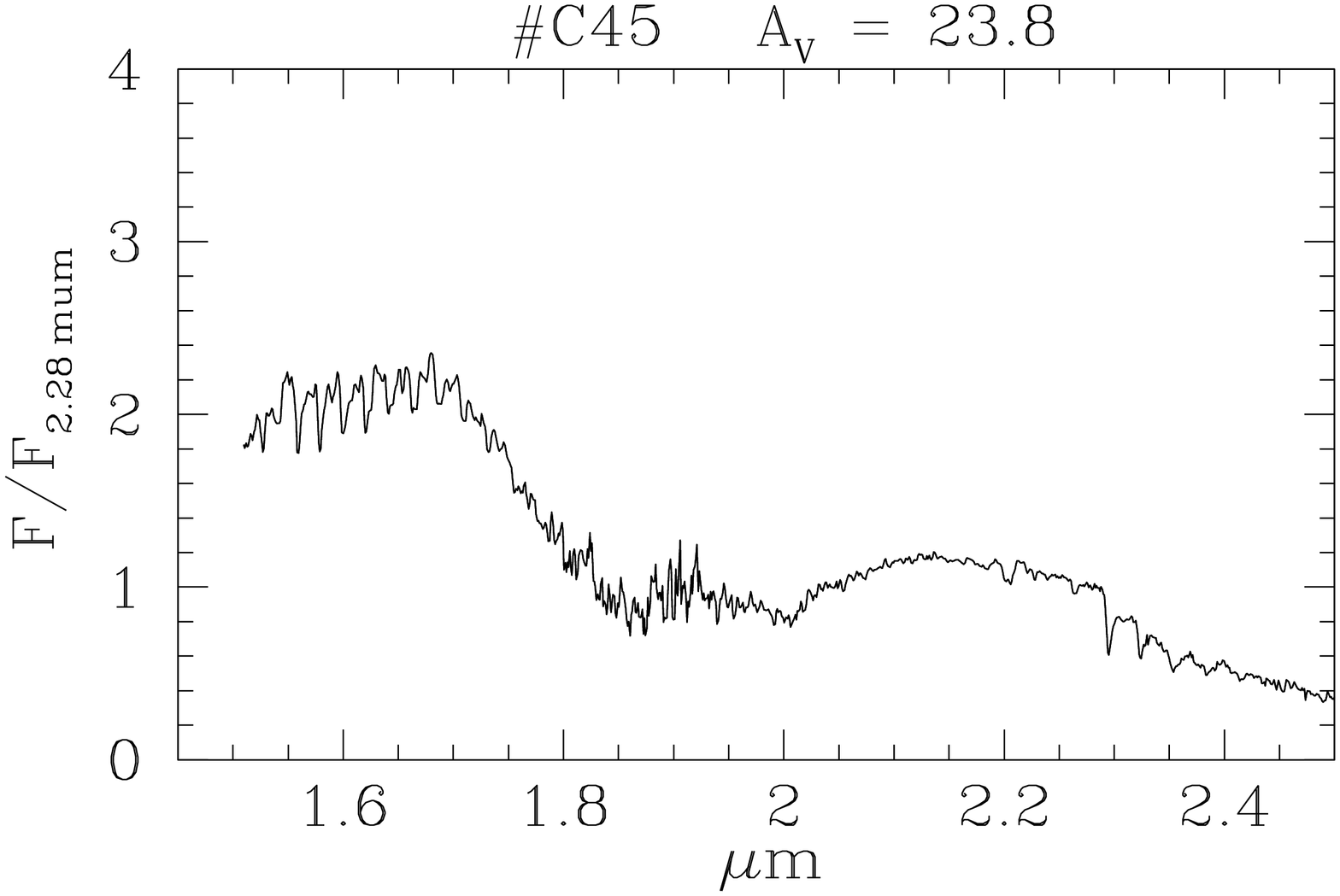,width=4.3cm,angle=270}   \epsfig{file=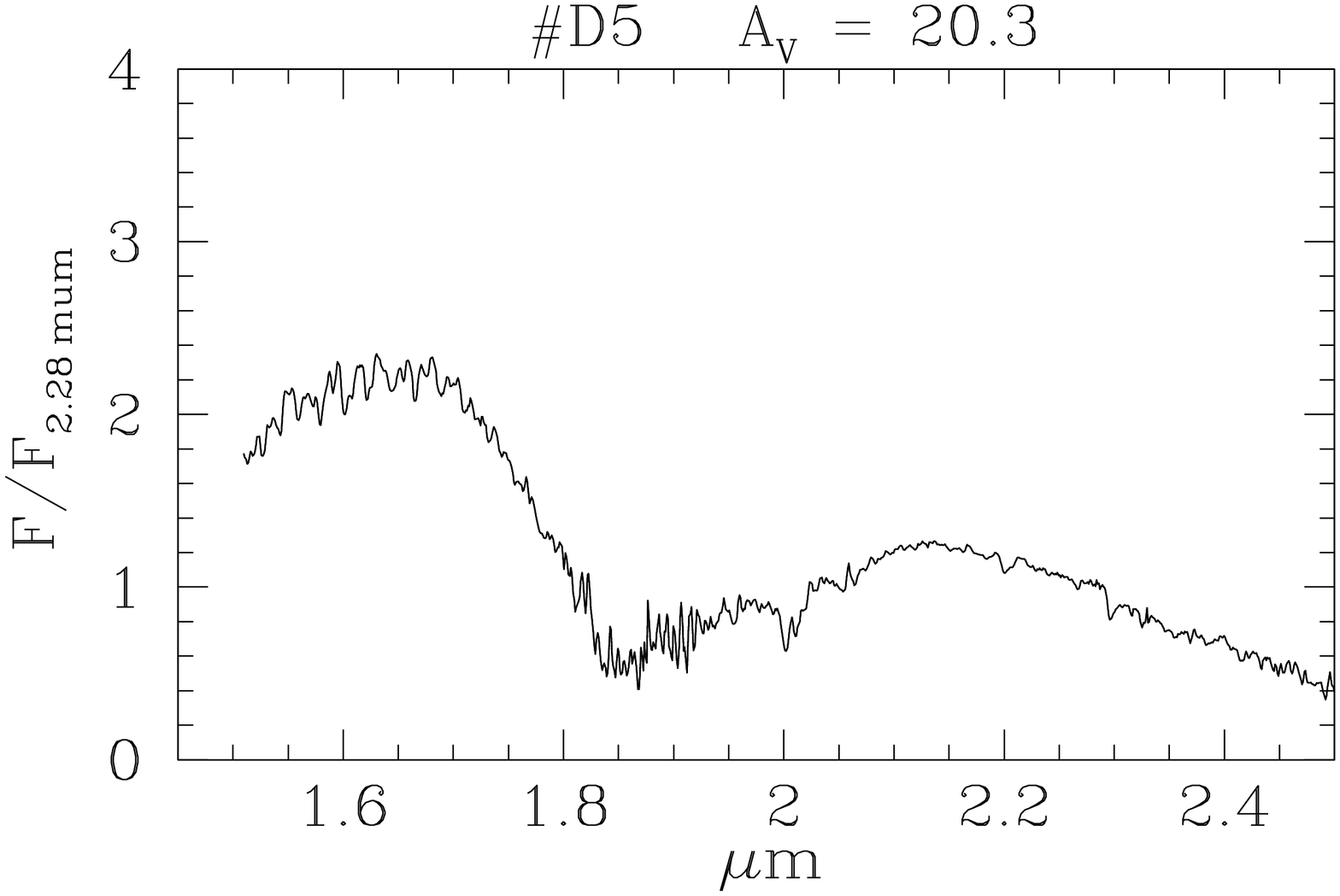,width=4.3cm,angle=270}

\epsfig{file=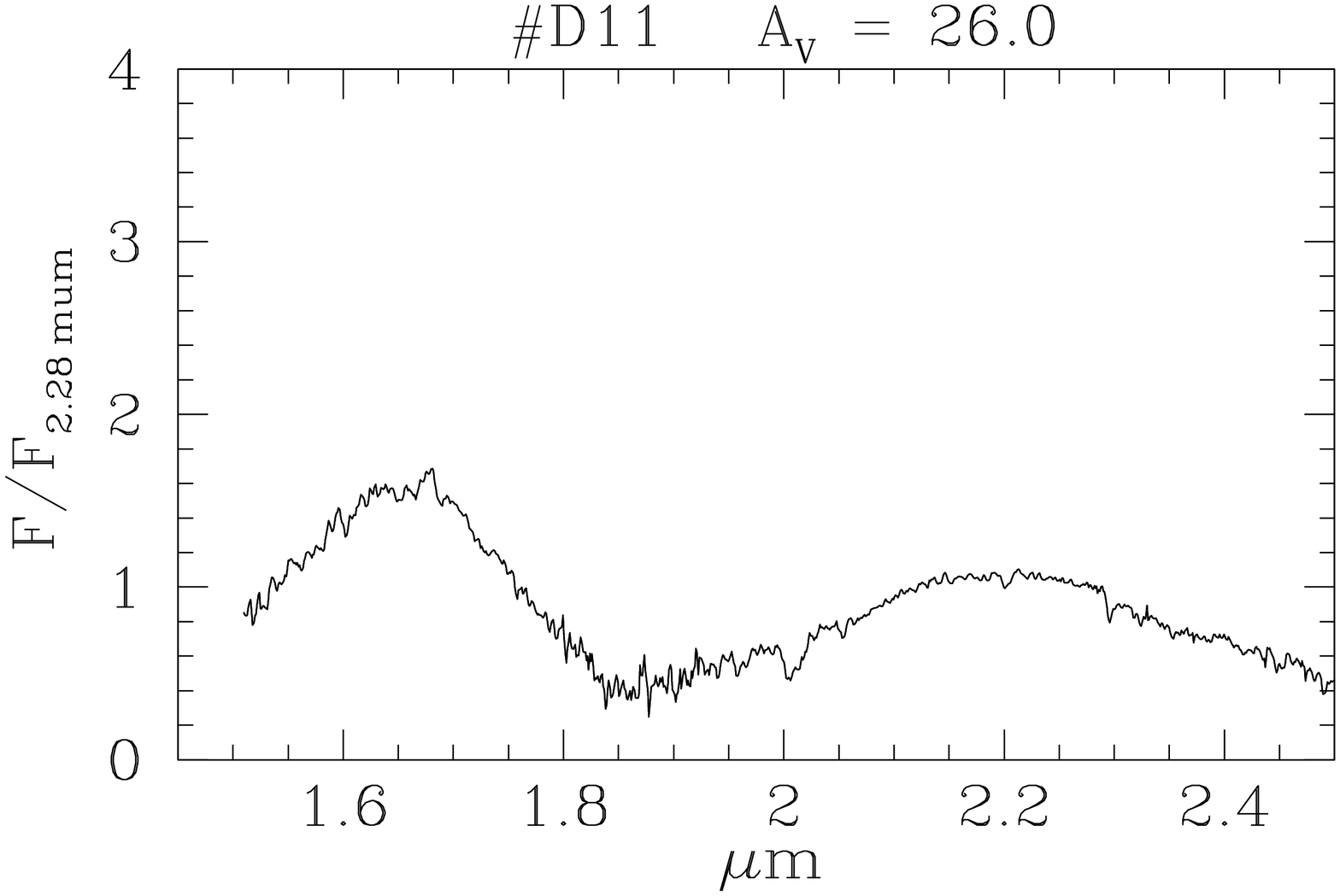,width=4.3cm,angle=270} \epsfig{file=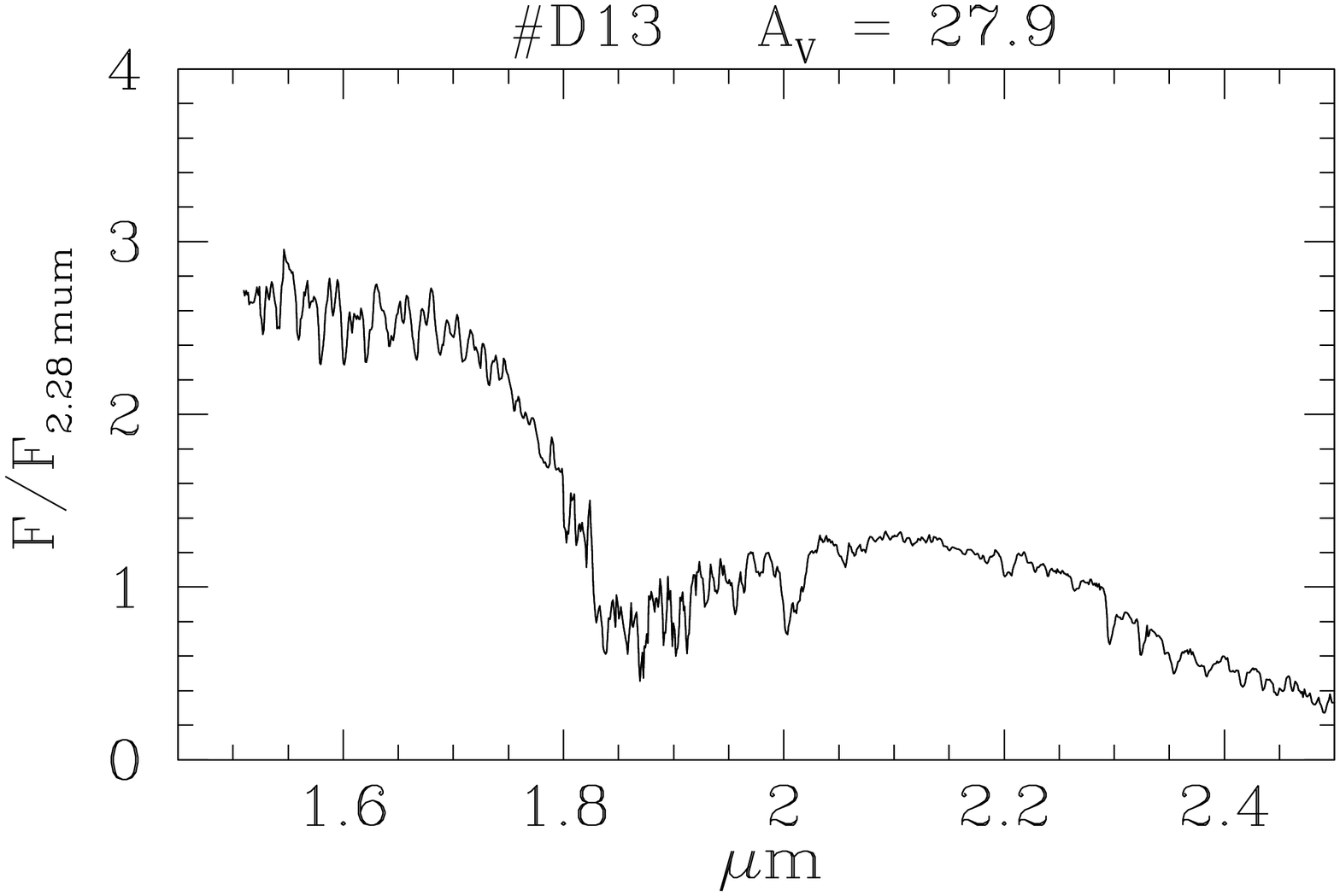,width=4.3cm,angle=270}   \epsfig{file=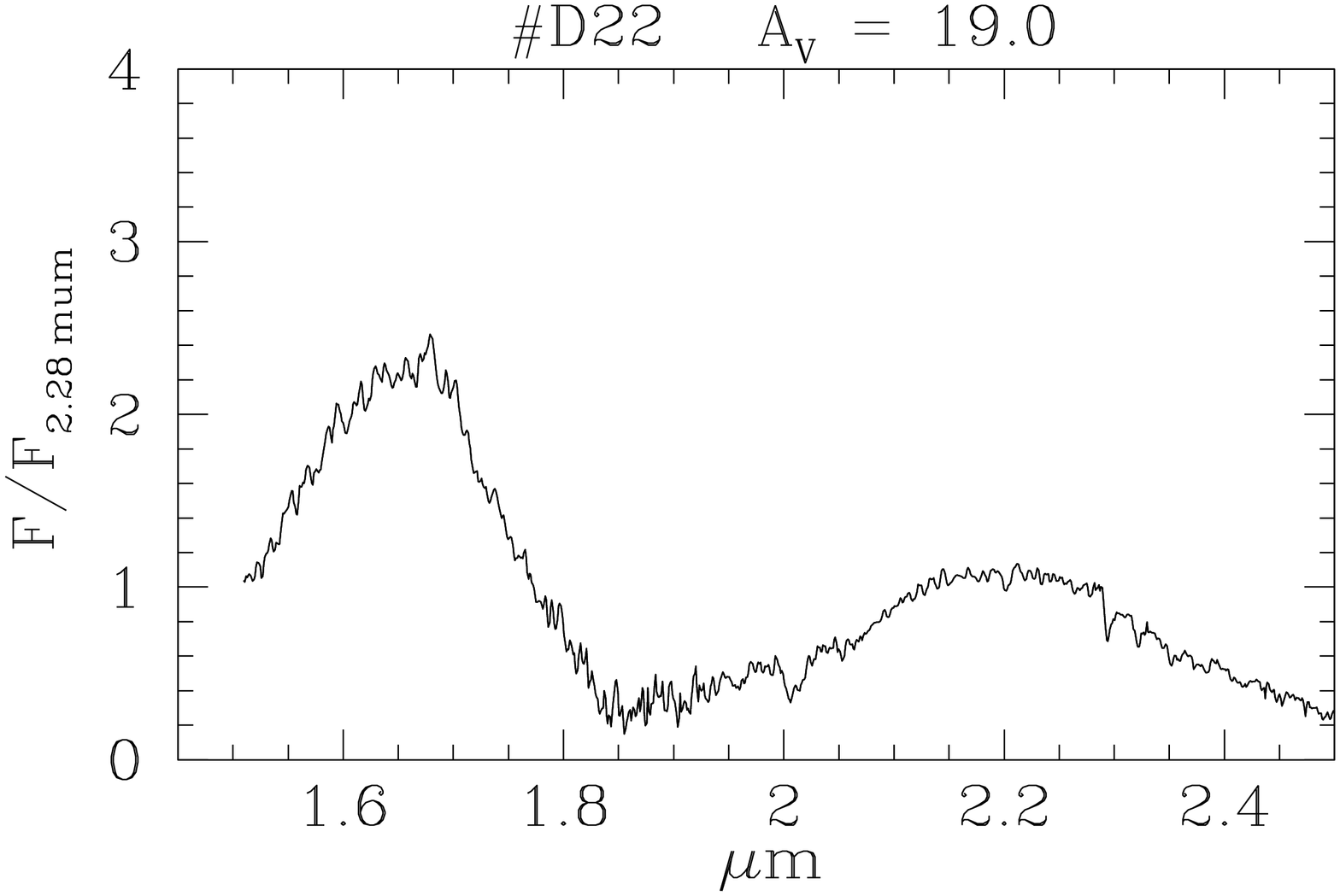,width=4.3cm,angle=270}}

\end{figure*}

\begin{figure*}[H!]
{\bf{Fig.~B.5. (continued)}}

{\epsfig{file=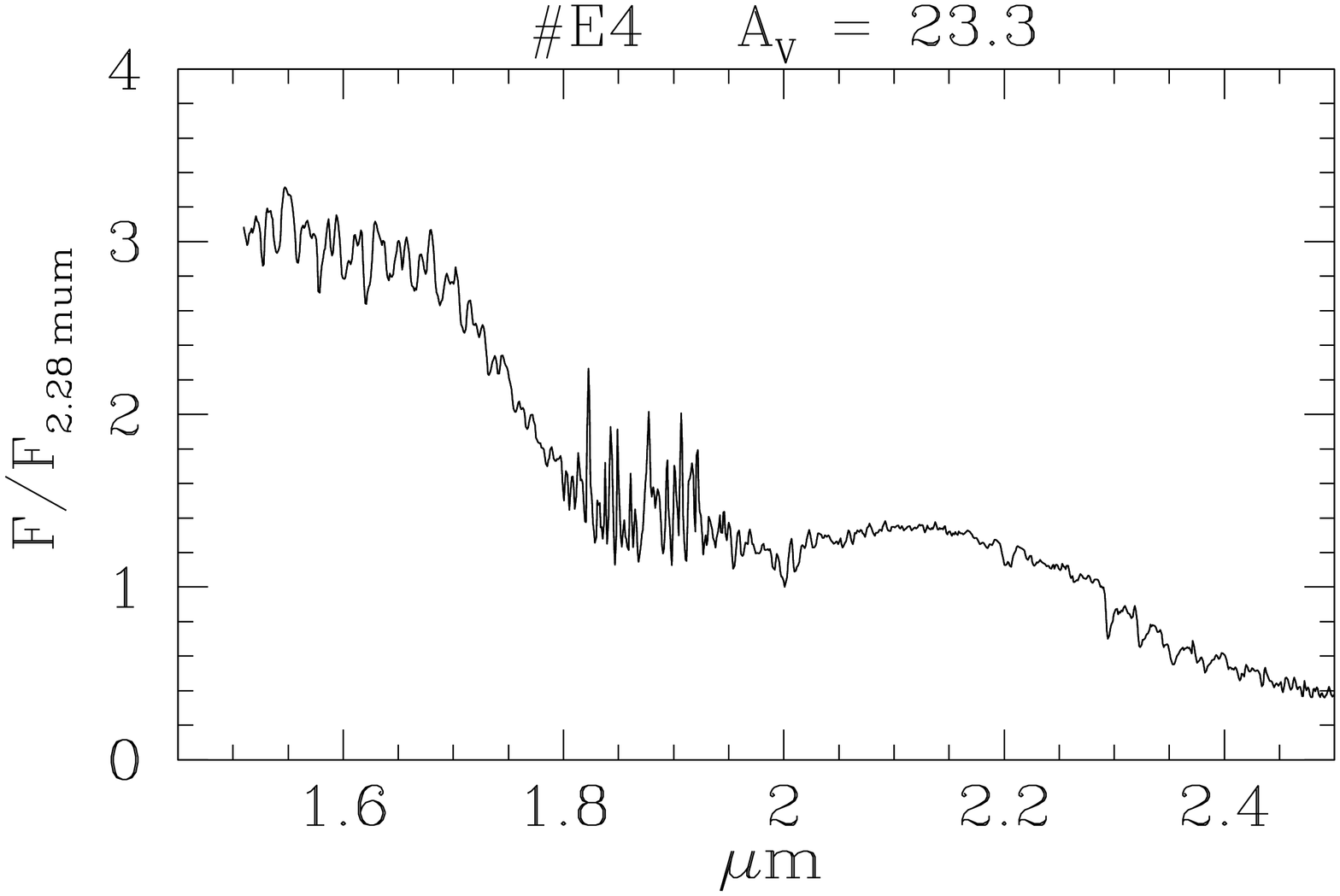,width=4.3cm,angle=270} \epsfig{file=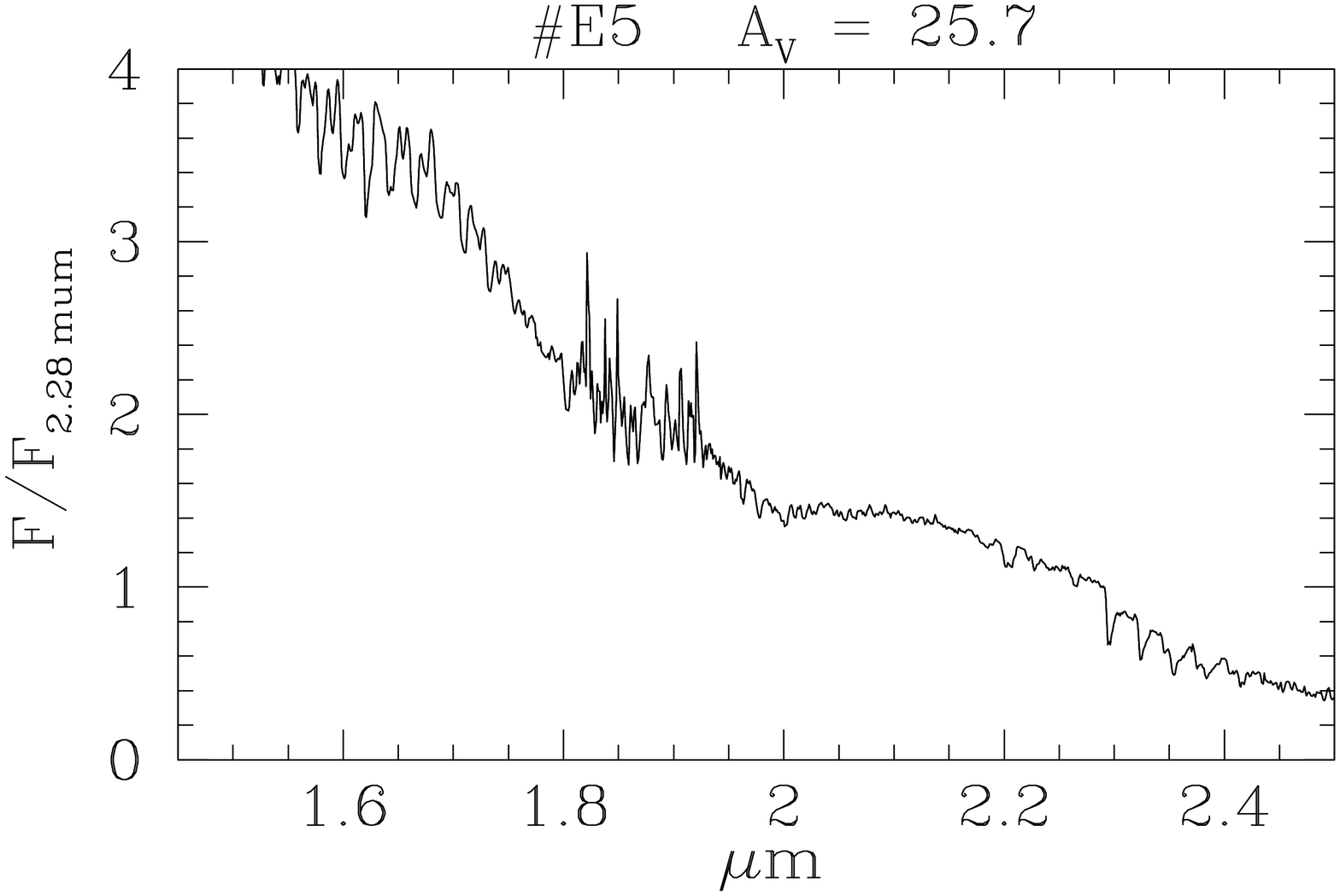,width=4.3cm,angle=270}   \epsfig{file=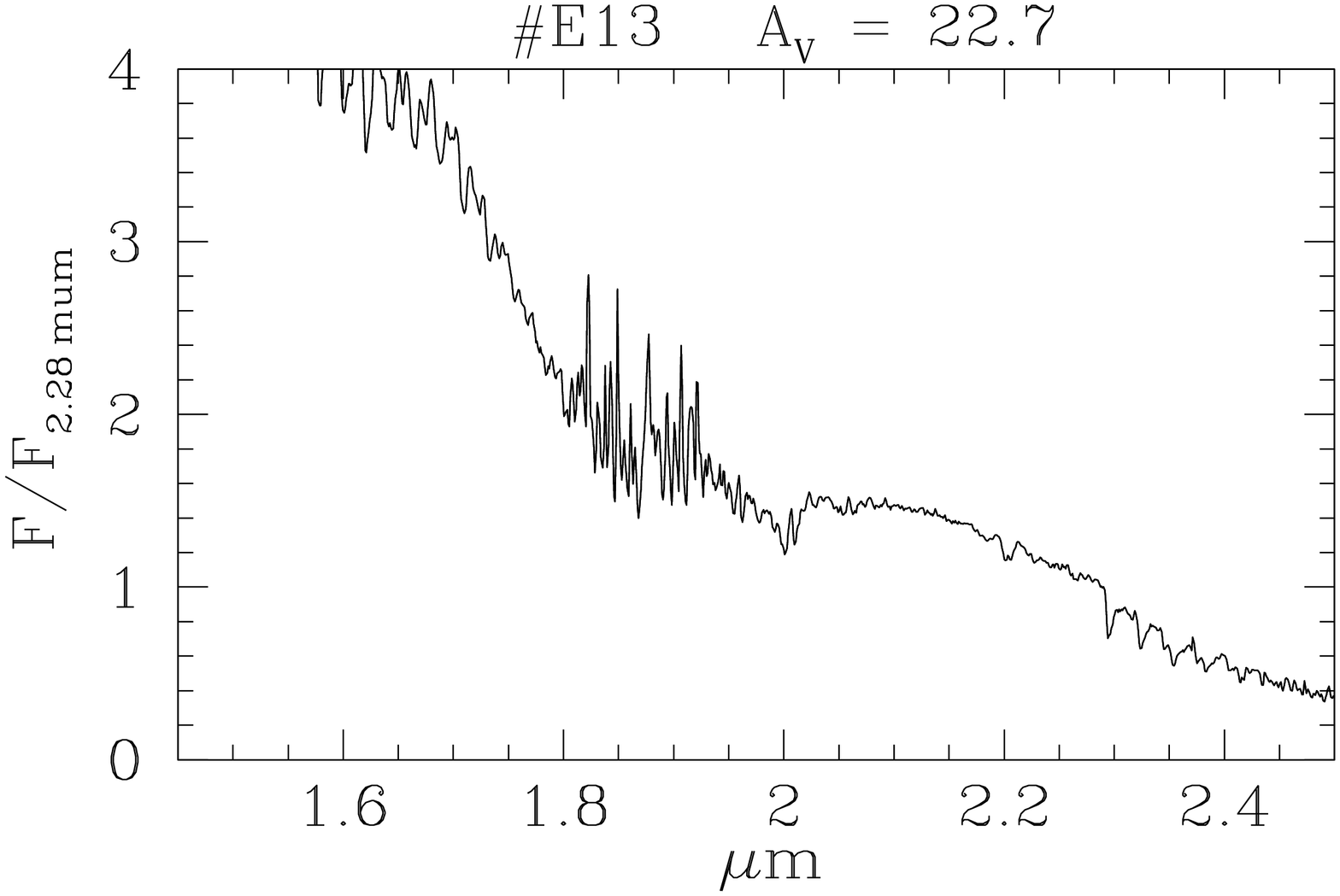,width=4.3cm,angle=270}

\epsfig{file=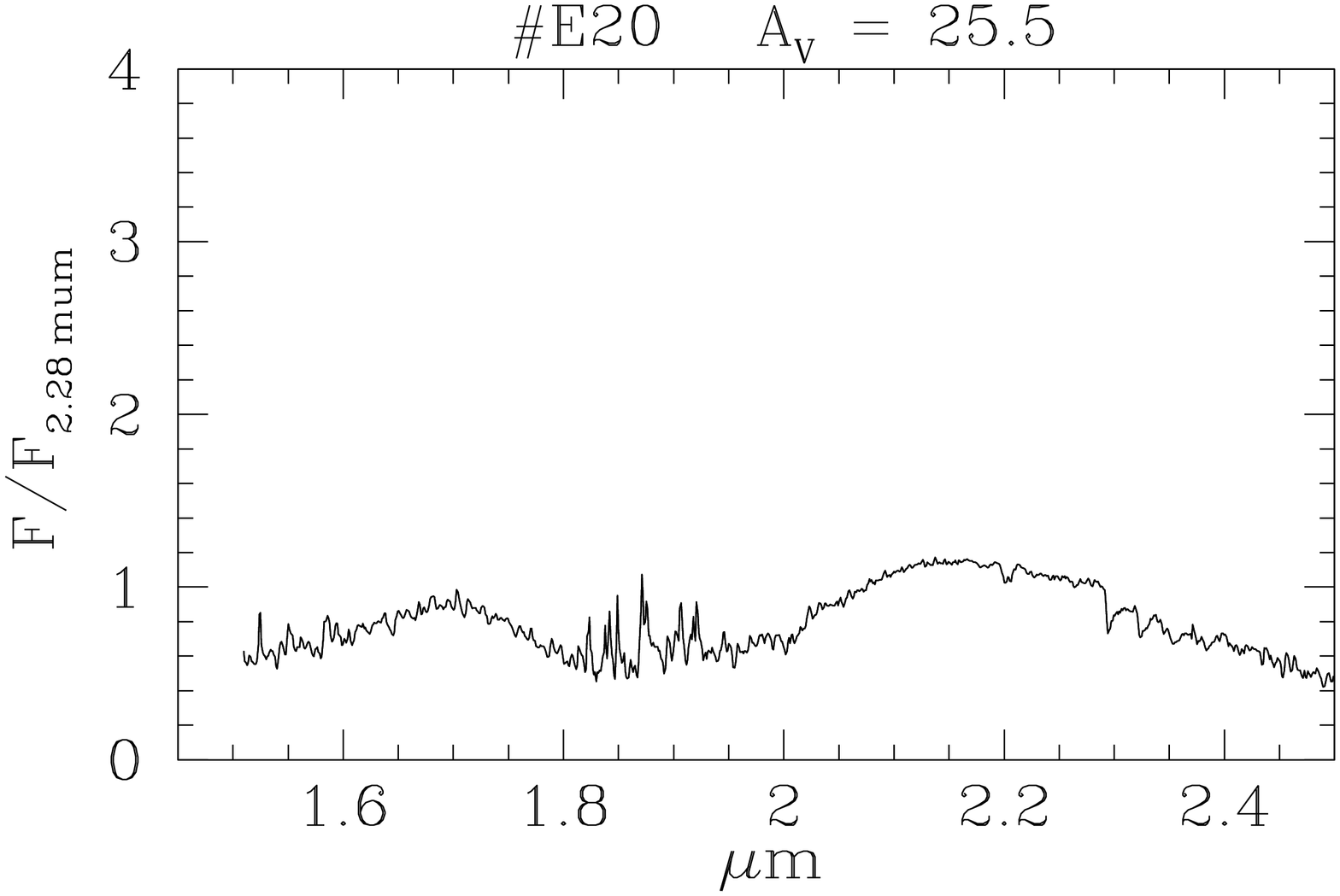,width=4.3cm,angle=270} \epsfig{file=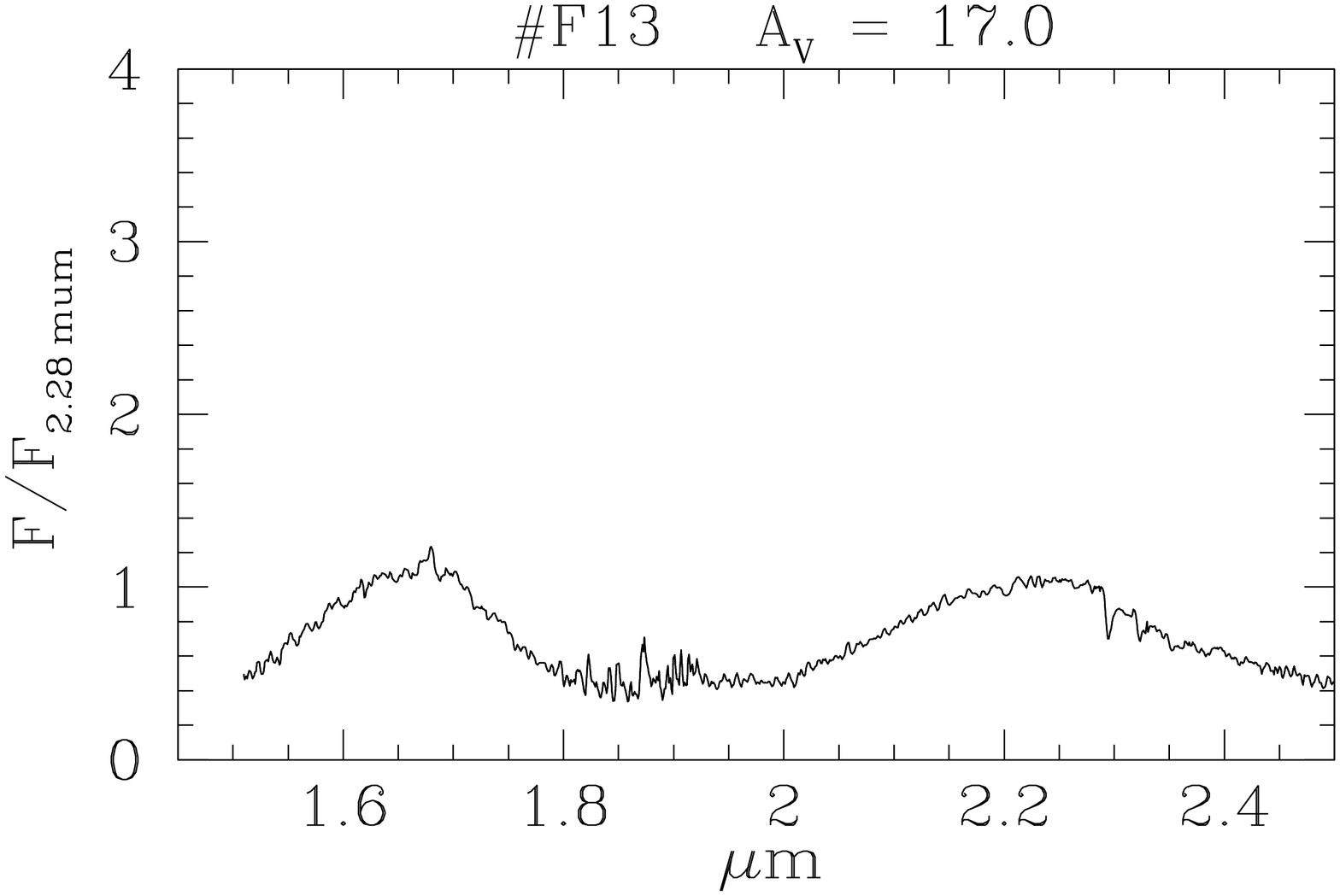,width=4.3cm,angle=270}   \epsfig{file=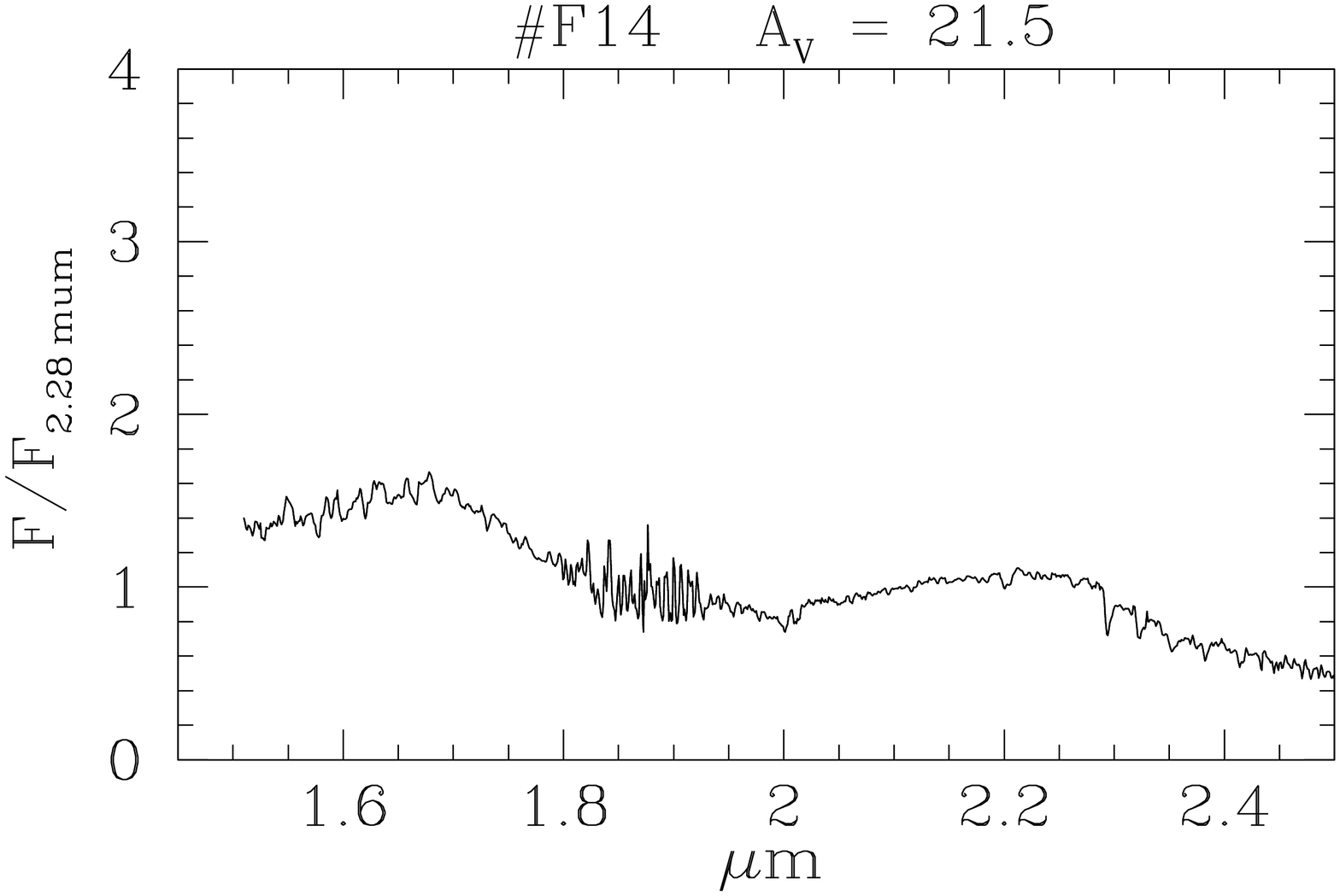,width=4.3cm,angle=270}

\epsfig{file=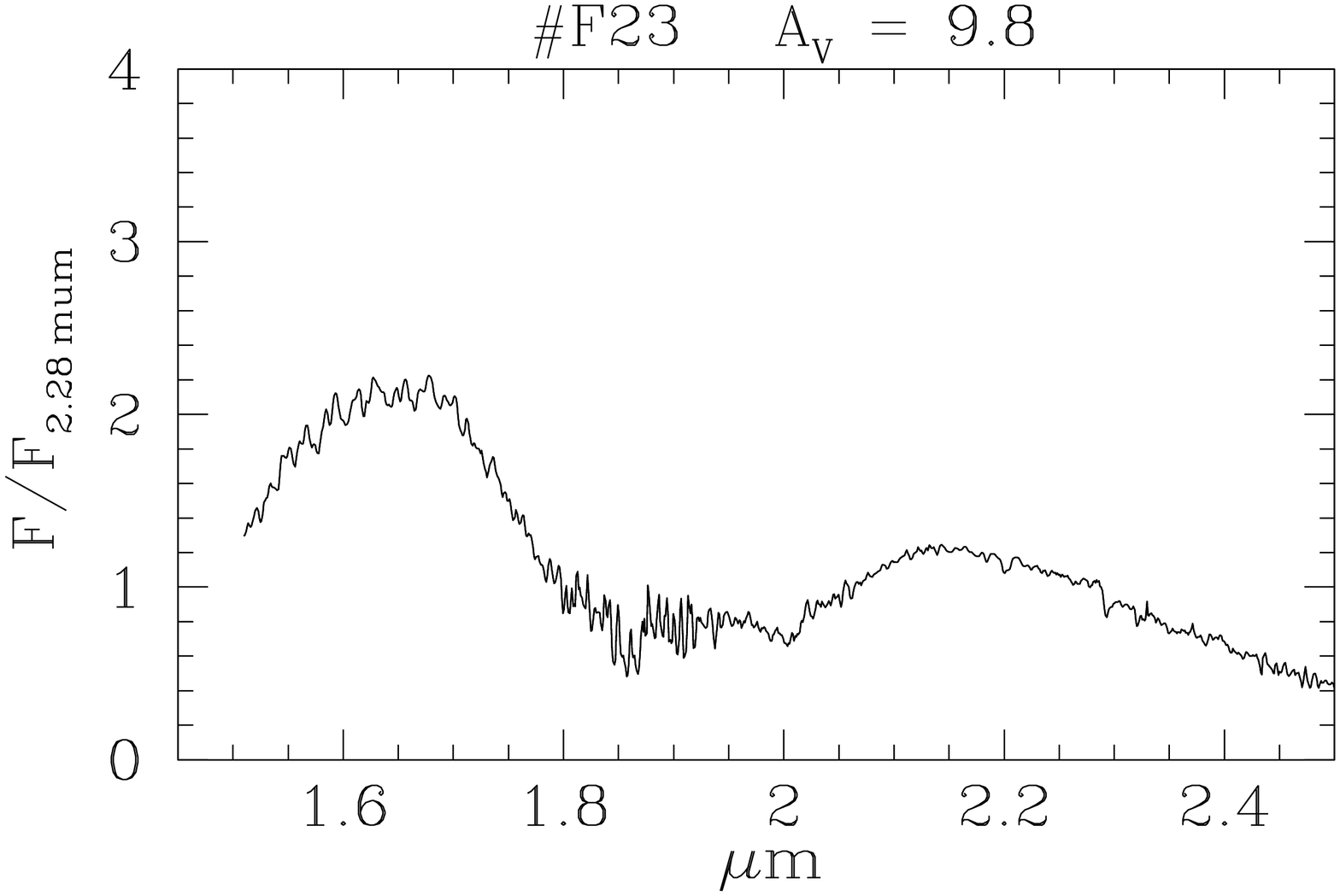,width=4.3cm,angle=270} \epsfig{file=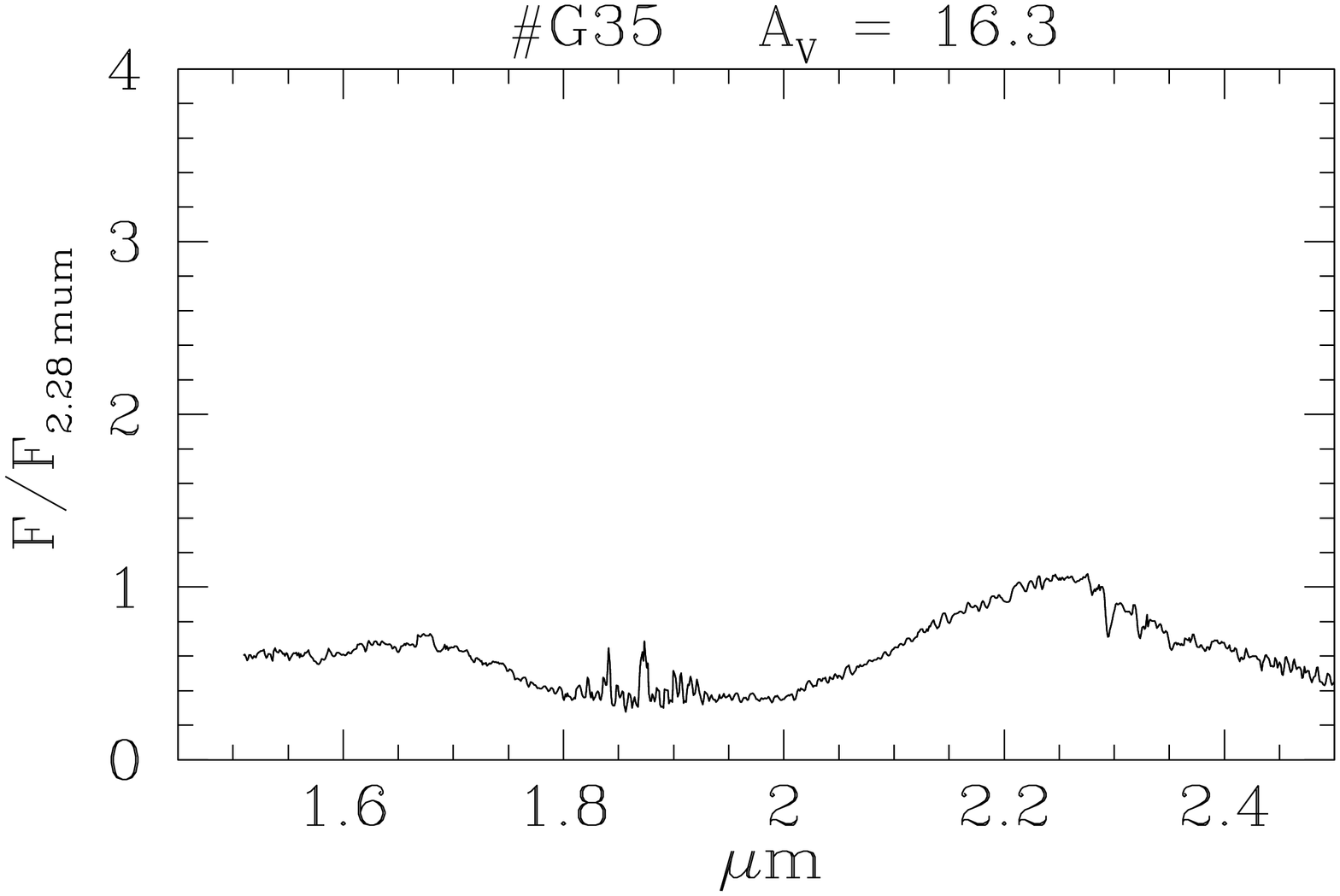,width=4.3cm,angle=270}   \epsfig{file=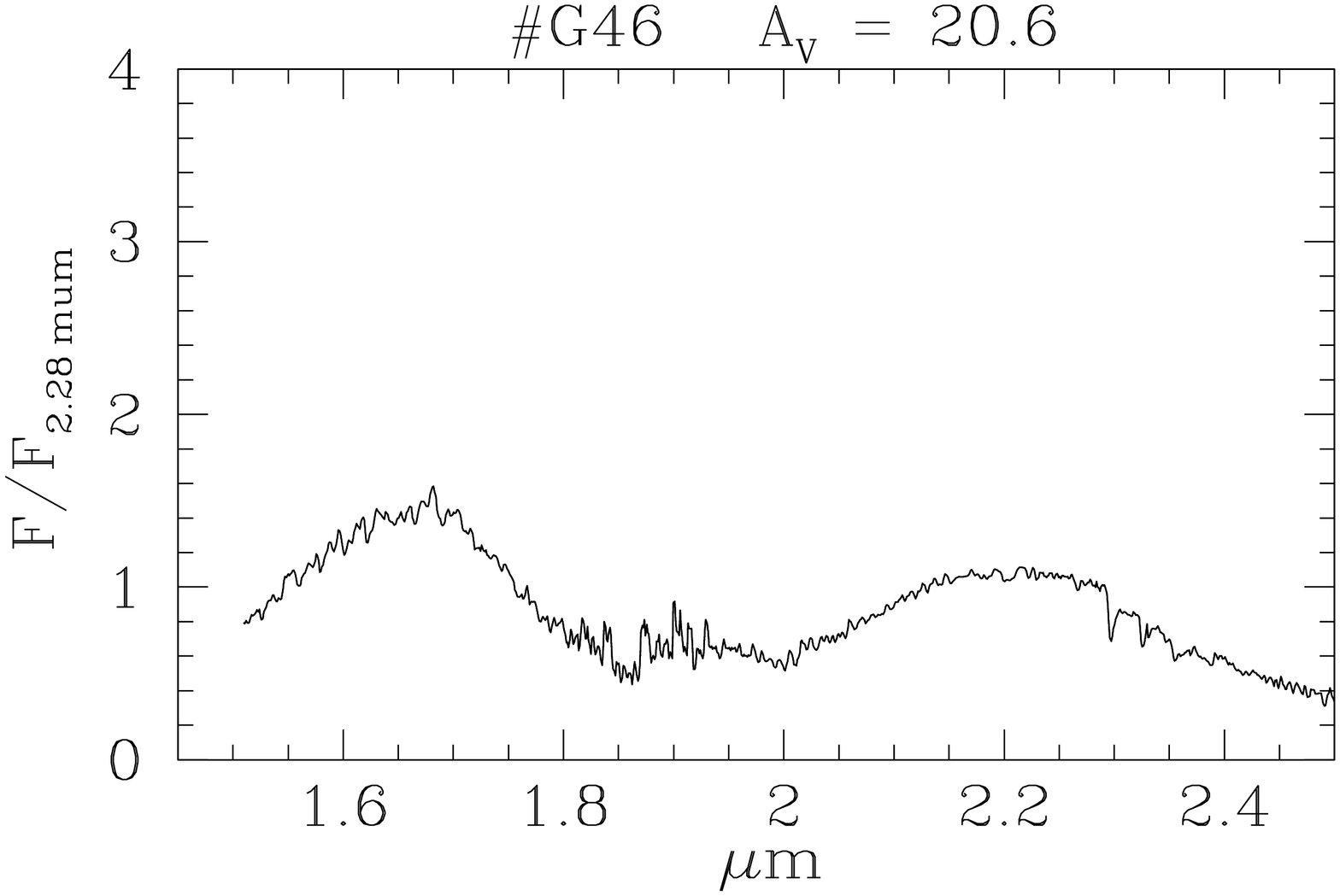,width=4.3cm,angle=270}

\epsfig{file=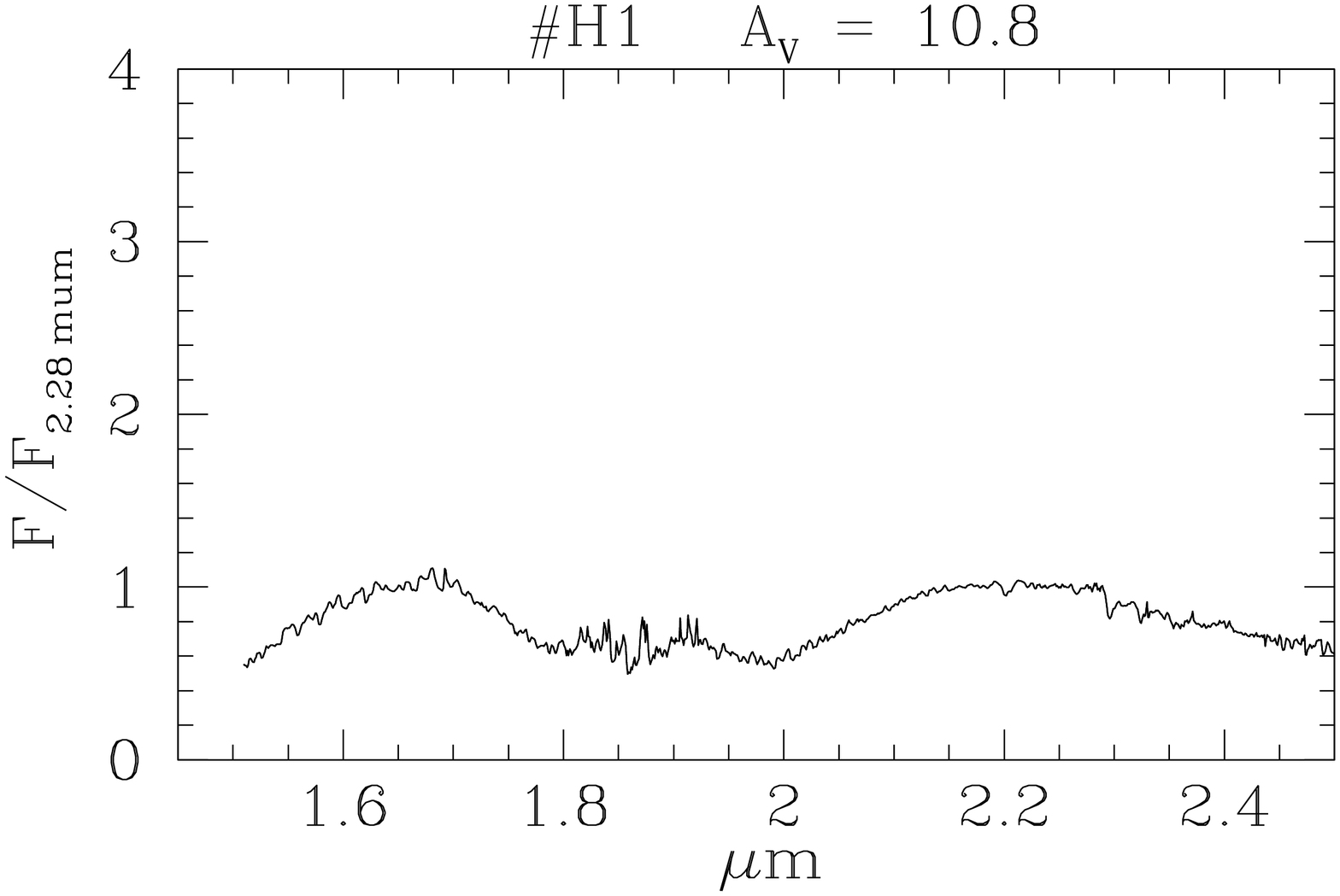,width=4.3cm,angle=270}}

\end{figure*}

\begin{figure*}[H!]
\caption{Red giant candidates}
{\epsfig{file=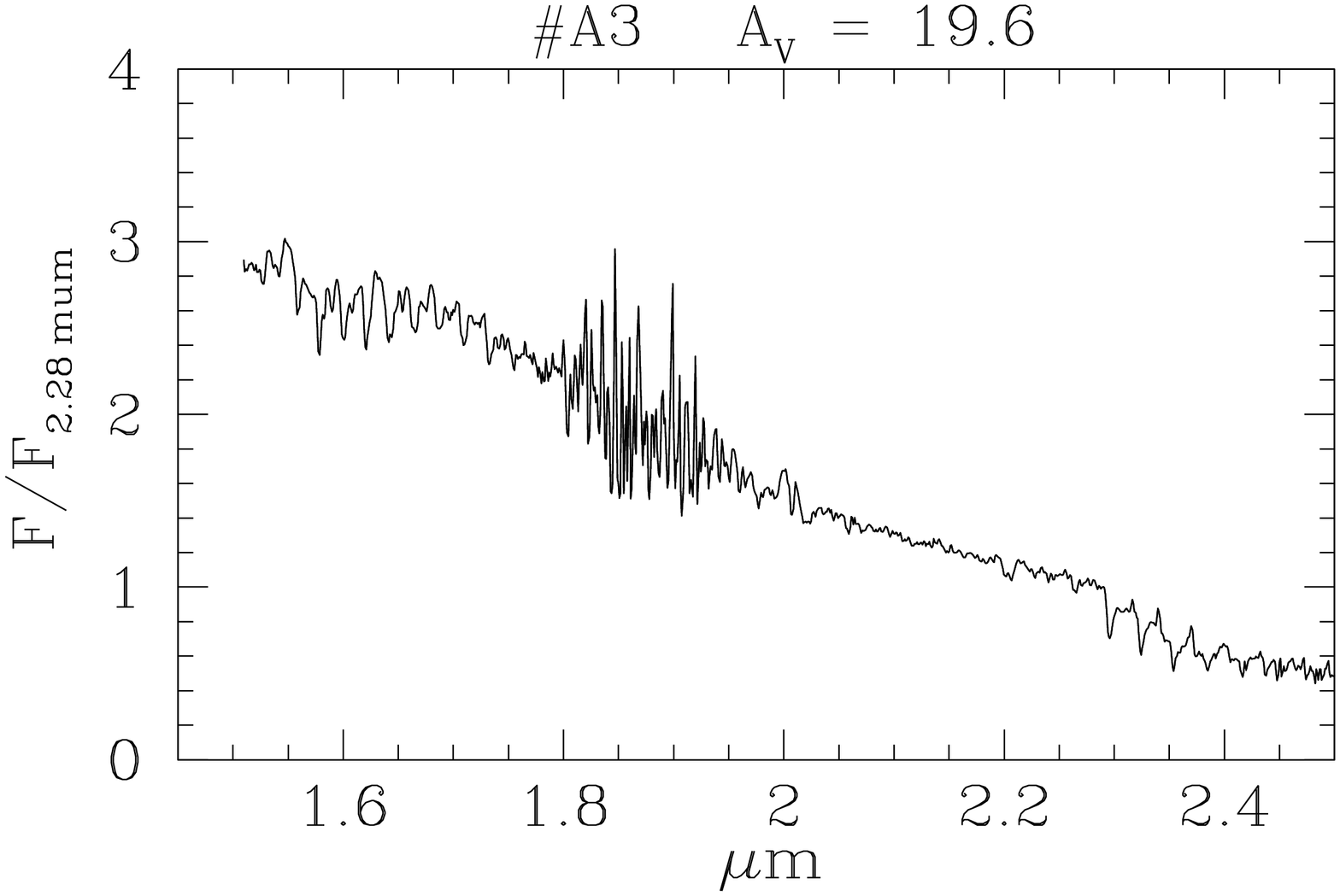,width=4.3cm,angle=270} \epsfig{file=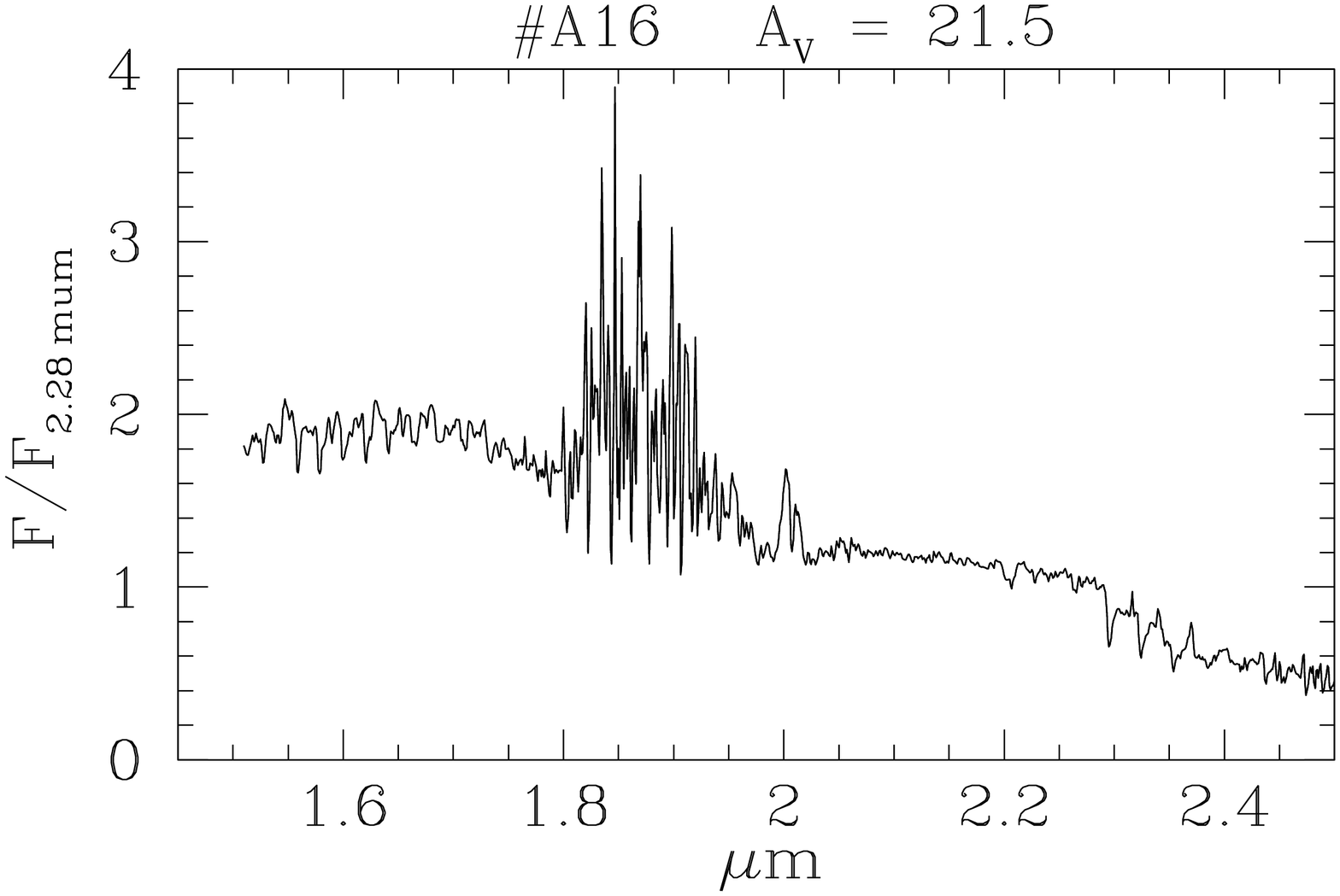,width=4.3cm,angle=270}   \epsfig{file=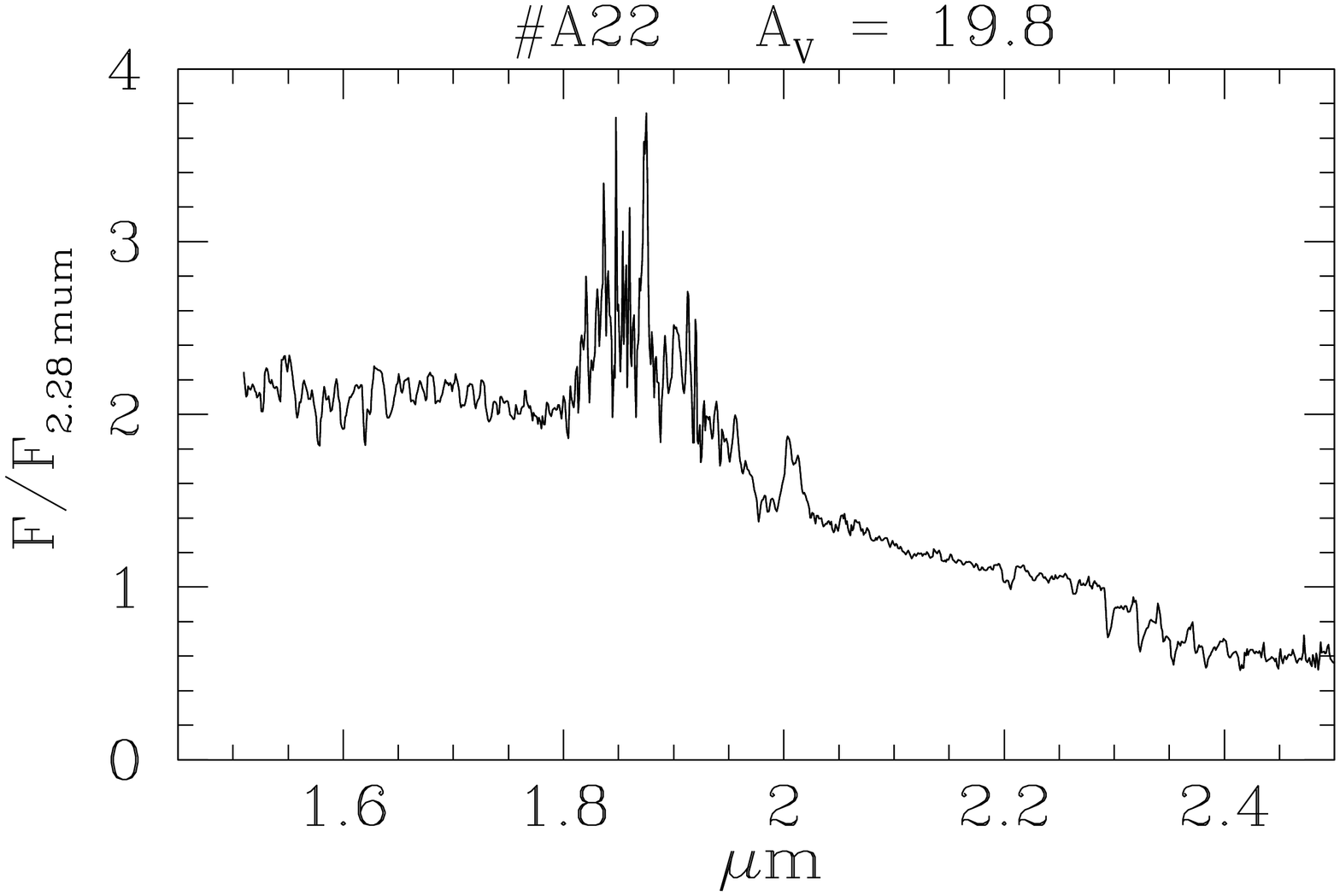,width=4.3cm,angle=270}

\epsfig{file=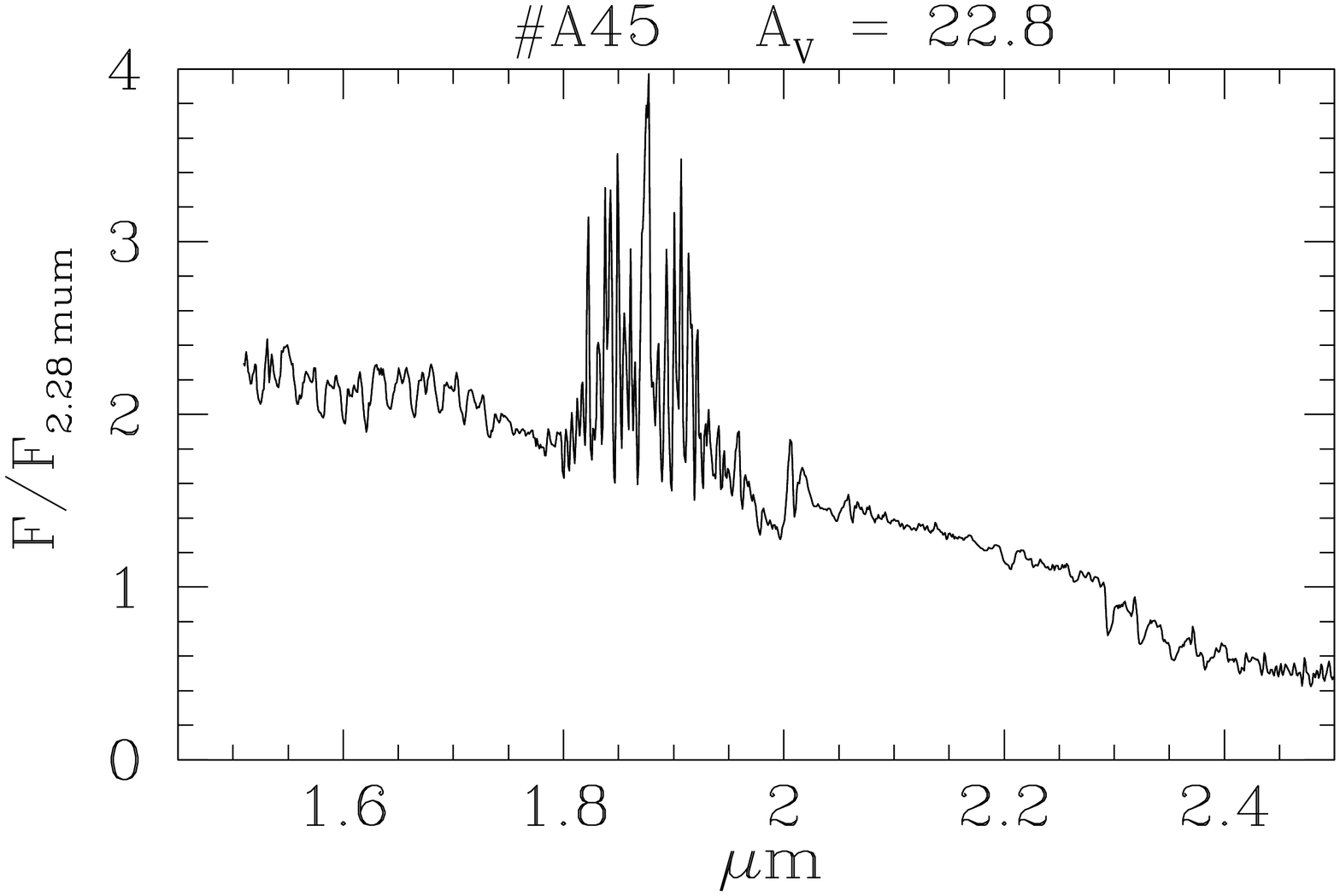,width=4.3cm,angle=270} \epsfig{file=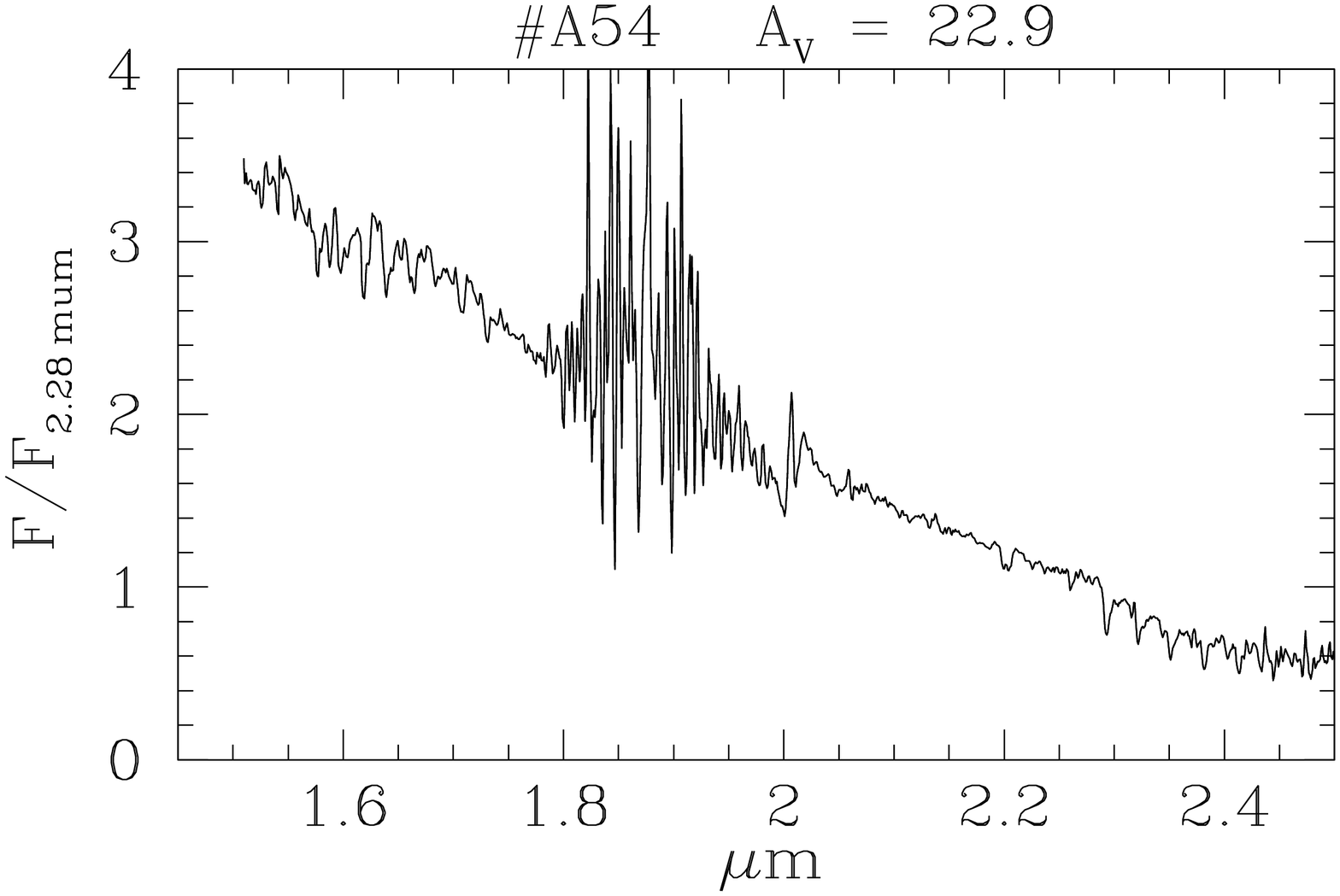,width=4.3cm,angle=270}   \epsfig{file=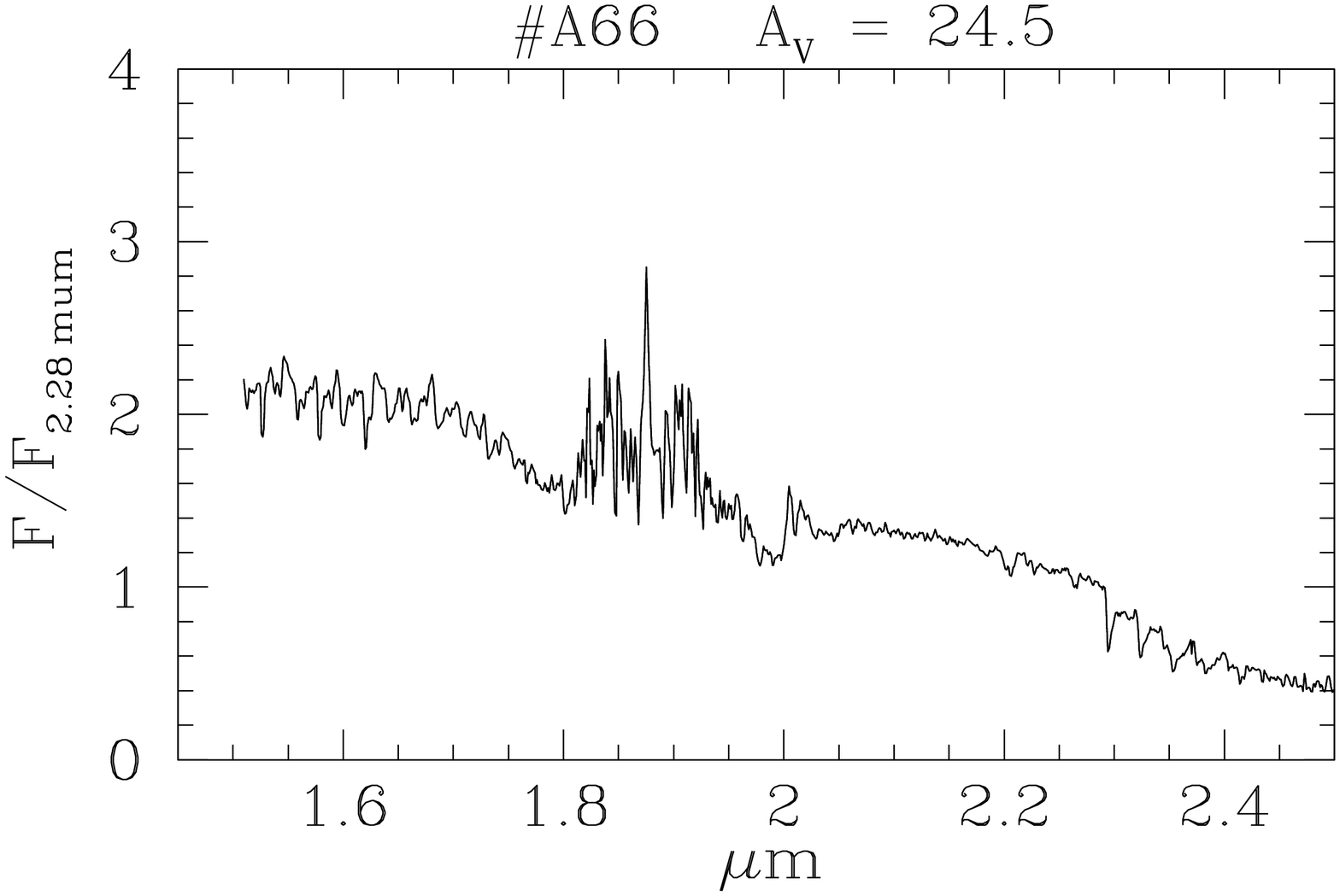,width=4.3cm,angle=270}

\epsfig{file=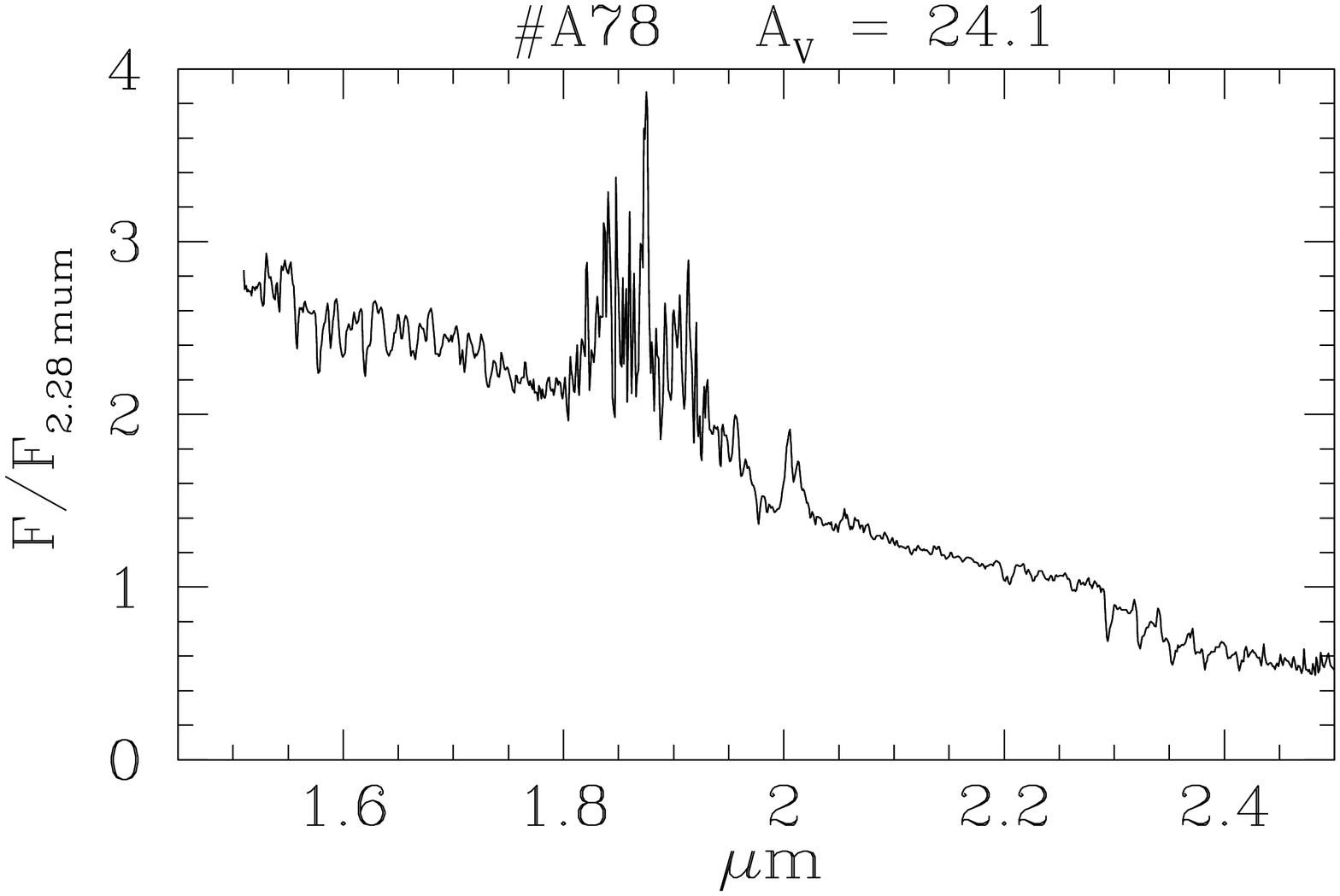,width=4.3cm,angle=270} \epsfig{file=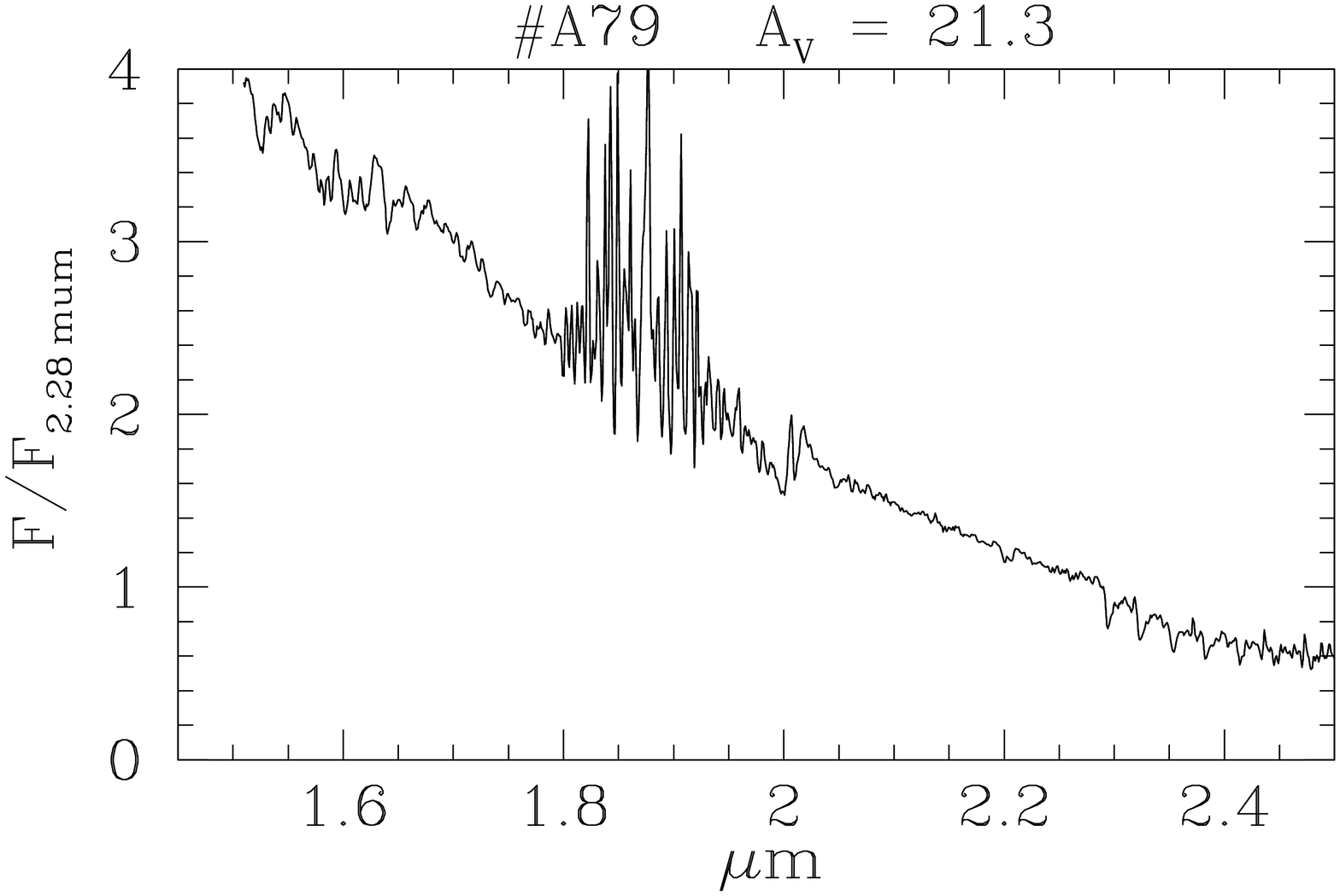,width=4.3cm,angle=270}   \epsfig{file=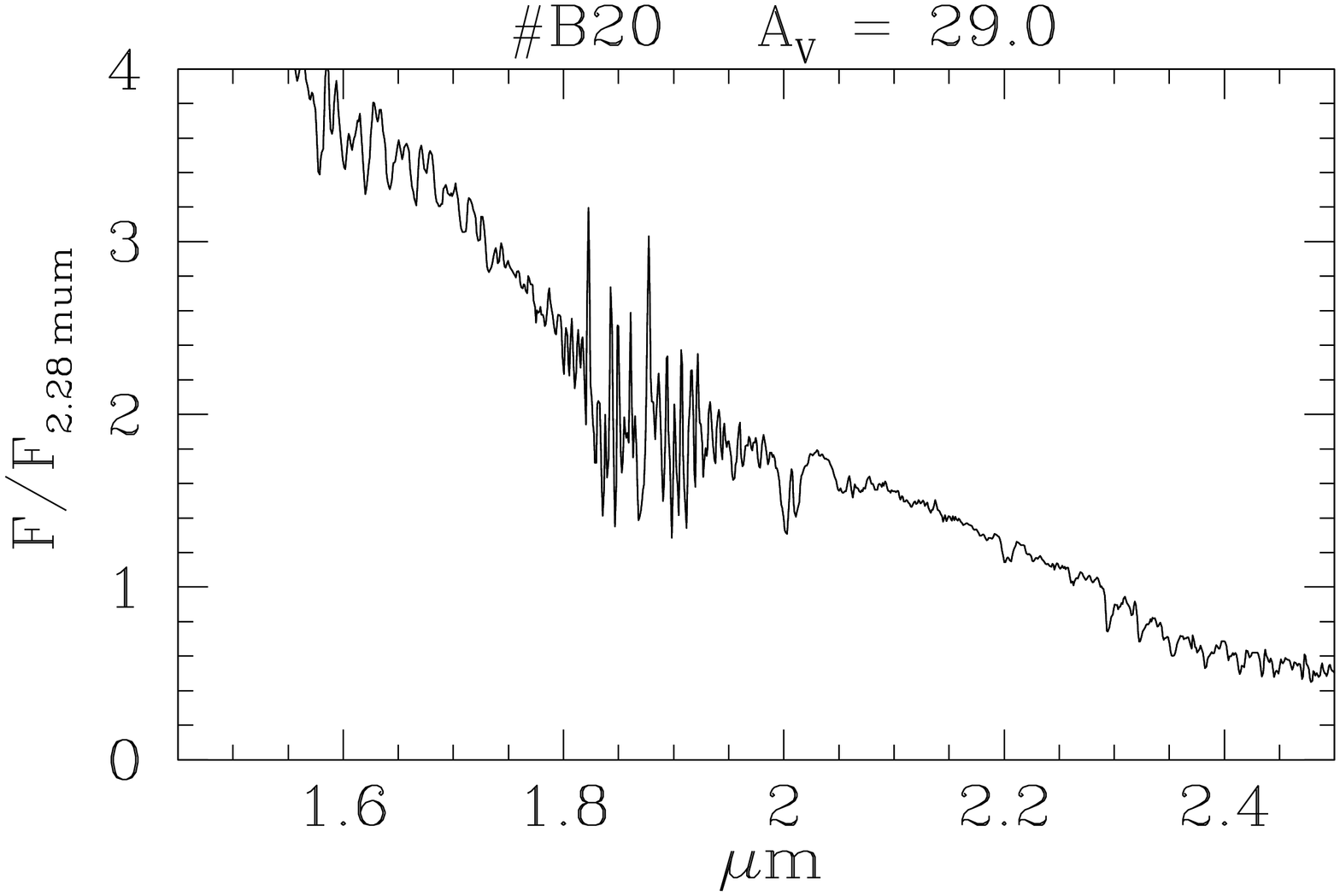,width=4.3cm,angle=270}

\epsfig{file=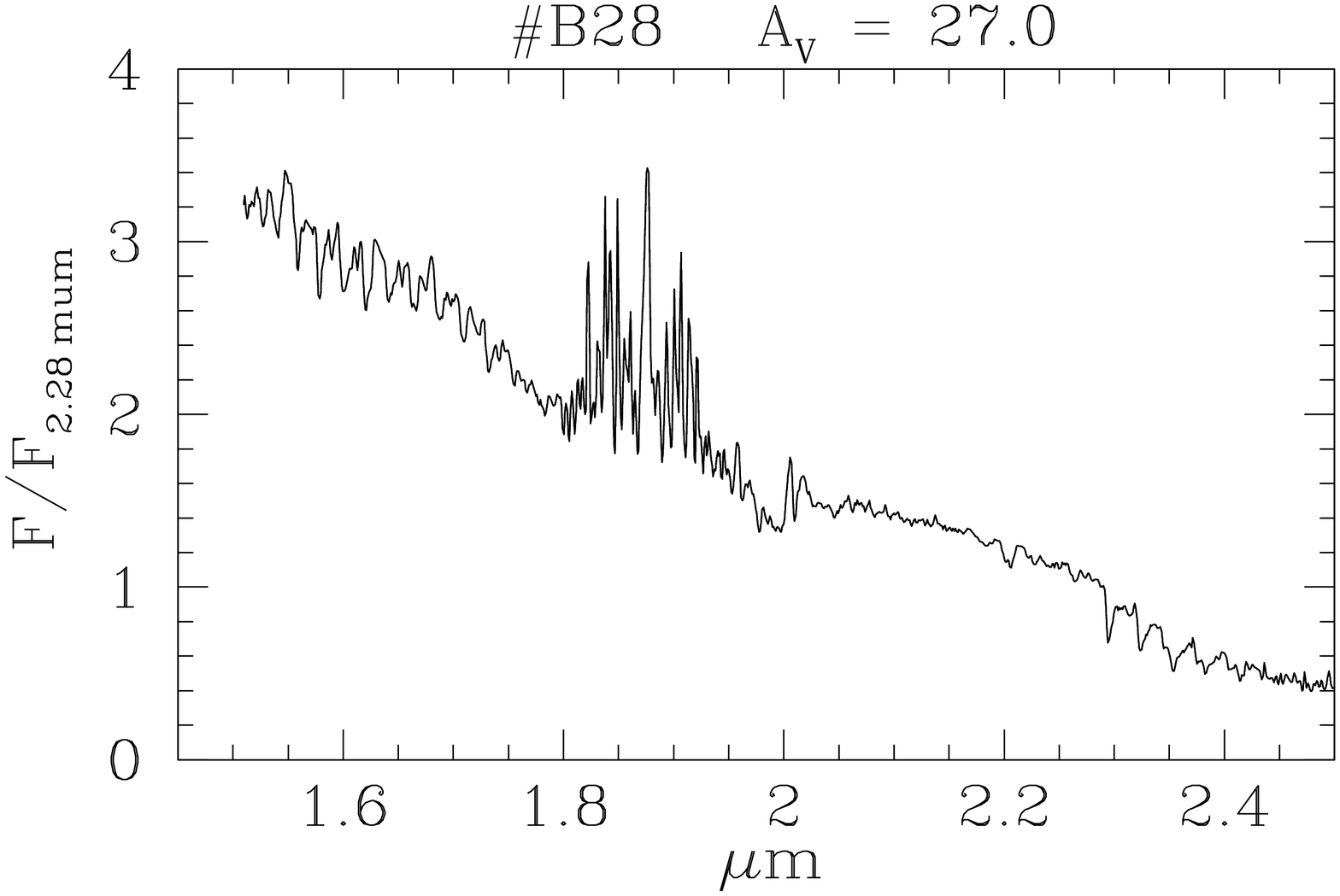,width=4.3cm,angle=270} \epsfig{file=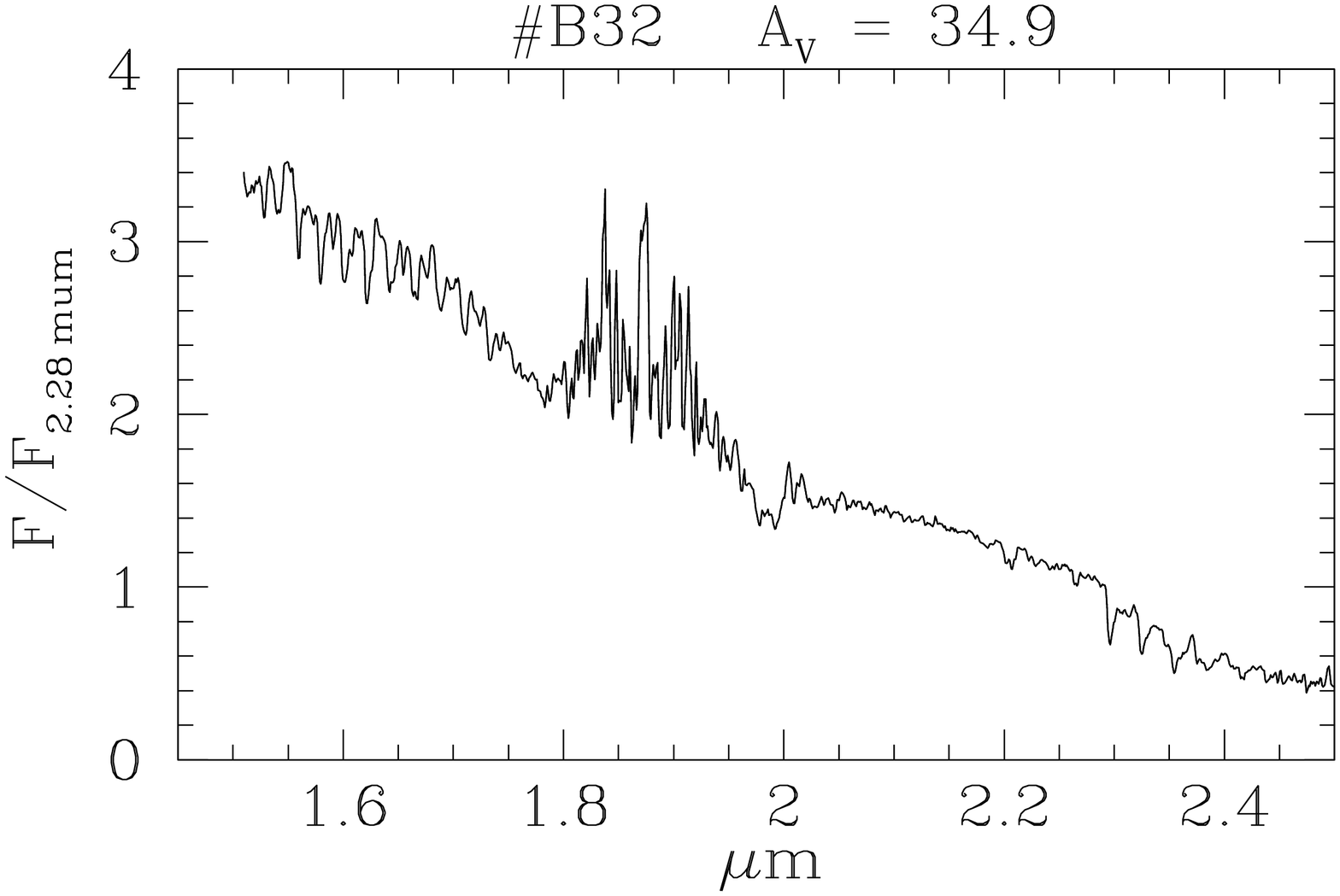,width=4.3cm,angle=270}   \epsfig{file=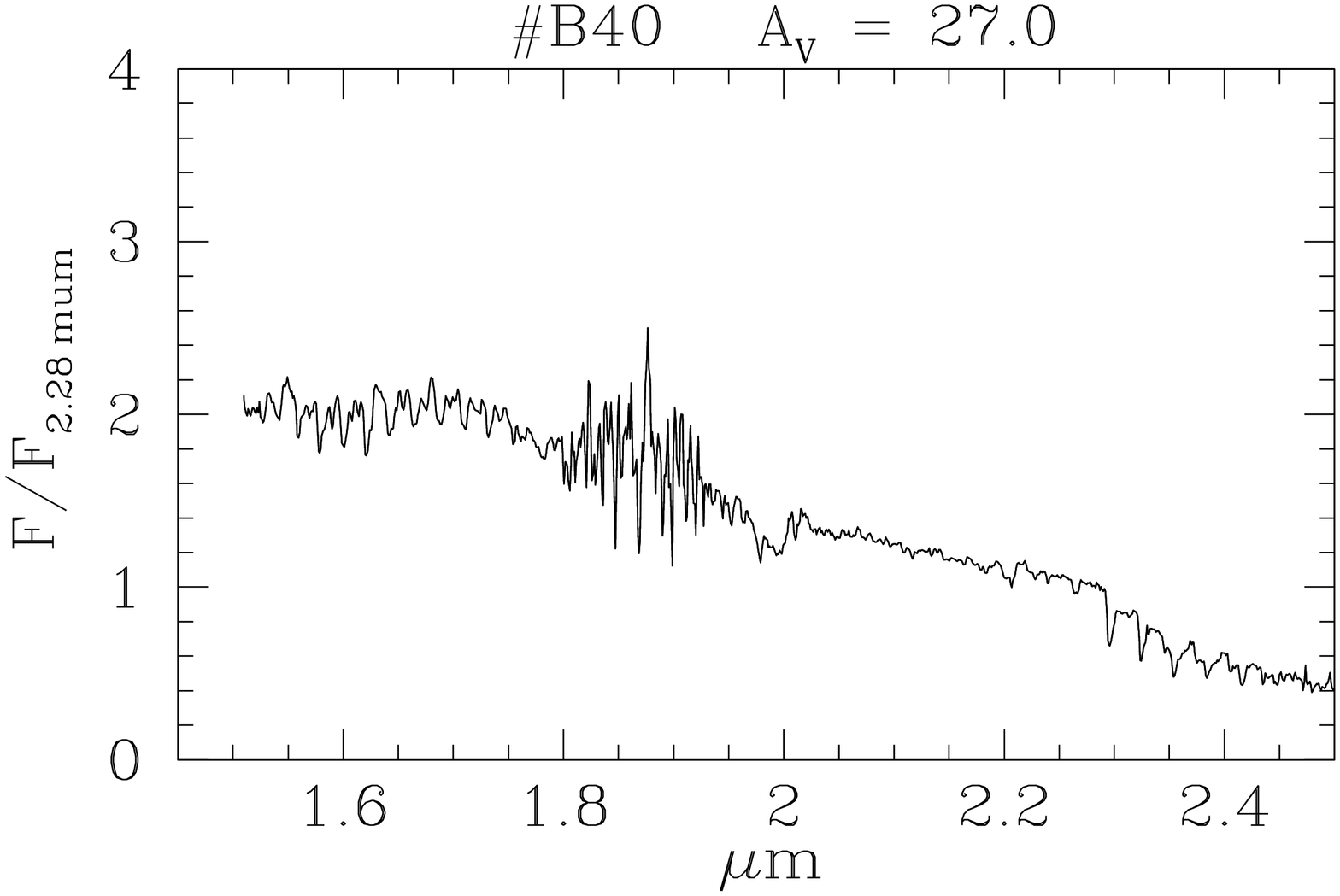,width=4.3cm,angle=270}}

\end{figure*}

\vfill \eject

\end{document}